\documentclass[12pt]{article}

\newcommand{\blind}{0}

\usepackage{setspace}

\usepackage[english]{babel}
\usepackage[utf8]{inputenc}
\usepackage[T1]{fontenc}
\usepackage{natbib}

\usepackage{changepage}
\usepackage[section]{placeins}

\usepackage{enumitem}
\setlist{noitemsep}  

\usepackage[colorinlistoftodos,color=orange!60]{todonotes}
\usepackage{regexpatch}
\makeatletter
\xpatchcmd{\@todo}{\setkeys{todonotes}{#1}}{\setkeys{todonotes}{inline,#1}}{}{}
\makeatother

\usepackage{amscd,amsmath,amssymb,amsfonts,amsthm,bbm,bm,latexsym,mathrsfs,mathtools,bigints}
\usepackage[subnum]{cases}	
\usepackage{arydshln}   
\usepackage{dsfont}
\def\I {\mathbb{I}}
\def\R {\mathds{R}}

\def\K {\mathbf{k}}
\def\C {\mathbf{C}}
\def\z {\mathbf{z}}
\def\x {\mathbf{x}}
\def\bmu {\boldsymbol\mu}
\def\bbeta {\boldsymbol\beta}
\def\bdelta {\boldsymbol\delta}
\def\X {\mathbf{X}}
\def\Z {\mathbf{Z}}
\makeatletter
\newcommand*{\distas}[1]{\mathbin{\overset{#1}{\kern\z@\sim}}}	
\makeatother
\newcommand*\abs[1]{\left|#1\right|}		
\newcommand{\norm}[1]{\left\lVert#1\right\rVert} 		
\allowdisplaybreaks
\def\bigtimes{\mathop{\mathchoice{
   \vcenter{\hbox to10bp{\vrule height15bp width0pt \pdfliteral{
   q 1 J .8 w 0 1 m 10 14 l S 0 14 m 10 1 l S Q
}\hss}}}{
   \vcenter{\hbox to10bp{\kern1bp\vrule height10bp width0pt \pdfliteral{
   q 1 J .65 w 0 0 m 8 10 l S 0 10 m 8 0 l S Q
}\hss}}}{\times}{\times}
}}

\newtheorem{theorem}{Theorem}
\newtheorem{lemma}{Lemma}[section]

\newtheorem{corollary}{Corollary}

\theoremstyle{remark}
\newtheorem{rem}{Remark}
\theoremstyle{plain}

\RequirePackage[colorlinks=true, citecolor=blue, urlcolor=blue]{hyperref}

\usepackage{xcolor,subfigure,epsfig,epstopdf,graphics,graphicx} 
\graphicspath{{./figures/}}
\usepackage[width=1.00\linewidth, font={footnotesize}]{caption}

\usepackage{rotating}
\usepackage{booktabs,longtable,float,array,multirow,colortbl,adjustbox}
\usepackage[flushleft]{threeparttable}
\usepackage{comment}

\newcolumntype{C}[1]{>{\centering\arraybackslash}p{#1}}

\makeatother

\begin{document}

\def\spacingset#1{\renewcommand{\baselinestretch}%
{#1}\small\normalsize} \spacingset{1}


\if0\blind
{
  \title{\bf Static and Dynamic BART for Rank-Order Data}
  \author{Matteo Iacopini\thanks{
    The authors gratefully acknowledge the seminar participants at Erasmus University Rotterdam, Queen Mary University of London, Catholic University of Milan, NESG 2023, and the 6th Annual Workshop on Financial Econometrics at \"Orebro for their useful feedback.
    The authors also thank Xinran Li for providing the replication code for their article.
    This research used the Computational resources provided by the Core Facility INDACO, which is a project of High-Performance Computing at the University of Milan.
    }\hspace{.2cm}\\
    Department of AI, Data and Decision Sciences, Luiss University\\
    and \\
    Eoghan O'Neill \\
    School of Economics, University College Dublin\\
    and \\
    Luca Rossini \\
    Department of Economics, Management and Quantitative Methods, \\University of Milan and Fondazione Eni Enrico Mattei
    }
  \maketitle
} \fi

\if1\blind
{
  \bigskip
  \bigskip
  \bigskip
  \begin{center}
    {\LARGE\bf Static and Dynamic BART for Rank-Order Data}
  \end{center}
  \medskip
} \fi

\bigskip

\begin{abstract}
Ranking lists are often provided at regular time intervals in a range of applications, including economics, sports, marketing, and politics.
Most popular methods for rank-order data postulate a linear specification for the latent scores, which determine the observed ranks, and ignore the temporal dependence of the ranking lists.
To address these issues, novel nonparametric static (ROBART) and autoregressive (ARROBART) models are developed, with latent scores defined as nonlinear Bayesian additive regression tree functions of covariates.

To make inferences in the dynamic ARROBART model, closed-form filtering, predictive, and smoothing distributions for the latent time-varying scores are derived. These results are applied in a Gibbs sampler with data augmentation for posterior inference. 

The proposed methods are shown to outperform existing competitors in simulation studies, static data applications to electoral data, stated preferences for sushi and movies, and dynamic data applications to economic complexity rankings of countries and weekly pollster rankings of NCAA football teams.
\end{abstract}

\noindent%
\vfill

\newpage
\spacingset{1.8} 

\section{Introduction}
\label{sec:introduction}

Rank-order data emerge when an expert is asked to rank a finite set of $N$ items according to a certain criterion, such as personal preference or a predetermined statistic.
The study of rank-order data is becoming popular in several fields, including political science \citep{gormley2008exploring}, music preferences \citep{bradlow2001bayesian}, economics \citep{FokPaapROL2012}, and sports analytics \citep{graves2003hierarchical}.
In practice, the rankers are likely heterogeneous in their qualities or opinions, and it is often of interest to investigate how the rankings may be dependent on covariates of the ranked items (e.g., summary statistics).
For example, for sports leagues such as the NCAA, NFL, or NBA, many experts provide ranking lists of teams by relying on their experiences and team-specific statistics, like the win percentage and the margin of victory.
Moreover, reflecting the inherently dynamic nature of competitors' abilities and human preferences, the experts' rankings may change over time \citep[e.g.,][]{graves2003hierarchical}.
Analogous considerations hold for rank-order data from other fields, such as marketing, politics, and economics. 

In this article, we focus on models for rank-order data \citep{alvo2014statistical,liu2019model}. Specifically, we consider the family of Thurstone models, which assumes the existence of an unobserved evaluation score for each of the $N$ items whose noisy realizations determine the rankings provided by the $M$ experts.
However, most of the existing models in this class are designed to study cross-sectional data and are inherently static, thus ignoring any temporal dependence.

\begin{figure}[t!h]
\centering
\hspace*{-8ex}
\captionsetup{width=0.93\linewidth}
\begin{tabular}{cc}
    \includegraphics[scale = 0.35]{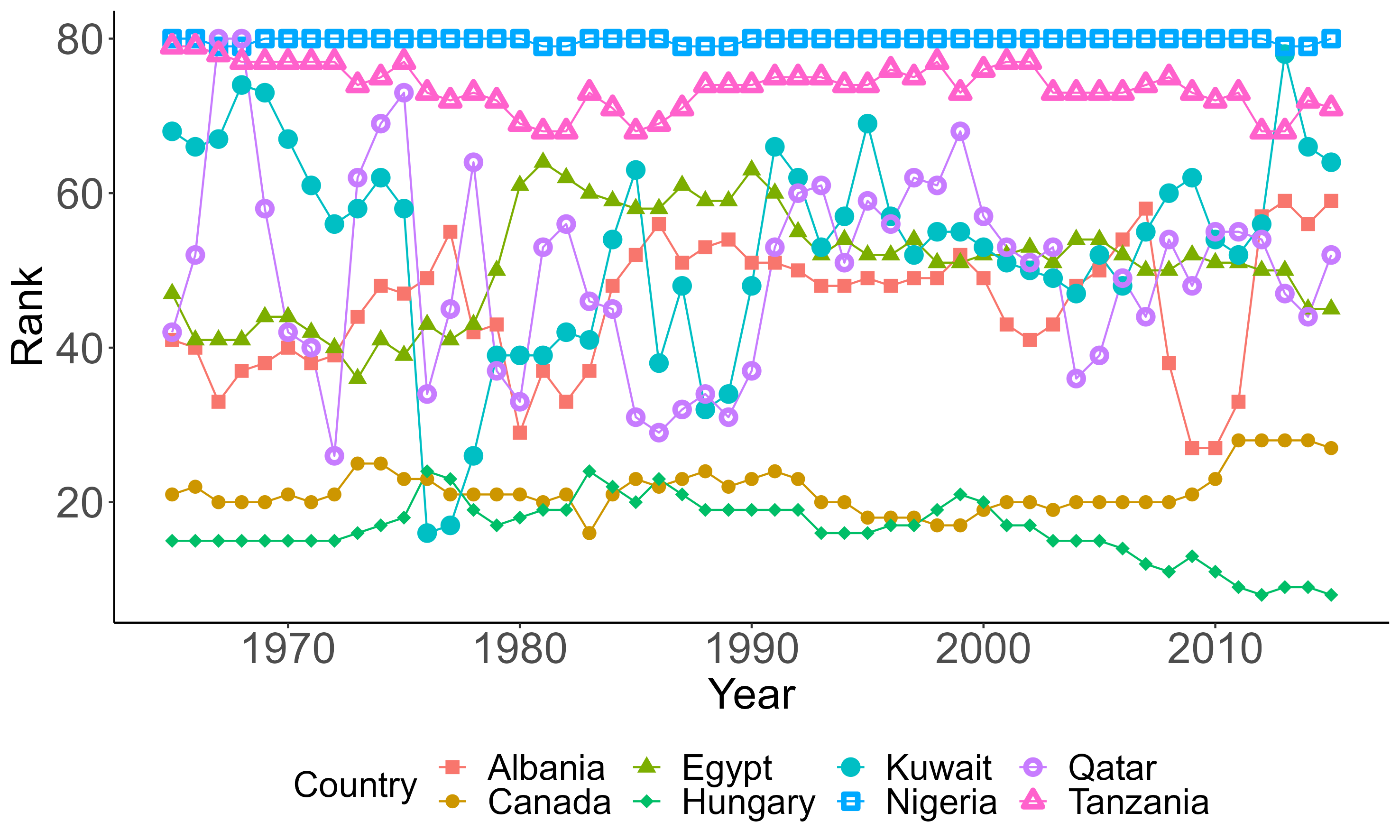} & 
    \includegraphics[scale = 0.35]{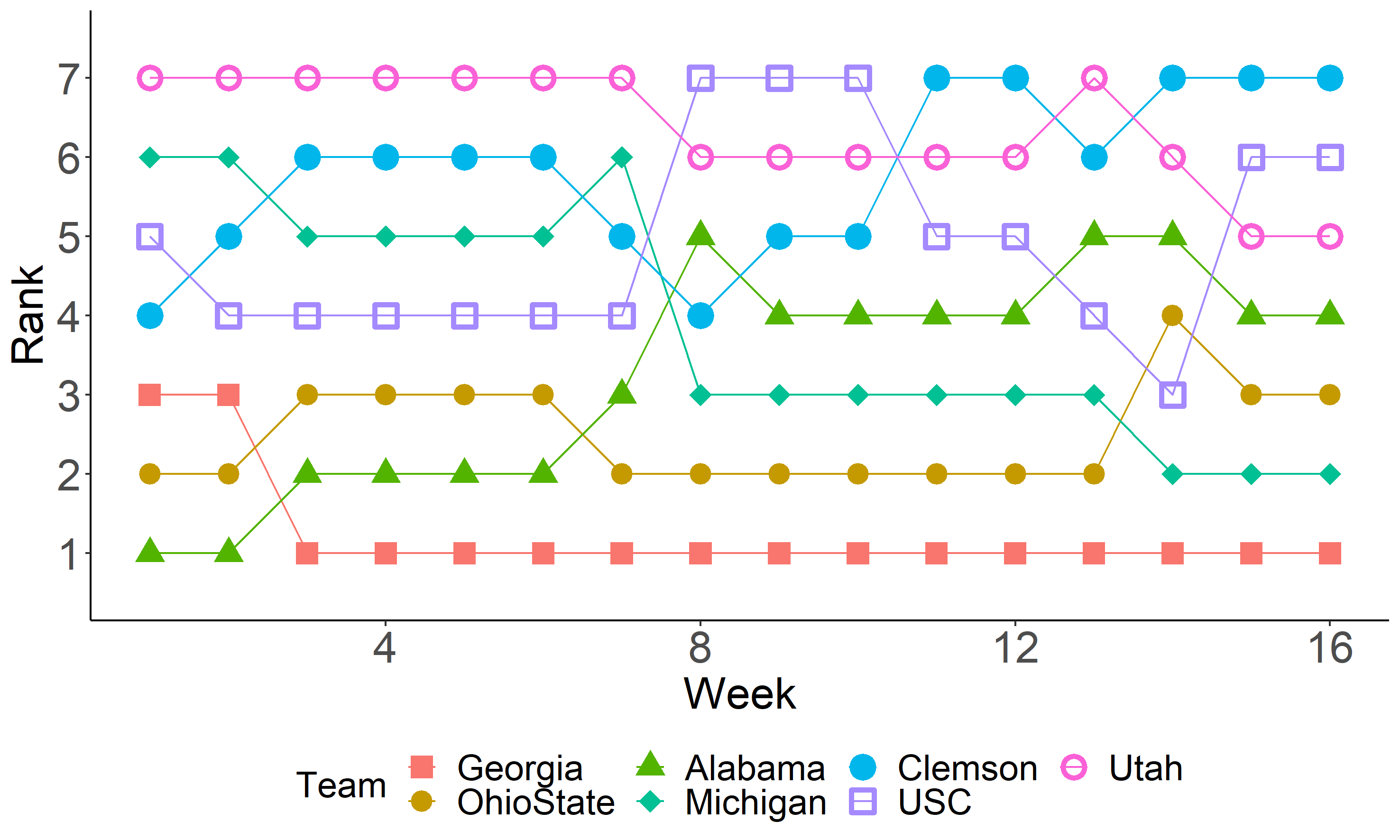}
\end{tabular}
\caption{Left panel: Economic Complexity Index rankings for a selection of countries by year.
Right panel: observed relative ranks for $N=7$ teams of the 2022 NCAA Division I Football season by AP Pollster Brian Howell.}
\label{fig:Econ_NCAA_data}
\end{figure}

Figure~\ref{fig:Econ_NCAA_data} shows the time-varying rankings of the Economic Complexity Index (ECI) \citep{simoes2011economic} for 8 countries from 1965 to 2015, and the 2022 weekly ranking of NCAA teams by a single pollster.
The changes in rankings over time are heterogeneous in magnitude, with both large and small shifts, as well as a few series having a stable ranking over time. Overall, these features highlight both slow evolution and large breaks due to events such as global crises or small events at a country level. The NCAA dataset shows similar features, with  smoother dynamics.
Section~\ref{sec:application} includes an application to these datasets among the others.

The main challenge in investigating and forecasting this class of data is the constraint among the items in a ranking list, which prevents the use of equation-by-equation modeling and calls for specific statistical tools. However, the existing rank-order models fall short of the flexibility necessary (i) to address the highly nonlinear relationship between the item's features and observed rankings and (ii) to forecast time series characterized by possibly large changes.

This article contributes to the literature on rank-order models by proposing two new methods for addressing these issues. Specifically, we leverage the Bayesian additive regression trees \citep[BART, see][]{chipman2010bart} framework to define a flexible nonparametric static model for rank-order data, named rank-order Bayesian additive regression trees (ROBART). {This model is an extension of the linear rank-order model of \cite{li2016bayesian} to a setting with non-linearities and covariates that are specific to both items and rankers.} Then, we extend this framework to a dynamic model called autoregressive rank-order Bayesian additive regression trees (ARROBART), which accounts for the persistence in time series of rank-order data (e.g., see Fig.~\ref{fig:Econ_NCAA_data}).

The most widely used families of models for rank data are the Thurstone \citep[][]{thurstone1927law} and Plackett-Luce \citep[][]{luce1959individual,plackett1975analysis} models, which differ in the distribution of the noise in the random utility representations (normal and Gumbel, respectively) and in the ordering interpretation (joint or sequential).
Relying on normal regression models makes the Thurstone family easier to interpret and extend in several directions. However, the estimation of these models is generally difficult, especially for large numbers of ranked items, due to the high-dimensional integral form of the likelihood function.
Within the Thurstone family, the Bayesian approach coupled with data augmentation methods provides an advantageous computational alternative that also allows for the introduction of covariates \citep{yu2000bayesian}.
\cite{johnson2002bayesian} developed a hierarchical Bayesian Thurstone model for rank-order data from known groups of evaluators, whereas \cite{li2016bayesian} introduced ranker-specific error variances and a Dirichlet process mixture prior for ranker-specific effects of covariates. Recently, \cite{gu2023social} introduced a Bayesian spatial autoregressive model to learn ranking data in social networks and to account for social dependencies among individuals.
We contribute to this literature by proposing two new Thurstone models, a static and a dynamic one, where a Bayesian nonparametric approach is used to capture the possibly nonlinear impact of covariates.
Moreover, we extend the proposed methods to address incomplete data (i.e., partial rankings) in the form of \textit{top-$q$} rankings, which frequently happens when rankers assess many items.

The study of \textit{time series} of rankings is less developed but has attracted increasing attention in recent years. 
Most existing dynamic ranking models are generalizations of the Plackett-Luce model \citep{bradlow2001bayesian, grewal2008university, glickman2015stochastic, henderson2018comparison}, whereas this article provides a dynamic extension to the Thurstone family.
We propose a novel time-varying nonparametric approach that considers an autoregressive process for the latent scores and uses a specification based on regression trees to flexibly model the impact of the lagged response and the external covariates. Its linear counterpart is also introduced as a benchmark.

The possibility of having large deviations as well as small changes in the time series of ranking lists, as shown in Fig.~\ref{fig:Econ_NCAA_data}, calls for the use of flexible tools to model the dynamics of the latent ``scores'' driving the ranking of each item.
To address these issues, we introduce the BART specification to model the conditional mean of latent scores. We leverage the good performance of this approach for fitting and forecasting regression models \citep[e.g., see][]{hill2020bayesian} in order-statistics models to design static and dynamic nonparametric frameworks for rank-order data.
Despite its advantages, BART has rarely been used in time series or state space models. Recent advances include Bayesian additive vector autoregressive trees for multiple time series \citep{huber2021inference}, and an extension to a mixed-frequency context for nowcasting in the presence of extreme observations \citep{huber2023nowcasting} .
Our contribution to this literature concerns the derivation of new closed-form filtering, predictive, {backward sampling,} and smoothing distributions for the latent time-varying scores in the ARROBART model. Furthermore, we show the proportionality of these distributions to finite mixtures.

The proposed ROBART and ARROBART models are tested on static and dynamic synthetic datasets, respectively, highlighting the superior performance against their linear counterparts as well as other existing methods.
The proposed static models are then applied to three datasets about sushi and movie preference data and election results, followed by two applications of the dynamic model to the rankings from the Economic Complexity Index (ECI) rankings and the 2022 NCAA pollsters displayed in Fig.~\ref{fig:Econ_NCAA_data}.
We find evidence supporting the superior performance of ROBART and ARROBART in the static and dynamic settings, respectively, relative to their linear counterparts, which suggests that it is important to account for both temporal persistence and nonlinearities.

The remainder of the article is organized as follows:
Section~\ref{sec:model} provides an overview of the Thurstone model and illustrates the ROBART and ARROBART models. It also provides a theorem describing the filtering and smoothing distributions for ARROBART.
Section~\ref{sec:inference} presents a Bayesian approach to inference. Section~\ref{sec:simulations} investigates the performance of our methods on simulated data. Section~\ref{sec:application} includes applications to three static ranking data sets and a time series of Economic Complexity Index rankings.
Section~\ref{sec:conclusion} concludes the article.


\section{Nonparametric Thurstone Models for Rank-Order Data}   \label{sec:model}

This section introduces notation, then presents a new class of nonparametric models for rank-order data that extends the Thurstone model \citep{thurstone1927law} by allowing for flexible, possibly nonlinear functions of the covariates. Then, we extend the framework to account for temporal dependence and persistence for investigating and forecasting time series data.

A full ranking of $N$ items is a map $\tau: \mathds{A}\to \mathcal{P}_N$ from a finite, ordered set of items, $\mathds{A} = \{ a_1,\ldots,a_N \}$ to the space of $N$-dimensional permutations, $\mathcal{P}_N$. Let $\#$ denote the cardinality of a set.
With a slight abuse of notation, we denote a full ranking of $N$ items by the vector $\tau = (\tau_1,\ldots,\tau_N)' \in\mathcal{P}_N$, where $\tau_i = \tau(i)$ is the rank of item $a_i \in\mathds{A}$, for $i = 1,\ldots, N$, and $\tau_i < \tau_j$ means that item $a_i$ is preferred to $a_j$, written $a_i \succ a_j$ since it is assigned a lower rank. Rankings and pairwise comparisons are related, as, given a full ranking $\tau$, it holds $[a_i \succ a_j] \iff \tau_i < \tau_j$ for each $i,j \in \{ 1,\ldots, N \}$.
Therefore, a full ranking $\tau$ is an ordered $N$-tuple representing a set of non-contradictory pairwise relationships in $\mathds{A}$, and as such, it encodes an ordered preference for all the items in $\mathds{A}$.

In this article, we consider a setting where $N$ items are ordered by $M$ rankers. For each item $i=1,\ldots,N$ and ranker $j=1,\ldots,M$, the $K_x$-dimensional vector of covariates is denoted by $\mathbf{X}_{ij} \in \R^{K_x}$.
To account for external information that might affect the rankers' decision process, it is possible to include item- and ranker-specific covariates, $\mathbf{x}_{i}^a \in\R^{K_a}$ and $\mathbf{x}_{j}^r \in\R^{K_r}$, as well as covariates specific to both items and rankers, $\mathbf{w}_{ij} \in\R^{K_w}$.
In the dynamic case, this setup is replicated for each of the $T$ time periods.
Finally, for any vector $\z = (z_1,\ldots,z_N)' \in \R^N$, we use $\operatorname{rank}(\z) = [z_{i_1} \succ \ldots \succ z_{i_N}]$ to denote the full ranking of $z_i$'s in decreasing order of preference, that is, $z_{i_1} < \ldots < z_{i_N}$, { i.e. lower latent variable scores correspond to higher rankings (more preferred or ``better'' items).
}

\subsection{Overview of Bayesian Additive Regression Trees}   \label{sec:preliminaries}


A Bayesian additive regression tree (BART) specification for a function $f: \R^{K_x} \to \R$ is an approximation of $f$ by means of a sum of regression trees.
Let us denote by $\mathcal{T}$ a binary tree consisting of a set of interior node decision rules and a set of terminal nodes (or leaves), and let $\bmu = \{ \mu_1,\ldots,\mu_b \}$ be a set of parameter values associated with each of the $b$ terminal nodes of $\mathcal{T}$.
A regression tree takes covariate values as input and produces a fitted outcome value given by the parameter associated with a terminal node of the tree.
This outcome is obtained by dividing the domain of the covariate $\mathbf{x}_i = (x_{i1},\ldots,x_{iK_x})' \in \R^{K_x}$ into disjoint regions $\{B_{\ell}\}_{\ell=1}^b$ using a sequence of binary rules of the form $\{ \mathbf{x}_i\in A_h \}$ versus $\{\mathbf{x}_i \notin A_h \}$, where $h \in \mathcal{I}$ indexes the internal nodes of the tree. Most commonly, $A_h$ is defined by a splitting variable $k_h$ and splitting point $c_h$, such as $A_h = \{\mathbf{x}_i : x_{ik_h} \le c_h \}$.
Then, $\mathbf{x}_i$ is assigned the value $\mu_\ell$, with $\ell \in\{ 1,\ldots,b \}$, associated with a single terminal node $B_{\ell}$ of $\mathcal{T}$ defined by a sequence of decision rules.

A tree $\mathcal{T}$ and a collection of leaf parameters $\bmu$ define a step function $g(\mathbf{x}_i | \mathcal{T},\bmu)$ that assigns value $\mu_\ell \in \bmu$ to $\mathbf{x}_i$, thus resulting in:
\begin{equation*}
    g(\mathbf{x}_i | \mathcal{T},\bmu) = \sum_{\ell=1}^b \mu_\ell \mathbb{I}(\mathbf{x}_i \in B_\ell),
\end{equation*}
where $\mathbb{I}(\cdot)$ is the indicator function and $\{ B_\ell \}_{\ell=1}^b$ is the partition of the domain of the covariate $\mathbf{x}_i$ associated to the tree $\mathcal{T}$. 
The BART specification consists in approximating the function $f$ through a sum of $S$ trees, as follows:
\begin{equation}
    f(\mathbf{x}_i) \approx \sum_{s=1}^S g_s(\mathbf{x}_i | \mathcal{T}_s, \bmu_s) = \sum_{s=1}^S \sum_{\ell=1}^{b_s} \mu_{\ell,s} \mathbb{I}(\mathbf{x}_i \in B_{\ell,s}),
\label{eq:BART_approx_f}
\end{equation}
where $g_s$ is a tree function that assigns $\mu_{\ell,s} \in \bmu_s$ to $\mathbf{x}_i$ for each binary regression tree $\mathcal{T}_s$.

\subsection{Static Model: Rank-Order BART}
\label{subsec:ROBART}

The Thurstone model can be described as a random utility model \citep[e.g., see][]{walker2002generalized}, where the full ranking $\tau$ is driven by a latent score, $\z$, such that for any two items $i_1,i_2$, it holds $z_{i_1} < z_{i_2}$ if and only if $i_1 \succ i_2$. The latent vector $\z$ is assumed to follow a multivariate normal distribution with mean $\boldsymbol{\gamma} = (\gamma_1,\ldots,\gamma_N)'$ representing the score of each item:
\begin{equation}
    \tau  = \operatorname{rank}(\z),  \qquad z_{i}  = \gamma_{i} + \varepsilon_{i}, \qquad \varepsilon_{i} \sim \mathcal{N}(0, \sigma_\varepsilon^2).
\label{eq:model_TMD}
\end{equation}
The model in Eq.~\eqref{eq:model_TMD} poses an identification problem since the observations only provide information about the ordering of the elements in $\z$, which implies that the parameters $(\boldsymbol{\gamma},\sigma_\varepsilon^2)$ are not separately identifiable from the likelihood.
Therefore, to ensure the identification of the parameters, we fix $\sigma_\varepsilon^2 = 1$.\footnote{If one is interested in interpreting the coefficients of some covariates, then it would be necessary to impose the additional constraint the $\boldsymbol{\gamma} \in \Gamma = \{ \boldsymbol{\gamma} \in\R^N: \mathbf{1}_N' \boldsymbol{\gamma} = 0 \}$, where $\mathbf{1}_N$ is an $N$-dimensional vector with all entries equal to 1.}

Consider a framework where no temporal observations are available or the time dependence is irrelevant to the analysis. 
Moreover, we assume each individual in a collection of $M$ professionals or rankers provides a ranking list of the same set of $N$ items, as is common in many real-world situations such as sports analytics and customer reviews.
To investigate data without temporal dependence, we define a new static model that extends the Thurston approach in Eq.~\eqref{eq:model_TMD} by including the BART specification from Eq.~\eqref{eq:BART_approx_f} for the conditional mean of the latent scores.
Specifically, we generalise Eq.~\eqref{eq:model_TMD} by assuming ranker-specific scores, $\z_j$, and, for each pair $(i,j)$, we replace the intercept $\gamma_{ij}$ with an unknown function of the covariates, $f$. Moreover, to cope with potential nonlinearities, interactions, and discontinuities in the rankers' dynamic scores, we adopt a nonparametric approach based on a BART specification for the latent regression function, $f$.

For each item $i=1,\ldots, N$ and ranker $j=1,\ldots,M$, let $\mathbf{X}_{ij} = (\mathbf{x}_i^{a\prime}, \mathbf{x}_j^{r\prime}, \mathbf{w}_{ij}')' \in \R^{K_x}$ be a covariate vector.
The Rank-Order BART model (ROBART) is defined as:\footnote{Assuming that the latent scores are item-ranker specific implicitly means assuming each ranker may make a different assessment of the items. Instead, when a single score is postulated for each item, it would imply postulating that each ranker makes a (noisy) evaluation of a unique and objective ranking of the items.
Moreover, the former approach allows us to include ranker-specific covariates, $\mathbf{x}_j^{r}$, and object-ranker-specific covariates, $\mathbf{w}_{ij}$, as factors possibly affecting the latent scores.}
\begin{align}
\begin{split}
\tau_{j}  = \operatorname{rank}(\z_{j}), \qquad z_{ij}   & = f(\mathbf{X}_{ij}) + \varepsilon_{ij}, \qquad \varepsilon_{ij} \distas{iid} \mathcal{N}(0,1), \\
f(\mathbf{X}_{ij}) & = \sum_{s=1}^S \sum_{\ell=1}^{b_s} \mu_{\ell,s} \mathbb{I}(\mathbf{X}_{ij} \in B_{\ell,s}),
\end{split}
\label{eq:model_ROBART}
\end{align}
where $f:\R^{K_x} \to \R$ is a regression function for which we assume a BART specification.
Denoting with $\delta_{\z}(\tau)$ the Dirac mass at $\z$, $\norm{\cdot}$ the $L_2$ norm, and $\boldsymbol{\theta}$ a vector collecting all the static parameters (ranker- and item-specific), one obtains the likelihood:
\begin{equation}
\begin{split}
    L(\tau_1,\ldots,\tau_M | \boldsymbol{\theta}) 
    & = \prod_{j=1}^M \int_{\R^N} (2\pi)^{-\frac{N}{2}} \exp\Bigg( -\frac{\big\|\z_{j} - f(\mathbf{X}_{ij}) \big\|^2}{2} \Bigg) \delta_{\operatorname{rank}(\z_{j})}(\tau_{j}) \: \mathrm{d}\z_j.
\end{split}
\end{equation}

\subsection{Dynamic Model: AutoRegressive Rank-Order BART}   \label{subsec:ARROBART}

In many real-world cases, multiple professionals continuously evaluate the items and provide their ranking lists periodically over time.
The latent scores that a ranker assigns to each item, which ultimately determine the professional's ranking lists observed by the researcher, evolve over time to reflect the change in the information set and in the features of the items.
Nonetheless, most of the existing models for rank-order data disregard the temporal dimension of the data, thus missing information relating to the serial correlation in the rankers' latent scores and observed ranks.
We aim to address this issue by proposing a modification of the Thurstone model that accounts for the temporal persistence of the ranking lists as implied by an evolution of the underlying rankers' scores.

Suppose that each ranker provides a ranking of items at each time $t=1,\ldots, T$, and let $\tau_{j,t}$ denote the observed ranking list of ranker $j$ in period $t$.
Moreover, let $\z_{j,t} = (z_{1j,t},\ldots,z_{Nj,t})'$, where $z_{ij,t}$ denotes the unobserved ranker $j$'s score for item $i$, such that $z_{i_1j,t} > z_{i_2j,t}$ if and only if $i_1 \prec i_2$ for ranker $j$ at time $t$.
In this setting, each latent score $z_{ij,t}$ can be modeled as an unknown function $f$ of its first lag  and the external covariates. This is done by considering the covariate vector $\mathbf{X}_{ij,t} = (z_{ij,t-1},
\mathbf{x}_{i,t}^{a\prime}, \mathbf{x}_{j,t}^{r\prime}, \mathbf{w}_{ij,t}')' \in\R^{K_x}$, where $\mathbf{x}_{i,t}^a$, $\mathbf{x}_{j,t}^r$, and $\mathbf{w}_{ij,t}$ are the item-, ranker-, and item/ranker-specific variables.

Therefore, in the presence of time series rank-order data where temporal persistence is a relevant feature, we extend the ROBART model in Eq.~\eqref{eq:model_ROBART} to a dynamic framework by defining an AutoRegressive Rank-Order BART model with exogenous covariates (ARROBARTX) as follows:
\begin{align}
\begin{split}
    \tau_{j,t}  = \operatorname{rank}(\z_{j,t}), \qquad
    z_{ij,t}  & = f(\mathbf{X}_{ij,t}) + \varepsilon_{ij,t}, \qquad \varepsilon_{ij,t} \distas{iid} \mathcal{N}(0,1), \\
    f(\mathbf{X}_{ij,t}) & = \sum_{s=1}^S \sum_{\ell=1}^{b_s} \mu_{\ell,s} \mathbb{I}(\mathbf{X}_{ij,t} \in B_{\ell,s}),
\end{split}
\label{eq:model_ARROBARTX}
\end{align}
where the regression function $f$ is approximated by a BART specification.
A special case of the ARROBARTX in Eq.~\eqref{eq:model_ARROBARTX} is an AutoRegressive Rank-Order BART model without covariates (ARROBART) and with one lag, which is obtained by assuming $\mathbf{X}_{ij,t} = \z_{j,t-1}$.
The ARROBART model can be interpreted as a hidden Markov model \citep[HMM, see][]{cappe2005hidden} with measurement and transition density given by, respectively:
\begin{align*}
    \tau_{j,t} | \z_{j,t}  \sim \delta_{\operatorname{rank}(\z_{j,t})}(\tau_{j,t}), \text{ and }    \z_{j,t} | \z_{j,t-1}  \sim \mathcal{N}_N(\z_{j,t} | f(\z_{j,t-1}), I_N),
\end{align*}
where $f$ is approximated by a BART and $I_N$ denotes the $N$-dimensional identity matrix.
Let $\boldsymbol{\tau} = \{ \tau_{j,t} : j \in \{1,\dots,M\}, t \in \{1,\dots,T\} \}$ denote the collection of all the observations and $\boldsymbol{\theta}$ be the collection of all parameters. The ARROBART likelihood function is:  
\begin{equation}
    L(\boldsymbol{\tau} | \boldsymbol{\theta}) = \prod_{j=1}^M  \underbrace{\int_{\R^N} \dots \int_{\R^N}}_{\text{T+1 integrals}}  \left( \prod_{t=1}^T p(\tau_{j,t} | \z_{j,t}) p(\z_{j,t} |\z_{j,t-1}, \boldsymbol{\theta} ) \right) p(\z_{j,0} | \boldsymbol{\theta})  \: \mathrm{d}\z_{j,0} \dots \mathrm{d}\z_{j,T},
\end{equation}
where $\z_{j,0}$ is the initial value of the latent vector. 
As the ARROBART model postulates independence across rankers $j$, hereafter, without loss of generality, we focus on the case $M=1$ and drop the corresponding index.

The temporal dependence among the latent variables introduces a filtering problem in the estimation of the ARROBART model, as making inferences on the latent vector $\z_t$ requires the derivation of appropriate filtering distributions and {appropriate sampling strategies}.

Let us first introduce some additional notation.
For a variable $x \in \mathcal{D} \subseteq \R$, the BART specification induces a partition $\{ C_k \}_{k=1}^K$ of the domain $\mathcal{D}$,
such that the (approximated) function has constant value $\tilde{\mu}_k$ over each region $C_k$. We refer to the Supplement for proof and give a sketch below.
First, notice that, for each tree $s \in \{1,\ldots,S \}$ defined in eq.~\eqref{eq:BART_approx_f}, the collection of intervals $\{ B_{\ell,s} \}_{\ell=1}^{b_s}$ is a partition of $\mathcal{D}$ and the $\ell$th leaf of the $s$th tree is associated to the scalar coefficient $\mu_{\ell,s} \in \R$.
Moreover, we can associate each tuple $(j_1,\ldots,j_S) \in \bigtimes_{s=1}^S \{1,\ldots,b_s\}$ one-to-one to an index $k \in \{ 1,\ldots,K \}$, where $K = \prod_{s=1}^S b_s$ is the total number of the tuples.
Then, for each tuple $(j_1,\ldots,j_S)$ and associated $k$ index, define $C_k = B_{j_1,1} \cap \ldots \cap B_{j_S,S}$ and let $\tilde{\mu}_k = \sum_{s=1}^S \mu_{j_s,s}$ be the associated coefficient.
Note that the intervals $C_k$ can be empty ($C_k = \emptyset$) or unbounded, that is $C_k = (-\infty, a)$ or $C_k = (b,\infty)$ for some $a,b \in \mathcal{D}$.

For simplicity, we introduce some shorthand notation.
For a vector $\tau_i$ and a set $A_i$, $i=1,\ldots,t$, we define the collections $\tau_{1:t} = \{ \tau_1,\ldots,\tau_t \}$ and $A_{1:t} = \{ A_1,\ldots,A_t \}$.
Let $\K = (k_1,\ldots,k_N)' \in \{1,\ldots,K\}^N$ denote a tuple or multi-index and $\boldsymbol{\tilde{\mu}}_{\K} = (\tilde{\mu}_{k_1}, \ldots, \tilde{\mu}_{k_N})'$ be the vector of leaf parameters indexed by $\K$. Then, a summation over multiple indices is written as
\begin{equation*}
\sum_{\K} a_{\K} = \sum_{\K \in \{1,\ldots,K\}^N} a_{\K} = \sum_{k_1=1}^K \dots \sum_{k_N=1}^K a_{k_1,\ldots,k_N}.
\end{equation*}
We define $A_t$ as the set of vectors $\z_t$ whose elements follow the same ordering as in the raking list $\tau_t$, that is:
\begin{equation*}
A_t = \big\{ \z_t\in\R^N : z_{j,t} < z_{i,t} \iff \tau_{j,t} < \tau_{i,t}, \;\;\forall \, i,j=1,\ldots,N \big\}.
\end{equation*}
For partial ranking data, if missingness of a rank implies absence of knowledge of an inequality between elements of $\bm{z}_t$, then the set $A_t$ is not constrained by the unknown inequality.
In detail, partial ranking occurs when each assessor provides the (full) ranking of a subset $A \subset \mathbb{A} = \{a_1,\ldots,a_N\}$ consisting of $k \leq N$ items, while leaving the other items $a_\ell \in A^c$ not ranked.
Let us denote this ranking of subset $A$ by $\tilde{\bm{\tau}}$, then we define a partial ranking as $\boldsymbol{\tau}: \mathbb{A} \to \mathcal{S} \subset \mathcal{P}_N$, where $\mathcal{S} = \{ \boldsymbol{\tau} \in \mathcal{P}_N : \text{ if } \tilde\tau_{i_1} < \tilde\tau_{i_2}, \text{ with } a_{i_1}, a_{i_2} \in A \text{ then } \tau_{i_1} < \tau_{i_2} \}$ is the set of possible augmented random permutation vectors, which coincides with the original partially ranked items together with the allowable values of the missing ranks.
This means that the order between items $a_i \in A$ in the latent rank vector $\boldsymbol{\tau}$ must be preserved as in the observed rank vector $\tilde{\boldsymbol{\tau}}$, whereas the ranks of the remaining items $a_j \in A^c$ are left unspecified.

Finally, we define $C_{\K} = (\bigtimes_{i=1}^N C_{k_i})$ and $C_{\K,t} = C_{\K} \cap A_t$, for $t=1,\ldots,T$. Therefore, for any $\z_t \in C_{\K,t}$, all the element-wise constraints $\{ C_{k_i} \}_{i=1}^N$ and the ordering restriction  $A_t$ are satisfied.

Theorem~\ref{theorem:filt_smoot} below derives the filtering, predictive, and smoothing distributions for the latent vector $\z_{j,t}$, then Corollary~\ref{corollary:mixture} shows that they are proportional to finite mixtures with time-varying weights and components.
We refer to the Supplement for the proofs.

\begin{theorem}    \label{theorem:filt_smoot}
The filtering distribution is given by
\begin{align}
p(\z_t | \tau_{1:t}) \propto \mathbb{I}(\z_t \in A_t) \sum_{\K \in \{1,\ldots,K\}^N} \mathcal{N}_N(\z_t | \boldsymbol{\tilde{\mu}}_{\K}, I_N) q_{\K, t}(A_{1:t-1}),
\label{eq:filtering}
\end{align}
where $A_t$ is the set of possible vectors $\z_t $ that agree with the rank-order given by $\tau_t$, $\boldsymbol{\tilde{\mu}}_{\K}$ are mean vectors corresponding to possible combinations of regions $C_{k_1},\ldots,C_{k_N}$ in which the latent variable $\z_{t-1}$ can be located, and for $t > 1$,
\begin{align*}
q_{\K,t}(A_{1:t-1}) & = \sum_{\mathbf{m} \in \{1,\ldots,K\}^N} q_{\mathbf{m},t-1}(A_{1:t-2}) \int_{\C_{\K,t-1}} \mathcal{N}_N(\z_{t-1} | \boldsymbol{\tilde{\mu}}_{\mathbf{m}}, I_N) \: \mathrm{d}\z_{t-1},
\end{align*}
whereas, at time $t=1$ there is no dependence on $A_t$ and we have:
\begin{align*}
q_{\K, 1} & = \int_{\C_{\K}} \mathcal{N}_N(\underline{\z}_0 | \underline{\z}_{prior}, I_N) \: \mathrm{d}\underline{\z}_0.
\end{align*}
The one-step-ahead predictive distribution, for each $t=1,\ldots,T-1$, is
\begin{align}
p(\z_{t+1} | \tau_{1:t}) \propto \sum_{\K \in \{1,\ldots,K\}^N} \mathcal{N}_N(\z_{t+1} | \boldsymbol{\tilde{\mu}}_{\K}, I_N) q_{\K,t+1}(A_{1:t}).
\label{eq:predictive}
\end{align}
The smoothing distribution is $p(\z_t | \tau_{1:T}) \propto p(\z_t | \tau_{1:t}) r_t(\z_t)$,
where $r_t(\z_t)$ embeds all information from the periods $t' = t+1,\ldots,T$ and is defined recursively for each $t < T$ as
\begin{align*}
    r_t(\z_t) & = \int_{A_{t+1}} r_{t+1}(\z_{t+1}) \mathcal{N}_N\Big(\z_{t+1} \big| \sum_{s=1}^S g_s(\z_t), I_N \Big) \: \mathrm{d}\z_{t+1} \\
     & = \int_{A_{t+1}} r_{t+1}(\z_{t+1}) \mathcal{N}_N\Big(\z_{t+1} \big| \sum_{\K \in\{1,\ldots,K\}^N} \boldsymbol{\tilde{\mu}}_{\K} \mathbb{I}(\z_t \in C_{\K}), I_N \Big) \: \mathrm{d}\z_{t+1},
\end{align*}
where tree function $g_s$ is defined as $g_s(\x | \mathcal{T}_s, \bmu_s) = \sum_{\ell=1}^{b_s} \mu_{\ell,s} \mathbb{I}(\x \in B_{\ell,s})$ and it holds $\sum_{s=1}^S g_s(\x | \mathcal{T}_s, \bmu_s) = \sum_{\K \in\{1,\ldots,K\}^N} \boldsymbol{\tilde{\mu}}_{\K} \mathbb{I}(\x \in C_{\K})$. At the final time $t=T$, it is given by $r_T(\z_T) = 1$.
\end{theorem}

Corollary~\ref{corollary:mixture} shows that both the filtering and smoothing distributions are proportional to finite mixtures.
{The alternative representation for the filtering  distribution and mixture distributions for backward sampling (see the Supplement)} can be used to design a more efficient algorithm for posterior inference based on a data augmentation approach coupled with a Metropolis-Hastings step (see Section~\ref{sec:inference}).

\begin{corollary}[Mixture representations]  \label{corollary:mixture}
At each time $t=1,\ldots,T$, the filtering distribution is proportional to the finite mixture:
\begin{align}
p(\z_t | \tau_{1:t}) & \propto \sum_{\K \in \{1,\ldots,K\}^N} w^F_{\K, t} u^F_{\K, t}(\z_t),
\end{align}
with weights $w^F_{\K, t}$ and components $u^F_{\K, t}(\z_t) = \bar{u}^F_{\K, t}(\z_t) / n_{\K, t}$, where
\begin{align*}
\bar{u}^F_{\K, t}(\z_t) & = \mathbb{I}(\z_t \in A_t) \mathcal{N}_N(\z_t | \boldsymbol{\tilde{\mu}}_{\K}, I_N), \\
n_{\K, t} & = \int_{\R^N} \bar{u}^F_{\K, t}(\z_t) \: \mathrm{d}\z_t = \int_{A_t} \mathcal{N}_N(\z_t | \boldsymbol{\tilde{\mu}}_{\K}, I_N) \: \mathrm{d}\z_t, \\
w^F_{\K, t} & = \frac{q_{\K,t}(A_{1:t-1}) n_{\K,t}}{\sum_{\mathbf{m} \in \{1,\ldots,K\}^N} q_{\mathbf{m},t}(A_{1:t-1}) n_{\mathbf{m},t}}.
\end{align*}
At each time $t=1,\ldots,T$, the smoothing distribution is proportional to the finite mixture:
\begin{align}
p(\z_t | \tau_{1:T}) &\propto \sum_{\K \in \{1,\ldots,K\}^N} w^S_{\K, t} u^S_{\K, t}(\z_t),
\end{align}
with weights $w^S_{\K,t}$ and components $u^S_{\K,t}(\z_t) = \bar{u}^S_{\K,t}(\z_t) / m_{\K,t}$, where
\begin{align*}
\bar{u}^S_{\K, t}(\z_t) & = r_t(\z_t) \mathbb{I}(\z_t \in A_t) \mathcal{N}_N(\z_t | \boldsymbol{\tilde{\mu}}_{\K}, I_N), \\
m_{\K, t} & = \int_{\R^N} \bar{u}^S_{\K, t}(\z_t) \: \mathrm{d}\z_t, \\
w^S_{\K, t} & = \frac{q_{\K, t} (A_{1:t-1}) m_{\K, t}}{\sum_{\boldsymbol{\ell} \in \{1,\ldots,K\}^N}  q_{\boldsymbol{\ell}, t} (A_{1:t-1}) m_{\boldsymbol{\ell},t}}, \qquad \K \in \{1,\ldots,K\}^N.
\end{align*}
\end{corollary}

Notice that $\bar{u}^F_{\K, t}(\z_t)$ is a multivariate normal distribution restricted such that the elements satisfy the specific ordering encoded by $A_t$.

{Furthermore, $p(\z_{1:T} | \tau_{1:T})  = p(\z_T | \tau_{1:T}) \prod_{t=1}^{T-1} p(\z_t | \z_{t+1}, \tau_{1:(T-1)})$ and backward sampling from $  p(\mathbf{z}_{t} | \tau_{1:T},  \mathbf{z}_{(t+1):T})$, consists of samples from mixtures of truncated Gaussian distributions. Details are provided in the Supplement.}


\section{Bayesian Inference}  \label{sec:inference}

\subsection{Prior on the BART Parameters} \label{subsec:PriorBART}

The BART specification is completed by choosing a prior for the parameters of the sum-of-trees model, namely, the binary trees $\mathcal{T}_s$ and the terminal node parameters $\bm\mu_s$. We assume:
\begin{equation}
p(\mathcal{T}_1,\bmu_1,\ldots,\mathcal{T}_S,\bmu_S) = \prod_{s=1}^S p(\bmu_s | \mathcal{T}_s) p(\mathcal{T}_s) = \prod_{s=1}^S \Big[ \prod_{\ell=1}^{b_s} p(\mu_{\ell,s} | \mathcal{T}_s) \Big] p(\mathcal{T}_s),
\end{equation}
where $\mu_{\ell,s} \in \bmu_s$.
To deal with the potential overfitting of the nonparametric BART specification, we follow \cite{chipman2010bart} and choose a regularization prior for the tree structure and terminal nodes that constrains the size and fit of each tree so that each one contributes only a small part to the overall fit. This prior strongly encourages each tree to be a weak learner, with leaf parameters close to zero.

The marginal prior for each tree structure, $p(\mathcal{T}_s)$, is specified by three parts. First, the probability that a node at depth $d=0,1,\ldots$ is nonterminal, given by $\underline{\alpha} (1+d)^{-\underline{\beta}}$, for $\underline{\alpha}\in(0,1)$, $\underline{\beta} \in [0,\infty)$. Smaller (larger) values of $\underline{\alpha}$ ($\underline{\beta}$) impose a larger penalty on more complex tree structures. We use the default choice proposed by \cite{chipman2010bart}, that is $\underline{\alpha} = 0.95$ and $\underline{\beta} = 2$.
Second, the distribution of the splitting variable at each interior node is assumed to be discrete uniform over all the covariates. Third, the distribution on the splitting rule assignment in each interior node, $c_h$, conditional on the splitting variable, is uniform on the discrete set of available splitting values.

The conditional prior on the terminal leaf parameters, $p(\mu_{\ell,s}| \mathcal{T}_s)$, is assumed to be the conjugate normal distribution $\mathcal{N}(\mu_\mu, \sigma^2_\mu)$, where the hyperparameters are chosen such that $\mathcal{N}(S \mu_\mu, S\sigma^2_\mu)$ assigns substantial probability to the interval $(z_{min}, z_{max})$.
In line with \cite{chipman2010bart}, the prior on $\mu_{\ell,s}$ is constructed with $\mu_\mu=0$ and $\sigma_\mu = (z_{max} - z_{min})/(2 k_{\mu} \sqrt{S})$, where $k_{\mu}$ is a hyperparameter set to 2, and $z_{max}$ and $z_{min}$ are the maximum and minimum of initial values of the latent outcome.

\subsection{Posterior Sampling for ARROBART}   \label{subsec:posterior_ARROBART}

To draw samples from the joint posterior distribution, a Markov Chain Monte Carlo (MCMC) algorithm is provided. The Gibbs sampler iterates over the following steps:
\begin{enumerate}[label=\arabic*)]
    \item {given the static parameters, sample the latent states $\z = \{ \z_1,\ldots,\z_T \}$;}
    \item {given the latent states,} sample the tree and leaf parameters from $p(\mathcal{T}_s, \bmu_s | \mathcal{T}_{-s}, \bmu_{-s}, \z)$, for $s=1,\ldots,S$:
    \begin{enumerate}[label=\alph*.]
        \item compute the $NJT$-dimensional vector of partial residuals, $R_s$, as
        \begin{equation*}
        R_{s,ijt} = {z}_{ijt} - \sum_{\substack{k=1\\ k \neq s}}^S g_k(\mathbf{X}_{ij,t} | \mathcal{T}_k, \bmu_k),
        \end{equation*}
        and notice that $p(\mathcal{T}_s, \bmu_s | \mathcal{T}_{-s}, \bmu_{-s}, \z) = p(\mathcal{T}_s, \bmu_s | R_s)$;
        \item since $p(\mathcal{T}_s | R_s)$ has a closed form, sample the trees structure from $p(\mathcal{T}_s | R_s)$ using a Metropolis-Hastings algorithm;
        \item sample the terminal nodes from $p(\bmu_s | \mathcal{T}_s, R_s)$, which is a set of independent draws from a normal distribution.
    \end{enumerate}
\end{enumerate}

{The latent states can be sampled through forward filtering backward sampling, using the filtering recursions in Theorem~\ref{theorem:filt_smoot} and a backward sampler described in the Supplement. However, this sampler is computationally intensive.} This is due to (i) evaluation of many Gaussian integrals over very small subspaces of $\R^N$ and (ii) drawing from an $N$-dimensional Gaussian distribution truncated such that its elements satisfy a given ordering constraint.
Both issues worsen as the number of items, $N$, increases.
To overcome these computational issues, for each period $t=1,\ldots, T$, we approximate the joint posterior distribution of the vector $\z_t$ and derive the element-wise posterior full conditional distribution for each $i=1,\ldots, N$, conditioning on all the other $j\neq i$ elements. Therefore, we replace sampling from a joint distribution by sampling from a collection of full conditional distributions, which results in a full Gibbs sampler of the latent scores.
This approach is similar to the standard technique used in rank-order data models \citep[e.g.,][Ch. 12]{hoff2009first}.
We refer to the Supplement for a detailed description of the sampling method.

Steps 2b and 2c concerning the static parameters of the tree structure are standard in the BART literature and follow directly from the algorithm in \cite{chipman2010bart}. The tree structure is updated by proposing one of four moves: growing a terminal node, pruning a pair of terminal nodes, changing a nonterminal rule, and swapping a rule between parent and child. Then, given the tree structure, the terminal node parameters are sampled directly from their normal conditional posterior distribution.


\section{Simulation Studies}
\label{sec:simulations}

This section investigates the performance of the proposed methods on synthetic data. In particular, Section~\ref{subsec:sim_ROBART} compares the ROBART model to the methods for rank-order data introduced by \cite{li2016bayesian}.
Then, Section~\ref{subsec:sim_ARROBART} presents a comparative study between ARROBART and a range of models that account for (or ignore) the possible temporal persistence of the data.

\subsection{Comparison of ROBART and Linear Models}  \label{subsec:sim_ROBART}

\subsubsection{Static Simulation Study with Full Rankings}

The performance of the ROBART model defined in Section~\ref{subsec:ROBART} is compared to the methods for rank-order data introduced by \cite{li2016bayesian} by using the normalized Kendall tau distance \citep{kendall1938new}. Given two ranking lists $\hat{\tau}$ and $\tau$, the normalized Kendall tau distance is defined as the proportion of pairwise disagreements between them:
\begin{equation}
K_n(\hat{\tau},\tau) = \frac{\# \big\{ (i,j) : i<j, \: [\hat{\tau}_i < \hat{\tau}_j \wedge \tau_i > \tau_j] \vee [\hat{\tau}_i > \hat{\tau}_j \wedge \tau_i < \tau_j] \big\} }{N(N-1)/2} \in [0,1].
\label{eq:scaled_Kendall_tau_distance}
\end{equation}
The three synthetic datasets used in this simulation are generated as in \cite{li2016bayesian}, and we report briefly their structures. For each $i = 1,\ldots, N$, item $i$ has a true score $\gamma_i$, while the covariate vectors $\mathbf{x}_i = (x_{i1},\ldots,x_{iK_x})'$ are drawn independently from a Normal distribution with zero mean and covariance equal to $\mathbb{C}ov(x_{il}, x_{im}) = \rho^{\abs{l-m}}$, for $1 \leq l, m \leq v$ and $\abs{\rho} < 1$. The true score vector $\bm{\gamma}$ is generated by considering the different roles of the covariates as:
\begin{enumerate}
	\item $\gamma_i = \mathbf{x}_i' \boldsymbol{\beta}$, where $\boldsymbol{\beta} = (3,2,-1,-0.5)'$, $K_x=4$, and $\rho=0$;
	\item $\gamma_i = \mathbf{x}_i' \boldsymbol{\beta} + \norm{\mathbf{x}_i}^2 $, where $\boldsymbol{\beta} = (3,2,1)'$, $K_x=3$, and $\rho=0.5$;
	\item $\gamma_i = \norm{\mathbf{x}_i}^2 $, where $K_x=4$, and $\rho=0.5$.
\end{enumerate}
The covariates are linearly related to the score in Scenario 1, have a nonlinear relationship in Scenario 3, and a combination of both in Scenario 2. 
Given the scores, $M$ full ranking lists $\{ \tau_j \}_{j=1}^M$ are generated as $\tau_j = \operatorname{rank}(\z_j)$ with $\z_j \distas{i.i.d.} \mathcal{N}_N(\boldsymbol{\gamma}, \sigma^2 I_N)$.

For each scenario and each value of $\sigma \in \{1,5,10,20,40\}$, we generate $100$ synthetic datasets with $N=50$ and $M=10$. 
We compare the results for ROBART against several competitors: the Borda Count, the Markov-Chain methods \citep[MC1, MC2, MC3 -][]{dwork2001rank}, the Cross-Entropy Monte Carlo method \citep[CEMC -][]{lin2009integration}, the Plankett-Luce (PL) model, the BARC model with and without covariates, and the BARC model with Mixture of rankers with different opinions \citep[BARCM -][]{li2016bayesian}. Further details on these methods are available in the Supplement.

We use the default values for the ROBART model as in \cite{chipman2010bart} and follow \cite{li2016bayesian} in choosing the hyperparameters for all the remaining models.

\begin{table}[ht]
\centering
\footnotesize
\setlength{\tabcolsep}{4pt}
\begin{adjustbox}{max width=\textwidth}
\begin{threeparttable}
\captionsetup{width=0.93\linewidth}
\caption{\label{tab:li_table} \small Comparison between ROBART, BARC, and other ranking methods. Full rankings.}
\hspace*{-2ex}
\begin{tabular}{l *{13}{c}}
		\toprule
		 & $\sigma$ & Borda & MC1 & MC2 & MC3 & CEMC & PL & BAR & $\text{BARC}_1$ & $\text{BARC}_2$ & $\text{BARC}_3$ & BARCM & ROBART \\ 
		\midrule
		\multirow{5}{1em}{
		\begin{rotate}{90} \hspace{-22pt} Scenario 1 \end{rotate}
		} & 1 & (0.03) & 1.31 & 1.08 & 1.01 & 2.55 & 5.62 & 0.99 & \textbf{0.65} & \textbf{0.65} & 0.98 & \textbf{0.65} & 0.89 \\ 
		& 5 & (0.13) & 1.27 & 1.02 & 1.01 & 1.13 & 1.09 & 0.99 & \textbf{0.65} & \textbf{0.65} & 0.97 & \textbf{0.65} & 0.68 \\
		& 10 & (0.22) & 1.57 & 1.01 & 1.00 & 1.05 & 1.07 & 1.00 & 0.71 & 0.71 & 0.99 & 0.71 & \textbf{0.69} \\ 
		& 20 & (0.32) &  1.47 & 1.00 & 1.00 & 1.02 & 1.04 & 0.99 & 0.78 & 0.78 & 0.99  & 0.78 & \textbf{0.75}\\ 
		& 40 & (0.41) & 1.23 & 1.00 & 1.00 & 1.01 & 1.01 & 0.99 & 0.88 & 0.88 & 0.99  & 0.88 & \textbf{0.87}\\ 
		\midrule
		\multirow{5}{1em}{
		\begin{rotate}{90} \hspace{-22pt} Scenario 2 \end{rotate}
		} & 1 & (0.03) & 1.33 & 1.04 & 1.00 & 2.65 & 1.06 & 0.99 & 0.97 & 0.97 & 0.98 & 0.97 & \textbf{0.86}  \\ 
		& 5 & (0.12) & 1.30 & 1.01 & 1.00 & 1.12 & 1.10 & 0.99 & 0.83 & 0.83 & 0.97 & 0.83 & \textbf{0.69} \\ 
		& 10 & (0.20) & 1.36 & 1.01 & 1.00 & 1.05 & 1.07 & 1.00 & 0.77 & 0.77 & 0.99 & 0.77 & \textbf{0.67} \\ 
		& 20 & (0.29) & 1.48 & 1.01 & 1.00 & 1.03 & 1.04 & 0.99 & 0.80 & 0.80 & 0.99 & 0.80 & \textbf{0.72} \\ 
		& 40 & (0.39) & 1.25 & 1.00 & 1.00 & 1.01 & 1.03 & 1.00 & 0.89 & 0.89 & 1.00 & 0.89 & \textbf{0.85}  \\ 
		\midrule
		\multirow{5}{1em}{
		\begin{rotate}{90} \hspace{-22pt} Scenario 3 \end{rotate}
		} & 1 & (0.05) & 1.28 & 1.03 & 1.00 & 1.60 & 1.11 & 1.00 & 1.00 & 1.00 & 1.00 & 1.00 & \textbf{0.88}  \\ 
		& 5 & (0.18) & 1.32 & 1.01 & 1.00 & 1.06 & 1.06 & 0.99 & 1.00 & 1.00 & 0.99 & 1.00 & \textbf{0.73} \\ 
		& 10 &  (0.29) & 1.38 & 1.00 & 1.00 & 1.03 & 1.04 & 0.99 & 1.00 & 1.00 & 0.99 & 1.00 & \textbf{0.78} \\ 
		& 20 & (0.36) & 1.34 & 1.00 & 1.00 & 1.02 & 1.03 & 1.00 & 1.00 & 1.00 & 1.00  & 1.00 & \textbf{0.85}\\ 
		& 40 & (0.42) & 1.18 & 1.00 & 1.00 & 1.01 & 1.01 & 1.00 & 1.00 & 1.00 & 1.00 & 1.00 & \textbf{0.91}  \\ 
		\bottomrule
\end{tabular}
\begin{tablenotes}
\footnotesize
\item \!\!\!\! Notes: Scaled Kendall tau distance of rank predictions for the scenarios specified by \cite{li2016bayesian}. The Borda column with parentheses shows the average Kendall tau distances between estimated and true ranking lists for the Borda Count predictions, while the remaining columns show tau distances relative to the tau distance attained by the Borda Count method. Bold denotes the best-performing model.
\end{tablenotes}
\end{threeparttable}
\end{adjustbox}
\end{table}

Table~\ref{tab:li_table} reports the Kendall tau distances between the true and estimated rank lists averaged over the $100$ replications. For each column, we provide the ratios of the average Kendall tau distances over the corresponding values for the Borda Count, which serves as the baseline.
In Scenarios 2 and 3, for any $\sigma$, the proposed ROBART model outperforms all the competing models. In particular, there is a strong gain in Scenario 3, for any $\sigma$, moving from around 10\% (for $\sigma = 1$ or $40$) to 27\% (for $\sigma = 5$). Similarly in Scenario 2, irrespective of $\sigma$, ROBART performs better than BARC and BARCM, with a gain of 10\% (for $\sigma$ equal to 1) to 20\% (for $\sigma$ equal to 5). Increasing the value of $\sigma$ leads to similar conclusions, except for $\sigma$ equal to 40, where the ROBART model provides slightly better predictions than BARCM and BARC.
These better performances are less evident in Scenario 1, where for small values of $\sigma$ (equal to 1 and 5), the ROBART is outperformed by the BARC and BARCM models, while it beats the other models of 1\% and 3\% when $\sigma$ is bigger than 10. As expected, nonlinear relationships in the data-generating process are correctly estimated by the BART specification. Unsurprisingly, the proposed model does not  yield improvements for estimates of linear functions of covariates when $\sigma$ is small.

\subsubsection{Static Simulation Study with Partial Rankings}

ROBART and ARROBART are extendable to partial ranking data by replacing the constraints on the latent scores with the less restrictive constraints implied by partial rankings. The Supplement contains additional details for the model and sampler for partial ranking data. We demonstrate the ability of our model to predict rankings using partial ranking information through a simulation study described by \cite{li2016bayesian}. 
The comparison also included two additional specifications: the RODART, which is ROBART with a Dirichlet hyperprior on covariates' splitting probabilities, and the SoftROBART, which is ROBART with soft and sparse splitting rules \citep{linero2018dart}.

We divide the $N=80$ items into $K = 1,2,4,8,10,16$ groups, of $N/K$ items. Pairwise comparison information is only available within groups. The DGP is otherwise the same as scenario 2 above.
Table~\ref{tab:li_tablePartial_sigma1} contains the correlations of rank predictions averaged over 10 repetitions of the simulation study with latent error standard deviations $\sigma\in\{1,5\}$, as in \cite{li2016bayesian}. The results confirm that ROBART still provides a considerable improvement over linear models when only partial rankings are available.
Finally, the SoftROBART is found to provide further improvements in ranking predictions.

\begin{table}[t!h]
\centering
\footnotesize
\setlength{\tabcolsep}{4pt}
\begin{adjustbox}{max width=\textwidth}
\begin{threeparttable}
\captionsetup{width=0.93\linewidth}
\caption{\label{tab:li_tablePartial_sigma1} \small Comparison between ROBART and other ranking methods. Partial rankings.}
\hspace*{-2ex}
\begin{tabular}{ll *{8}{c}}
  \toprule
& & K & BAR & BARC1 & BARC2 & BARC3 & ROBART & RODART & SoftROBART \\ 
  \midrule
\multirow{5}{1em}{
		\begin{rotate}{90} \hspace{-22pt} $\sigma=1$ \end{rotate}
		} & \multirow{5}{1em}{
		\begin{rotate}{90} \hspace{-22pt} Scenario 2 \end{rotate}
		} & 1 & (0.03) & 1.00 & 1.00 & 1.00 & 0.83 & 0.84 & \textbf{0.67} \\ 
& & 2 & (0.07) & 0.61 & 0.60 & 0.59 & 0.32 & 0.34 & \textbf{0.28} \\ 
& & 4 & (0.08) & 0.59 & 0.62 & 0.64 & 0.35 & 0.39 & \textbf{0.28} \\ 
& & 8 & (0.13) & 0.51 & 0.49 & 0.56 & 0.24 & 0.27 & \textbf{0.21} \\ 
& & 10 & (0.12) & 0.59 & 0.59 & 0.65 & 0.26 & 0.33 & \textbf{0.22} \\ 
& & 16 & (0.17) & 0.55 & 0.56 & 0.58 & 0.23 & 0.36 & \textbf{0.21} \\ 
   \midrule
\multirow{5}{1em}{
		\begin{rotate}{90} \hspace{-22pt} $\sigma=5$ \end{rotate}
		} & \multirow{5}{1em}{
		\begin{rotate}{90} \hspace{-22pt} Scenario 2 \end{rotate}
		} & 1 & (0.12) & 0.83 & 0.83 & 0.98 & 0.60 & 0.68 & \textbf{0.46} \\ 
& & 2 & (0.13) & 0.80 & 0.79 & 1.04 & 0.54 & 0.64 & \textbf{0.46} \\ 
& & 4 & (0.14) & 0.78 & 0.78 & 0.95 & 0.56 & 0.73 & \textbf{0.46} \\ 
& & 8 & (0.17) & 0.63 & 0.64 & 0.84 & 0.46 & 0.58 & \textbf{0.36} \\ 
& & 10 & (0.16) & 0.65 & 0.65 & 0.88 & 0.45 & 0.57 & \textbf{0.37} \\ 
& & 16 & (0.19) & 0.60 & 0.60 & 0.78 & 0.42 & 0.58 & \textbf{0.35} \\ 
\bottomrule
\end{tabular}
\begin{tablenotes}
\footnotesize
\item \!\!\!\! Notes: Scaled Kendall tau distance of rank predictions for the scenarios specified by  \cite{li2016bayesian}. The BAR column with parentheses shows the average Kendall tau distances between estimated and true ranking lists for the BAR predictions, while the remaining columns show tau distances relative to the tau distance attained by BAR. \newline RODART is ROBART with a Dirichlet hyperprior on covariates' splitting probabilities. SoftROBART is ROBART with soft and sparse splitting rules.
\end{tablenotes}
\end{threeparttable}
\end{adjustbox}
\end{table}

\subsubsection{Static Simulation Study with Ranker-specific Covariates}

Table ~\ref{tab:li_table)rankerspecific} contains the results for a variation on the static simulation study described by \cite{li2016bayesian} in which the covariates are ranker-specific. For each $ (i,m) \in  \{1,\ldots, N\} \times \{1,\dots,M\}$, item $(i,m)$ has a true score $\gamma_{i,m}$, while the covariate vectors $\mathbf{x}_{i,m} = (x_{i,m,1},\ldots,x_{i,m,K_x})'$ are drawn independently from a Normal distribution with zero mean and covariance equal to $\mathbb{C}ov(x_{i,m,l}, x_{i,m,s}) = \rho^{\abs{l-s}}$, for $1 \leq l, s \leq v$ and $\abs{\rho} < 1$. The true score vector $\bm{\gamma}$ is generated by considering the different roles of the covariates as:
\begin{enumerate}
	\item $\gamma_{i,m} = \mathbf{x}_{i,m}' \boldsymbol{\beta}$, where $\boldsymbol{\beta} = (3,2,-1,-0.5)'$, $K_x=4$, and $\rho=0$;
	\item $\gamma_{i,m} = \mathbf{x}_{i,m}' \boldsymbol{\beta} + \norm{\mathbf{x}_{i,m}}^2 $, where $\boldsymbol{\beta} = (3,2,1)'$, $K_x=3$, and $\rho=0.5$;
	\item $\gamma_{i,m} = \norm{\mathbf{x}_{i,m}}^2 $, where $K_x=4$, and $\rho=0.5$.
\end{enumerate}

ROBART provides much better predictions than other methods for all scenarios and values of $\sigma$. However, the benchmark methods were not originally intended for data with ranker-specific covariates.

\begin{table}[ht]
\centering
\footnotesize
\setlength{\tabcolsep}{4pt}
\begin{adjustbox}{max width=\textwidth}
\begin{threeparttable}
\captionsetup{width=0.93\linewidth}
\caption{\label{tab:li_table)rankerspecific} \small Comparison between ROBART, BARC, and other ranking methods. Full rankings with ranker-specific covariates.}
\hspace*{-2ex}
\begin{tabular}{l *{13}{c}}
		\toprule
		 & $\sigma$ & Borda & MC1 & MC2 & MC3 & CEMC & PL & BAR & $\text{BARC}_1$ & $\text{BARC}_2$ & $\text{BARC}_3$ & ROBART \\ 
		\midrule
		\multirow{5}{1em}{
		\begin{rotate}{90} \hspace{-22pt} Scenario 1 \end{rotate}
		} & 1 & (0.40) & 1.27 & 1.00 & 1.00 & 1.00 & 1.03 & 1.27 & 1.27 & 1.27 & 1.27 & \textbf{0.09} \\ 
		& 5 &  (0.44) & 1.15 & 1.00 & 1.00 & 1.01 & 1.02 & 1.15 & 1.15 & 1.15 & 1.15 & \textbf{0.20} \\ 
		& 10 & (0.47) & 1.08 & 1.00 & 1.00 & 1.00 & 1.00 & 1.08 & 1.08 & 1.08 & 1.08 & \textbf{0.31} \\ 
		& 20 & (0.48) & 1.03 & 1.00 & 1.00 & 1.00 & 1.00 & 1.03 & 1.03 & 1.03 & 1.03 & \textbf{0.51} \\ 
		& 40 & (0.49) & 1.02 & 1.00 & 1.00 & 1.00 & 1.00 & 1.02 & 1.02 & 1.02 & 1.02 & \textbf{0.72} \\ 
		\midrule
		\multirow{5}{1em}{
		\begin{rotate}{90} \hspace{-22pt} Scenario 2 \end{rotate}
		} & 1 & (0.40) & 1.27 & 1.00 & 1.00 & 1.00 & 1.03 & 1.27 & 1.27 & 1.27 & 1.27 & \textbf{0.10} \\ 
		& 5 &  (0.43) & 1.16 & 1.00 & 1.00 & 1.01 & 1.03 & 1.16 & 1.16 & 1.16 & 1.16 & \textbf{0.17} \\ 
		& 10 & (0.46) & 1.09 & 1.00 & 1.00 & 1.00 & 1.02 & 1.09 & 1.09 & 1.09 & 1.09 & \textbf{0.26} \\ 
		& 20 & (0.48) & 1.05 & 1.00 & 1.00 & 1.00 & 1.01 & 1.05 & 1.05 & 1.05 & 1.05 & \textbf{0.41} \\ 
		& 40 & (0.49) & 1.03 & 1.00 & 1.00 & 1.00 & 1.00 & 1.03 & 1.03 & 1.03 & 1.03 & \textbf{0.62} \\ 
		\midrule
		\multirow{5}{1em}{
		\begin{rotate}{90} \hspace{-22pt} Scenario 3 \end{rotate}
		} & 1 & (0.41) & 1.25 & 1.00 & 1.00 & 1.01 & 1.04 & 1.25 & 1.25 & 1.25 & 1.25 & \textbf{0.13} \\ 
		& 5 & (0.45) & 1.10 & 1.00 & 1.00 & 1.01 & 1.02 & 1.10 & 1.10 & 1.10 & 1.10 & \textbf{0.27} \\ 
		& 10 &  (0.47) & 1.05 & 1.00 & 1.00 & 1.00 & 1.01 & 1.05 & 1.05 & 1.05 & 1.05 & \textbf{0.44} \\ 
		& 20 & (0.49) & 1.03 & 1.00 & 1.00 & 1.00 & 1.00 & 1.03 & 1.03 & 1.03 & 1.03 & \textbf{0.64} \\ 
		& 40 & (0.49) & 1.02 & 1.00 & 1.00 & 1.00 & 1.00 & 1.02 & 1.02 & 1.02 & 1.02 & \textbf{0.79} \\ 
		\bottomrule
\end{tabular}
\begin{tablenotes}
\footnotesize
\item \!\!\!\! Notes: Scaled Kendall tau distance of rank predictions for modified scenarios in which the covariates are ranker-specific. The Borda column with parentheses shows the average Kendall tau distances between estimated and true ranking lists for the Borda Count predictions, while the remaining columns show tau distances relative to the tau distance attained by the Borda Count method. Bold denotes the best-performing model.
\end{tablenotes}
\end{threeparttable}
\end{adjustbox}
\end{table}

\FloatBarrier

\subsection{Comparison of ARROBART and Linear Models} \label{subsec:sim_ARROBART}

We investigate the performance of the proposed ARROBART against several competitors by means of the normalized Kendall tau distance defined in Eq.~\eqref{eq:scaled_Kendall_tau_distance}.
In particular, the synthetic datasets are generated according to three main scenarios:
\begin{enumerate}
    \item $\gamma_{ij,t} = 0.1 z_{ij,t-1}^2$,
    \item $\gamma_{ij,t} = 0.05 z_{ij,t-1} + 0.1 z_{ij,t-1}^2$,
    \item $\gamma_{ij,t} = 0.1z_{ij,t-1}x_{ij,t-1,1}$, \ $K_x=3$, \ $\rho=0.5$,
\end{enumerate} 
where in the last scenario the $K_x$ covariates $\mathbf{x}_{ij,t} = (x_{ij,t,1},\ldots,x_{ij,t, K_x})'$ are drawn independently from a multivariate normal distribution with mean zero and $\mathbb{C}ov(x_{ij,t,l}, x_{ij,t,m}) = \rho^{\abs{l-m}}$, for $1 \leq l, m \leq v$ and $\abs{\rho} < 1$. In detail, each scenario describes a nonlinear relationship, where $i=1,\ldots, N$, $j=1,\ldots, M$, and $t=1,\ldots, T$. We fix the number of rankers, $M$, the number of items, $N$ and the period, $T$, to $5$, $20$ and $52$, respectively.

For each scenario, the latent scores and the full ranking lists are generated as $\z_{j,t} \distas{iid} \mathcal{N}_N(\boldsymbol{\gamma}_{jt}, \sigma^2 I_N)$ and $\tau_{j,t} = \operatorname{rank}(\z_{j,t})$, respectively.
Moreover, for each scenario, we generate three datasets varying the signal-to-noise ratio; specifically, the data are simulated using a value of $\sigma \in \{ 0.1, 0.5, 1.0, 1.5, 2.0 \}$.

We compare the results for the full and the partial ranking against different competitors, such as (i) an autoregressive linear model for the latent score (ARROLinear); (ii) an ARROBART specification with exogenous covariates (ARROBARTX);  (iii) an autoregressive linear model with exogenous covariates (ARROLinearX). Moreover, we extend the ARROBART model to include the lag-1 of the observed rank (ARROBART-lag) and compare it with (i) an analogous linear model (ARROLinear-lag); (ii) a ROBART model using the first lag of the observed rank as a covariate (ROBART-lag); (iii) a linear rank order model (BARC) with the first lag of the {observed rank} (ROLinear-lag).

\begin{table}[t!h]
\centering
\scriptsize
\setlength{\tabcolsep}{1pt}
\begin{adjustbox}{max width=\textwidth}
\begin{threeparttable}
\captionsetup{width=0.93\linewidth}
\caption{\label{tab:ARROSimu}\small Comparison between ARROBART and ARROLinear with and without covariates for the full (top panel) and the partial (bottom panel) ranking.}
\hspace*{-2ex}
\begin{tabular}{l *{5}{c} c *{5}{c} c *{5}{c}}
    \toprule
     & \multicolumn{5}{c}{Scenario 1} & \;\; & \multicolumn{5}{c}{Scenario 2} & \;\; & \multicolumn{5}{c}{Scenario 3} \\
    \multicolumn{1}{c}{$\sigma$} & 0.1 & 0.5 & 1.0 & 1.5 & 2.0 &   &0.1 & 0.5 & 1.0 & 1.5 & 2.0 &  & 0.1 & 0.5 & 1.0 & 1.5 & 2.0 \\
    \midrule
    \multicolumn{2}{c}{\textbf{Full ranking}} & \\
    ARROBART    & (\textbf{0.38}) & (\textbf{0.23}) & (\textbf{0.22}) & (\textbf{0.16}) & (\textbf{0.15}) & 
                & (\textbf{0.13}) & (\textbf{0.18}) & (\textbf{0.17}) & (\textbf{0.15}) & (\textbf{0.13}) &
                & (0.54) & (0.52) & (0.49) & (0.50) &  (0.52)\\
    ARROLinear  & 1.35 &  2.10 & 2.27 & 2.90 & 2.95 & 
                & 2.54 & 2.21 & 2.42 & 2.65 & 2.87 & 
                & 0.95 & 0.94 & 0.99 & 1.02 & 0.96\\
    ARROBARTX   & -- & -- & -- & -- & -- &
                & -- & -- & -- & -- & -- & 
                & \textbf{0.89} & \textbf{0.92} & \textbf{0.94} & \textbf{0.99} & \textbf{0.94} \\
    ARROLinearX & -- & -- & -- & -- & -- &
                & -- & -- & -- & -- & -- & 
                & 0.92 & 0.97 & 1.01 & 1.01 & 0.99 \\
    \midrule
    \multicolumn{2}{c}{\textbf{Partial ranking}} & \\
	ARROBART    & (\textbf{0.45}) & (\textbf{0.36}) & (\textbf{0.32}) & (\textbf{0.29}) & (\textbf{0.30}) &
               & (\textbf{0.19}) & (\textbf{0.29}) & (\textbf{0.26}) & (\textbf{0.26}) & (\textbf{0.27}) &
                & (0.49) & (0.49) & (\textbf{0.49}) & (0.50) & (0.51) \\
    ARROLinear  &  1.13 & 1.31 & 1.55 & 1.71 & 1.65 &
                & 2.06 & 1.62 & 1.83 & 1.90 & 1.70 &
                &  1.01 & 1.02 & 1.01 & 0.96 & 1.00 \\
    ARROBARTX   & -- & -- & -- & -- & -- &
                & -- & -- & -- & -- & -- &
                & \textbf{0.97} & 1.02 & \textbf{1.00} & \textbf{0.95} & \textbf{0.95} \\
    ARROLinearX & -- & -- & -- & -- & -- &
                & -- & -- & -- & -- & -- &
                & 1.00 & \textbf{0.99} & 1.03 & 0.97 & 0.99  \\
	\bottomrule
\end{tabular}
\begin{tablenotes}
\footnotesize
\item \!\!\!\! Notes: The ARROBART row shows in brackets the average Kendall's tau distances between estimated and true ranking lists using the ARROBART method. The remaining rows report the ratio of average Kendall's tau for the model in the row to average Kendall's tau for ARROBART. Values greater than one mean that the benchmark outperforms the competitor. Bold denotes the best-performing model.
\end{tablenotes}
\end{threeparttable}
\end{adjustbox}
\end{table}

\begin{table}[t!h]
\centering
\scriptsize
\setlength{\tabcolsep}{1pt}
\begin{adjustbox}{max width=\textwidth}
\begin{threeparttable}
\captionsetup{width=0.93\linewidth}
\caption{\label{tab:ARROlagSimu}\small Comparison between ARROBART, ARROLinear, ROBART, and ROLinear with one lag of the observed ranks as a covariate for the full (top panel) and the partial (bottom panel) ranking.}
\hspace*{-2ex}
\begin{tabular}{l c *{5}{c} c *{5}{c} c *{5}{c}}
    \toprule
     & \;\; & \multicolumn{5}{c}{Scenario 1} & \;\; & \multicolumn{5}{c}{Scenario 2} & \;\; & \multicolumn{5}{c}{Scenario 3} \\
    \multicolumn{2}{c}{$\sigma$} & 0.1 & 0.5 & 1.0 & 1.5 & 2.0 &   & 0.1 & 0.5 & 1.0 & 1.5 & 2.0 &   & 0.1 & 0.5 & 1.0 & 1.5 & 2.0 \\
    \midrule
    \multicolumn{2}{c}{\textbf{Full ranking}} & \\
    ARROBART-lag    & & (\textbf{0.45}) & ({0.34}) & (0.27) & (\textbf{0.19}) & (\textbf{0.18}) & 
                & (\textbf{0.19}) & ({0.29}) & ({0.25}) & (\textbf{0.17}) & (\textbf{0.15}) & 
                & (0.48) & (0.42) & (0.42) & (0.41) & (0.44) \\
    ARROLinear-lag  & & 1.14 & 1.43 & 1.84 & 2.42 & 2.44 & 
                & 1.57 & 1.34 & 1.62 & 2.28 & 2.35 & 
                & 1.06 & 1.19 & 1.17 & 1.25 & 1.16 \\
    ROBART-lag      & & 1.04 & 1.03 & {0.99} & 1.06 & \textbf{1.00} &
                & 1.47 & 1.07 & 1.01 & 1.24 & 1.11 & 
                & \textbf{0.68} & \textbf{0.95} & \textbf{0.80} & \textbf{0.85} &  \textbf{0.80} \\
    ROLinear-lag    & & 1.14 & 1.45 & 1.82 & 2.40 & 2.43 & 
                & 1.53 & 1.33 &  1.62 & 2.29 & 2.37 & 
                & 1.06 & 1.21 & 1.17 & 1.25 & 1.16\\
    \multicolumn{2}{c}{\textbf{With Irrelevant X}} & \\
    ARROBART-lag & & 1.07 & 1.00 & 1.03 & 1.20 & 1.15  & & 1.49 &  \textbf{0.98} &  \textbf{0.93} & 1.24 & 1.25 \\ 
    ARROLinear-lag & & 1.10 & 1.42 & 1.84 & 2.40 & 2.48  & & 1.76 & 1.31 & 1.62 & 2.29 & 2.44  \\ 
    ROBART-lag & & 1.03 &  \textbf{0.99} &  \textbf{0.97} & 1.17 & 1.13 & & 1.62 & 1.03 & 0.98 & 1.34 & 1.17 \\ 
    ROLinear-lag & &  1.10 & 1.44 & 1.84 & 2.38 & 2.49 & & 1.73 & 1.29 & 1.62 & 2.27 & 2.41  \\
    \midrule 
    \multicolumn{2}{c}{\textbf{Partial ranking}} & \\
	ARROBART-lag    & & (\textbf{0.47}) & (\textbf{0.42}) & (\textbf{0.33}) & ({0.32}) & (\textbf{0.30}) &
                    & ({0.30}) & (\textbf{0.35}) & (\textbf{0.30}) & (\textbf{0.29}) & (\textbf{0.28}) & 
                    & ({0.45}) & ({0.47}) & ({0.46}) & ({0.48}) & ({0.49})  \\
    ARROLinear-lag  & & 1.09 & 1.09 & 1.42 & 1.46 & 1.60 &
                    & 1.21 & 1.34 & 1.62 & 1.63 & 1.63 & 
                    & 1.09 & 1.03 & 1.10 & 1.01 & 1.03 \\
    ROBART-lag      & & 1.04 & 1.01 & 1.05 & \textbf{0.94} & 1.01 &
                    & \textbf{0.96} & 1.03 & 1.07 & 1.01 & \textbf{1.00} & 
                    & \textbf{0.84} & \textbf{0.82} & \textbf{0.81} & \textbf{0.80} & \textbf{0.75} \\
    ROLinear-lag    & & 1.08 & 1.08 & 1.44 & 1.47 & 1.64 &
                    & 1.22 & 1.37 & 1.65 & 1.68 & 1.68 & 
                    & 1.10 & 1.03 & 1.11 & 1.01 & 1.04 \\
	\bottomrule
\end{tabular}
\begin{tablenotes}
\footnotesize
\item \!\!\!\! Notes: The ARROBART-lag row shows in brackets the average Kendall's tau distances between estimated and true ranking lists using the ARROBART-lag method. The remaining rows report the ratio of average Kendall's tau for the model in the row to average Kendall's tau for ARROBART-lag. Values greater than one mean that the benchmark outperforms the competitor. Bold denotes the best-performing model.
\end{tablenotes}
\end{threeparttable}
\end{adjustbox}
\end{table}

The results are shown in Table~\ref{tab:ARROSimu}, where the normalized Kendall tau distance is reported for the benchmark ARROBART model, whereas for the competitors the ratio of their performance over the ARROBART is shown.
Instead, Table~\ref{tab:ARROlagSimu} illustrates the outcome for all the models estimated using also the lag of the observed rank as a covariate, where we use the ARROBART-lag model as a benchmark.
For the benchmark models, the closer the normalized Kendall tau distance is to zero, the better the performance. Therefore, ratios of Kendall's tau distances for the competitors greater than 1 mean that the benchmark is outperforming them (and vice versa).
The results suggest several interesting insights.
First, the ARROBART (ARROBART-lag) benchmark outperforms the competitors for almost all the noise levels in scenarios 1 and 2. 
In scenarios 1 and 2, ARROBART-lag is almost uniformly better than all methods. 
The improvement over ARROLinear  and Kendall's tau for ARROBART are worse in the presence of a strong signal ($\sigma = 0.1$); this is possibly due to the sub-optimal performance of the BART specification in almost the absence of noise and is in line with previous results in the literature \citep{chipman2010bart}.
Second, in the third scenario, the ARROBARTX (Tab.~\ref{tab:ARROSimu}) and ROBART-lag (Tab.~\ref{tab:ARROlagSimu}) are the best models. This demonstrates the ability of the proposed methods to leverage exogenous covariates when they are relevant to the regression. 
Third, the results suggest that the nonlinear specifications, that is the ARROBART and ROBART models, almost always outperform their respective linear counterparts. 
Finally, the results from Tab.~\ref{tab:ARROSimu} and ~\ref{tab:ARROlagSimu} are confirmed even in the presence of partial ranking. 


\FloatBarrier

\section{Real Data Applications}
\label{sec:application}

The usefulness of our static data models is demonstrated by an application to three standard label ranking benchmark datasets \citep{de2018discovering}.
We include in these results a variation on ROBART, named SoftROBART, with soft and sparse splitting rules \citep{linero2018dart}.
Then, we compare static and dynamic model specifications on the two time series datasets presented in Section~\ref{sec:introduction}, about country-rankings on the economic complexity index and the 2022 NCAA rankings.

\subsection{Static Label Rankings}

The \texttt{sushi} dataset includes 5000 rankings of 10 items: shrimp, sea eel, tuna, squid, sea urchin, salmon roe, egg fatty tuna, tuna roll and cucumber roll \citep{kamishima2010survey}. The variables include age, gender, answer duration, a binary variable indicating if the respondent moved to another city, and categorical variables for current and past region and prefecture. The categorical variables were re-coded as dummy variables resulting in 128 predictors.

The \texttt{Top7Movies} dataset is a subset of the MovieLens dataset \citep{harper2016movielens} processed by \cite{de2018discovering}. The rankings were constructed from ratings of 7 movies submitted by 602 individuals. Predictors include gender, age, latitude and longitude of zipcode, city, state, and occupation. The categorical variables for city, state, and occupation were coded as dummy variables, resulting in a total of 446 predictors. This dataset contains many ties, and therefore allows for a comparison of methods in an application to partial ranking data.

The \texttt{german2009} dataset was constructed from 2009 electoral results for 5 political parties across 413 districts. The predictors include an urban dummy variable, state, region, latitude, longitude, electoral participation rates, and a number of demographic and economic variables. The state and region variables were re-coded as dummy variables, resulting in 51 predictors in total. These rankings contain a small number of ties.

It can be observed in Table \ref{tab:staticdataapplications} below that ROBART and SoftROBART produce much more accurate ranking forecasts for the \texttt{Top7Movies} data than the benchmark linear models, and there is a smaller improvement for the \texttt{sushi} data. For the  \texttt{german2009} data, standard ROBART does not  outperform linear models, whereas SoftROBART produces a comparable improvement to that achieved for the \texttt{Top7Movies} data. This suggests that there may be some smoothness or sparsity in the data generating process.

\begin{table}[t!h]
\centering
\begin{threeparttable}
\captionsetup{width=0.93\linewidth}
\caption{\label{tab:staticdataapplications} \small Kendall's tau distance between one-step ahead out-of-sample forecasted rankings and true rankings.}
\begin{tabular}{l | ccccc}
  \toprule
 & BARC1 & BARC2 & BARC3 & ROBART & SoftROBART \\ 
  \midrule
 Sushi &  (0.34) & 1.00 & 1.00 & \textbf{0.97} & \textbf{0.97} \\ 
Top-7 Movies & (0.44) & 1.07 & 1.03 & 0.79 & \textbf{0.78} \\ 
German Election 2009 & (0.06) & 0.94 & 0.95 & 1.09 & \textbf{0.76} \\
   \bottomrule
\end{tabular}
\begin{tablenotes}
\footnotesize
\item \!\!\!\! Notes: All methods were evaluated using 5-fold cross-validation. All results except the BARC1 results are relative to BARC1.
\end{tablenotes}
\end{threeparttable}
\end{table}

\subsection{Dynamic Label Ranking: Economic Complexity Index}

The ECI dataset \citep{alvo2014statistical,liu2019model} includes data for 80 countries from 1965 to 2015. The dependent variable is the country-year specific ranking in an index defined by the Observatory of Economic Complexity. The independent variables include the lag of the observed rank, GDP per capita, population, population growth, cereal production in tonnes, the dependency ratio, and growth in the dependency ratio.
These are not necessarily the most relevant variables for the index of interest. However, this exercise demonstrates the ability of the new methods to select relevant variables.

Table~\ref{tab:ECIktauavg} shows that standard ROBART and ARROBART do not improve on a static linear model. However, RODART and ARRODART, variations on ROBART and ARROBART for which a Dirichlet hyperprior distribution is specified for sparse splitting rules \citep{linero2018bayesian}, improve substantially on the linear models.
Specifically, the dynamic ARRODART model provides better forecasts than the static RODART model.
We refer to the Supplement for the data sources and further results.

\begin{table}[t!h]
\centering
\begin{threeparttable}
\captionsetup{width=0.93\linewidth}
\caption{\label{tab:ECIktauavg} \small Kendall's tau distance between one-step ahead out-of-sample forecasted rankings and true rankings.}
\begin{tabular}{cccccc}
\toprule
  ROLinear & ARROLinear & ROBART & RODART & ARROBART & ARRODART \\ 
\midrule
 1.00 & 1.41 & 0.99 & 0.93 & 0.99 & \textbf{0.81} \\ 
\bottomrule
\end{tabular}
\begin{tablenotes}
\footnotesize
\item \!\!\!\! Notes: All results except for the ROLinear result report the ratio of average Kendall's tau for the model in the row to average Kendall's tau for ROLinear. Values greater than one mean that ROLinear outperforms the competitor. Bold denotes the best-performing model.
\end{tablenotes}
\end{threeparttable}
\end{table}

\subsection{Panel of Label Rankings: NCAA Pollsters}

We apply the proposed methods to a real dataset of pollster-specific rankings of American football teams from the Associated Press weekly poll for the 2022 NCAA Division I Football season (see Fig.~\ref{fig:Econ_NCAA_data}). We refer to the Supplement for the details of the data sources and further results.
The dataset includes the ranking lists of $M=15$ pollsters across $T=16$ weeks. As many teams are not listed in the top 20 by at least one pollster in at least one week, we consider a subset of $N=7$ teams to obtain a time series of full rankings.

\begin{figure}[t!h]
\centering
\captionsetup{width=0.93\linewidth}
\includegraphics[scale = 0.8]{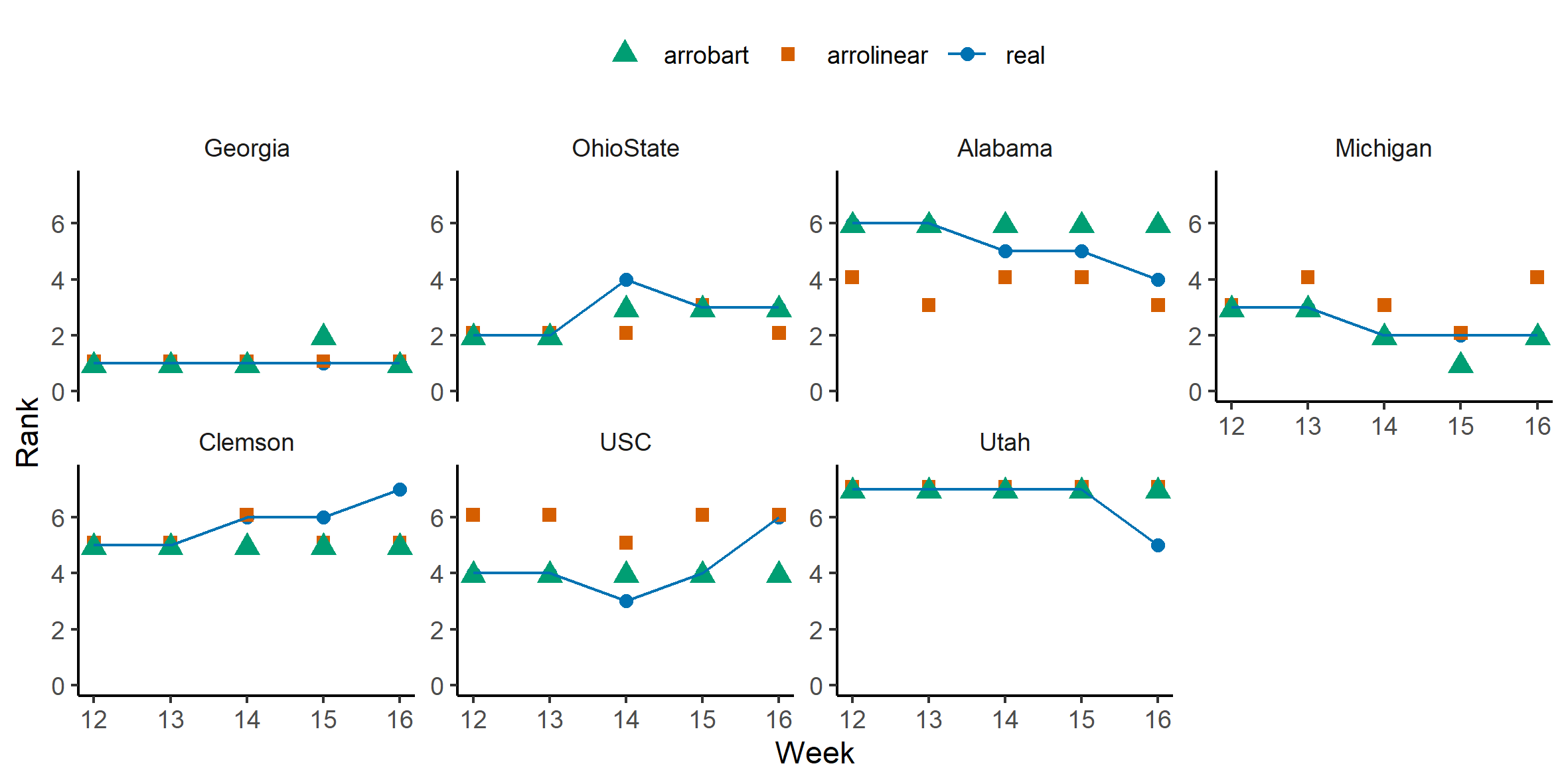}
\caption{AP Pollster Brian Howell: observed poll ranks (blue dots and line) and ranks forecasted using the ARROBART (green triangles) and ARROLinear (orange squares) models.}
\label{fig:realranks_onegraph}
\end{figure}

To inspect in more detail the advantages of the BART specification compared to the linear one, we report the point forecasts from the best ARROBART and ARROLinear models in Fig.~\ref{fig:realranks_onegraph} against the observed ranks, for $N=7$ teams and $h=5$ test periods. As the rankers are assumed to be independent, we present the results for a single pollster.
The ARROBART model is able to correctly predict the rank for several teams, with minor deviations for Alabama and Clemson. Conversely, the ARROLinear forecasts are generally more distant from the actual ranks. 
In particular, it can be observed that while both methods often produce similar forecasts, ARROBART produces more accurate forecasts for Alabama, USC, and Michigan in weeks 12 to 14.

The point forecast falls short of quantifying the uncertainty about the predicted value. To tackle this issue, we investigate in deeper detail the posterior predictive distribution of the best-performing model.
Figure~\ref{fig:postpredictive_michigan_ranker11} reports the posterior predictive distribution of the rank for Alabama and Michigan, whose observed ranks have changed multiple times during the testing sample (see Fig.~\ref{fig:realranks_onegraph}).
For both teams, the predictive density of the rank is unimodal and changes over time in terms of its mode and dispersion, suggesting a horizon-specific uncertainty of predictions. Furthermore, the comparison of Fig.~\ref{fig:realranks_onegraph} and Fig.~\ref{fig:postpredictive_michigan_ranker11}  highlights that in those periods where the point forecast differs from the observed one, the latter falls within the second-most frequent rank of the posterior predictive distribution.

\begin{figure}[t!h]
\centering
\setlength{\abovecaptionskip}{-2pt}
\captionsetup{width=0.93\linewidth}
\hspace*{-4.5ex}
\begin{tabular}{cc}
\includegraphics[scale=0.35]{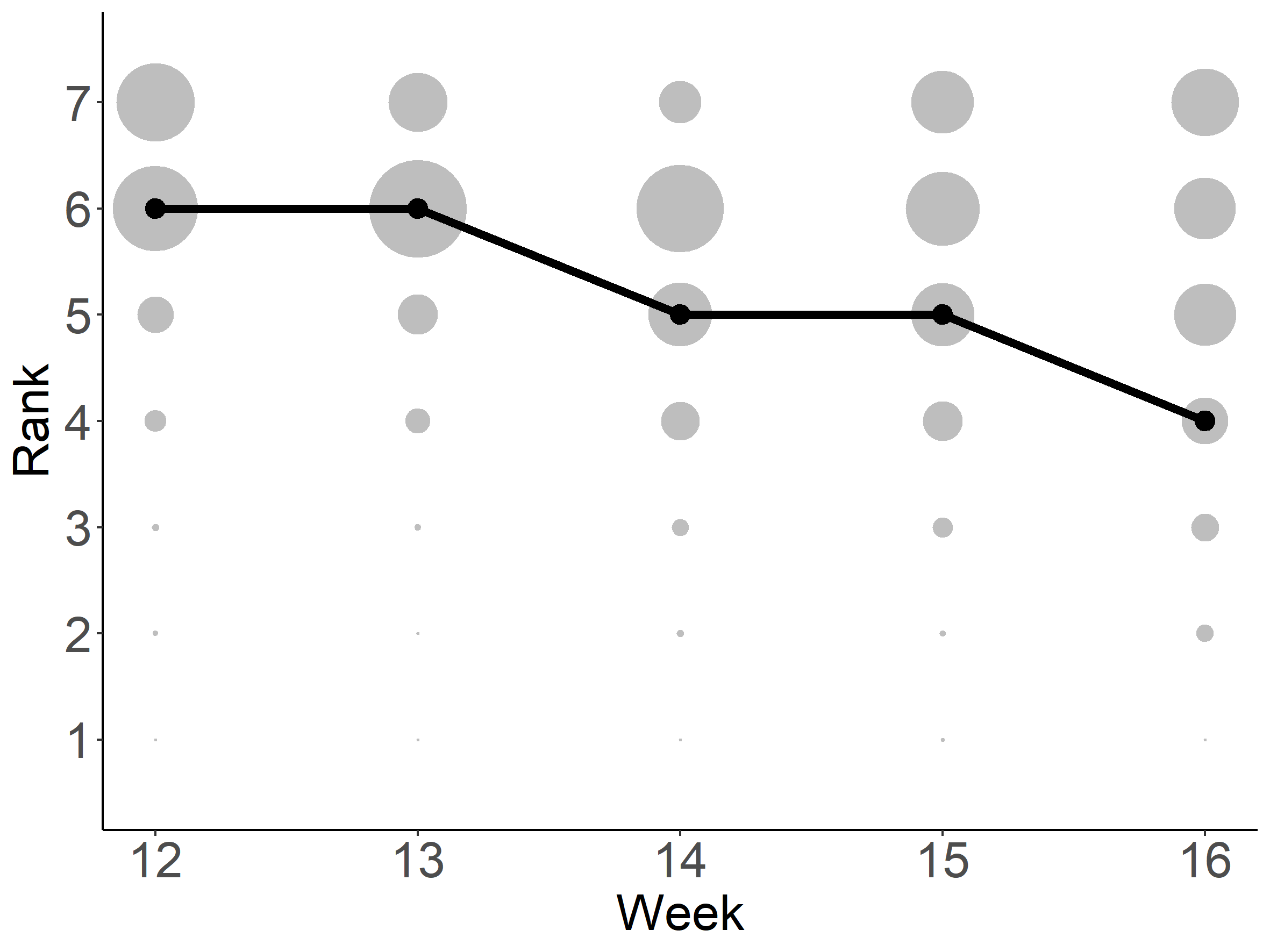} &
\includegraphics[scale=0.35]{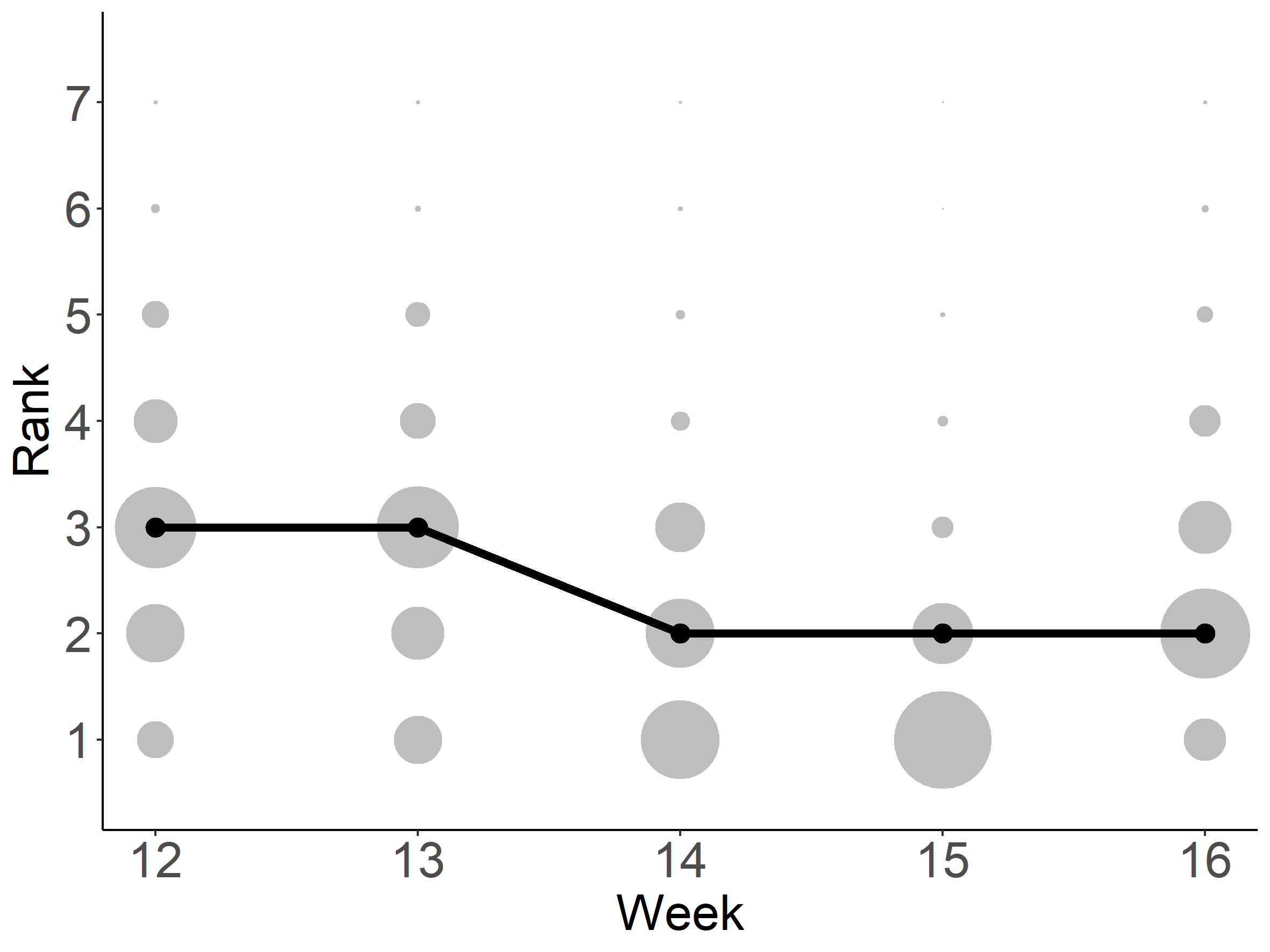}
\end{tabular}
\caption{ARROBART posterior predictive distribution (grey balls, with size proportional to the probability mass) against the observed ranks (black dots and line) for Alabama (left) and Michigan (right) ranks of AP Pollster Brian Howell.}
\label{fig:postpredictive_michigan_ranker11}
\end{figure}

\FloatBarrier


\section{Discussion}
\label{sec:conclusion}

This article introduces a novel class of nonparametric Thurstone models for static and dynamic regressions for rank-order data called ROBART and ARROBART.
The proposed frameworks use a BART specification for the regression function of the latent variable representation to allow for flexible, possibly nonlinear relationships between the latent scores and their lagged values as well as exogenous covariates.
The filtering and smoothing recursions for the dynamic model are derived, followed by a discussion {of Gibbs sampling methods}.
A simulation study illustrates that the proposed ROBART and ARROBART outperform the competitors in forecasting. Then, ROBART is applied to rankings of sushi, movies, and political parties, and ARROBART is applied to a time series of economic complexity and NCAA rankings.
The results highlight the better performance of the ROBART and ARROBART models compared to linear and static counterparts, suggesting the importance of accounting for temporal persistence and the usefulness of BART in accounting for the nonlinearities driving the evolution of the latent item-specific scores.

The proposed methods can be applied to a wide range of datasets.
Examples include the analysis of the main drivers of customers' preferences, university rankings, and political ratings, in addition to sports data.
{{Based on the results from both simulations and real data applications, we encourage the use of the ROBART model or its variants to investigate static (full or partial) ranking data. Conversely, the ARROBART model or its variants are applicable to dynamic ranking data. Moreover, if the data shows sparsity, (AR)RODART should be used, and in the absence of item-specific covariates, (AR)ROBART/(AR)RODART with item-specific trees shall be preferred.}}

The proposed ARROBART model can be generalized to $p > 1$ lags in the presence of more complex dynamics of the latent scores.
Other interesting possible extensions of the proposed methods include investigating the performance of more general BART specifications. For instance, the introduction of a varying-coefficient BART \citep{deshpande2020vcbart} or partially linear BART model for the latent scores \citep{zeldow2019semiparametric}. 
The proposed autoregressive BART methods can also be adapted to investigate binary, ordered categorical, unordered categorical data, or censored outcome data. These models involve an autoregressive Gaussian process for the latent scores and are therefore similar to the rank-order model described in this article. The samplers can be derived in a similar way to the sampler in this work, and have simpler constraints on the latent variables than rank-order outcome models.
Finally, it would be interesting to extend the proposed dynamic model to investigate the problem of aggregation of ranking lists \citep{deng2014bayesian,zhu2023partition} in a time series setting.

\section*{Data Availability Statement}

The data that support the findings of this study are openly available. These data were derived from the following resources: 

\noindent \url{https://collegepolltracker.com}

\noindent \url{https://www.teamrankings.com/}

\noindent \url{https://doi.org/10.17632/3mv94c8jpc.2}

\noindent \url{https://ourworldindata.org}

\section*{Conflict of Interest Statement}

The authors report there are no competing interests to declare. 

\bibliographystyle{chicago}
\bibliography{EoLuMa_refs}

\appendix

\noindent \textbf{Supplementary Appendix}

\bigskip 

This Supplement contains the proofs of the results in the main article in Section~\ref{sec:apdx_proofs}. Section~\ref{sec:apdx_MCMC} contains extra details for the MCMC algorithm, including sampling of the path of the latent states in the ARROBART and ARROlinear models, and the sampler for models of partial rankings. 
Section~\ref{sec:apdx_simualtion_details} contains descriptions of the models compared in the simulation study and real data application, and results for new simulation studies with partial rankings, ranker-specific covariates, and dynamic rankings forecasted by ranker-specific models. Section \ref{MCMCdiag_sec} contains MCMC diagnostic results. Section \ref{sec:apdx_NCAA_application} contains a dynamic data application to a panel of pollster-specific rankings of NCAA football teams.  
Section \ref{ECI_sec} provides additional results for the application to the time series of Economic Complexity Index rankings presented in the main text. Section \ref{sec:apdx_tempdepend_sec} summarizes the temporal dependence of predicted and true rankings from the dynamic simulation study presented in the main text.

\newpage
\spacingset{1.8} 

\renewcommand{\thesection}{S.\arabic{section}}
\renewcommand{\theequation}{S.\arabic{equation}}
\renewcommand{\thefigure}{S.\arabic{figure}}
\renewcommand{\thetable}{S.\arabic{table}}
\setcounter{section}{0}
\setcounter{table}{0}
\setcounter{figure}{0}
\setcounter{equation}{0}
\setcounter{page}{1}

\section{Additional results and proofs}  \label{sec:apdx_proofs}

This Section contains the proof of Theorem 1 in the main article and additional results.
First of all, let us recall some notations from Section 2 of the main article.

For a variable $x \in \mathcal{D} \subseteq \R$, the BART specification induces a partition $\{ C_k \}_{k=1}^K$ of the domain $\mathcal{D}$ (see Section~\ref{sec:apdx_proofs_S11}).
Notice that $\{ B_{\ell,s} \}_{\ell=1}^{b_s}$ is a partition of $\mathcal{D}$ for each tree $s \in \{1,\ldots,S \}$, where the $\ell$th leaf of the $s$th tree is associated to a scalar coefficient $\mu_{\ell,s} \in \R$.
Let us define a one-to-one map that associates each tuple $(j_1,\ldots,j_S) \in \bigtimes_{s=1}^S \{1,\ldots,b_s\}$ to an index $k \in \{ 1,\ldots,K \}$, where $K = \sum_{s=1}^S b_s$ is the total number of the tuples.
Then, for each tuple $(j_1,\ldots,j_S)$ and associated $k$ index, define $C_k = B_{j_1,1} \cap \ldots \cap B_{j_S,S}$ and let $\tilde{\mu}_k = \sum_{s=1}^S \mu_{j_s,s}$ be the associated coefficient.
Note that the intervals $C_k$ can be empty ($C_k = \emptyset$) or unbounded, that is $C_k = (-\infty, a)$ and $C_k = (b,\infty)$ for some $a,b \in \mathcal{D}$.

We denote by $\K = (k_1,\ldots,k_N)' \in \{1,\ldots,K\}^N$ a tuple or multi-index and define $\boldsymbol{\tilde{\mu}}_{\K} = (\tilde{\mu}_{k_1}, \ldots, \tilde{\mu}_{k_N})'$ as a vector of leaf parameters indexed by $\K$. Then, we use the shorthand notation
\begin{equation*}
\sum_{\K} a_{\K} = \sum_{k_1=1}^K \dots \sum_{k_N=1}^K a_{k_1,\ldots,k_N}.
\end{equation*}
In the presence of summations across different tuples, we use the notation $\K^{(m)} = (k_1^{(m)},\ldots,k_N^{(m)})' \in \{1,\ldots, K\}^N$, for $m \in \mathbb{N}$.
Given an $N$-dimensional vector of ranks at time $t$, $\tau_t$, we introduce $A_t$ as the set of vectors $\z\in\R^N$ whose elements follow the same ordering as in $\tau_t$, that is:
\begin{equation*}
A_t = \big\{ \z\in\R^N : z_{j,t} < z_{i,t} \iff \tau_{j,t} < \tau_{i,t}, \;\forall \, i,j=1,\ldots,N \big\}.
\end{equation*}
We use the notation $\C_{\K} = (\bigtimes_{i=1}^N C_{k_i})$ and $\C_{\K,t} = \C_{\K} \cap A_t$, for $t=1,\ldots,T$. 
For a vector $\tau$, we denote by $\tau_{1:t}$ the collection $\{\tau_1,\ldots,\tau_t\}$, and by $A_{1:t}$ the collection of sets $\{ A_1,\ldots,A_t \}$.

\subsection{Additional Lemmas} \label{sec:apdx_proofs_S11}

The following Lemmas show that the sets $\{ C_k \}_{k=1}^K$ form a partition of the space $\mathcal{D} = \R$, and that the intersection of a Cartesian product of $\{ C_k \}_{k=1}^K$ and a set induced by the observed ranks is convex.

\begin{lemma} \label{lemma:Partition}
The collection of sets $\{C_k\}_{k=1}^K$ is a partition of $\R$.
\end{lemma}
\smallskip

\noindent\textbf{Proof of Lemma~\ref{lemma:Partition}:}

\smallskip
\noindent Start by proving that the collection of sets $\{C_k\}_{k=1}^K$ consists of disjoint sets.
For any two sets $C_m$, $C_{\ell}$, with $m,\ell \in \{1,\ldots,K\}$ and $m\neq \ell$, by definition of $C_k$ there exists an $s \in \{1,\ldots,S\}$ such that $C_m$ and $C_{\ell}$ are constructed from distinct leafs of tree $s$.
This means that there exists $j_m, j_{\ell}$ such that $j_m \neq j_{\ell}$ and
\begin{align*}
C_m \subseteq B_{j_m, s} & \: \text{ by definition of } C_m, \\
C_\ell \subseteq B_{j_\ell, s} & \: \text{ by definition of } C_\ell.
\end{align*}
Also, as $j_m \neq j_\ell$, the sets $ B_{j_m, s}$ and $B_{j_\ell, s}$ are disjoint.
Since $C_m$ and $C_\ell$ are subsets of disjoint sets, they are disjoint.

Now, prove that the collection of sets $\{C_k\}_{k=1}^K$ covers the space $\R$.
First, note that $\cup_{k=1}^K C_k \subseteq \R$ by definition.
It is necessary to show $\cup_{k=1}^K C_k \supseteq \R$.
Assume that $z$ is the only splitting variable.
Since each tree $s \in \{1,\ldots,S\}$ induces a partition of $\R$, for each $z \in \R$ and each $s \in \{1,\ldots,S\}$, there exists an index $j_s$ such that $z \in B_{j_s}$.
Therefore, there exists $j_1,\ldots,j_S$ such that $z \in B_{j_1} \cap \ldots \cap B_{j_S} \in \{ C_k \}_{k=1}^K$.

\qed 

\vspace*{1.5ex}

\noindent The set $\C_{\K,t} = \C_{\K} \cap A_t$ is formed by the intersection of $A_t$, which imposes an ordering on $\z$, and $\C_{\K} = (\bigtimes_{i=1}^K C_{k_i})$, which is the Cartesian product of $N$ intervals in $(a,b) \subset \R$ (where $a,b$ may be $\pm\infty$).
Notice that all these sets are convex and open.

It is important to remark that $\C_{\K,t}$ can be expressed by means of a set of linear inequality constraints of the type $A \z -\mathbf{b} >0$.

\begin{lemma}\label{lemma:convex}
For any $\K = (k_1,\ldots,k_N) \in \{1,\ldots,K\}^N$ and any time $t \in \{1,\ldots,T\}$, the set $\C_{\K,t} = ( \bigtimes_{i=1}^N C_{k_i} ) \cap A_t$ is convex.
\end{lemma}
\smallskip

\noindent\textbf{Proof of Lemma~\ref{lemma:convex}:}

\smallskip
\noindent By definition, $\{ C_k \}_{k=1}^K$ is a partition of $\R$, where each $C_k$ is segment, hence a convex set.
Therefore the Cartesian product $\C_{\K} = (\bigtimes_{i=1}^N C_{k_i})$ is convex for each tuple $\K = (k_1,\ldots,k_N) \in \{1,\ldots,K\}^N$.

Since the intersection of convex sets is convex, it remains to show that $A_t$ is convex.
Recall that $A_t$ is the subspace of $\R^N$ constructed by restricting each coordinate value to agree with the observed rankings $\tau_t$. In words, if $\tau_{t,1} > \tau_{t,2}$, then $A_t$ imposes a restriction $x_1 > x_2$ on $\R^2$, that is the second coordinate to be smaller than the first one.
Therefore, as $A_t$ is defined by hyperplanes in $\R^N$, which are all convex sets, it follows that $A_t$ is convex.

\qed

\subsection{Proof of Theorem 1}

Following the previous Lemmas, the BART sum-of-trees regression function can then be rewritten using the partition $\{ C_k \}_k$ and the associated coefficients $\{ \tilde{\mu}_k \}_{k=1}^K$ as follows:
\begin{align}
\sum_{s=1}^S g_s(x_i) = \sum_{k=1}^K \tilde{\mu}_k \I(x_i \in C_k).
\label{eq:sumG}
\end{align}
We are now ready to prove Theorem 1.

\vspace{1.6ex}

\noindent\textbf{Proof of Theorem 1}

\smallskip

\noindent \underline{\textit{Filtering distribution.}}
The proof is made by induction. Therefore, it starts by deriving the filtering distribution for the first two time periods, then the result is extended for a generic $t \geq 1$.
In the following, denote by $\tau_{1:t} = (\tau_1',\ldots,\tau_t')'$ the collection of vectors $\tau_{\tilde{t}}$ for each $\tilde{t}=1,\ldots,t$. 

\noindent For $t=1$, recall that
\begin{align*}
p(\z_1 | \tau_1, \underline{\z}_0) & \propto p(\tau_1 | \z_1, \underline{\z}_0) p(\z_1 | \underline{\z}_0) = p(\tau_1 | \z_1) p(\z_1 | \underline{\z}_0) \\
 & \propto \I(\z_1 \in A_1) \mathcal{N}_N\Big(\z_1 \big|  \sum_s g_s(\underline{\z}_0), I_N \Big),
\end{align*}
where $A_1$ is a set identified through the order of the elements of the observed rank vector $\tau_1$, $\z_t$ denotes the $N$-dimensional vector of latent variables in period $t$, and $\sum_s g_s(\underline{\z}_0)$ is an $N$-dimensional vector of leaf mean values corresponding to the regions associated with each element of $\underline{\z}_0$.

\noindent After integrating out $\underline{\z}_0$, the density of the latent vector $\z_1$ given the observed ranks at the same time, $\tau_1$, is proportional to
{
$$p(\z_1 | \tau_1)  \propto p(\tau_1 | \z_1) p(\z_1 ) = p(\tau_1 | \z_1) \int p(\z_1, \z_0) d \z_0 = p(\tau_1 | \z_1) \int p(\z_1 | \z_0) p(\z_0) d \z_0 $$
}
$$
p(\z_1 | \tau_1) \propto \delta_{rank(\z_1)}(\tau_1)  \int_{\R^N} \mathcal{N}_N\Big(\z_1 \big| \sum_s g_s(\underline{\z}_0), I_N \Big) \mathcal{N}_N\left(\underline{\z}_0 | \underline{\z}_{prior}, I_N \right) \: \mathrm{d}\underline{\z}_0,
$$
where ${\sum_s g_s(\underline{\z}_0 ) }$ is a vector of values determined by the terminal nodes of the BART specification associated with each element of $\underline{\z}_0$.
By Eq.~\eqref{eq:sumG}, one gets
\begin{align*}
p(\z_1 | \tau_1) & \propto \mathbb{I}(\z_1 \in A_1) \\
 & \quad \times \bigint_{\R^N} \mathcal{N}_N\left(\z_1 \Bigg| 
\begin{bmatrix}
    \sum_{k_1=1}^K  \tilde{\mu}_{k_1} \mathbb{I}(\underline{\z}_{0,1} \in C_{k_1}) \\
    \sum_{k_2=1}^K  \tilde{\mu}_{k_2} \mathbb{I}(\underline{\z}_{0,2} \in C_{k_2}) \\
    \vdots \\
    \sum_{k_N=1}^K  \tilde{\mu}_{k_N} \mathbb{I}(\underline{\z}_{0,N} \in C_{k_N}) 
\end{bmatrix},
I_N \right) \mathcal{N}_N\left(\underline{\z}_0 | \underline{\z}_{prior}, I_N \right) \: \mathrm{d}\underline{\z}_0.
\end{align*}
As $\{ C_k \}_{k=1}^K$ forms a partition of $\R$, one can define a partition of $\R^N$ by collecting the regions $\C_{\K} = (\bigtimes_{i=1}^N C_{k_i})$, for each tuple $\K = (k_1,\ldots,k_N) \in \{1,\ldots,K\}^N = \mathcal{K}$. In particular, these regions partition $\R^N$ into hypercuboids, with the boundaries given by a grid that is symmetrical in all $N$ dimensions.
This partition allows to separate out the integral over $\R^N$ into a sum of integrals, one on each region $\C_{\K} \in \mathcal{K}$.

Therefore, exploiting the partition $\{ \C_{\K} \}_{\K}$ and the indicator functions appearing in the conditional mean of the first normal density, the integral can be simplified as
\begin{align*}
    & p(\z_1 | \tau_1) \propto \mathbb{I}(\z_1 \in A_1) \\
    & \times \sum_{\K} \bigint_{\C_{\K}} \!\!\! \mathcal{N}_N\left(\z_1 \Bigg| 
    \begin{bmatrix}
    \sum_{k_1=1}^K  \tilde{\mu}_{k_1} \mathbb{I}(\underline{\z}_{0,1} \in C_{k_1}) \\
    \sum_{k_2=1}^K  \tilde{\mu}_{k_2} \mathbb{I}(\underline{\z}_{0,2} \in C_{k_2}) \\
    \vdots \\
    \sum_{k_N=1}^K  \tilde{\mu}_{k_N} \mathbb{I}(\underline{\z}_{0,N} \in C_{k_N}) 
    \end{bmatrix},
    I_N \right) \mathcal{N}_N(\underline{\z}_0 | \underline{\z}_{prior}, I_N) \: \mathrm{d}\underline{\z}_0 \\
    & \propto \mathbb{I}(\z_1 \in A_1)  \sum_{\K} \int_{\C_{\K}} \mathcal{N}_N(\z_1 \big| \boldsymbol{\tilde{\mu}}_{\K}, I_N)  \mathcal{N}_N(\underline{\z}_0 | \underline{\z}_{prior}, I_N) \: \mathrm{d}\underline{\z}_0 \\
    & \propto \mathbb{I}(\z_1 \in A_1)  \sum_{\K} \mathcal{N}_N(\z_1 | \boldsymbol{\tilde{\mu}}_{\K}, I_N)  \int_{\C_{\K}}  \mathcal{N}_N(\underline{\z}_0 | \underline{\z}_{prior}, I_N) \: \mathrm{d}\underline{\z}_0,
\end{align*}
where $\boldsymbol{\tilde{\mu}}_{\K} = (\tilde{\mu}_{k_1}, \tilde{\mu}_{k_2}, \ldots, \tilde{\mu}_{k_N})'$ is a vector of parameters indexed by the tuple $\K = (k_1,\ldots,k_N)$.
To further simplify the notation, let us define $q_{\K, 1} =  \int_{\C_{\K}} \mathcal{N}_N(\underline{\z}_0 | \underline{\z}_{prior}, I_N) \mathrm{d}\underline{\z}_0$.
Then, the density of the latent outcomes at time $t=1$, given the ranks observed at the same time, is proportional to
\begin{align*}
    p(\z_1 | \tau_1) \propto \mathbb{I}(\z_1 \in A_1) \sum_{\K} \mathcal{N}_N(\z_1 | \boldsymbol{\tilde{\mu}}_{\K}, I_N)  q_{\K, 1}.
\end{align*}
Notice that $q_{\K,1} \in (0,1)$ for each tuple $\K$ and $\sum_{\K} q_{\K,1} = 1$.

Consider now the second time period ($t=2$). The filtered distribution has the following form:
\begin{align*}
    p(\z_2 | \tau_{1:2}) & \propto \int_{\R^N} p(\tau_2| \z_2) p(\z_2 | \z_1) p(\z_1 | \tau_1) \: \mathrm{d}\z_1 \\
    & \propto \mathbb{I}(\z_2 \in A_2) \int_{\R^N} p(\z_2 | \z_1) p(\z_1 | \tau_1) \: \mathrm{d}\z_1 \\
    & \propto \mathbb{I}(\z_2 \in A_2) \bigint_{\R^N} \mathcal{N}_N\left(\z_2 \Bigg| 
    \begin{bmatrix}
    \sum_{k_1=1}^K  \tilde{\mu}_{k_1} \mathbb{I}(\z_{1,1} \in C_{k_1}) \\
    \vdots \\
    \sum_{k_N=1}^K  \tilde{\mu}_{k_N} \mathbb{I}(\z_{1,N} \in C_{k_N})
    \end{bmatrix},
    I_N \right) \mathbb{I}(\z_1 \in A_1) \\
    & \quad \times \sum_{\mathbf{m}} \mathcal{N}_N(\z_1 |  \boldsymbol{\tilde{\mu}}_{\mathbf{m}}, I_N)  q_{\mathbf{m}, 1} \: \mathrm{d}\z_1,
\end{align*}
where $\mathbf{m} = (m_1,\ldots,m_N) \in \{1,\ldots,K\}^N$.
Separating the integral over regions $\C_{\K}$, for $\K \in \{1,\ldots,K\}^N$, one gets
\begin{align*}
    p(\z_2 | \tau_{1:2}) & \propto \mathbb{I}(\z_2 \in A_2) \sum_{\K} \int_{\C_{\K}} \mathcal{N}_N(\z_2 | \boldsymbol{\tilde{\mu}}_{\K}, I_N) \mathbb{I}(\z_1 \in A_1)\sum_{\mathbf{m}} \mathcal{N}_N(\z_1 | \boldsymbol{\tilde{\mu}}_{\mathbf{m}}, I_N)  q_{\mathbf{m}, 1} \: \mathrm{d}\z_1\\
    & \propto \mathbb{I}(\z_2 \in A_2) \sum_{\K} \mathcal{N}_N(\z_2 | \boldsymbol{\tilde{\mu}}_{\K}, I_N) \int_{\C_{\K}} \mathbb{I}(\z_1 \in A_1) 
    \sum_{\mathbf{m}}  \mathcal{N}_N(\z_1 | \boldsymbol{\tilde{\mu}}_{\mathbf{m}}, I_N)  q_{\mathbf{m}, 1} \: \mathrm{d}\z_1.
\end{align*}
Restricting the domain of integration to the space where the integrand is non-zero yields
\begin{align*}
    p(\z_2 | \tau_{1:2}) & \propto \mathbb{I}(\z_2 \in A_2) \sum_{\K} \mathcal{N}_N(\z_2 | \boldsymbol{\tilde{\mu}}_{\K}, I_N) 
    \int_{\C_{\K,1}} \sum_{\mathbf{m}}  \mathcal{N}_N(\z_1 | \boldsymbol{\tilde{\mu}}_{\mathbf{m}}, I_N)  q_{\mathbf{m}, 1} \: \mathrm{d}\z_1,
\end{align*}
where $\C_{\K,1} = \C_{\K} \cap A_1$.
The integrals and the summations can be switched, leading to
\begin{align*}
    p(\z_2 | \tau_{1:2}) & \propto \mathbb{I}(\z_2 \in A_2) \sum_{\K} \mathcal{N}_N(\z_2 | \boldsymbol{\tilde{\mu}}_{\K}, I_N) 
    \sum_{\mathbf{m}} q_{\mathbf{m}, 1}  \int_{\C_{\K,1}} \mathcal{N}_N(\z_1 | \boldsymbol{\tilde{\mu}}_{\mathbf{m}}, I_N) \: \mathrm{d}\z_1.
\end{align*}
Let us use the following notation:
\begin{equation*}
    q_{\K, 2}(A_1) = \sum_{\mathbf{m}} q_{\mathbf{m}, 1}  \int_{\C_{\K,1}}  \mathcal{N}_N(\z_1 | \boldsymbol{\tilde{\mu}}_{\mathbf{m}}, I_N) \: \mathrm{d}\z_1.
\end{equation*}
More generally, for each $t>1$, we denote by $A_{1:t-1}$ the collection $\{ A_1,\ldots,A_{t-1} \}$ and
\begin{align}
    q_{\K, t}(A_{1:t-1}) & = \sum_{\mathbf{m}} q_{\mathbf{m}, t-1}(A_{1:t-2})  \int_{\C_{\K,t-1}} \mathcal{N}_N(\z_{t-1} | \boldsymbol{\tilde{\mu}}_{\mathbf{m}}, I_N) \: \mathrm{d}\z_{t-1}
\label{eq_qkt}
\end{align}
Therefore, the filtering distribution for the latent vector of outcomes at time $t=2$ is proportional to
\begin{align*}
    p(\z_2 | \tau_{1:2}) \propto \mathbb{I}(\z_2 \in A_2) \sum_{\K} \mathcal{N}_N(\z_2 | \boldsymbol{\tilde{\mu}}_{\K}, I_N) q_{\K, 2}(A_1).
\end{align*}

Then, having proved the result for $t=1$ and $t=2$, the proof is completed by assuming that the result holds for a generic $t-1$ and proving that it holds for $t$.
The filtering distribution at time $t$ is given by:
\begin{align*}
    p(\z_t | \tau_{1:t}) & \propto \int_{\R^N} p(\tau_t| \z_t) p(\z_t | \z_{t-1}) p(\z_{t-1} | \tau_{1:t-1}) \: \mathrm{d}\z_{t-1} \\
    & \propto \mathbb{I}(\z_t \in A_t) \int_{\R^N} p(\z_t | \z_{t-1}) p(\z_{t-1} | \tau_{1:t-1}) \: \mathrm{d}\z_{t-1} \\ 
    & \propto \mathbb{I}(\z_t \in A_t) \bigint_{\R^N} \mathcal{N}_N\left(\z_t \Bigg|
    \begin{bmatrix}
    \sum_{k_1=1}^K \tilde{\mu}_{k_1} \mathbb{I}(\z_{t-1,1} \in C_{k_1}) \\
    \vdots \\
    \sum_{k_N=1}^K \tilde{\mu}_{k_N} \mathbb{I}(\z_{t-1,N} \in C_{k_N})
    \end{bmatrix},
    I_N \right) \mathbb{I}(\z_{t-1} \in A_{t-1}) \\ 
    & \quad \times \sum_{\mathbf{m}} \mathcal{N}_N(\z_{t-1} | \boldsymbol{\tilde{\mu}}_{\mathbf{m}}, I_N)  q_{\mathbf{m}, t-1}(A_{1:t-2}) \: \mathrm{d}\z_{t-1},
\end{align*}
where $q_{\textbf{m}, t-1}(A_{1:t-2})$ is defined in Eq.~\eqref{eq_qkt}.

\noindent Separating the integral over regions $\C_{\K}$, for $\K \in \{1,\ldots,K\}^N$, one gets
\begin{align*}
    p(\z_t | \tau_{1:t}) & \propto \mathbb{I}(\z_t \in A_t) \sum_{\K} \int_{\C_{\K}} \mathcal{N}_N(\z_t | \boldsymbol{\tilde{\mu}}_{\K}, I_N) \mathbb{I}(\z_{t-1} \in A_{t-1}) \\
    & \quad \times \sum_{\mathbf{m}} \mathcal{N}_N(\z_{t-1} | \boldsymbol{\tilde{\mu}}_{\mathbf{m}}, I_N)  q_{\mathbf{m}, t-1}(A_{1:t-2}) \: \mathrm{d}\z_{t-1} \\
    & \propto \mathbb{I}(\z_t \in A_t) \sum_{\K} \mathcal{N}_N(\z_t | \boldsymbol{\tilde{\mu}}_{\K}, I_N) \int_{\C_{\K}} \mathbb{I}(\z_{t-1} \in A_{t-1}) \\
    & \quad \times \sum_{\mathbf{m}} \mathcal{N}_N(\z_{t-1} | \boldsymbol{\tilde{\mu}}_{\mathbf{m}}, I_N)  q_{\mathbf{m}, t-1}(A_{1:t-2}) \: \mathrm{d}\z_{t-1}.
\end{align*}
Restricting the range of integration to the region where the integrand is non-zero yields
\begin{align*}
    p(\z_t | \tau_{1:t}) & \propto \mathbb{I}(\z_t \in A_t) \sum_{\K} \mathcal{N}_N(\z_t | \boldsymbol{\tilde{\mu}}_{\K}, I_N)  \int_{\C_{\K,t-1}} \sum_{\mathbf{m}} \mathcal{N}_N(\z_{t-1} |  \boldsymbol{\tilde{\mu}}_{\mathbf{m}}, I_N)  q_{\mathbf{m}, t-1}(A_{1:t-2}) \: \mathrm{d}\z_{t-1},
\end{align*}
where $\C_{\K,t-1} = \C_{\K} \cap A_{t-1}$.
By swapping the summations and the integrals one gets
\begin{align*}
    p(\z_t | \tau_{1:t}) & \propto \mathbb{I}(\z_t \in A_t) \sum_{\K} \mathcal{N}_N(\z_t | \boldsymbol{\tilde{\mu}}_{\K}, I_N) \\
    & \quad \times \sum_{\mathbf{m}} q_{\mathbf{m}, t-1}(A_{1:t-2}) \int_{\C_{\K,t-1}}  \mathcal{N}_N(\z_{t-1} | \boldsymbol{\tilde{\mu}}_{\mathbf{m}}, I_N) \: \mathrm{d}\z_{t-1} \\
    & \propto \mathbb{I}(\z_t \in A_t) \sum_{\K} \mathcal{N}_N(\z_t | \boldsymbol{\tilde{\mu}}_{\K}, I_N) q_{\K, t}(A_{1:t-1}),
\end{align*}
where
\begin{equation*}
    q_{\K, t}(A_{1:t-1}) = \sum_{\mathbf{m}} q_{\mathbf{m}, t-1}(A_{1:t-2}) \int_{\C_{\K,t-1}} \mathcal{N}_N(\z_{t-1} |   \boldsymbol{\tilde{\mu}}_{\mathbf{m}}, I_N) \: \mathrm{d}\z_{t-1} =  \sum_{\mathbf{m}} q_{\mathbf{m}, t-1}(A_{1:t-2})  r_{ \K,  \mathbf{m}}^{(t-1)}.
\end{equation*}
This completes the first part of the proof.

\vspace*{2.5ex}

\noindent \underline{\textit{Predictive distribution.}}
The one-step-ahead predictive distribution is given by
\begin{align}
p(\z_{t+1} | \tau_{1:t}) = \int_{\R^N} p(\z_{t+1}, \z_t | \tau_{1:t}) \: \mathrm{d}\z_t = \int_{\R^N} p(\z_{t+1} | \z_t) p(\z_t | \tau_{1:t}) \: \mathrm{d}\z_t,
\label{eq:supp_Pred}
\end{align}
where $p(\z_{t+1} |\z_t)$ is the transition density and $p(\z_t | \tau_{1:t})$ is the filtered distribution.
Substituting the latter in Eq.~\eqref{eq:supp_Pred}, one gets
\begin{align*}
p(\z_{t+1} | \tau_{1:t}) & \propto \int_{\R^N}  \mathcal{N}_N\left(\z_{t+1} \Bigg|
\begin{bmatrix}
    \sum_{m_1=1}^K \tilde{\mu}_{m_1} \mathbb{I}(\z_{t,1} \in C_{m_1}) \\
    \vdots \\
    \sum_{m_N=1}^K \tilde{\mu}_{m_N} \mathbb{I}(\z_{t,N} \in C_{m_N}) 
\end{bmatrix},
I_N \right) \\
 & \quad \times \mathbb{I}(\z_t \in A_t) \sum_{\K} \mathcal{N}_N\left(\z_t | \boldsymbol{\tilde{\mu}}_{\K}, I_N \right) q_{\K, t}(A_{1:t-1}) \: \mathrm{d}\z_t.
\end{align*}
Then, restricting the integral to the space where the integrand is non-zero and swapping the integral and summations yields
\begin{align*}
p(\z_{t+1} | \tau_{1:t}) & \propto \sum_{\K} q_{\K, t}(A_{1:t-1}) \\
 & \times \bigint_{A_t} \mathcal{N}_N\left(\z_{t+1} \Bigg|
\begin{bmatrix}
    \sum_{m_1=1}^K  \tilde{\mu}_{m_1} \mathbb{I}(\z_{t,1} \in C_{m_1}) \\
    \vdots \\
    \sum_{m_N=1}^K  \tilde{\mu}_{m_N} \mathbb{I}(\z_{t,N} \in C_{m_N}) 
\end{bmatrix},
I_N \right) \mathcal{N}_N(\z_t | \boldsymbol{\tilde{\mu}}_{\K}, I_N) \: \mathrm{d}\z_t.
\end{align*}
By separating the integral over regions $(\bigtimes_{i=1}^N C_{m_i})$, for $(m_1,\ldots,m_N ) \in \{1,\ldots,K\}$ and reordering terms, one gets
\begin{align*}
p(\z_{t+1} | \tau_{1:t}) & \propto \sum_{\K} q_{\K, t}(A_{1:t-1}) 
 \sum_{\mathbf{m}} \int_{\C_{\mathbf{m},t}} \mathcal{N}_N(\z_{t+1} | \boldsymbol{\tilde{\mu}}_{\mathbf{m}}, I_N) \mathcal{N}_N(\z_t |\boldsymbol{\tilde{\mu}}_{\K}, I_N) \: \mathrm{d}\z_t \\
 & \propto \sum_{\mathbf{m}} \mathcal{N}_N(\z_{t+1} | \boldsymbol{\tilde{\mu}}_{\mathbf{m}}, I_N) 
 \sum_{\K} q_{\K, t}(A_{1:t-1})   \int_{\C_{\mathbf{m},t}}  \mathcal{N}_N(\z_t | \boldsymbol{\tilde{\mu}}_{\K}, I_N) \: \mathrm{d}\z_t,
\end{align*}
where $\C_{\mathbf{m},t} = \C_{\mathbf{m}} \cap A_t$.
Concluding, the one-step-ahead predictive distribution is proportional to
\begin{align*}
p(\z_{t+1} | \tau_{1:t}) \propto \sum_{\mathbf{m}}  \mathcal{N}_N(\z_{t+1} | \boldsymbol{\tilde{\mu}}_{\mathbf{m}}, I_N) q_{\mathbf{m},t+1}(A_{1:t}),
\end{align*}
where
\begin{equation*}
q_{\mathbf{m},t+1}(A_{1:t}) = \sum_{\K} q_{\K, t}(A_{1:t-1})  \int_{\C_{\mathbf{m},t}}  \mathcal{N}_N(\z_t | \boldsymbol{\tilde{\mu}}_{\K}, I_N) \: \mathrm{d}\z_t = \sum_{\K} q_{\K, t}(A_{1:t-1})  r_{\mathbf{m}, \K}^{(t)}.
\end{equation*}

The predictive distribution just provided is an approximation, but we are also able to compute the exact one-step-ahead predictive distribution. First of all, the normalising constant is calculated as follows:
\begin{align*}
    & \int \sum_{\mathbf{m}}  \mathcal{N}_N(\z_{t+1} | \boldsymbol{\tilde{\mu}}_{\mathbf{m}}, I_N) q_{\mathbf{m},t+1}(A_{1:t}) \mathrm{d}\z_{t+1}  \\
    &= \sum_{\mathbf{m}}  q_{\mathbf{m},t+1}(A_{1:t}) \int  \mathcal{N}_N(\z_{t+1} | \boldsymbol{\tilde{\mu}}_{\mathbf{m}}, I_N) \mathrm{d}\z_{t+1} = \sum_{\mathbf{m}}  q_{\mathbf{m},t+1}(A_{1:t}). 
\end{align*}
This normalising constant can also be expressed as follows
\begin{align*}
    &\sum_{\mathbf{m}} q_{\mathbf{m},t+1}(A_{1:t}) = \sum_{\mathbf{m}}  \sum_{\K} q_{\K, t}(A_{1:t-1})  \int_{\C_{\mathbf{m},t}}  \mathcal{N}_N(\z_t | \boldsymbol{\tilde{\mu}}_{\K}, I_N) \: \mathrm{d}\z_t \\
&= \sum_{\K} q_{\K, t}(A_{1:t-1}) \sum_{\mathbf{m}} \int_{\C_{\mathbf{m},t}}  \mathcal{N}_N(\z_t | \boldsymbol{\tilde{\mu}}_{\K}, I_N) \: \mathrm{d}\z_t  \\
&= \sum_{\K} q_{\K, t}(A_{1:t-1}) \int_{A_{t}}  \mathcal{N}_N(\z_t | \boldsymbol{\tilde{\mu}}_{\K}, I_N) \: \mathrm{d}\z_t = \sum_{\K} q_{\K, t}(A_{1:t-1}) \xi_{\K, t},
\end{align*}
where $\xi_{\K, t} = \int_{A_{t}}  \mathcal{N}_N(\z_t | \boldsymbol{\tilde{\mu}}_{\K}, I_N) \: \mathrm{d}\z_t$.

Thus the one-step-ahead predictive distribution is equal to
\begin{align*}
    p(\z_{t+1} | \tau_{1:t}) &= \int_{\R^N}  \mathcal{N}_N\left(\z_{t+1} \Bigg|
    \begin{bmatrix}
    \sum_{m_1=1}^K \tilde{\mu}_{m_1} \mathbb{I}(\z_{t,1} \in C_{m_1}) \\
    \vdots \\
    \sum_{m_N=1}^K \tilde{\mu}_{m_N} \mathbb{I}(\z_{t,N} \in C_{m_N}) 
    \end{bmatrix},
    I_N \right) \\
    & \quad \times \frac{1}{\sum_{\boldsymbol{\ell}}  q_{\boldsymbol{\ell}, t}(A_{1:t-1}) \xi_{\boldsymbol{\ell},\mathbf{t}} } \mathbb{I}(\z_t \in A_t) \sum_{\K} \mathcal{N}_N\left(\z_t | \boldsymbol{\tilde{\mu}}_{\K}, I_N \right) q_{\K, t}(A_{1:t-1}) \: \mathrm{d}\z_t.
\end{align*}
Then, restricting the integral to the space where the integrand is non-zero and swapping the integral and summations yields
\begin{align*}
    p(\z_{t+1} | \tau_{1:t}) & = \frac{1}{\sum_{\boldsymbol{\ell}}  q_{\boldsymbol{\ell}, t}(A_{1:t-1}) \xi_{\boldsymbol{\ell},\mathbf{t}} } \sum_{\K} q_{\K, t}(A_{1:t-1}) \\
    & \times \bigint_{A_t} \mathcal{N}_N\left(\z_{t+1} \Bigg| 
    \begin{bmatrix}
    \sum_{m_1=1}^K  \tilde{\mu}_{m_1} \mathbb{I}(\z_{t,1} \in C_{m_1}) \\
    \vdots \\
    \sum_{m_N=1}^K  \tilde{\mu}_{m_N} \mathbb{I}(\z_{t,N} \in C_{m_N}) 
    \end{bmatrix}, 
    I_N \right) \mathcal{N}_N(\z_t | \boldsymbol{\tilde{\mu}}_{\K}, I_N) \: \mathrm{d}\z_t.
\end{align*}
By separating the integral over regions $(\bigtimes_{i=1}^N C_{m_i})$, for $(m_1,\ldots,m_N ) \in \{1,\ldots,K\}$ and reordering terms, one gets
\begin{align*}
    p(\z_{t+1} | \tau_{1:t}) & = \frac{1}{\sum_{\boldsymbol{\ell}}  q_{\boldsymbol{\ell}, t}(A_{1:t-1}) \xi_{\boldsymbol{\ell},\mathbf{t}} }  \sum_{\K} q_{\K, t}(A_{1:t-1}) 
    \sum_{\mathbf{m}} \int_{\C_{\mathbf{m},t}} \mathcal{N}_N(\z_{t+1} | \boldsymbol{\tilde{\mu}}_{\mathbf{m}}, I_N) \mathcal{N}_N(\z_t |\boldsymbol{\tilde{\mu}}_{\K}, I_N) \: \mathrm{d}\z_t \\
    & = \frac{1}{\sum_{\boldsymbol{\ell}}  q_{\boldsymbol{\ell}, t}(A_{1:t-1}) \xi_{\boldsymbol{\ell},\mathbf{t}} } \sum_{\mathbf{m}} \mathcal{N}_N(\z_{t+1} | \boldsymbol{\tilde{\mu}}_{\mathbf{m}}, I_N) 
    \sum_{\K} q_{\K, t}(A_{1:t-1})   \int_{\C_{\mathbf{m},t}}  \mathcal{N}_N(\z_t | \boldsymbol{\tilde{\mu}}_{\K}, I_N) \: \mathrm{d}\z_t,
\end{align*}
where $\C_{\mathbf{m},t} = \C_{\mathbf{m}} \cap A_t$.
Concluding, the one-step-ahead predictive distribution is equal to
\begin{align*}
p(\z_{t+1} | \tau_{1:t}) = \frac{1}{\sum_{\boldsymbol{\ell}}  q_{\boldsymbol{\ell}, t}(A_{1:t-1}) \xi_{\boldsymbol{\ell},\mathbf{t}} } \sum_{\mathbf{m}}  \mathcal{N}_N(\z_{t+1} | \boldsymbol{\tilde{\mu}}_{\mathbf{m}}, I_N) q_{\mathbf{m},t+1}(A_{1:t}),
\end{align*}
where
\begin{equation*}
q_{\mathbf{m},t+1}(A_{1:t}) = \sum_{\K} q_{\K, t}(A_{1:t-1})  \int_{\C_{\mathbf{m},t}}  \mathcal{N}_N(\z_t | \boldsymbol{\tilde{\mu}}_{\K}, I_N) \: \mathrm{d}\z_t.
\end{equation*}

\vspace*{2.5ex}

\noindent \underline{\textit{Smoothing distribution.}}
The smoothing distribution at the final time $t=T$ coincides with the filtered distribution that is
\begin{align*}
    p(\z_T | \tau_{1:T}) \propto \mathbb{I}(\z_T \in A_T) \sum_{\K} \mathcal{N}_N(\z_T | \boldsymbol{\tilde{\mu}}_{\K}, I_N) q_{\K, T}(A_{1:T-1}).
\end{align*}
The smoothing distribution at a generic time $t$ can be derived by developing a backwards recursion for $t=T-1,\ldots,1$, based on the relationship:
\begin{align*}
    p(\z_t | \tau_{1:T}) = p(\z_t | \tau_{1:t}) \int_{\R^N} \frac{ p(\z_{t+1} | \tau_{1:T}) p(\z_{t+1} | \z_{t}) }{ p(\z_{t+1} | \tau_{1:t})} \: \mathrm{d}\z_{t+1}.
\end{align*}

\noindent Start by considering time $t=T-1$, where the smoothed distribution is
\begin{align*}
    p(\z_{T-1} | \tau_{1:T}) = p(\z_{T-1} | \tau_{1:T-1}) \int_{\R^N} \frac{ p(\z_{T} | \tau_{1:T}) p(\z_{T} | \z_{T-1}) }{ p(\z_{T} | \tau_{1:T-1})} \: \mathrm{d}\z_{T}.
\end{align*}
In this equation, $p(\z_T | \tau_{1:T})$ is the smoothing distribution at the following period, which is already available, $p(\z_{T} | \z_{T-1})$ is the transition density, and $p(\z_{T} | \tau_{1:T-1})$ is the predictive distribution. Since all these distributions are known, by direct substitution one gets:
\begin{align*}
    p(\z_{T-1} | \tau_{1:T}) & = p(\z_{T-1} | \tau_{1:T-1}) \int_{\R^N} \frac{ \mathbb{I}(\z_T \in A_T) \sum_{\K} \mathcal{N}_N(\z_T | \boldsymbol{\mu}_{\K}, I_N) q_{\K, T}(A_{1:T-1}) }{ \sum_{\mathbf{m}} \mathcal{N}_N(\z_{T} | \boldsymbol{\mu}_{\mathbf{m}}, I_N) q_{\mathbf{m},T}(A_{1:T-1}) } \\
    & \quad \times \mathcal{N}_N\Big( \z_T | \sum_s g_s(\z_{T-1}), I_N \Big) \: \mathrm{d}\z_T,
\end{align*}
where $\sum_s g_s(\z_{T-1})$ is a vector of values determined by the terminal nodes associated with each individual $\z_{T-1}$ value.
Thus, the smoothing distribution at time $T-1$ is given by
\begin{align*}
    p(\z_{T-1} | \tau_{1:T}) = p(\z_{T-1} | \tau_{1:T-1}) \int_{\R^N} \mathbb{I}(\z_T \in A_T) \mathcal{N}_N\Big(\z_T \big| \sum_s g_s(\z_{T-1}), I_N \Big) \: \mathrm{d}\z_T.
\end{align*}
Restricting the integral to the space where the integrand is non-zero yields
\begin{align*}
    p(\z_{T-1} | \tau_{1:T}) & = p(\z_{T-1} | \tau_{1:T-1}) \int_{A_T} \mathcal{N}_N\Big(\z_T \big| \sum_s g_s(\z_{T-1}), I_N \Big) \: \mathrm{d}\z_T \\
    & = p(\z_{T-1} | \tau_{1:T-1}) r_{T-1}(\z_{T-1}),
\end{align*}
where
\begin{align*}
    r_{T-1}(\z_{T-1}) & = \int_{A_T} \mathcal{N}_N\Big( \z_T \big| \sum_s g_s(\z_{T-1}), I_N \Big) \: \mathrm{d}\z_{T} \\
    & = \bigint_{A_T} \mathcal{N}_N\left(\z_T \Bigg| \begin{bmatrix}
    \sum_{k_1=1}^K \tilde{\mu}_{k_1} \mathbb{I}(\z_{T-1,1} \in C_{k_1}) \\
    \vdots \\
    \sum_{k_N=1}^K \tilde{\mu}_{k_N} \mathbb{I}(\z_{T-1,N} \in C_{k_N}) 
\end{bmatrix}, I_N \right) \: \mathrm{d}\z_T.
\end{align*}

\noindent At time $t= T-2$, the smoothing distribution has the following form
\begin{align*}
p(\z_{T-2} | \tau_{1:T}) & \propto p(\z_{T-2} | \tau_{1:T-2}) \\
 & \quad \times \int_{\R^N} \frac{\mathbb{I}(\z_{T-1} \in A_{T-1}) \sum_{\K} \mathcal{N}_N(\z_{T-1} | \boldsymbol{\tilde{\mu}}_{\K}, I_N) q_{\K, T-1}(A_{1:T-2}) r_{T-1}(\z_{T-1}) }{ \sum_{\mathbf{m}}  \mathcal{N}_N(\z_{T-1} | \boldsymbol{\tilde{\mu}}_{\mathbf{m}}, I_N) q_{\mathbf{m},T-1}(A_{1:T-2}) } \\
 & \quad \times \mathcal{N}_N(\z_{T-1} | \sum_s g_s(\z_{T-2}), I_N) \:  \mathrm{d}\z_{T-1}.
\end{align*}
Restricting the range of the integral to where the integrand is non-zero and simplifying the terms in the fraction, one obtains
\begin{align*}
p(\z_{T-2} | \tau_{1:T}) & \propto p(\z_{T-2} | \tau_{1:T-2}) \int_{A_{T-1}} \!\!\! r_{T-1}(\z_{T-1}) \mathcal{N}_N\Big(\z_{T-1} \big| \sum_s g_s(\z_{T-2}), I_N \Big) \: \mathrm{d}\z_{T-1} \\
 & \propto p(\z_{T-2} | \tau_{1:T-2}) r_{T-2}(\z_{T-2}),
\end{align*}
where
\begin{align*}
r_{T-2}(\z_{T-2}) & = \int_{A_{T-1}} \!\!\! r_{T-1}(\z_{T-1}) \mathcal{N}_N\Big(\z_{T-1} \big| \sum_s g_s(\z_{T-2}), I_N \Big) \: \mathrm{d}\z_{T-1}.
\end{align*}
Given the smoothing distribution at time $t = T-1$ and $t=T-2$, now assume that the result holds for a generic $t+1$ and prove that it also holds for $t$.
Using the previous results and induction, one obtains the smoothing distribution at time $t$:
\begin{align*}
    p(\z_t | \tau_{1:T}) & \propto p(\z_t | \tau_{1:t}) r_t(\z_t) \\
    & \propto r_t(\z_t) \mathbb{I}(\z_t \in A_t) \sum_{\K} \mathcal{N}_N(\z_{t} | \boldsymbol{\tilde{\mu}}_{\K}, I_N) q_{\K, t}(A_{1:t-1}),
\end{align*}
where $r_t(\z_t)$ embeds all information from the periods $t'=t+1,\ldots,T$ and, starting from $r_T(\z_T) = 1$, it is defined by the recursion
\begin{align*}
    r_t(\z_t) & = \int_{A_{t+1}} r_{t+1}(\z_{t+1}) \mathcal{N}_N\Big( \z_{t+1} \big| \sum_{s} g_s(\z_t), I_N \Big) \: \mathrm{d}\z_{t+1}.
\end{align*}

\qed

\vspace*{2.5ex}

\noindent\textbf{Proof of Corollary 1}
\smallskip

\noindent Concerning the filtering distribution, one has:
\begin{align*}
p(\z_t | \tau_{1:t}) & \propto \mathbb{I}(\z_t \in A_t) \sum_{\K} \mathcal{N}_N(\z_t | \boldsymbol{\tilde{\mu}}_{\K}, I_N) q_{\K,t}(A_{1:t-1}) \\
 & \propto \sum_{\K} q_{\K,t}(A_{1:t-1}) \bar{u}^F_{\K,t}(\z_t) \\
 & \propto \sum_{\K} q_{\K,t}(A_{1:t-1}) n_{\K, t} u^F_{\K,t}(\z_t),
\end{align*}
where $u^F_{\K,t}(\z_t) = \bar{u}^F_{\K,t}(\z_t) / n_{\K, t}$ is a probability density function, with
\begin{align*}
\bar{u}^F_{\K,t}(\z_t) & = \mathbb{I}(\z_t \in A_t) \mathcal{N}_N(\z_t | \boldsymbol{\tilde{\mu}}_{\K}, I_N), \\
n_{\K, t} & = \int_{\R^N} \bar{u}^F_{\K,t}(\z_t) \: \mathrm{d}\z_t = \int_{A_t} \mathcal{N}_N(\z_t | \boldsymbol{\tilde{\mu}}_{\K}, I_N) \: \mathrm{d}\z_t,
\end{align*}
and the normalizing weights are given by
\begin{align*}
w^F_{\K, t} = \frac{q_{\K,t}(A_{1:t-1}) n_{\K, t}}{\sum_{\mathbf{m}} q_{\mathbf{m},t}(A_{1:t-1}) n_{\mathbf{m}, t}}.
\end{align*}
Combining these quantities, one gets the mixture representation:
\begin{align*}
p(\z_t | \tau_{1:t}) & = \sum_{\K} w^F_{\K, t} u^F_{\K,t}(\z_t).
\end{align*}

\noindent Concerning the predictive distribution, one has:
\begin{align*}
p(\z_{t+1} | \tau_{1:t}) & \propto  \sum_{\K} \mathcal{N}_N(\z_{t+1} | \boldsymbol{\tilde{\mu}}_{\K}, I_N) q_{\K,t+2}(A_{1:t}) \\
 & \propto \sum_{\K} q_{\K,t}(A_{1:t}) \bar{u}^P_{\K,t+1}(\z_{t+1}) \\
\end{align*}
where $ \bar{u}^P_{\K,t+1}(\z_{t+1})  = \mathcal{N}_N(\z_{t+1} | \boldsymbol{\tilde{\mu}}_{\K}, I_N) $, and the normalizing weights are given by
\begin{align*}
w^P_{\K, t+1} = \frac{q_{\K,t+1}(A_{1:t}) }{\sum_{\mathbf{m}} q_{\mathbf{m},t+1}(A_{1:t}) }.
\end{align*}
Combining these quantities, one gets the mixture representation:
\begin{align*}
p(\z_{t+1} | \tau_{1:t}) & = \sum_{\K} w^P_{\K, t+1} \bar{u}^P_{\K,t+1}(\z_{t+1}).
\end{align*}

\noindent Concerning the smoothing distribution, one has:
\begin{align*}
\bar{p}(\z_t | \tau_{1:T}) & \propto r_t(\z_t) \mathbb{I}(\z_t \in A_t) \sum_{\K} \mathcal{N}_N(\z_t | \boldsymbol{\tilde{\mu}}_{\K}, I_N) q_{\K,t}(A_{1:t-1}) \\
 & \propto \sum_{\K} q_{\K,t}(A_{1:t-1}) \bar{u}^S_{\K,t}(\z_t) \\
 & \propto \sum_{\K} q_{\K,t}(A_{1:t-1}) m_{\K, t} u^S_{\K,t}(\z_t),
\end{align*}
where $u^S_{\K,t}(\z_t) = \bar{u}^S_{\K,t}(\z_t) / m_{\K, t}$ is a probability density function, with
\begin{align*}
\bar{u}^S_{\K,t}(\z_t) & = \mathbb{I}(\z_t \in A_t) r_t(\z_t) \mathcal{N}_N(\z_t | \boldsymbol{\tilde{\mu}}_{\K}, I_N), \\
m_{\K} & = \int_{\R^N} \bar{u}^S_{\K,t}(\z_t) \: \mathrm{d}\z_t,
\end{align*}
and the normalizing weights are given by
\begin{align*}
    w^S_{\K, t} = \frac{q_{\K,t}(A_{1:t-1}) m_{\K, t}}{\sum_{\boldsymbol{\ell}} q_{\boldsymbol{\ell}, t}(A_{1:t-1}) m_{\boldsymbol{\ell}, t}},
\end{align*}
where $\boldsymbol{\ell} = (\ell_1,\ldots,\ell_N) \in \{ 1,\ldots,K \}^N$.
Combining these quantities, one gets the mixture representation:
\begin{align*}
    p(\z_t | \tau_{1:T}) & = \sum_{\K} w^S_{\K, t} u^S_{\K,t}(\z_t).
\end{align*}

\qed

\vspace*{2.5ex}

\noindent The following Remark provides a compact way to express the functions $r_t(\z_t)$ in the smoothing distribution.

\smallskip

\begin{rem} \label{remark:smoothing}
Let
\begin{align*}
    r^{(T)}(\z_{T-1}) & = \int_{A_T} \mathcal{N}_N\Big( \z_T | \sum_s g_s(\z_{T-1}), I_N \Big) \: \mathrm{d}\z_T, \\
    r_{\K}^{(T)} & = \int_{A_T} \mathcal{N}_N(\z_T | \boldsymbol{\tilde{\mu}}_{\K}, I_N) \: \mathrm{d}\z_T, \\
    r_{\K, \mathbf{m}}^{(t)} & = \int_{\C_{\K,t}} \mathcal{N}_N(\z_{t} | \boldsymbol{\tilde{\mu}}_{\mathbf{m}}, I_N) \: \mathrm{d}\z_{t}, \qquad \text{ for } t < T, \\
    r_{\K}^{(t)}(\z_{t-1}) & = \int_{\C_{\K,t}} \mathcal{N}_N\Big( \z_{t} | \sum_s g_s(\z_{t-1}), I_N \Big) \: \mathrm{d}\z_{t}, \qquad \text{ for } t < T.
\end{align*}
Then, the term $r_t(\z_t)$ in the smoothing distribution in Eq. (9) in the main article can be simplified as follows:
\begin{align*}
    r_t(\z_t) & = \begin{cases}
    1 & \text{ if } t = T, \\
    r^{(T)}(\z_{T-1}) & \text{ if } t = T-1, \\
    \displaystyle\sum_{\K^T} r_{\K^T}^{(T)} \sum_{\K^{T-1}} r_{\K^T,\K^{T-1}}^{(T-1)} \dots \sum_{\K^{t+2}} r_{\K^{t+3},\K^{t+2}}^{(t+2)} r_{\K^{t+2}}^{(t+1)}(\z_t) & \text{ if } t < T-1.
\end{cases}
\end{align*}
where $\K^t = (k_1^t,\ldots,k_N^t)$ is a multi-index for each $t$. \footnote{We have written $\K$ in most of this document for simplicity of exposition. However, in the presence of other covariates, the sum-of-trees defines a different partition of the possible values of the lag of the latent score in each time period, conditional on the time-period specific values of other covariates. Therefore, in each time period, the tuples can index elements of different partitions and can take different ranges of values (because the number of elements of the partition can vary). The results are unaffected apart from editing of indices to make them time-specific. See also the paragraph below beginning with ``Including Covariates as Splitting Variables''.}
\end{rem}

\vspace*{2.5ex}

\noindent\textbf{Proof of Remark~\ref{remark:smoothing}:}

\smallskip

\noindent Let us start by defining
\begin{align*}
    r^{(T)}(\z_{T-1}) & = \int_{A_T} \mathcal{N}_N\Big( \z_T | \sum_s g_s(\z_{T-1}), I_N \Big) \: \mathrm{d}\z_T, \\
    r_{\K}^{(T)} & = \int_{A_T} \mathcal{N}_N(\z_T | \boldsymbol{\tilde{\mu}}_{\K}, I_N) \: \mathrm{d}\z_T, \\
    r_{\K, \mathbf{m}}^{(t)} & = \int_{\C_{\K,t}} \mathcal{N}_N(\z_{t} | \boldsymbol{\tilde{\mu}}_{\mathbf{m}}, I_N) \: \mathrm{d}\z_{t}, \qquad \text{ for } t < T, \\
    r_{\K}^{(t)}(\z_{t-1}) & = \int_{\C_{\K,t}} \mathcal{N}_N\Big( \z_{t} | \sum_s g_s(\z_{t-1}), I_N \Big) \: \mathrm{d}\z_{t}, \qquad \text{ for } t < T,
\end{align*}
and recall  that $r_{T}(\z_{T}) = 1$. Substituting in the recursion for $r_t(\z_t)$ with time $t=T-1$, one obtains
\begin{align*}
    r_{T-1}(\z_{T-1}) & = \int_{A_{T}} r_{T}(\z_{T}) \mathcal{N}_N\Big( \z_{T} \big| \sum_s g_s(\z_{T-1}), I_N \Big) \: \mathrm{d}\z_{T} \\
    & = \int_{A_{T}}  \mathcal{N}_N\Big( \z_{T} \big| \sum_s g_s(\z_{T-1}), I_N \Big) \: \mathrm{d}\z_{T} \\
    & = r^{(T)}(\z_{T-1}).
\end{align*}
In the same way at time $t=T-2$, one gets
\begin{align*}
    r_{T-2}(\z_{T-2}) & = \int_{A_{T-1}} r_{T-1}(\z_{T-1}) \mathcal{N}_N\Big( \z_{T-1} \big| \sum_s g_s(\z_{T-2}), I_N \Big) \: \mathrm{d}\z_{T-1} \\
    & = \int_{A_{T-1}} \left[ \int_{A_{T}}  \mathcal{N}_N\Big( \z_{T} \big| \sum_s g_s(\z_{T-1}), I_N \Big) \: \mathrm{d}\z_{T} \right] \mathcal{N}_N\Big( \z_{T-1} \big| \sum_s g_s(\z_{T-2}), I_N \Big) \: \mathrm{d}\z_{T-1} \\
    & = \sum_{\K} \int_{\C_{\K,T-1}} \left[ \int_{A_T}  \mathcal{N}_N( \z_{T} \big| \boldsymbol{\tilde{\mu}}_{\K}, I_N) \: \mathrm{d}\z_{T} \right] \mathcal{N}_N\Big( \z_{T-1} \big| \sum_s g_s(\z_{T-2}), I_N \Big) \: \mathrm{d}\z_{T-1} \\
    & =   \sum_{\K} \int_{\C_{\K,T-1}} r_{\K}^{(T)} \mathcal{N}_N\Big( \z_{T-1} \big| \sum_s g_s(\z_{T-2}), I_N \Big) \: \mathrm{d}\z_{T-1} \\
    & =   \sum_{\K}  r_{\K}^{(T)} \int_{\C_{\K,T-1}} \mathcal{N}_N\Big( \z_{T-1} \big| \sum_s g_s(\z_{T-2}), I_N \Big) \: \mathrm{d}\z_{T-1} \\
    & =   \sum_{\K}  r_{\K}^{(T)} r_{\K}^{(T-1)}(\z_{T-2}).
\end{align*}
Moving to time $t=T-3$, one has
\begin{align*}
    r_{T-3}(\z_{T-3}) & = \int_{A_{T-2}} r_{T-2}(\z_{T-2}) \mathcal{N}_N\Big( \z_{T-2} \big| \sum_s g_s(\z_{T-3}), I_N \Big) \: \mathrm{d}\z_{T-2} \\
    & = \int_{A_{T-2}} \Big[ \sum_{\K}  r_{\K}^{(T)} \int_{\C_{\K,T-1}} \!\!\!\! \mathcal{N}_N\Big( \z_{T-1} \big| \sum_s g_s(\z_{T-2}), I_N \Big) \: \mathrm{d}\z_{T-1} \Big] \\
    & \quad \times \mathcal{N}_N\Big( \z_{T-2} \big| \sum_s g_s(\z_{T-3}), I_N \Big) \: \mathrm{d}\z_{T-2} \\
    & = \sum_{\mathbf{m}} \int_{\C_{\mathbf{m},T-2}} \Big[ \sum_{\K}  r_{\K}^{(T)} \int_{\C_{\K,T-1}} \!\!\!\! \mathcal{N}_N( \z_{T-1} \big| \boldsymbol{\tilde{\mu}}_{\mathbf{m}}, I_N) \: \mathrm{d}\z_{T-1} \Big] \\
    & \quad \times \mathcal{N}_N\Big( \z_{T-2} \big| \sum_s g_s(\z_{T-3}), I_N \Big) \: \mathrm{d}\z_{T-2} \\
    & = \sum_{\mathbf{m}}\int_{\C_{\mathbf{m},T-2}} \Big[ \sum_{\K}  r_{\K}^{(T)} r_{\K,\mathbf{m}}^{(T-1)} \Big] \mathcal{N}_N\Big( \z_{T-2} \big| \sum_s g_s(\z_{T-3}), I_N \Big) \: \mathrm{d}\z_{T-2} \\
    & = \sum_{\mathbf{m}}\Big[ \sum_{\K}  r_{\K}^{(T)} r_{\K,\mathbf{m}}^{(T-1)} \Big] \int_{\C_{\mathbf{m},T-2}} \mathcal{N}_N\Big( \z_{T-2} \big| \sum_s g_s(\z_{T-3}), I_N \Big) \: \mathrm{d}\z_{T-2} \\
    & = \sum_{\K}  r_{\K}^{(T)} \sum_{\mathbf{m}} r_{\K,\mathbf{m}}^{(T-1)}  r_{\mathbf{m}}^{(T-2)}(\z_{T-3}).
\end{align*}
Let us now introduce a compact notation for the multi-indices by defining $\K^t = (k_1^t,\ldots,k_N^t)$ for each $t=1,\ldots, T$. In the last equation above, this corresponds to the use of $\K^T = \K$ and $\K^{T-1} = \mathbf{m}$.
Using this notation to rewrite the previous results yields the following.
\begin{align*}
    r_{T-1}(\z_{T-1}) & = r^{(T)}(\z_{T-1}), \\
    r_{T-2}(\z_{T-2}) & = \sum_{\K^T} r_{\K^T}^{(T)} r_{\K^{T}}^{(T-1)}(\z_{T-2}), \\
    r_{T-3}(\z_{T-3}) & = \sum_{\K^T} r_{\K^T}^{(T)} \sum_{\K^{T-1}}  r_{\K^{T},\K^{T-1}}^{(T-1)} r_{\K^{T-1}}^{(T-2)}(\z_{T-3}).
\end{align*}
Iterating this procedure by induction on $t$, the result follows.

\qed

\vspace*{2.5ex}

Based on Remark~\ref{remark:smoothing}, we can provide proof of the exact smoothing distribution as a mixture representation from Corollary 1 in the main article.  
\bigskip

\noindent\textbf{Alternative Proof of Corollary 1}

\smallskip

Concerning the smoothing distribution, one has:
\begin{align*}
    \bar{p}(\z_t | \tau_{1:T})  & \propto \sum_{\K} q_{\K,t}(A_{1:t-1}) m_{\K, t} \frac{\bar{u}^S_{\K,t}(\z_t)}{m_{\K, t}}.
\end{align*}
Note that the normalizing constant can be represented as 
\begin{align*}
    \int \sum_{\boldsymbol{\ell}} q_{\boldsymbol{\ell},t}(A_{1:t-1}) m_{\boldsymbol{\ell}, t} \frac{\bar{u}^S_{\boldsymbol{\ell},t}(\z_t)}{m_{\boldsymbol{\ell}, t}} \: \mathrm{d}\z_t =  \sum_{\boldsymbol{\ell}} q_{\boldsymbol{\ell},t}(A_{1:t-1})  \int \bar{u}^S_{\boldsymbol{\ell},t}(\z_t)  \: \mathrm{d}\z_t =  \sum_{\boldsymbol{\ell}} q_{\boldsymbol{\ell},t}(A_{1:t-1})  m_{\boldsymbol{\ell}, t},
\end{align*}
therefore the smoothing distribution is
\begin{align*}
    \bar{p}(\z_t | \tau_{1:T}) & = \sum_{\K} \frac{q_{\K,t}(A_{1:t-1}) m_{\K, t}}{ \sum_{\boldsymbol{\ell}} q_{\boldsymbol{\ell},t}(A_{1:t-1}) m_{\boldsymbol{\ell}, t}} \frac{\bar{u}^S_{\K,t}(\z_t)}{m_{\K, t}}.
\end{align*}
In particular, $m_{\K, t}$ must be calculated in order to be sampled from this distribution, and it can be calculated $\forall \K$ as follows:
\begin{align*}
    m_{\K, t} = \int \bar{u}^S_{\boldsymbol{\ell},t}(\z_t)  \mathrm{d}\z_t  = \int  \mathbb{I}(\z_t \in A_t) r_t(\z_t) \mathcal{N}_N(\z_t | \boldsymbol{\tilde{\mu}}_{\K}, I_N) \: \mathrm{d}\z_t,
\end{align*}
where 
\begin{align*}
    r_t(\z_t) & = \begin{cases}
    1 & \text{ if } t = T, \\
    r^{(T)}(\z_{T-1}) & \text{ if } t = T-1, \\
    \displaystyle\sum_{\K^T} r_{\K^T}^{(T)} \sum_{\K^{T-1}} r_{\K^T,\K^{T-1}}^{(T-1)} \dots \sum_{\K^{t+2}} r_{\K^{t+3},\K^{t+2}}^{(t+2)} r_{\K^{t+2}}^{(t+1)}(\z_t) & \text{ if } t < T-1.
\end{cases}
\end{align*}
Thus, we have 
\begin{align*}
    m_{\K, T} = \int  \mathbb{I}(\z_T \in A_T)  \mathcal{N}_N(\z_T | \boldsymbol{\tilde{\mu}}_{\K}, I_N) \: \mathrm{d}\z_T  =  r_{\K}^{(T)} = \xi_{\K, T}.
\end{align*}
Moving to $T-1$, we have 
\begin{align*}
    m_{\K, T-1} &= \int  \mathbb{I}(\z_{T-1} \in A_{T-1})  r^{(T)}(\z_{T-1}) \mathcal{N}_N(\z_{T-1} | \boldsymbol{\tilde{\mu}}_{\K}, I_N) \: \mathrm{d}\z_{T-1} \\
    & = \int  \mathbb{I}(\z_{T-1} \in A_{T-1})  \int_{A_T} \mathcal{N}_N\Big( \z_T | \sum_s g_s(\z_{T-1}), I_N \Big) \: \mathrm{d}\z_T \mathcal{N}_N(\z_{T-1} | \boldsymbol{\tilde{\mu}}_{\K}, I_N) \: \mathrm{d}\z_{T-1}.
\end{align*}
Then separate out integral over $\mathrm{d}\z_{T-1}$ into regions $\C_{\K^{T},T-1}$, one obtains
\begin{align*}
    m_{\K, T-1} & = \sum_{\K^{T}} \int_{ \C_{\K^{T},T-1} }  \int_{A_T} \mathcal{N}_N\Big( \z_T | \boldsymbol{\tilde{\mu}}_{\K^{T}}, I_N \Big) \: \mathrm{d}\z_T \mathcal{N}_N(\z_{T-1} | \boldsymbol{\tilde{\mu}}_{\K}, I_N) \: \mathrm{d}\z_{T-1} \\
    & = \sum_{\K^{T}}  \int_{A_T} \mathcal{N}_N\Big( \z_T | \boldsymbol{\tilde{\mu}}_{\K^{T}}, I_N \Big) \: \mathrm{d}\z_T \int_{ \C_{\K^{T},T-1} }  \mathcal{N}_N(\z_{T-1} | \boldsymbol{\tilde{\mu}}_{\K}, I_N) \: \mathrm{d}\z_{T-1}   \\
    & = \sum_{\K^{T}} \bigg(  \int_{A_T} \mathcal{N}_N\Big( \z_T | \boldsymbol{\tilde{\mu}}_{\K^{T}}, I_N \Big) \: \mathrm{d}\z_T \bigg) r_{\K^{T-1}, \K}^{(T-1)} = \sum_{\K^{T}} r_{\K^{T}, \K}^{(T-1)}  \int_{A_T} \mathcal{N}_N\Big( \z_T | \boldsymbol{\tilde{\mu}}_{\K^{T}}, I_N \Big) \: \mathrm{d}\z_T  \\
    & = \sum_{\K^{T}} r_{\K^{T}, \K}^{(T-1)} r_{\K^{T}}^{(T)}.
\end{align*}
Extending the result for $t < T-1$, one has
\begin{align*}
    m_{\K, t}  =  \int \mathbb{I}(\z_t \in A_t)  \sum_{\K^T} r_{\K^T}^{(T)} \sum_{\K^{T-1}} r_{\K^T,\K^{T-1}}^{(T-1)} \dots \sum_{\K^{t+2}} r_{\K^{t+3},\K^{t+2}}^{(t+2)} r_{\K^{t+2}}^{(t+1)}(\z_t)  \mathcal{N}_N(\z_t | \boldsymbol{\tilde{\mu}}_{\K}, I_N) \: \mathrm{d} \z_t,  
\end{align*}
and note that 
\begin{align*}
r_{\K^{t+2}}^{(t+1)}(\z_{t}) & = \int_{\C_{\K^{t+2},t+1}} \mathcal{N}_N\Big( \z_{t+1} | \sum_s g_s(\z_{t}), I_N \Big) \: \mathrm{d}\z_{t+1}. 
\end{align*}
Thus rearranging the previous equation leads to 
\begin{align*}
     m_{\K, t}  &=  \int_{A_t}   \sum_{\K^T} r_{\K^T}^{(T)} \sum_{\K^{T-1}} r_{\K^T,\K^{T-1}}^{(T-1)} \dots \sum_{\K^{t+2}} r_{\K^{t+3},\K^{t+2}}^{(t+2)} \\
     &\quad \times \int_{\C_{\K^{t+2},t+1}} \mathcal{N}_N\Big( \z_{t+1} | \sum_s g_s(\z_{t}), I_N \Big) \: \mathrm{d}\z_{t+1}   \mathcal{N}_N(\z_t | \boldsymbol{\tilde{\mu}}_{\K}, I_N) \: \mathrm{d} \z_t \\
     &=  \sum_{\K^T} r_{\K^T}^{(T)} \sum_{\K^{T-1}} r_{\K^T,\K^{T-1}}^{(T-1)} \dots \sum_{\K^{t+2}} r_{\K^{t+3},\K^{t+2}}^{(t+2)}  \\
     &\quad \times  \int_{\C_{\K^{t+2},t+1}} \int_{A_t} \mathcal{N}_N\Big( \z_{t+1} | \sum_s g_s(\z_{t}), I_N \Big)    \mathcal{N}_N(\z_t | \boldsymbol{\tilde{\mu}}_{\K}, I_N) \: \mathrm{d} \z_t   \: \mathrm{d}\z_{t+1}.
\end{align*}
Then separating the integral over  $\z_t$ into regions $\K^{t+1}$ and taking the sum over $\K^{t+1}$ of integrals restricted to ranges $\C_{\K^{t+1}}$, one restricts the range of the integral over $\z_t$ to $A_T$ and the resulting integral will be restricted to $\C_{\K^{t+1},t}$. 

Furthermore, note that $ \sum_s g_s(\z_{t})$ can be replaced by $\boldsymbol{\tilde{\mu}}_{\K^{t+1}}$, one obtains
\begin{align*}
    m_{\K, t}  & =    \sum_{\K^T} r_{\K^T}^{(T)} \sum_{\K^{T-1}} r_{\K^T,\K^{T-1}}^{(T-1)} \dots \sum_{\K^{t+2}} r_{\K^{t+3},\K^{t+2}}^{(t+2)} \\
    &\quad \times \int_{\C_{\K^{t+2},t+1}} \sum_{\K^{t+1}} \int_{\C_{\K^{t+1},t}} \mathcal{N}_N\Big( \z_{t+1} | \boldsymbol{\tilde{\mu}}_{\K^{t+1}}, I_N \Big)    \mathcal{N}_N(\z_t | \boldsymbol{\tilde{\mu}}_{\K}, I_N) \: \mathrm{d} \z_t   \: \mathrm{d}\z_{t+1} \\
    &=  \sum_{\K^T} r_{\K^T}^{(T)} \sum_{\K^{T-1}} r_{\K^T,\K^{T-1}}^{(T-1)} \dots \sum_{\K^{t+2}} r_{\K^{t+3},\K^{t+2}}^{(t+2)} \\
    &\quad \times \sum_{\K^{t+1}}   \int_{\C_{\K^{t+2},t+1}} \mathcal{N}_N\Big( \z_{t+1} | \boldsymbol{\tilde{\mu}}_{\K^{t+1}}, I_N \Big)  \int_{\C_{\K^{t+1},t}}   \mathcal{N}_N(\z_t | \boldsymbol{\tilde{\mu}}_{\K}, I_N) \: \mathrm{d} \z_t   \: \mathrm{d}\z_{t+1} \\
    &=   \sum_{\K^T} r_{\K^T}^{(T)} \sum_{\K^{T-1}} r_{\K^T,\K^{T-1}}^{(T-1)} \dots \sum_{\K^{t+2}} r_{\K^{t+3},\K^{t+2}}^{(t+2)} \\
    &\quad \times \sum_{\K^{t+1}}   \int_{\C_{\K^{t+2},t+1}} \mathcal{N}_N\Big( \z_{t+1} | \boldsymbol{\tilde{\mu}}_{\K^{t+1}}, I_N \Big) \: \mathrm{d} \z_{t+1}   \int_{\C_{\K^{t+1},t}}   \mathcal{N}_N(\z_t | \boldsymbol{\tilde{\mu}}_{\K}, I_N)   \: \mathrm{d}\z_{t}.
\end{align*}
Then, since $r_{\K^{t+2}, \K^{t+1}}^{(t+1)} = \int_{\C_{\K^{t+2},t+1}} \mathcal{N}_N\Big( \z_{t+1} | \boldsymbol{\tilde{\mu}}_{\K^{t+1}}, I_N \Big) \: \mathrm{d} \z_{t+1}$ and $ r_{\K^{t+1}, \K}^{(t)} = \int_{\C_{\K^{t+1},t}}   \mathcal{N}_N(\z_t | \boldsymbol{\tilde{\mu}}_{\K}, I_N) \: \mathrm{d}\z_{t} $, one concludes that
\begin{align*}
    m_{\K, t} = \sum_{\K^T} r_{\K^T}^{(T)} \sum_{\K^{T-1}} r_{\K^T,\K^{T-1}}^{(T-1)} \dots \sum_{\K^{t+2}} r_{\K^{t+3},\K^{t+2}}^{(t+2)} \sum_{\K^{t+1}}   r_{\K^{t+2}, \K^{t+1}}^{(t+1)}   r_{\K^{t+1}, \K}^{(t)}.
\end{align*}

\qed

\section{MCMC implementation details}\label{sec:apdx_MCMC}

\subsection{Sampler for ROBART}
\label{sec:apdx_posterior_ROBART}

In this Section, we provide an MCMC algorithm for the ROBART posterior sampler. The steps are structured as follows:
\begin{enumerate}[label=\arabic*)]
\item sample the latent variables $z_{ij}$, for $i=1,\ldots,N$ and $j=1,\ldots,J$, from the truncated Gaussian distribution
\begin{equation}
p(z_{ij} | \z_{-i,j}, \z_{-j}, \mathcal{T}_1,\boldsymbol\mu_1,\ldots,\mathcal{T}_S,\boldsymbol\mu_S) \propto \mathcal{N}\Big( z_{ij} \big| \sum_{s=1}^S g_s(\mathbf{X}_{ij}|\mathcal{T}_s,\boldsymbol\mu_s), 1 \Big) \I_{A_{ij}}(z_{ij}),
\end{equation}
where the set $A_{ij}$ is determined by $z_{-i,j}$ and $\tau_{j}$ such that $\operatorname{rank}(\z_{j}) = \tau_{j}$, that is\footnote{See also \citet[][Ch. 12]{hoff2009first}.}
\begin{equation*}
A_{ij} = \Big\{ z_{ij} : \max\{z_{kj} : \tau_{kj} < \tau_{ij} \} < z_{ij} < \min\{z_{kj} : \tau_{ij} < \tau_{kj} \} \Big\};
\end{equation*}
\item sample the tree structure and leaf parameters from $p(\mathcal{T}_s, \boldsymbol\mu_s | \mathcal{T}_{-s}, \boldsymbol\mu_{-s}, \z)$, for $s=1,\ldots,S$. This is done similarly to standard BART, with $\mathbf{y}$ replaced by $\z = (\z_{1,1}',\ldots,\z_{N,J}')'$, as follows:
\begin{enumerate}[label=\alph*.]
\item compute the $NJ$-dimensional vector of partial residuals, $R_s$ as 
\begin{equation*}
R_{s, ij} = \z_{i,j} - \sum_{\substack{k=1\\k \neq s}}^S g_k(\mathbf{X}_{ij}| \mathcal{T}_k, \boldsymbol\mu_k),
\end{equation*}
and notice that $p(\mathcal{T}_s, \boldsymbol\mu_s | \mathcal{T}_{-s}, \boldsymbol\mu_{-s}, \z) = p(\mathcal{T}_s, \boldsymbol\mu_s | R_s)$;
\item since $p(\mathcal{T}_s | R_s)$ has a closed form, sample the trees structure from $p(\mathcal{T}_s | R_s)$ using a Metropolis-Hastings algorithm;
\item sample the terminal nodes from $p(\boldsymbol\mu_s | \mathcal{T}_s, R_s)$, which is a set of independent draws from a Gaussian distribution.
\end{enumerate}
\end{enumerate}


\subsection{Sampling the path of the latent score}
\label{sec:apdx_sampling_filter_smooth}

\subsubsection{ARROBART model}

{
\paragraph{Backward Sampling}
The marginal smoothing distribution derived in the main text cannot be applied in a Gibbs sampler because we require samples of $\bm{z}_{1:T}$ instead of marginal distribution samples. Instead, we can in principal (although potentially not in practice due to computational complexity) sample as follows. First, note that ideally, we would sample the full block of latent scores for all time periods and items simultaneously. This is not feasible, although the distribution can be decomposed as follows
\begin{align*}
    p(\z_{1:T}| \tau_{1:T}) & = \sum_{ \mathbf{k}_1, \dots, \mathbf{k}_T } p(\z_{1:T}, \mathbf{k}_1,\dots,\mathbf{k}_T | \tau_{1:T} ) \propto \sum_{ \mathbf{k}_1, \dots, \mathbf{k}_T } p( \tau_{1:T} | \z_{1:T}, \mathbf{k}_1,\dots,\mathbf{k}_T) p(\z_{1:T}, \mathbf{k}_1,\dots,\mathbf{k}_T) \\
    & \propto \sum_{ \mathbf{k}_1, \dots, \mathbf{k}_T } \underbrace{p( \tau_{1:T} | \z_{1:T}) p(  \z_{1:T} | \mathbf{k}_1,\dots,\mathbf{k}_T) }_{\text{NT-dim Truncated Gaussian}} p( \mathbf{k}_1,\dots,\mathbf{k}_T) \\
    & \propto \sum_{ \mathbf{k}_1, \dots, \mathbf{k}_T } \underbrace{p( \tau_{1:T} | \z_{1:T}) p(  \z_{1:T} | \mathbf{k}_1,\dots,\mathbf{k}_T) }_{\text{NT-dim Truncated Gaussian}} p( \mathbf{k}_1)\prod_{t=2}^T p( \mathbf{k}_t | \mathbf{k}_{t-1})
\end{align*}
with
\begin{align*}
    p(\mathbf{k}_1) & = q_{\K, 1} = \int_{\C_{\K_1}} \mathcal{N}_N(\underline{\z}_0 | \underline{\z}_{prior}, I_N) \mathrm{d}\underline{\z}_0 \\
    p( \mathbf{k}_t | \mathbf{k}_{t-1}) & = r_{\K_t, \mathbf{k}_{t-1}}^{(t)}  = \int_{\C_{\K_t}} \mathcal{N}_N(\z_{t} | \boldsymbol{\tilde{\mu}}_{\mathbf{k}_{t-1}}, I_N) \: \mathrm{d}\z_{t}.
\end{align*}
Differently, we may write $p( \mathbf{k}_1,\dots,\mathbf{k}_T) $ as an $NT$ dimensional Gaussian integral
\begin{align*}
    p( \mathbf{k}_1,\dots,\mathbf{k}_T) = \int_{\C_{\mathbf{k}_1} \times \dots \times  \C_{\mathbf{k}_T} } \mathcal{N}_N(\z_{t} | (\boldsymbol{\tilde{\mu}}_{\mathbf{k}_{1}}^T,\dots, \boldsymbol{\tilde{\mu}}_{\mathbf{k}_{T}}^T)^T, I_{TN}) \: \mathrm{d}\z_{1:T}.
\end{align*}
The number of elements of the above mixture is exponential in $T$ and it is unlikely that sampling from the above mixture would be computationally feasible. 
A computationally feasible approach, based on the filtering results above, would be to sample backwards, that is, forward filtering backward sampling. Starting from the decomposition:
\begin{align*}
    p(\z_{1:T} | \tau_{1:T}) & = p(z_T | \tau_{1:T}) p(\z_{1:(T-1)} | \z_{T}, \tau_{1:T}) = p(\z_T | \tau_{1:T}) \prod_{t=1}^{T-1} p(\z_t | \z_{t+1},\ldots,\z_T, \tau_{1:T}) \\
    &  = p(\z_T | \tau_{1:T}) \prod_{t=1}^{T-1} p(\z_t | \z_{t+1}, \tau_{1:(T-1)}).
\end{align*}
Therefore, we may sample from $p(\z_{T} | \tau_{1:T})$, then sequentially from $p(\z_{(T-1)} | \z_T, \tau_{1:T})$, then $p(\z_{(T-2)} | \z_{(T-1)},  \z_T, \tau_{1:T})$,   and so on until $p(\z_1 | \z_{2:T}, \tau_{1:T})$.
To derive a sampler for  \newline $p(\z_{T-\ell} | \tau_{1:T}, \mathbf{z}_{(T-\ell+1):T}) = p(\z_{T-\ell} | \tau_{1:(T-\ell)}, \mathbf{z}_{(T-\ell+1)})$, first note that
\begin{align*}
    & p(\mathbf{z}_{T-\ell} | \tau_{1:(T-\ell)},  \mathbf{z}_{T-\ell+1}) \propto 
    p(\mathbf{z}_{T-\ell+1} | \mathbf{z}_{T-\ell} ) p(\mathbf{z}_{T-\ell} | \tau_{1:(T-\ell)}).
\end{align*}
Then, by substituting for the latent variable and the previously derived filtering distribution, we obtain
\begin{align*}
    p(\mathbf{z}_{T-\ell} | \tau_{1:(T-\ell)},  \mathbf{z}_{T-\ell+1})  & \propto \mathcal{N}_N \Big(\z_{T-\ell+1} | \sum_{\mathbf{k}_{T-\ell+1}} \tilde{\bmu}_{\mathbf{k}_{T-\ell+1}} \mathbb{I}(\z_{T-\ell} \in C_{\mathbf{k}_{T-\ell+1}}), I_N \Big)   \\
    & \quad \times \I(\z_{T-\ell} \in A_{T-\ell}) \sum_{\mathbf{k}_{T-\ell}}  \mathcal{N}_N \left(\z_{T-\ell}  | \tilde{\bmu}_{\mathbf{k}_{T-\ell}}, I_N \right) q_{\mathbf{k}_{T-\ell}} (A_{1:(t-\ell-1)}),
\end{align*}
where the mean of the first Gaussian distribution is a sum over $N$ dimensional tuples $\mathbf{k}_{T-\ell+1}$, and $\tilde{\bmu}_{\mathbf{k}_{T-\ell+1}} \mathbb{I}(\z_{T-\ell} \in C_{\mathbf{k}_{T-\ell+1}})$ is a vector of terminal node parameters for a vector $\z_{T-\ell}$ that is in the $N$-dimensional region $C_{\mathbf{k}_{T-\ell+1}}$.
This can be re-written as
\begin{align*}
    p(\mathbf{z}_{T-\ell} &| \tau_{1:(T-\ell)}, \mathbf{z}_{T-\ell+1})  \propto \I(\z_{T-\ell} \in A_{T-\ell}) \sum_{\mathbf{k}_{T-\ell+1}} \mathbb{I}(\z_{T-\ell} \in C_{\mathbf{k}_{T-\ell+1}}) \mathcal{N}_N \left(\z_{T-\ell+1} |  \tilde{\bmu}_{\mathbf{k}_{T-\ell+1}}, I_N \right) \\
    & \quad \times \sum_{\mathbf{k}_{T-\ell}}  \mathcal{N}_N \left(\z_{T-\ell}   |    \tilde{\bmu}_{\mathbf{k}_{T-\ell}}, I_N \right) q_{\mathbf{k}_{T-\ell}} (A_{1:(t-\ell-1)}) \\
    & \propto \sum_{\mathbf{k}_{T-\ell+1}} \I(\z_{T-\ell} \in \tilde{C}_{\mathbf{k}_{T-\ell+1}}) \mathcal{N}_N \left(\z_{T-\ell+1} | \tilde{\bmu}_{\mathbf{k}_{T-\ell+1}}, I_N \right) \sum_{\mathbf{k}_{T-\ell}}  \mathcal{N}_N \left(\z_{T-\ell} | \tilde{\bmu}_{\mathbf{k}_{T-\ell}}, I_N \right) q_{\mathbf{k}_{T-\ell}} (A_{1:(t-\ell-1)}) \\
    & \propto \sum_{\mathbf{k}_{T-\ell+1}} \sum_{\mathbf{k}_{T-\ell}} \Big[ \mathcal{N}_N(\z_{T-\ell+1} | \tilde{\bmu}_{\mathbf{k}_{T-\ell+1}}, I_N) \I(\z_{T-\ell} \in \tilde{C}_{\mathbf{k}_{T-\ell+1}}) \mathcal{N}_N(\z_{T-\ell} | \tilde{\bmu}_{\mathbf{k}_{T-\ell}}, I_N) q_{\mathbf{k}_{T-\ell}}(A_{1:(t-\ell-1)}) \Big].
\end{align*}
It can be observed that this is proportional to a mixture of truncated Gaussian distributions
\begin{align*}
    & \propto \sum_{\mathbf{k}_{T-\ell+1}} \sum_{\mathbf{k}_{T-\ell}} \Big[ \mathcal{N}_N(\z_{T-\ell+1} | \tilde{\bmu}_{\mathbf{k}_{T-\ell+1}}, I_N) \I(\z_{T-\ell} \in \tilde{C}_{\mathbf{k}_{T-\ell+1}}) \mathcal{N}_N(\z_{T-\ell} | \tilde{\bmu}_{\mathbf{k}_{T-\ell}}, I_N) q_{\mathbf{k}_{T-\ell}}(A_{1:(t-\ell-1)}) \Big] \\
    & \propto \sum_{\mathbf{k}_{T-\ell+1}} \sum_{\mathbf{k}_{T-\ell}} \mathcal{TN}_{\tilde{C}_{\mathbf{k}_{T-\ell+1}}}(\z_{T-\ell} | \tilde{\bmu}_{\mathbf{k}_{T-\ell}}, I_N) \times \mathcal{N}_N(\z_{T-\ell+1} | \tilde{\bmu}_{\mathbf{k}_{T-\ell+1}}, I_N) q_{\mathbf{k}_{T-\ell}}(A_{1:(t-\ell-1)})
\end{align*}
with each component of the mixture corresponding to a $2N$-dimensional tuple $(\mathbf{k}_{T-\ell+1}, \mathbf{k}_{T-\ell})$. The normalizing constant for each truncated Gaussian distribution in the mixture is:
\begin{align*}
\tilde{n}_{\mathbf{k}_{T-\ell+1}, \mathbf{k}_{T-\ell}} & \coloneqq \int_{\R} \I(\z_{T-\ell} \in \tilde{C}_{\mathbf{k}_{T-\ell+1}})  \mathcal{N}_N(\z_{T-\ell} | \tilde{\bmu}_{\mathbf{k}_{T-\ell}}, I_N) \mathrm{d} \z_{T-\ell} \\
    & = \int_{\tilde{C}_{\mathbf{k}_{T-\ell+1}}}  \mathcal{N}_N(\z_{T-\ell} | \tilde{\bmu}_{\mathbf{k}_{T-\ell}}, I_N) \mathrm{d} \z_{T-\ell}.
\end{align*}
Therefore the mixture may be expressed as:
\begin{align*}
    p(\mathbf{z}_{T-\ell} | \tau_{1:(T-\ell)},  \mathbf{z}_{T-\ell+1})  & \propto  \sum_{\mathbf{k}_{T-\ell+1}} \sum_{\mathbf{k}_{T-\ell}} \mathcal{N}_N(\z_{T-\ell+1} | \tilde{\bmu}_{\mathbf{k}_{T-\ell+1}}, I_N) q_{\mathbf{k}_{T-\ell}} (A_{1:(t-\ell-1)}) \\
    & \quad \times \tilde{n}_{\mathbf{k}_{T-\ell+1}, \mathbf{k}_{T-\ell}} \frac{\I(\z_{\mathbf{k}_{T-\ell}} \in \tilde{C}_{\mathbf{k}_{T-\ell+1}})  \mathcal{N}_N(\z_{T-\ell} | \tilde{\bmu}_{\mathbf{k}_{T-\ell}}, I_N)}{\tilde{n}_{ \mathbf{k}_{T-\ell+1}, \mathbf{k}_{T-\ell}}}.
\end{align*}
Let us now define the normalized weights of the mixture as
\begin{align*}
    \tilde{w}_{\mathbf{k}_{T-\ell+1}, \mathbf{k}_{T-\ell}} & = \frac{ \mathcal{N}_N(\z_{T-\ell+1} | \tilde{\bmu}_{\mathbf{k}_{T-\ell+1}}, I_N) q_{\mathbf{k}_{T-\ell}} (A_{1:(t-\ell-1)}) \tilde{n}_{\mathbf{k}_{T-\ell+1}, \mathbf{k}_{T-\ell}} }{ \sum_{\mathbf{m}_{T-\ell+1}} \sum_{\mathbf{m}_{T-\ell}} \mathcal{N}_N(\z_{T-\ell+1} | \tilde{\bmu}_{\mathbf{m}_{T-\ell+1}}, I_N) q_{\mathbf{m}_{T-\ell}} (A_{1:(t-\ell-1)}) \tilde{n}_{\mathbf{m}_{T-\ell+1}, \mathbf{m}_{T-\ell}} }.
\end{align*}
Therefore we obtain
\begin{align*}
    p(\mathbf{z}_{T-\ell} | \tau_{1:(T-\ell)},  \mathbf{z}_{T-\ell+1})  & \propto \sum_{\mathbf{k}_{T-\ell+1}} \sum_{\mathbf{k}_{T-\ell}}  \tilde{w}_{\mathbf{k}_{T-\ell+1}, \mathbf{k}_{T-\ell}}  \mathcal{TN}_{\tilde{C}_{\mathbf{k}_{T-\ell+1}}} \left(\z_{T-\ell}   |    \tilde{\bm{\mu}}_{\mathbf{k}_{T-\ell}}, I_N \right),
\end{align*}
where $\mathcal{TN}_{\tilde{C}_{\mathbf{k}_{T-\ell+1}}} $ is an $N$ dimensional truncated multivariate Gaussian distribution restricted to $\tilde{C}_{\mathbf{k}_{T-\ell+1}}$.
Recall that at $T-\ell=1$, instead of $q_{\mathbf{m}_{T-\ell}} (A_{1:(t-\ell-1)})$ we have $q_{\K,1} = \int_{\C_{\K}} \mathcal{N}_N(\underline{\z}_0 | \underline{\z}_{prior}, I_N) \mathrm{d}\underline{\z}_0$.

\bigskip

To summarize, we can sample from the joint smoothing distribution $p(\z_{1:T} | \tau_{1:T})$ as follows (although computationally not feasible):
\begin{enumerate}
    \item Sample $p(\z_T | \tau_{1:T})$ from the previously derived filtering distribution $p(\z_T | \tau_{1:T}) \propto \I(\z_T \in A_T) \sum_{\K} \mathcal{N}_N(\z_T | \boldsymbol{\tilde{\mu}}_{\K}, I_N) q_{\K,T}(A_{1:T-1})$.
    
    \item From $\ell = 1$ to $\ell = T-1$, sample from $p(\mathbf{z}_{T-\ell} | \tau_{1:(T-\ell)},  \mathbf{z}_{T-\ell+1}) $ by iterating the following steps:
    \begin{enumerate}
        \item Calculate weights $\tilde{w}_{ \mathbf{k}_{T-\ell+1}, \mathbf{k}_{T-\ell} }$.
        \item Draw a $2N$-tuple $\mathbf{k}_{T-\ell+1}, \mathbf{k}_{T-\ell}$ with probabilities given by the calculated weights.
        \item Given the drawn tuple, sample from $ \mathcal{TN}_{\tilde{C}_{\mathbf{k}_{T-\ell+1}}} (\z_{T-\ell} | \tilde{\bmu}_{\mathbf{k}_{T-\ell}}, I_N)$.
    \end{enumerate}
\end{enumerate}
}

\paragraph{Full Gibbs Sampler}
{
Evaluating many potentially small-volume Gaussian integrals is computationally costly and sampling $\z_t$ from a large mixture of multivariate Truncated Gaussian distribution with many constraints is computationally costly.} We provide a method following \cite{hoff2009first} that approximates the path of a latent score conditional on the other components of the vector of latent scores. This provides a full Gibbs sampler of the latent scores.

Initially, we derive the probability density for the conditional distribution for different time periods. For 
$T- \ell = T$, we have a simple case since the density for the conditional distribution is a truncated Gaussian. Hence, for a specific $i$, we have
\begin{align*}
    p(z_{T,i} | \tau_{1:T}, \z_{1:T,-i}, \z_{1:(T-1),i} ) = \mathcal{TN}_{ [\max(z_{T,j} : \tau_{T,j} < \tau_{T,i}), \min(z_{T,j} : \tau_{T,j} > \tau_{T,i})]  } \big( z_{T,i} | g(z_{T-1,i}), 1 \big), 
\end{align*}
where ${\z}_{1:T,-i} = (\z_{1,-i},\ldots, \z_{T,-i})$ is the collection of vectors where the $i$-th element is removed (e.g., $\z_{1,-i} = (z_{1,1},\ldots, z_{1,i-1}, z_{1,i+1},\ldots, z_{1,M})$) and $\tau_{1:t} = (\tau_1',\ldots,\tau_t')'$ is the collection of vectors $\tau_{\tilde{t}}$ for each $\tilde{t}=1,\ldots,t$.

Moving to the previous time ($T-1$), we have that the probability density for the $i$-th latent score at time $T-1$ given all the other latent scores can be represented as 
\begin{align*}
    p(z_{(T-1),i} & | \tau_{1:T}, \z_{1:T,-i}, z_{T,i}, \z_{1:(T-2),i}) \\
    & \propto p(\tau_{T},{\z}_{T} | z_{(T-1),i}, \tau_{1:(T-1)}, \z_{1:(T-1),-i}, \z_{1:(T-2),i} ) \\
    & \quad \times p( z_{(T-1),i} | \tau_{1:(T-1)}, \z_{1:(T-1),-i}, \z_{1:(T-2),i} ) \\
    & \propto p(\tau_{T} | \z_{T}) \times p({\z}_{T} | z_{(T-1),i}, \tau_{1:(T-1)}, \z_{1:(T-1),-i},{\z}_{1:(T-2),i}) \\
    & \quad \times p( z_{(T-1),i} | \tau_{1:(T-1)}, \z_{1:(T-1),-i}, \z_{1:(T-2),i} ).
\end{align*}
Once the conditional distribution is represented as above, we can exploit the conditional independence relationships and get rid of terms not depending on  $z_{T-1,i}$ to obtain:
\begin{align*}
    p( z_{(T-1),i} & | \tau_{1:T}, \z_{1:T,-i}, z_{T,i}, \z_{1:(T-2),i})  \\
    & \propto p({\z}_{T} | z_{(T-1),i}, \tau_{1:(T-1)}, \z_{(T-1),-i}) \times p( z_{(T-1),i} | \tau_{1:(T-1)}, \z_{(T-1),-i}, z_{(T-2),i} ) \\
    & \propto p(z_{T,i} | z_{(T-1),i}, \tau_{1:(T-1)}, \z_{(T-1),-i}) \times \prod_{j \neq i} p(z_{T,j} | z_{(T-1),i}, \tau_{1:(T-1)}, \z_{(T-1),-i}) \\
    & \quad \times p( z_{(T-1),i} | \tau_{1:(T-1)}, \z_{(T-1),-i}, z_{(T-2),i} ) \\
    & \propto p(z_{T,i} | z_{(T-1),i}, \tau_{1:(T-1)}, \z_{(T-1),-i}) \times p( z_{(T-1),i} | \tau_{1:(T-1)}, \z_{(T-1),-i}, z_{(T-2),i} ).\\
    & \propto \mathcal{N} \Big(z_{T,i} | g(z_{T-1,i} ), 1 \Big) \\
    & \quad \times \mathcal{TN}_{ [\max(z_{(T-1),j} : \tau_{(T-1),j} < \tau_{(T-1),i}), \min(z_{(T-1),j} : \tau_{(T-1),j} > \tau_{(T-1),i})]  } \Big(z_{(T-1),i} | g(z_{(T-2),i} ), 1 \Big).
\end{align*}
The second term can be interpreted as an approximation of the filter that results from using a ``Gibbs'' approach by conditioning on $\z_{T-1,-i}$, and from conditioning also on the past $z_{T-2,i}$.

For the time period $T-2$, the conditional distribution of the $i$th latent score given the future and the past  time period of the  latent scores is equal to 
\begin{align*}
    p(z_{(T-2),i} & | \tau_{1:T}, \z_{(T-1):T,i}, \z_{1:T,-i}, \z_{1:(T-3),i}) = \\
    p(z_{(T-2),i} & | \tau_{1:T}, \z_{(T-1):T}, \z_{1:(T-1),-i}, \z_{1:(T-3),i}) \propto \\
    & \propto  p( \tau_T, \tau_{T-1}, \z_{(T-1):T} | z_{(T-2),i}, \tau_{1:(T-2)}, \z_{(T-2),-i}, \z_{1:(T-3)}) \\
    & \quad \times p( z_{(T-2),i} | \tau_{1:(T-2)}, \z_{(T-2),-i}, \z_{1:(T-3)}) \\
    & \propto p(\tau_T,  \z_{T} | \tau_{T-1}, \z_{(T-1)} z_{(T-2),i}, \tau_{1:(T-2)}, \z_{(T-2),-i}, \z_{1:(T-3)}) \\
    & \quad \times p(\tau_{T-1}, \z_{(T-1)} | z_{(T-2),i}, \tau_{1:(T-2)}, \z_{(T-2),-i}, \z_{1:(T-3)}) \\
    & \quad \times p( z_{(T-2),i}  | \tau_{1:(T-2)}, \z_{(T-2),-i}, \z_{1:(T-3)}).
\end{align*}
Conditioning on ${\z}_{(T-2)} $ removes dependence on $z_{T-1,i}$, therefore we write the conditional distribution as
\begin{align*}
    p(z_{(T-2),i} & | \tau_{1:T}, \z_{(T-1):T}, \z_{1:(T-1),-i}, z_{(T-1),i}, \z_{1:(T-3),-i}) \propto \\
    & \propto p( \tau_{T-1}, \z_{(T-1)} | z_{(T-2),i}, \tau_{1:(T-2)}, \z_{(T-2),-i}, \z_{1:(T-3)}) \\
    & \quad \times p( z_{T-2,i}  | \tau_{1:(T-2)}, \underline{\z}_{(T-2),-i}, \underline{\z}_{1:(T-3)}) \\ 
    & \propto p(\tau_{T-1}| \z_{(T-1)} ) \times p(  \z_{(T-1)} | z_{(T-2),i}, \tau_{1:(T-2)}, \z_{(T-2),-i}, \z_{1:(T-3)}) \\
    & \quad \times p( z_{(T-2),i}  | \tau_{1:(T-2)}, \z_{(T-2),-i}, \z_{1:(T-3)}). 
\end{align*}
Then, by conditional independence, one obtains
\begin{align*}
    p(z_{(T-2),i} & | \tau_{1:T}, \z_{(T-1):T}, \z_{1:(T-1),-i}, z_{(T-1),i}, \z_{1:(T-3),-i}) \propto \\
    & \propto p(z_{(T-1),i} | z_{(T-2),i}, \tau_{1:(T-2)}, \z_{(T-2),-i}) \times  \prod_{j \neq i} p(  z_{(T-1),j} | z_{(T-2),i}, \tau_{1:(T-2)}, \z_{(T-2),-i}) \\
    & \quad \times p( z_{(T-2),i}  | \tau_{1:(T-2)}, \z_{(T-2),-i}, \z_{1:(T-3)}) \\   
    & \propto p(z_{(T-1),i} | z_{(T-2),i}, \tau_{1:(T-2)},{\z}_{(T-2),-i}) \times  p( z_{(T-2),i}  | \tau_{1:(T-2)}, \z_{(T-2),-i}, \z_{1:(T-3)}) \\
    & \quad = \mathcal{N} \Big(z_{(T-1),i}|  g(z_{(T-2),i} ), 1 \Big) \\
    & \quad \times \mathcal{TN}_{ [\max(z_{(T-2),j} : \tau_{(T-2),j} < \tau_{(T-2),i}), \min(z_{(T-2),j} : \tau_{(T-2),j} > \tau_{(T-2),i})] } \big(z_{(T-2),i}| g(z_{(T-3),i} ), 1 \big),
\end{align*}
which is the product of a Gaussian distribution and a truncated Gaussian distribution over the interval $ [\max(z_{(T-2),j}: \tau_{(T-2),j} < \tau_{(T-2),i}), \min(z_{(T-2),j}: \tau_{(T-2),j} > \tau_{(T-2),i})]$.

Regarding the time period $T-3$, the conditional distribution of the latent score can be represented as
\begin{align*}
    p(z_{(T-3),i} & | \tau_{1:T}, \z_{(T-2):T,i}, \z_{1:T,-i}, \z_{1:(T-4),i}) \propto \\
    & \propto p(\tau_{(T-2): T}, \z_{(T-2):T} | z_{(T-3),i}, \tau_{1:(T-3)}, \z_{(T-3),-i}, \z_{1:(T-4)}) \\
    & \quad \times p( z_{(T-3),i}  | \tau_{1:(T-3)}, \z_{(T-3),-i}, \z_{1:(T-4)}) \\
    & \propto p(\tau_{(T-1):T},  \z_{(T-1):T} | \tau_{T-2}, \z_{(T-2)} z_{(T-3),i}, \tau_{1:(T-3)}, \z_{(T-3),-i}, \z_{1:(T-4)}) \\
    & \quad \times p(\tau_{T-2}, \z_{(T-2)} | z_{(T-3),i}, \tau_{1:(T-3)}, \z_{(T-3),-i}, \z_{1:(T-4)}) \\
    & \quad \times p( z_{(T-3),i}  | \tau_{1:(T-3)}, \z_{(T-3),-i}, \z_{1:(T-4)}).
\end{align*}
Conditioning on ${\z}_{(T-3)} $ removes dependence on $z_{(T-2),i}$, therefore we get
\begin{align*}
    p(z_{(T-3),i} & | \tau_{1:T}, \z_{(T-2):T,i}, \z_{1:T,-i}, \z_{1:(T-4),i}) \propto \\
    & \propto p(\tau_{T-2}, \z_{(T-2)} | z_{(T-3),i}, \tau_{1:(T-3)}, \z_{(T-3),-i}, \z_{1:(T-4)}) \\
    & \quad \times p( z_{(T-3),i}  | \tau_{1:(T-3)}, \z_{(T-3),-i}, \z_{1:(T-4)}) \\ 
    & \propto p(\tau_{T-2}| \z_{(T-2)} ) \times p(  \z_{(T-2)} | z_{(T-3),i}, \tau_{1:(T-3)}, \z_{(T-3),-i}, \z_{1:(T-4)}) \\
    & \quad \times p( z_{(T-3),i}  | \tau_{1:(T-3)}, \z_{(T-3),-i},{\z}_{1:(T-4)}).
\end{align*}
Since conditional independence appears, we have
\begin{align*}
    p(z_{(T-3),i} & | \tau_{1:T}, \z_{(T-2):T,i}, \z_{1:T,-i}, \z_{1:(T-4),i}) \propto \\
    & \propto p(z_{(T-2),i} | z_{(T-3),i}, \tau_{1:(T-3)}, \z_{(T-3),-i}) \times \prod_{j \neq i} p( z_{(T-2),j} | z_{(T-3),i}, \tau_{1:(T-3)}, \z_{(T-3),-i}) \\
    & \quad \times p( z_{(T-3),i}  | \tau_{1:(T-3)}, \z_{(T-3),-i},{\z}_{1:(T-4)}) \\   
    & \propto p(z_{(T-2),i} | z_{(T-3),i}, \tau_{1:(T-3)},{\z}_{(T-3),-i}) \times p( z_{T-3,i}  | \tau_{1:(T-3)},{\z}_{(T-3),-i},{\z}_{1:(T-4)}) \\
    & = \mathcal{N}\Big(z_{(T-2),i} | g(z_{(T-3),i} ), 1 \Big) \\
    & \quad \times \mathcal{TN}_{ [\max(z_{(T-3),j} : \tau_{(T-3),j} < \tau_{(T-3),i}), \min(z_{(T-3),j} : \tau_{(T-3),j} > \tau_{(T-3),i})] } \big(z_{(T-3),i} | g(z_{(T-4),i} ), 1 \big).
\end{align*}
Once we have provided the conditional distribution for different $T$, we can provide the general case,  $T-\ell$, where $\ell>3$. In this scenario, the probability density for the conditional distribution of the latent score is
\begin{align*}
    p(z_{(T-\ell),i} & | \tau_{1:T}, \z_{(T-\ell+1):T,i}, \z_{1:T,-i}, \z_{1:(T-\ell-1),i}) \propto \\
    & \propto p(\tau_{(T-\ell+1): T}, \z_{(T-\ell+1):T} | z_{(T-\ell),i}, \tau_{1:(T-\ell)}, \z_{(T-\ell),-i}, \z_{1:(T-\ell-1)}) \\
    & \quad \times p(z_{(T-\ell),i} | \tau_{1:(T-\ell)}, \z_{(T-\ell),-i}, \z_{1:(T-\ell-1)}) \\
    & \propto p(\tau_{(T-\ell+2):T}, \z_{(T-\ell+2):T} | \tau_{T-\ell+1}, \z_{(T-\ell+1)}, \tau_{1:(T-\ell)}, \z_{(T-\ell),-i}, \z_{1:(T-\ell-1)}) \\
    & \quad \times p(\tau_{T-\ell+1}, \z_{(T-\ell+1)} | z_{(T-\ell),i}, \tau_{1:(T-\ell)}, \z_{(T-\ell),-i}, \z_{1:(T-\ell-1)}) \\
    & \quad \times p(z_{(T-\ell),i} | \tau_{1:(T-\ell)}, \z_{(T-\ell),-i}, \z_{1:(T-\ell-1)}).
\end{align*}
Conditioning on ${\z}_{(T-\ell)} $ removes dependence on $z_{(T-\ell+1),i}$, one obtains
\begin{align*}
    p(z_{(T-\ell),i} & | \tau_{1:T}, \z_{(T-\ell+1):T,i}, \z_{1:T,-i}, \z_{1:(T-\ell-1),i}) \propto \\
    & \propto p(\tau_{T-\ell+1}, \z_{(T-\ell+1)} | z_{(T-\ell),i}, \tau_{1:(T-\ell)}, \z_{(T-\ell),-i}, \z_{1:(T-\ell-1)}) \\
    & \quad \times p( z_{(T-\ell),i}  | \tau_{1:(T-\ell)}, \z_{(T-\ell),-i}, \z_{1:(T-\ell-1)}) \\ 
    & \propto p(\tau_{T-\ell+1}| \z_{(T-\ell+1)} ) \times p({\z}_{(T-\ell+1)} | z_{(T-\ell),i}, \tau_{1:(T-\ell)}, \z_{(T-\ell),-i}, \z_{1:(T-\ell-1)}) \\
    & \quad \times  p( z_{(T-\ell),i}  | \tau_{1:(T-\ell)}, \z_{(T-\ell),-i}, \z_{1:(T-\ell-1)}).
\end{align*}
Then, by conditional independence, it is possible to rewrite the conditional distribution as
\begin{align*}
    p(z_{(T-\ell),i} & | \tau_{1:T}, \z_{(T-\ell+1):T,i}, \z_{1:T,-i}, \z_{1:(T-\ell-1),i}) \propto \\
    & \propto p(z_{(T-\ell+1),i} | z_{(T-\ell),i}, \tau_{1:(T-\ell)},{\z}_{(T-\ell),-i}) \times \prod_{j \neq i} p(z_{(T-\ell+1),j} | z_{(T-\ell),i}, \tau_{1:(T-\ell)}, \z_{(T-\ell),-i}) \\
    & \quad \times p( z_{(T-\ell),i}  | \tau_{1:(T-\ell)}, \z_{(T-\ell),-i}, \z_{1:(T-\ell-1)}) \\ 
    & \propto p(z_{(T-\ell+1),i} | z_{(T-\ell),i}, \tau_{1:(T-\ell)}, \z_{(T-\ell),-i}) \times p( z_{(T-\ell),i} | \tau_{1:(T-\ell)}, \z_{(T-\ell),-i}, \z_{1:(T-\ell-1)}) \\ 
    & = \mathcal{N} \Big(z_{(T- \ell + 1),i} | g(z_{(T- \ell),i} ), 1 \Big)  \\
    & \quad \times \mathcal{TN}_{ [\max(z_{(T- \ell),j} : \tau_{(T- \ell),j} < \tau_{(T- \ell),i}), \min(z_{(T- \ell),j} : \tau_{(T- \ell),j} > \tau_{(T- \ell),i})]  } \Big(z_{(T- \ell),i} |g(z_{(T- \ell -1),i} ), 1 \Big).
\end{align*}
Similar results hold for $T-\ell = 2$, while for $T-\ell = 1$, we derive the conditional distribution of the latent score as
\begin{align*}
    p(z_{1,i} & | \tau_{1:T}, \z_{2:T,i}, \z_{1:T,-i}) \propto  p( \tau_{2: T}, \z_{2:T} | z_{1,i}, \tau_{1}, \z_{1,-i}) \times p( z_{1,i} | \tau_{1}, \z_{1,-i}) \\
    & \propto  p( \tau_{3:T},  \z_{3:T} | \tau_{2}, \z_{2} z_{1,i}, \tau_{1}, \z_{1,-i}) \times p( \tau_{2}, \z_{2} | z_{1,i}, \tau_{1}, \z_{1,-i}) \times p( z_{1,i}  | \tau_{1}, \z_{1,-i}).
\end{align*}
Conditioning on $\z_{2}$ removes dependence on $z_{1,i}$, and therefore we write
\begin{align*}
    p(z_{1,i} & | \tau_{1:T}, \z_{2:T,i}, \z_{1:T,-i}) \propto p( \tau_{2}, \z_{2} | z_{1,i}, \tau_{1}, \z_{1,-i}) \times p( z_{1,i}  | \tau_{1}, \z_{1,-i}) \\ 
    & \propto p(\tau_2| \z_{2} )p(  \z_{2} | z_{1,i}, \tau_{1}, \z_{1,-i}) \times p( z_{1,i}  | \tau_{1},\z_{1,-i}). 
\end{align*}
Then, by conditional independence
\begin{align*}
    p(z_{1,i} & | \tau_{1:T}, \z_{2:T,i}, \z_{1:T,-i}) \propto  p(  {z}_{2,i} | z_{1,i}, \tau_{1}, \z_{1,-i}) \times  \prod_{j \neq i} p(  z_{2,j} | z_{1,i}, \tau_{1}, \z_{1,-i}) \times p( z_{2,i}  | \tau_{1}, \z_{1,-i}) \\ 
    & \propto p(z_{2,i} | z_{1,i}, \tau_{1}, \z_{1,-i}) \times p( z_{1,i}  | \tau_{1}, \z_{1,-i}),
\end{align*}
where $p( z_{2,i} | \tau_{1}, \z_{1,-i})$ is a Gaussian density and, for $t=1$, by integrating out $\underline{\z}_0$, we have
\begin{align*}
    p(z_{1,i} | \tau_1, \z_{1,-i} ) \propto \mathbb{I}\Big( \max_{j \neq i} \{ z_{j,1} : \tau_{j,1} < \tau_{i,1} \} < z_{i,1} < \min_{j \neq i} \{ z_{j,1} : \tau_{j.1} > \tau_{i,1} \}  \Big)   \sum_{k_i}^K  \mathcal{N}_1 \Big(z_{1,i} |  \tilde{\mu}_{k_i} , 1  \Big) q_{k_i, 1}.
\end{align*}
In conclusion, we obtain the conditional density for $T-\ell = 1$ as
\begin{align*}
    p(z_{1,i} &| \tau_{1:T}, \z_{1:T,-i}, \z_{2:T,i} ) = \mathcal{N}\Big(z_{2,i} | g(z_{1,i} ), 1 \Big) \\
    & \quad \times  \mathbb{I}\Big( \max_{j \neq i} \{ z_{j,1} : \tau_{j,1} < \tau_{i,1} \} < z_{i,1} < \min_{j \neq i} \{ z_{j,1} : \tau_{j.1} > \tau_{i,1} \}  \Big)   \sum_{k_i}^K  \mathcal{N}_1 \Big(z_{1,i} |  \tilde{\mu}_{k_i} , 1  \Big) q_{k_i, 1}.
\end{align*}
Below, we re-express the full conditional distributions derived above as mixture distributions for which sampling is straightforward. Initially, we start with the case $T- \ell = T$, where the full conditional distribution is defined as
\begin{align*}
    p(z_{T,i} | \tau_{1:T}, \z_{1:T,-i}, \z_{1:(T-1),i} ) = \mathcal{TN}_{ [\max(z_{T,j} : \tau_{T,j} < \tau_{T,i}), \min(z_{T,j} : \tau_{T,j} > \tau_{T,i})]  } \Big(z_{T,i} | g(z_{(T-1),i} ), 1 \Big),
\end{align*}
which is a truncated Gaussian distribution.

Moving to the general case $T-\ell$, by using the previously derived results, we obtain
\begin{align*}
    p(z_{(T-\ell),i} & | \tau_{1:T}, \z_{(T-\ell+1):T,i}, \z_{1:T,-i}, \z_{1:(T-\ell-1),i}) \propto \\
    & \propto p(z_{(T-\ell+1),i} | z_{(T-\ell),i}, \tau_{1:(T-\ell)}, \z_{(T-\ell),-i}) \times p( z_{(T-\ell),i}  | \tau_{1:(T-\ell)}, \z_{(T-\ell),-i}, \z_{1:(T-\ell-1)}) \\ 
    & = p(  z_{(T-\ell+1),i} | z_{(T-\ell),i}) \times  p( z_{(T-\ell),i}  | \tau_{1:(T-\ell)}, \z_{(T-\ell),-i}, \z_{1:(T-\ell-1)}).
  \end{align*}
Then we separate the probability $p(  z_{(T-\ell+1),i} | z_{(T-\ell),i}, \tau_{1:(T-\ell)}, \z_{(T-\ell),-i})$ into regions of $ z_{(T-\ell),i}$ defined by the sum-of-trees
\begin{align*}
    p(z_{(T-\ell+1),i} | z_{(T-\ell),i}, \tau_{1:(T-\ell)}, \z_{(T-\ell),-i}) = \mathcal{N} \Big(z_{(T- \ell + 1),i} | g(z_{(T- \ell),i} ), 1 \Big) \\
    = \sum_{\tilde{k}_{(T-\ell)} = 1}^{K_{(T-\ell)}}  \mathcal{N} \Big(z_{(T- \ell + 1),i} | \tilde{\mu}_{ \tilde{k}_{(T- \ell),i}}, 1 \Big) \mathbb{I}\big( z_{(T-\ell),i} \in C_{\tilde{k}_{(T-\ell)}} \big).
\end{align*}
This sum-of-tress is defined over the regions $z_{(T-\ell),i}$, denoted by $ z_{(T-\ell),i} \in C_{\tilde{k}_{(T-\ell)}}$, and note that $z_{(T-\ell),i}$ is in one of these regions, for which $ z_{(T-\ell+1),i}$ is normally distributed with constant mean.
Therefore, one obtains
\begin{align*}
    p(z_{(T-\ell),i} & | \tau_{1:T}, \z_{(T-\ell+1):T,i}, \z_{1:T,-i}, \z_{1:(T-\ell-1),i}) \propto \\
    & \propto p(z_{(T-\ell+1),i} | z_{(T-\ell),i}, \tau_{1:(T-\ell)}, \z_{(T-\ell),-i}) \times  p( z_{(T-\ell),i}  | \tau_{1:(T-\ell)}, \z_{(T-\ell),-i}, \z_{1:(T-\ell-1)}) \\ 
    & \propto \Bigg[ \sum_{\tilde{k}_{(T-\ell)} = 1}^{K_{(T-\ell)}}  \mathcal{N}\Big(z_{(T- \ell + 1),i} | \tilde{\mu}_{ \tilde{k}_{(T- \ell),i}}, 1 \Big) \mathbb{I}\big( z_{(T-\ell),i} \in C_{\tilde{k}_{(T-\ell)}} \big) \Bigg] \\
    & \quad \times \mathcal{TN}_{ [\max(z_{(T- \ell),j} : \tau_{(T- \ell),j} < \tau_{(T- \ell),i}), \min(z_{(T- \ell),j} : \tau_{(T- \ell),j} > \tau_{(T- \ell),i})]  } \Big(z_{(T- \ell),i} |g(z_{(T- \ell -1),i} ), 1 \Big) \\
    & = \sum_{\tilde{k}_{(T-\ell)} = 1}^{K_{(T-\ell)}}  \mathcal{N} \Big(z_{(T- \ell + 1),i} | \tilde{\mu}_{ \tilde{k}_{(T- \ell),i}}, 1 \Big) \times \mathcal{TN}_{ \left[ \tilde{C}_{\tilde{k}_{(T-\ell)}, i}^{(T-\ell),\text{min}},  \tilde{C}_{\tilde{k}_{(T-\ell)}, i}^{(T-\ell),\text{max}} \right]  } \Big(z_{(T- \ell),i} |g(z_{(T- \ell -1),i} ), 1 \Big).
\end{align*}

\noindent For the time period $t=1$, we define
\begin{align*}
    p(z_{1,i} & | \tau_{1:T}, \z_{2:T,i}, \z_{1:T,-i})  \propto p(  z_{2,i} | z_{1,i}, \tau_{1},{\z}_{1,-i}) \times p( z_{1,i}  | \tau_{1}, \z_{1,-i}).
\end{align*}
Then, we separate the probability $p(z_{2,i} | z_{1,i}, \tau_{1}, \z_{1,-i})$ into regions of $z_{1,i}$ defined by the sum-of-trees
\begin{align*}
    p(z_{2,i} | z_{1,i}, \tau_{1}, \z_{1,-i}) =  \sum_{\tilde{k}_{1} = 1}^{K_{1}} \mathcal{N} \Big(z_{2,i} | \tilde{\mu}_{ \tilde{k}_{1,i}}, 1 \Big) \mathbb{I}\Big(\tilde{C}_{\tilde{k}_{1}, i}^{(1),\text{min}} < z_{i,1} < \tilde{C}_{\tilde{k}_{1}, i}^{(1),\text{max}}  \Big), 
\end{align*}
which are defined over the regions $z_{1,i}$, denoted by $ z_{1,i} \in C_{\tilde{k}_{1}}$, and note that $z_{1,i}$ is in one of these regions, for which $ z_{2,i}$ is normally distributed with constant mean.

Therefore, we get the following full conditional distribution at time 1,
\begin{align*}
    p( z_{1,i} & | \tau_{1:T}, \z_{2:T,i}, \z_{1:T,-i}) \propto \\
    & \propto \sum_{\tilde{k}_{1} = 1}^{K_{1}} \mathcal{N} \Big(z_{2,i} | \tilde{\mu}_{ \tilde{k}_{1,i}}, 1 \Big) \mathbb{I}\Big(\tilde{C}_{\tilde{k}_{1}, i}^{(1),\text{min}} < z_{i,1} < \tilde{C}_{\tilde{k}_{1}, i}^{(1),\text{max}}  \Big) \times \sum_{k_{0,i}=1}^{K_0} \mathcal{N}_1 \Big(z_{1,i} |  \tilde{\mu}_{k_{0,i}} , 1  \Big) q_{k_{0,i}, 1} \\
    & = \sum_{\tilde{k}_{1} = 1}^{K_{1}} \sum_{k_{0,i}=1}^{K_0} \mathcal{N} \Big(z_{2,i} | \tilde{\mu}_{ \tilde{k}_{1,i}}, 1 \Big) \times q_{k_{0,i}, 1}  \times  \mathcal{TN}_{\big[ \tilde{C}_{\tilde{k}_{1}, i}^{min}, \tilde{C}_{\tilde{k}_{1}, i}^{max} \big]} \big(z_{1,i} | \tilde{\mu}_{k_{0,i}}, 1 \big).
\end{align*}

\paragraph{Including Covariates as Splitting Variables.}

If the trees can split on covariates, then the following steps must be taken into consideration. The mean of the latent score predicted by the sum of trees is now, $g(z_{T-\ell,i}, X_{T-\ell+1,i} )$. Furthermore, the relevant leaf at time period $T-\ell$ now depends on $X_{T-\ell+1,i}$. Depending on the values of $X_{T-\ell+1,i}$, the set of possible leaves corresponding to different values of $z_{T-\ell,i}$ will change. Therefore each tree's partition of possible values of $z_{T-\ell,i}$ is now specific to the time period $T-\ell$ and to $i$. The set of possible intersections of such regions across all trees in the sum is also specific to $T-\ell$ and to $i$. The filtering and smoothing distributions defined above would therefore involve $C_{\tilde{k}_{T-\ell+1, i}}$ with $\tilde{k}_{T-\ell+1, i} = 1,2,...,K_{T-\ell+1,i}$, and otherwise remain the same.

\subsubsection{ARROLinear model}
\label{sec:apdx_arrolinear_zsamples}

\paragraph{Case $0 < \ell < T-1$.}
First note that, as for ARROBART, we can generally obtain
\begin{align*}
p( z_{(T-\ell),i} & | \tau_{1:T}, \z_{(T-\ell+1):T}, \z_{1:(T-\ell-1)}, \z_{(T-\ell),-i}) = \\
p( z_{(T-\ell),i} & | \tau_{1:T}, z_{(T-\ell+1),i}, z_{(T-\ell-1),i}, \z_{(T-\ell),-i}) \propto \\
  & \propto p( z_{(T-\ell+1),i} | z_{(T-\ell),i}, \tau_{1:(T-\ell)}, \z_{(T-\ell),-i}) \times p( z_{(T-\ell),i}  | \tau_{1:(T-\ell)}, \z_{(T-\ell),-i}, \z_{1:(T-\ell-1)}).
\end{align*}
Then, applying the ARROLinear model we have
\begin{align*}
 & \propto \mathcal{N}(z_{T-\ell+1,i}| \x_{i, T-\ell+1} \bbeta + z_{T-\ell, i} \gamma,1) \times \mathcal{N}(z_{T-\ell, i} | \x_{i,T-\ell} \bbeta + z_{T-\ell-1, i}\gamma, 1) \\
 & \quad \times \mathbb{I}\big( \max(z_{(T-\ell), j} : \tau_{(T-\ell), j} < \tau_{(T-\ell), i}) < z_{T-\ell,i} < \min(z_{(T-\ell), j} : \tau_{(T-\ell), j} > \tau_{(T-\ell), i} ) \big) \\
 & \propto \exp\Big\{ \!-\frac{1}{2} \big[ (z_{T-\ell+1,i} - \x_{i, T-\ell+1} \bbeta - z_{T-\ell, i} \gamma )^2 \big] \Big\}  \exp\Big\{ -\frac{1}{2} \big[ (z_{T-\ell, i}- \x_{i,T-\ell} \bbeta - z_{T-\ell-1, i}\gamma)^2 \big] \Big\} \\
 & \quad \times \mathbb{I}\big( \max(z_{(T-\ell), j} : \tau_{(T-\ell), j} < \tau_{(T-\ell), i}) < z_{T-\ell,i} < \min(z_{(T-\ell), j} : \tau_{(T-\ell), j} > \tau_{(T-\ell), i} ) \big) \\
 & \propto \exp\Big\{ \! -\frac{1}{2} \big[ (z_{T-\ell+1,i} - \x_{i, T-\ell+1} \bbeta - z_{T-\ell, i} \gamma)^2 +( z_{T-\ell, i}- \x_{i,T-\ell} \bbeta - z_{T-\ell-1, i}\gamma)^2 \big] \Big\} \\
 & \quad \times \mathbb{I}\big( \max(z_{(T-\ell), j} : \tau_{(T-\ell), j} < \tau_{(T-\ell), i}) < z_{T-\ell,i} < \min(z_{(T-\ell), j} : \tau_{(T-\ell), j} > \tau_{(T-\ell), i} ) \big) \\
 & \propto \exp\Big\{ \! -\frac{1}{2} \big[ (z_{T-\ell+1,i} - \x_{i, T-\ell+1} \bbeta)^2 + (z_{T-\ell, i} \gamma )^2 - 2 (z_{T-\ell, i} \gamma)(z_{T-\ell+1,i} - \x_{i, T-\ell+1} \bbeta) \\
 & \quad + z_{T-\ell, i}^2 + (\x_{i,T-\ell} \bbeta + z_{T-\ell-1, i}\gamma)^2 - 2 (\x_{i,T-\ell} \bbeta + z_{T-\ell-1, i}\gamma) z_{T-\ell, i} \big] \Big\} \\
 & \quad \times \mathbb{I}\big( \max(z_{(T-\ell), j} : \tau_{(T-\ell), j} < \tau_{(T-\ell), i}) < z_{T-\ell,i} < \min(z_{(T-\ell), j} : \tau_{(T-\ell), j} > \tau_{(T-\ell), i} ) \big) \\
 & \propto \exp\{ \! -\frac{1}{2} \big[  (z_{T-\ell, i} \gamma )^2  - 2 (z_{T-\ell, i} \gamma )(z_{T-\ell+1,i} - \x_{i, T-\ell+1} \bbeta) \\
 & \quad + z_{T-\ell, i}^2 - 2 (\x_{i,T-\ell} \bbeta + z_{T-\ell-1, i}\gamma) z_{T-\ell, i} \big] \Big\} \\
 & \times \mathbb{I}\big( \max(z_{(T-\ell), j} : \tau_{(T-\ell), j} < \tau_{(T-\ell), i}) < z_{T-\ell,i} < \min(z_{(T-\ell), j} : \tau_{(T-\ell), j} > \tau_{(T-\ell), i} ) big) \\
 & \propto \exp\Big\{ \! -\frac{1}{2} \big[ (\gamma^2+1)z_{T-\ell, i}^2  - 2  (z_{T-\ell+1,i}\gamma - \x_{i, T-\ell+1} \bbeta \gamma + \x_{i,T-\ell} \bbeta  + z_{T-\ell-1, i}\gamma) z_{T-\ell, i} \big] \Big\} \\
 & \quad \times \mathbb{I}\big( \max(z_{(T-\ell), j} : \tau_{(T-\ell), j} < \tau_{(T-\ell), i}) < z_{T-\ell,i} < \min(z_{(T-\ell), j} : \tau_{(T-\ell), j} > \tau_{(T-\ell), i} ) \big).
\end{align*}
Then, we complete the square. In general, $ax^2+bx+c = a(x-h)^2+k$, where $h=-\frac{b}{2a}$ and $k = c - \frac{b^2}{4a}$. Here, we have $a = \gamma^2 + 1 $, $b= - 2  (z_{T-\ell+1,i}\gamma - \x_{i, T-\ell+1} \bbeta \gamma + \x_{i,T-\ell} \bbeta  + z_{T-\ell-1, i}\gamma) $, and $c=0$. Note also that $k$ does not depend on $z_{T-\ell, i}$. Therefore
\begin{align*}
 & \propto \exp\Bigg\{ \! - \frac{1}{2}\left[ (\gamma^2+1) \left( z_{T-\ell, i}^2 - \frac{2  (z_{T-\ell+1,i}\gamma - \x_{i, T-\ell+1} \bbeta \gamma + \x_{i,T-\ell} \bbeta  + z_{T-\ell-1, i}\gamma)}{2(\gamma^2+1)} \right)^2 + k \right] \Bigg\} \\
 & \quad \times \mathbb{I}\big( \max(z_{(T-\ell), j} : \tau_{(T-\ell), j} < \tau_{(T-\ell), i}) < z_{T-\ell,i} < \min(z_{(T-\ell), j} : \tau_{(T-\ell), j} > \tau_{(T-\ell), i}) \big) \\
 & \propto \exp\Bigg\{ \! - \frac{1}{2\frac{1}{(\gamma^2+1) }}\left(z_{T-\ell, i}^2   - \frac{   z_{T-\ell+1,i}\gamma - \x_{i, T-\ell+1} \bbeta \gamma + \x_{i,T-\ell} \bbeta  + z_{T-\ell-1, i}\gamma }{\gamma^2+1} \right)^2 \Bigg\} \\
 & \quad \times \mathbb{I}\big( \max(z_{(T-\ell), j} : \tau_{(T-\ell), j} < \tau_{(T-\ell), i})< z_{T-\ell,i} < \min(z_{(T-\ell), j} : \tau_{(T-\ell), j} > \tau_{(T-\ell), i}) \big),
\end{align*}
which gives a truncated Gaussian density with mean and variance
\begin{align*}
\mu & = \frac{z_{T-\ell+1,i}\gamma - \x_{i, T-\ell+1} \bbeta \gamma + \x_{i,T-\ell} \bbeta  + z_{T-\ell-1, i}\gamma}{\gamma^2+1}, \qquad
\sigma^2 = \frac{1}{\gamma^2+1}.
\end{align*}

\paragraph{Case $\ell=0$.}
For $T-\ell = T$ we have
\begin{align*}
p(z_{T,i} | \tau_{1:T}, \z_{1:T,-i}, \z_{1:(T-1),i} ) & = \mathcal{TN}_{ [\max(z_{T,j} : \tau_{T,j} < \tau_{T,i}), \min(z_{T,j} : \tau_{T,j} > \tau_{T,i})] } \Big(z_{T,i} | g(z_{T-1,i} ), 1 \Big) \\
 & = \mathcal{TN}_{ [\max(z_{T,j} : \tau_{T,j} < \tau_{T,i}), \min(z_{T,j} : \tau_{T,j} > \tau_{T,i})]  } \Big(z_{T,i} | \x_{i,T} \bbeta + z_{T-1, i}\gamma, 1 \Big).
\end{align*}

\paragraph{Case $\ell=T-1$.}
For $T-\ell = 1$ we have
\begin{align*}
    p( z_{1,i} | \tau_{1:T}, \z_{2:T,i}, \z_{1:T,-i}) & \propto p( z_{2,i} | z_{1,i}, \tau_{1}, \z_{1,-i}) \times p( z_{1,i} | \tau_{1}, \z_{1,-i}) \\
    & \propto  p( z_{2,i} | z_{1,i}, \tau_{1}, \z_{1,-i}) \times p( \tau_{1} |z_{1,i}, \z_{1,-i}) \times p( z_{1,i} | \z_{1,-i}).
\end{align*}
The first term above is a Gaussian density, the second is an indicator function, and it is necessary to derive the third term by considering the prior distribution over the unobserved latent value in the period before the first observation, $z_{0,i}$, as
\begin{align*}
    p( z_{1,i} | \z_{1,-i}) & = \int p( z_{1,i},z_{0,i} | \z_{1,-i}) \mathrm{d}z_{0,i} \\
    & = \int p( z_{1,i} |z_{0,i}, \z_{1,-i}) p( z_{0,i} | \z_{1,-i}) \mathrm{d} z_{0,i}.
\end{align*}
Then, note that $p(z_{1,i} | z_{0,i}, \z_{1,-i})$ and $p(z_{0,i} | \z_{1,-i})$ are both Gaussian densities with means $\x_{1,i}\bbeta + z_{0,i} \gamma$ and $z_{p,i}$, respectively, and variances equal to 1. Here $z_{p,i}$ denotes the prior mean of $z_{0,i}$ (while the prior variance is equal to 1)
\begin{align*}
    p( z_{1,i} | \z_{1,-i}) & = \int \frac{1}{\sqrt{2\pi}} \exp\Bigg\{ \!-\frac{1}{2} \Big( z_{1,i} - (\x_{1,i} \bbeta + z_{0,i} \gamma) \Big)^2 \Bigg\} \frac{1}{\sqrt{2\pi}} \exp\Bigg\{ \! -\frac{1}{2} \Big( z_{0,i} - z_{p,i} \Big)^2  \Bigg\}  \mathrm{d}z_{0,i} \\
    & \propto \int \exp\Bigg\{ \! -\frac{1}{2} \Bigg[ \Big( z_{1,i} - (\x_{1,i} \bbeta + z_{0,i} \gamma) \Big)^2 + \Big( z_{0,i} - z_{p,i} \Big)^2 \Bigg] \Bigg\}  \mathrm{d}z_{0,i} \\
    & \propto \int \exp\Bigg\{ \! -\frac{1}{2} \Bigg[ z_{0,i}^2 \gamma^2 + (z_{1,i} - \x_{1,i} \bbeta)^2  - 2 z_{0,i} \gamma (z_{1,i} - \x_{1,i} \bbeta) + z_{0,i}^2 + z_{p,i}^2 - 2 z_{0,i} z_{p,i} \Bigg] \Bigg\}  \mathrm{d}z_{0,i}.
\end{align*}
Taking out of the integral terms that are not functions of $z_{0,i}$ we obtain
\begin{align*}
    & \propto \exp \Bigg\{ -\frac{1}{2} \Bigg[  ( z_{1,i} - \x_{1,i} \bbeta )^2  + z_{p,i}^2  \Bigg] \Bigg\}  \int \exp\Bigg\{ \! -\frac{1}{2} \Bigg[ (\gamma^2+1) z_{0,i}^2 - 2 z_{0,i} \Big( \gamma (z_{1,i} - \x_{1,i} \bbeta) + z_{p,i} \Big) \Bigg] \Bigg\}  \mathrm{d}z_{0,i}.
\end{align*}
Then, we complete the square, $ax^2+bx+c = a(x-h)^2+k$, where $h=-\frac{b}{2a}$ and $k = c - \frac{b^2}{4a}$. Here, $a = \gamma^2+1 $, $b = - 2 \Big( \gamma ( z_{1,i} - \x_{1,i} \bbeta ) +z_{p,i} \Big)$, and $c=0$, thus
\begin{align*}
    & \propto \exp\Bigg\{ \! -\frac{1}{2} \Bigg[  ( z_{1,i} - \x_{1,i} \bbeta )^2 + z_{p,i}^2  \Bigg] \Bigg\} \\
    & \quad \times \int \exp \Bigg\{ -\frac{1}{2} \Bigg[ (\gamma^2+1) \Bigg(  z_{0,i} - \frac{2 ( \gamma (z_{1,i} - \x_{1,i} \bbeta) + z_{p,i})}{2(\gamma^2+1)} \Bigg)^2 - \frac{4 \Big(\gamma (z_{1,i} - \x_{1,i} \bbeta) + z_{p,i} \Big)^2}{4(\gamma^2+1)}  \Bigg] \Bigg\}  \mathrm{d}z_{0,i} \\
    & \propto \exp\Bigg\{ \! -\frac{1}{2} \Bigg[ (z_{1,i} - \x_{1,i} \bbeta)^2 + z_{p,i}^2 - \frac{ \Big( \gamma (z_{1,i} - \x_{1,i} \bbeta) + z_{p,i} \Big)^2 }{\gamma^2+1} \Bigg] \Bigg\} \\
    & \quad \times \int \exp\Bigg\{ \! -\frac{1}{2\frac{1}{\gamma^2+1}} \Bigg[ \Bigg( z_{0,i}    - \frac{ \gamma (z_{1,i} - \x_{1,i} \bbeta) + z_{p,i}}{\gamma^2+1} \Bigg)^2 \Bigg] \Bigg\}  \mathrm{d}z_{0,i}.
\end{align*}
The integrand above is proportional to the density of a Gaussian distribution with mean and variance equal to
\begin{align*}
\mu = \frac{\gamma (z_{1,i} - \x_{1,i} \bbeta) + z_{p,i}}{\gamma^2+1}, \qquad 
\sigma^2 = \frac{1}{\gamma^2+1}.
\end{align*}
Since there are no constraints on $z_{0,i}$ (no ranks are observed in period 0), the integral is over the range $(-\infty, \infty)$ and it is equal to the constant $\frac{2\pi}{\gamma^2+1}$, which is not a function of $z_{1,i}$. Therefore the integral is part of the constant of proportionality, and we may write
\begin{align*}
    & \propto \exp \Bigg\{ -\frac{1}{2} \Bigg[  ( z_{1,i} - \x_{1,i} \bbeta )^2 + z_{p,i}^2 - \frac{ \Big( \gamma ( z_{1,i} - \x_{1,i} \bbeta ) + z_{p,i} \Big)^2 }{\gamma^2+1} \Bigg] \Bigg\} \\
    & \propto \exp \Bigg\{ -\frac{1}{2} \Bigg[  ( z_{1,i} - \x_{1,i} \bbeta )^2   - \frac{ \Big( \gamma ( z_{1,i} - \x_{1,i} \bbeta ) +z_{p,i} \Big)^2 }{\gamma^2+1} \Bigg] \Bigg\}.
\end{align*}
Thus, it remains to rewrite the expression in the exponential as a quadratic function of $z_{1,i}$, and then complete the square to obtain a Gaussian density as
\begin{align*}
    & \propto \exp\Bigg\{ -\frac{1}{2} \Bigg[   z_{1,i}^2 + (\x_{1,i} \bbeta )^2  - 2 z_{1,i} (\x_{1,i} \bbeta ) - \frac{1 }{\gamma^2+1}  \Big( \gamma^2 ( z_{1,i} - \x_{1,i} \bbeta )^2 +z_{p,i}^2 + 2 \gamma ( z_{1,i} - \x_{1,i} \bbeta ) z_{p,i} \Big) \Bigg] \Bigg\} \\
    & \propto \exp\Bigg\{ -\frac{1}{2} \Bigg[   z_{1,i}^2 + (\x_{1,i} \bbeta )^2  - 2 z_{1,i} (\x_{1,i} \bbeta ) + \\
    & \quad - \frac{1}{\gamma^2+1}  \Big( \gamma^2 ( z_{1,i}^2 + (\x_{1,i} \bbeta)^2 - 2 z_{1,i} \x_{1,i} \bbeta )  + 2 \gamma  z_{1,i} z_{p,i} - 2 \gamma \x_{1,i} \bbeta  z_{p,i} \Big) \Bigg] \Bigg\} \\
    & \propto \exp\Bigg\{ -\frac{1}{2} \Bigg[ z_{1,i}^2 - 2 z_{1,i} (\x_{1,i} \bbeta) - \frac{1 }{\gamma^2+1}  \Big( \gamma^2 ( z_{1,i}^2 - 2 z_{1,i} \x_{1,i} \bbeta) + 2 \gamma  z_{1,i} z_{p,i}  \Big) \Bigg] \Bigg\} \\
    & \propto \exp\Bigg\{ -\frac{1}{2} \Bigg[ z_{1,i}^2 - 2 z_{1,i} (\x_{1,i} \bbeta ) - \frac{\gamma^2}{\gamma^2+1}  z_{1,i}^2  + 2 \frac{\gamma^2 \x_{1,i} \bbeta}{\gamma^2+1} z_{1,i} - \frac{2 \gamma z_{p,i}}{\gamma^2+1} z_{1,i} \Bigg] \Bigg\} \\
    & \propto \exp\Bigg\{ -\frac{1}{2} \Bigg[ \frac{1}{\gamma^2+1} z_{1,i}^2 + \Big( 2\frac{\gamma^2 \x_{1,i} \bbeta}{\gamma^2+1} - 2\x_{1,i} \bbeta - \frac{2 \gamma z_{p,i}}{\gamma^2+1} \Big) z_{1,i} \Bigg] \Bigg\} \\
    & \propto \exp\Bigg\{ -\frac{1}{2} \Bigg[ \frac{1}{\gamma^2+1}  z_{1,i}^2 + \Big( 2 \x_{1,i} \bbeta(\frac{\gamma^2}{\gamma^2+1}-1) - \frac{2 \gamma z_{p,i}}{\gamma^2+1} \Big) z_{1,i}   \Bigg] \Bigg\} \\
    & \propto \exp\Bigg\{ -\frac{1}{2} \Bigg[ \frac{1}{\gamma^2+1}  z_{1,i}^2 - \frac{2}{\gamma^2+1} \Big( \x_{1,i} \bbeta  + z_{p,i} \gamma \Big) z_{1,i} \Bigg] \Bigg\}.
\end{align*}
Then, we complete the square, $ax^2+bx+c = a(x-h)^2+k$, where $h=-\frac{b}{2a}$ and $k = c - \frac{b^2}{4a}$. Here $a = \frac{1}{\gamma^2+1} $, $b = - \frac{2}{\gamma^2+1} \Big( \x_{1,i} \bbeta  + z_{p,i} \gamma \Big)$, and $c=0$, thus
\begin{align*}
    & \propto \exp \Bigg\{ -\frac{1}{2} \Bigg[ \frac{1}{\gamma^2+1} \Big(  z_{1,i} -  \frac{2 (   \x_{1,i} \bbeta  + z_{p,i} \gamma )}{ 2\frac{\gamma^2+1}{\gamma^2+1} } \Big)^2  + k \Bigg] \Bigg\}.
\end{align*}
Rearranging, and noting that $k$ is not a function of $z_{1,i}$, we obtain
\begin{align*}
    p( z_{1,i} | \z_{1,-i}) & \propto \exp\Bigg\{ -\frac{1}{2} \frac{ 1 }{\gamma^2+1} \Bigg[       z_{1,i} -  (\x_{1,i} \bbeta  + z_{p,i} \gamma ) \Bigg]^2 \Bigg\}.
\end{align*}
This is proportional to a Gaussian density with mean $\x_{1,i} \bbeta + z_{p,i} \gamma$ and variance $\gamma^2+1 $. Returning to the overall expression to be derived, we have
\begin{align*}
    p( z_{1,i} & | \tau_{1:T}, \z_{2:T,i}, \z_{1:T,-i})  \propto  p(z_{2,i} | z_{1,i}, \tau_{1}, \z_{1,-i}) \times p( \tau_{1} |z_{1,i}, \z_{1,-i}) \times p( z_{1,i}  | \z_{1,-i}) \\
    & \propto \mathcal{N}(z_{2,i}| \x_{2,i} \bbeta  +  z_{1,i} \gamma, 1)  \mathbb{I}\big( \max(z_{1, j} : \tau_{1, j} < \tau_{1, i}) < z_{1,i} < \min(z_{1, j} : \tau_{1, j} > \tau_{1, i} ) \big) \\
    & \quad \times \mathcal{N}(z_{1,i}| \x_{1,i} \bbeta  + z_{p,i} \gamma, \gamma^2 + 1) \\
    & \propto  \mathbb{I}\big( \max(z_{1, j} : \tau_{1, j} < \tau_{1, i}) < z_{1,i} < \min(z_{1, j} : \tau_{1, j} > \tau_{1, i} ) \big) \\
    & \quad \times \frac{1}{\sqrt{2\pi}} \exp\Bigg\{ -\frac{1}{2} \Big( z_{2,i} - (\x_{2,i} \bbeta + z_{1,i} \gamma) \Big)^2 \Bigg\} \\
    & \quad \times \frac{1}{\sqrt{2\pi}} \exp\Bigg\{ -\frac{1}{2(\gamma^2 + 1)} \Big( z_{1,i} - (\x_{1,i} \bbeta + z_{p,i} \gamma)  \Big)^2  \Bigg\} \\
    & \propto \mathbb{I}\big( \max(z_{1, j} : \tau_{1, j} < \tau_{1, i}) < z_{1,i} < \min(z_{1, j} : \tau_{1, j} > \tau_{1, i} ) \big) \\
    & \quad \times \exp\Bigg\{ -\frac{1}{2} \Bigg[ \Big( z_{2,i} - (\x_{2,i} \bbeta + z_{1,i} \gamma)  \Big)^2 + \frac{1}{\gamma^2+1} \Big( z_{1,i} - (\x_{1,i} \bbeta + z_{p,i} \gamma)  \Big)^2  \Bigg]\Bigg\}.
\end{align*}
It remains to express the terms inside the exponential function as a quadratic function of $z_{1,i}$ and complete the square to obtain a Gaussian density, as
\begin{align*}
    & \propto  \mathbb{I}\big( \max(z_{1, j} : \tau_{1, j} < \tau_{1, i}) < z_{1,i} < \min(z_{1, j} : \tau_{1, j} > \tau_{1, i} ) \big) \\
    & \quad \times \exp\Bigg\{ -\frac{1}{2} \Bigg[z_{1,i}^2 \gamma^2  + (z_{2,i}- \x_{2,i} \bbeta)^2  - 2 (z_{2,i}- \x_{2,i} \bbeta) \gamma z_{1,i} \\
    & \quad \quad + \frac{1}{\gamma^2+1} \Big( z_{1,i}^2 + (\x_{1,i} \bbeta + z_{p,i} \gamma)^2-2 z_{1,i} (\x_{1,i} \bbeta + z_{p,i} \gamma)  \Big)  \Bigg]\Bigg\} \\
    & \propto \mathbb{I}\big( \max(z_{1, j} : \tau_{1, j} < \tau_{1, i}) < z_{1,i} < \min(z_{1, j} : \tau_{1, j} > \tau_{1, i} ) \big) \\
    & \quad \times \exp\Bigg\{ -\frac{1}{2} \Bigg[ z_{1,i}^2 \gamma^2 - 2 (z_{2,i}- \x_{2,i} \bbeta) \gamma z_{1,i}  + \frac{1}{\gamma^2+1} \Big( z_{1,i}^2 -2 z_{1,i} (\x_{1,i} \bbeta + z_{p,i} \gamma)  \Big) \Bigg] \Bigg\} \\
    & \propto \mathbb{I}\big( \max(z_{1, j} : \tau_{1, j} < \tau_{1, i}) < z_{1,i} < \min(z_{1, j} : \tau_{1, j} > \tau_{1, i}) \big)  \\
    & \quad \times \exp\Bigg\{ -\frac{1}{2} \Bigg[ (\gamma^2 + \frac{1}{\gamma^2+1}) z_{1,i}^2   -2  \Big( (z_{2,i}- \x_{2,i} \bbeta) \gamma  + \frac{1}{\gamma^2+1} (\x_{1,i} \bbeta + z_{p,i} \gamma)  \Big)z_{1,i}  \Bigg] \Bigg\}.
\end{align*}
Then, we complete the square, $ax^2+bx+c = a(x-h)^2+k$, where $h=-\frac{b}{2a}$ and $k = c - \frac{b^2}{4a}$. Here $a = (\gamma^2 + \frac{1}{\gamma^2+1}) $, $b = -2  \Big( (z_{2,i}- \x_{2,i} \bbeta) \gamma  + \frac{1}{\gamma^2+1} (\x_{1,i} \bbeta + z_{p,i} \gamma)  \Big)$, and $c=0$, thus
\begin{align*}
    & \propto \mathbb{I}\big( \max(z_{1, j} : \tau_{1, j} < \tau_{1, i}) < z_{1,i} < \min(z_{1, j} : \tau_{1, j} > \tau_{1, i} ) \big) \\
    & \quad \times \exp\Bigg\{ -\frac{1}{2} \Bigg[ (\gamma^2 + \frac{1}{\gamma^2+1}) \Bigg(   z_{1,i} - \frac{ (z_{2,i}- \x_{2,i} \bbeta) \gamma  + \frac{1}{\gamma^2+1} (\x_{1,i} \bbeta + z_{p,i} \gamma)  }{\gamma^2 + \frac{1}{\gamma^2+1}}  \Bigg)^2 +k \Bigg] \Bigg\} \\
    & \propto \mathbb{I}(\max(z_{1, j} : \tau_{1, j} < \tau_{1, i}) < z_{1,i} < \min(z_{1, j} : \tau_{1, j} > \tau_{1, i} ) \big) \\
    & \quad \times \exp\Bigg\{ -\frac{1}{2} (\gamma^2 + \frac{1}{\gamma^2+1}) \Bigg( z_{1,i} - \frac{ (z_{2,i}- \x_{2,i} \bbeta) \gamma  + \frac{1}{\gamma^2+1} (\x_{1,i} \bbeta + z_{p,i} \gamma) }{\gamma^2 + \frac{1}{\gamma^2+1}} \Bigg)^2 \Bigg\},
\end{align*}
which is a truncated Gaussian distribution with mean and variance equal to
\begin{align*}
    \mu = \frac{ (z_{2,i}- \x_{2,i} \bbeta) \gamma  + \frac{1}{\gamma^2+1} (\x_{1,i} \bbeta + z_{p,i} \gamma)  }{\gamma^2 + \frac{1}{\gamma^2+1}}, \qquad
    \sigma^2 = \frac{1}{\gamma^2 + \frac{1}{\gamma^2+1}}.
\end{align*}
Equivalently, this is a truncated Gaussian distribution with mean and variance
\begin{align*}
    \mu = \frac{ \gamma(\gamma^2 + 1) (z_{2,i}- \x_{2,i} \bbeta)  +  \x_{1,i} \bbeta + z_{p,i} \gamma}{(\gamma^2 + 1)\gamma^2 + 1}, \qquad
    \sigma^2 = \frac{\gamma^2 + 1}{(\gamma^2 + 1)\gamma^2 + 1}.
\end{align*}
Therefore, one has
\begin{align*}
    & p( z_{1,i}  | \tau_{1:T}, \z_{2:T,i}, \z_{1:T,-i}) \propto \\
    & \propto \mathcal{TN}_{[\max(z_{1, j} : \tau_{1, j} < \tau_{1, i}), \min(z_{1, j} : \tau_{1, j} > \tau_{1, i} )]} \Bigg(z_{1,i} \Bigg| \frac{\gamma(\gamma^2 + 1) (z_{2,i}- \x_{2,i} \bbeta) + \x_{1,i} \bbeta + z_{p,i} \gamma}{(\gamma^2 + 1)\gamma^2 + 1}, \frac{\gamma^2 + 1}{(\gamma^2 + 1)\gamma^2 + 1} \Bigg).
\end{align*}

\subsection{Samplers for Data with Partial Rankings}
\label{sec:apdx_sampler_partial_rankings}

\subsubsection{Sampler for Static Data with Partial Rankings}

Let $\tau = (\tilde\tau, \tau_\delta) \in \mathbb{N}^N$, where $\tilde\tau$ is the vector of observed ranks and $\tau_{\delta}$ is the collection of the remaining $N-k$ unobserved ranks.
Similarly, let $\z = (\tilde\z, \z_\delta) \in \R^N$,  where $\tilde\z$ is the latent score of the top-$k$ observed ranks and $\z_\delta$ is the score of the remaining $N-k$ unobserved ranks.
\begin{equation}
\begin{split}
    \tau & = \operatorname{rank}(\z),  \qquad
    \z  = \boldsymbol{\gamma} + \boldsymbol\varepsilon, \qquad \boldsymbol\varepsilon \sim \mathcal{N}_N(\mathbf{0}, \Sigma_\varepsilon) \equiv \mathcal{N}_N(\mathbf{0}, I_N).
\end{split}
\label{eq:model_TMD}
\end{equation}
Therefore, considering $M$ rankers, the likelihood is
\begin{align*}
    L(\tilde\tau_1,\ldots,\tilde\tau_M | \boldsymbol{\theta}) 
     & = \prod_{j=1}^M \int_{\R^N} p(\tilde\tau_{j}, \z_{j} | \boldsymbol{\theta}) \: \mathrm{d}\z_{j} 
       = \prod_{j=1}^M \int_{\R^N} p(\underbrace{\tilde\z_{j}, \z_{\delta,j}}_{\z_j} | \boldsymbol{\theta}) \times p(\tilde\tau_{j} | \z_{j}) \: \mathrm{d}\z_{j} \\
     & = \prod_{j=1}^M \int_{\R^N} \mathcal{N}_N\big( \z_j | f(\mathbf{X}_{ij}), \: I_N \big) \times \delta_{\operatorname{rank}(\z_{j})}(\tilde\tau_{j}) \: \mathrm{d}\z_j \\
    & = \prod_{j=1}^M \int_{\R^N} (2\pi)^{-\frac{N}{2}} \exp\Bigg( -\frac{\big\|\z_{j} - f(\mathbf{X}_{ij}) \big\|^2}{2} \Bigg) \delta_{\operatorname{rank}(\z_{j})}(\tilde\tau_{j}) \: \mathrm{d}\z_j,
\end{align*}
where $\delta_{\operatorname{rank}(\z_{j})}(\tilde\tau_{j})$ is a binary indicator function equal to 1 if the observed ranks agree with the latent variable vector $\operatorname{rank}(\z_{j})$ and 0 otherwise. This likelihood differs from that of the full ranking setting in so far as partial rankings imply fewer constraints. The exact constraints depend on the type of missingness of rankings. For example, in the top-$k$ ranking setting, it is known that the items outside of the top $k$, i.e. not one of the $k$ lowest (best) rankings, for which rankings are unobserved, must have latent outcome values above those of the items with observed rankings.

The sampler proceeds in the same way as for the full ranking case, except there are less constraints on the sampled latent variables in the step in which the latent variables are sampled in each MCMC iteration. 

Note that our extension to partial ranking data follows the approach of \cite{li2016bayesian}, and relies on a missingness at random assumption, which is satisfied by many common types of partial ranking data, including top-$k$ rankings.
Specifically, the assumption postulates
\begin{equation*}
    p(\bdelta_j| \tau, \Z_j, \mathcal{T}_1, \ldots, \mathcal{T}_S, \bmu_{1}, \ldots, \bmu_{S}, \X) = p(\bdelta_j | \tilde\tau, \tilde{\Z}_j,\mathcal{T}_1, \ldots, \mathcal{T}_S, \bmu_{1}, \ldots, \bmu_{S}, \X)
\end{equation*}
where $\tilde{\Z}_j$ denotes the scores of the items with observed rankings for ranker $j$. This assumption states that the probability of missingness is unrelated to the rankings or scores of the missing items conditional on the rankings and scores of the non-missing items. In other words, missingness provides no additional information about the rankings of the items for which rankings are missing, beyond that provided by the information relevant to the observed rankings.

\subsubsection{Sampler for Dynamic Data with Partial Rankings}

The ARROBART and ARROLinear models for data with partial rankings are adjusted in a similar way to the static case described above. Indicator functions for agreement of latent outcomes with rankings are re-defined so that they only relate to observe rankings and other inequalities implied by the type of missingness (e.g., the top-$k$ setting). The derivations of the samplers described in this supplement and in the main text proceed in the same way as for the full ranking case except the constrained spaces of latent outcome vectors, $A_t$, which are defined by different constraints (these constraints are less restrictive because fewer ranks are observed). The samplers are adjusted similarly by applying constraints defined by observed ranking information. 

\section{Model Specifications and  Simulation Study Details}
\label{sec:apdx_simualtion_details}

\subsection{Model specifications in Section 4.1}
The performance of the ROBART model defined in Section 4 is compared to the methods for rank-order data introduced by \cite{li2016bayesian}.
Specifically, we consider the following models.
The Borda Count uses the arithmetic means of the average ranking positions over all ranking lists, the Markov Chain-based methods are based on the stationary distributions of Markov chains whose transition matrices are constructed based on the ranking lists, the CEMC approach tries to find the ranking list minimising the average distance from all ranking lists using a stochastic search method. See the supplementary material of \cite{li2016bayesian} for a review of these methods. The MLE of the Plankett-Luce model is computed using the \texttt{R} package \texttt{StatRank}, which uses a minorization-maximization algorithm. See the supplementary material of \cite{li2016bayesian} for a review of these methods.
For the BARC, BARCM and ROBART, we retain 10,000 MCMC draws after 2,000 burn-in draws.

\subsection{Model specifications in Section 4.2}

For the methods compared in the main text, the latent scores are modelled as follows. All coefficients and intercepts in linear models gave Gaussian prior distributions. Here, for clarity, we use  $\x_{i,t}$ to denote covariates excluding the lag of observed ranks and the lag of the latent scores, whereas in the main text $\x_{i,t}$ can include the lag of observed ranks and the lag of the latent scores.

\begin{itemize}
    \item ROBART-lag. A ROBART model using the first lag of the response variable as a covariate:
    \begin{equation*}
        z_{i} = f(z_{i,t-1}, \tau_{i,t-1})  + \epsilon_{i},
    \end{equation*}
    where $f(\cdot, \cdot)$ has a BART specification. 
    \item ROBARTX-lag. A ROBART model using the first lag of the response variable as a covariate:
    \begin{equation*}
        z_{i,t} = f(z_{i,t-1}, \tau_{i,t-1}, \x_{i,t})  + \epsilon_{i,t},
    \end{equation*}
    where $\x_{i,t}$ refers to the covariates.    
    \item ROLinear-lag. A linear rank order model (BARC) with the first lag of the response variable:
    \begin{equation*}
        z_{i,t} = \beta_{0,i} +  \beta_{1} \tau_{i,t-1}  + \epsilon_{i,t}.
    \end{equation*}

    \item ARROLinear model. It replaces the unknown function $f$ with a linear specification, thus obtaining an autoregressive linear model for the latent score:
    \begin{equation*}
        z_{i,t} =  \alpha_{0,i} + \alpha_1 z_{i,t-1} + \epsilon_{i,t},
    \end{equation*}

    \item ARROLinearX model. Analogous to the ARROLinear, with the addition of exogenous covariates:
    \begin{equation*}
        z_{i,t} =  \alpha_{0,i} + \alpha_1 z_{i,t-1} + \boldsymbol\eta' \x_{i,t} + \epsilon_{i,t}.
    \end{equation*}

    \item ARROLinear-lag model. Linear specification, adding the lag-1 observed ranking as a covariate:
    \begin{equation*}
        z_{i,t} = \alpha_{0,i} + \alpha_{1} z_{i,t-1} + \beta \tau_{i,t-1} + \epsilon_{i,t},
    \end{equation*}
    
    \item ARROBARTX model. An ARROBART specification with the addition of exogenous covariates:
    \begin{equation*}
        z_{i,t} = f(z_{i,t-1},  \x_{i,t})   + \epsilon_{i,t}.
    \end{equation*}
    
    \item ARROBART-lag model. The transition equation for the latent scores includes the lagged latent score and the lag-1 value of the item's observed ranking:
    \begin{equation*}
        z_{i,t} = f(z_{i,t-1}, \tau_{i,t-1}) + \epsilon_{i,t},
    \end{equation*}
    where $f$ has a BART specification.
    \item ARROBARTX-lag model. The transition equation for the latent scores includes the lagged latent score and the lag-1 value of the item's observed ranking:
    \begin{equation*}
        z_{i,t} = f(z_{i,t-1}, \tau_{i,t-1}, \x_{i,t}) + \epsilon_{i,t}.
    \end{equation*}
\end{itemize}

\subsection{Static Simulation Study with Partial Rankings}

We demonstrate the ability of our model to predict rankings using partial ranking information through a simulation study described by \cite{li2016bayesian}. We divide the $N=80$ items into $K = 1,2,4,8,10,16$ groups, of $N/K$ items. Pairwise comparison information is only available within groups. The DGP is otherwise the same as scenario 2.  Table \ref{tab:li_tablePartial_sigma1} contains the tau correlations of rank predictions averaged over 10 repetitions of the simulations study with latent error standard deviations $\sigma=1$ and $\sigma=5$, as in \cite{li2016bayesian}. The results confirm that ROBART still provides a considerable improvement over linear models when only partial rankings are available. Moreover, ROBART with soft splitting rules introduced by \cite{linero2018dart} can provide further large improvements in ranking predictions.

\begin{table}[ht]
\centering
\footnotesize
\setlength{\tabcolsep}{4pt}
\begin{adjustbox}{max width=\textwidth}
\begin{threeparttable}
\captionsetup{width=0.93\linewidth}
\caption{\label{tab:li_tablePartial_sigma1} \small Partial ranking simulation results, $\sigma=1$. Comparison between ROBART, BARC, and other ranking methods.}
\hspace*{-2ex}
\begin{tabular}{ll *{8}{c}}
  \hline
& & K & BAR & BARC1 & BARC2 & BARC3 & ROBART & RODART & SoftROBART \\ 
  \hline
\multirow{5}{1em}{
		\begin{rotate}{90} \hspace{-22pt} $\sigma=1$ \end{rotate}
		} & \multirow{5}{1em}{
		\begin{rotate}{90} \hspace{-22pt} Scenario 2 \end{rotate}
		} & 1 & (0.03) & 1.00 & 1.00 & 1.00 & 0.83 & 0.84 & \textbf{0.67} \\ 
& & 2 & (0.07) & 0.61 & 0.60 & 0.59 & 0.32 & 0.34 & \textbf{0.28} \\ 
& & 4 & (0.08) & 0.59 & 0.62 & 0.64 & 0.35 & 0.39 & \textbf{0.28} \\ 
& & 8 & (0.13) & 0.51 & 0.49 & 0.56 & 0.24 & 0.27 & \textbf{0.21} \\ 
& & 10 & (0.12) & 0.59 & 0.59 & 0.65 & 0.26 & 0.33 & \textbf{0.22} \\ 
& & 16 & (0.17) & 0.55 & 0.56 & 0.58 & 0.23 & 0.36 & \textbf{0.21} \\ 
   \hline
\multirow{5}{1em}{
		\begin{rotate}{90} \hspace{-22pt} $\sigma=5$ \end{rotate}
		} & \multirow{5}{1em}{
		\begin{rotate}{90} \hspace{-22pt} Scenario 2 \end{rotate}
		} & 1 & (0.12) & 0.83 & 0.83 & 0.98 & 0.60 & 0.68 &\textbf{0.46} \\ 
& & 2 & (0.13) & 0.80 & 0.79 & 1.04 & 0.54 & 0.64 & \textbf{0.46} \\ 
& & 4 & (0.14) & 0.78 & 0.78 & 0.95 & 0.56 & 0.73 & \textbf{0.46} \\ 
& & 8 & (0.17) & 0.63 & 0.64 & 0.84 & 0.46 & 0.58 & \textbf{0.36} \\ 
& & 10 & (0.16) & 0.65 & 0.65 & 0.88 & 0.45 & 0.57 & \textbf{0.37} \\ 
& & 16 & (0.19) & 0.60 & 0.60 & 0.78 & 0.42 & 0.58 & \textbf{0.35} \\ 
\hline
\hline
\end{tabular}
\begin{tablenotes}
\footnotesize
\item \!\!\!\! Notes: scaled Kendall tau distance across the same scenarios as in \cite{li2016bayesian}. The BAR column with parentheses shows the average Kendall tau distances between estimated and true ranking lists for the BAR predictions, while the remaining columns show tau distances relative to the tau distance attained by BAR.
\end{tablenotes}
\end{threeparttable}
\end{adjustbox}
\end{table}

\FloatBarrier

\subsection{Static Simulation Study with Ranker-specific Covariates}

Table \ref{tab:li_table)rankerspecific} contains the results for a variation on the static simulation study described by \cite{li2016bayesian} in which the covariates are ranker-specific. For each $ (i,m) \in  \{1,\ldots, N\} \times \{1,\dots,M\}$, item $(i,m)$ has a true score $\gamma_{i,m}$, while the covariate vectors $\mathbf{x}_{i,m} = (x_{i,m,1},\ldots,x_{i,m,K_x})'$ are drawn independently from a Gaussian distribution with zero mean and covariance equal to $\mathbb{C}ov(x_{i,m,l}, x_{i,m,s}) = \rho^{\abs{l-s}}$, for $1 \leq l, s \leq v$ and $\abs{\rho} < 1$. The true score vector $\bm{\gamma}$ is generated by considering the different roles of the covariates as:
\begin{enumerate}
	\item $\gamma_{i,m} = \mathbf{x}_{i,m}' \boldsymbol{\beta}$, where $\boldsymbol{\beta} = (3,2,-1,-0.5)'$, $K_x=4$, and $\rho=0$;
	\item $\gamma_{i,m} = \mathbf{x}_{i,m}' \boldsymbol{\beta} + \norm{\mathbf{x}_{i,m}}^2 $, where $\boldsymbol{\beta} = (3,2,1)'$, $K_x=3$, and $\rho=0.5$;
	\item $\gamma_{i,m} = \norm{\mathbf{x}_{i,m}}^2 $, where $K_x=4$, and $\rho=0.5$.
\end{enumerate}

\begin{table}[ht]
\centering
\footnotesize
\setlength{\tabcolsep}{4pt}
\begin{adjustbox}{max width=\textwidth}
\begin{threeparttable}
\captionsetup{width=0.93\linewidth}
\caption{\label{tab:li_table)rankerspecific} \small Comparison between ROBART and other ranking methods. Ranker-specific covariates.}
\hspace*{-2ex}
\begin{tabular}{l *{13}{c}}
		\hline
		 & $\sigma$ & Borda & MC1 & MC2 & MC3 & CEMC & PL & BAR & $\text{BARC}_1$ & $\text{BARC}_2$ & $\text{BARC}_3$ & ROBART \\ 
		\hline
		\multirow{5}{1em}{
		\begin{rotate}{90} \hspace{-22pt} Scenario 1 \end{rotate}
		} & 1 & (0.40) & 1.27 & 1.00 & 1.00 & 1.00 & 1.03 & 1.27 & 1.27 & 1.27 & 1.27 & 0.09 \\ 
		& 5 &  (0.44) & 1.15 & 1.00 & 1.00 & 1.01 & 1.02 & 1.15 & 1.15 & 1.15 & 1.15 & 0.20 \\ 
		& 10 & (0.47) & 1.08 & 1.00 & 1.00 & 1.00 & 1.00 & 1.08 & 1.08 & 1.08 & 1.08 & 0.31 \\ 
		& 20 & (0.48) & 1.03 & 1.00 & 1.00 & 1.00 & 1.00 & 1.03 & 1.03 & 1.03 & 1.03 & 0.51 \\ 
		& 40 & (0.49) & 1.02 & 1.00 & 1.00 & 1.00 & 1.00 & 1.02 & 1.02 & 1.02 & 1.02 & 0.72 \\ 
		\hline
		\multirow{5}{1em}{
		\begin{rotate}{90} \hspace{-22pt} Scenario 2 \end{rotate}
		} & 1 & (0.40) & 1.27 & 1.00 & 1.00 & 1.00 & 1.03 & 1.27 & 1.27 & 1.27 & 1.27 & 0.10 \\ 
		& 5 &  (0.43) & 1.16 & 1.00 & 1.00 & 1.01 & 1.03 & 1.16 & 1.16 & 1.16 & 1.16 & 0.17 \\ 
		& 10 & (0.46) & 1.09 & 1.00 & 1.00 & 1.00 & 1.02 & 1.09 & 1.09 & 1.09 & 1.09 & 0.26 \\ 
		& 20 & (0.48) & 1.05 & 1.00 & 1.00 & 1.00 & 1.01 & 1.05 & 1.05 & 1.05 & 1.05 & 0.41 \\ 
		& 40 & (0.49) & 1.03 & 1.00 & 1.00 & 1.00 & 1.00 & 1.03 & 1.03 & 1.03 & 1.03 & 0.62 \\ 
		\hline
		\multirow{5}{1em}{
		\begin{rotate}{90} \hspace{-22pt} Scenario 3 \end{rotate}
		} & 1 & (0.41) & 1.25 & 1.00 & 1.00 & 1.01 & 1.04 & 1.25 & 1.25 & 1.25 & 1.25 & 0.13 \\ 
		& 5 & (0.45) & 1.10 & 1.00 & 1.00 & 1.01 & 1.02 & 1.10 & 1.10 & 1.10 & 1.10 & 0.27 \\ 
		& 10 &  (0.47) & 1.05 & 1.00 & 1.00 & 1.00 & 1.01 & 1.05 & 1.05 & 1.05 & 1.05 & 0.44 \\ 
		& 20 & (0.49) & 1.03 & 1.00 & 1.00 & 1.00 & 1.00 & 1.03 & 1.03 & 1.03 & 1.03 & 0.64 \\ 
		& 40 & (0.49) & 1.02 & 1.00 & 1.00 & 1.00 & 1.00 & 1.02 & 1.02 & 1.02 & 1.02 & 0.79 \\ 
		\hline
\end{tabular}
\begin{tablenotes}
\footnotesize
\item \!\!\!\! Notes: scaled Kendall tau distance across modified scenarios in which the covariates are ranker-specific. The Borda column with parentheses shows the average Kendall tau distances between estimated and true ranking lists for the Borda Count predictions, while the remaining columns show tau distances relative to the tau distance attained by the Borda Count method. 
\end{tablenotes}
\end{threeparttable}
\end{adjustbox}
\end{table}

\FloatBarrier

\subsection{Dynamic Simulation Study with Ranker-specific Models}
Table \ref{tab:dynsimstudy_sig1_rankerspecific} contains the results for an alternative approach in which a separate model is fitted to each ranker's time series of rankings. The DGP is scenario 1 of the dynamic simulation study described in the main text. The results show that all methods except ARROBART (without the lag of observed ranks) provide predictions that are uncorrelated with the true rankings. The ARROBART results are also somewhat disappointing. This result might be explained by the small number of observations per ranker for training of ranker-specific models. We only include results for $\sigma=0.1$ due to the high computational cost of estimation of separate models for each ranker, and the expectation that large improvements are unlikely to be observed for higher values of $\sigma$.

\begin{table}[t!h]
\centering
\begin{threeparttable}
\captionsetup{width=0.90\linewidth}
\caption{\label{tab:dynsimstudy_sig1_rankerspecific} \small Average Kendall's tau distance of 1 period ahead forecasts from true values for ranker-specific models, scenario 1, $\sigma=0.1$.}
\begin{tabular}{ccccc}
  \toprule
  ARRO \newline BART & RO \newline BART lag & ARRO \newline linear lag & ARRO \newline linear & RO \newline linear lag \\ 
  \midrule
  0.44 & 0.48 & 0.50 & 0.50 & 0.50 \\ 
 \bottomrule
\end{tabular}
\begin{tablenotes}
\footnotesize
\item \!\!\!\! 
\end{tablenotes}
\end{threeparttable}
\end{table}

\FloatBarrier

\section{MCMC Diagnostics and Computational Details}\label{MCMCdiag_sec}

\subsection{MCMC Diagnostics}
This subsection provides different MCMC diagnostics for the models considered in the simulation experiments. We present the results of the Geweke test applied to Kendall's tau averaged across rankers in each iteration for different values of $\sigma$ (see Tables~\ref{tab:gew_Ktau_avg_sig0p1_by_sim} and \ref{tab:gew_Ktau_first_sig0p1_by_sim}).

\begin{table}[t!h]
\centering
\begin{threeparttable}
\captionsetup{width=0.93\linewidth}
\caption{\label{tab:gew_Ktau_avg_sig0p1_by_sim} \small Z statistic from Geweke test for Kendall's tau averaged across 5 rankers in each iteration, scenario 2, $\sigma = 0.1$, 10 repetitions.}
\begin{tabular}{r|rrrrrrrrrr}
 Repetition  & 1 & 2 & 3 & 4 & 5 & 6 & 7 & 8 & 9 & 10 \\ 
  \hline
ARROBART & 0.85 & 0.05 & -0.14 & -1.53 & 0.73 & -0.03 & -2.00 & 0.56 & 0.56 & -0.80 \\ 
  ARROLinear Lag & -1.81 & -0.76 & 0.08 & 1.14 & -0.11 & 0.58 & 0.96 & -0.97 & 0.25 & -0.34 \\ 
  ARROLinear & 1.67 & -1.26 & 0.06 & -0.23 & 1.90 & -0.69 & 1.03 & 0.49 & 0.16 & 0.66 \\ 
  ROLinear & -1.80 & 0.38 & 0.69 & -0.73 & -1.01 & 0.20 & 1.60 & -2.90 & -0.15 & -0.01 \\ 
  ROBART & 0.17 & -1.60 & 0.44 & 0.86 & 1.36 & 2.87 & 0.16 & 2.77 & -2.22 & -4.39 \\ 
\end{tabular}
\begin{tablenotes}
\footnotesize
\item \!\!\!\! Notes: The variable to which the test is applied is the average of ranker-specific Kendall's tau values calculated for each of 5 ranker's predicted rankings in period $T+1$ in each MCMC iteration.
\end{tablenotes}
\end{threeparttable}
\end{table}

\begin{table}[t!h]
\centering
\begin{threeparttable}
\captionsetup{width=0.93\linewidth}
\caption{\label{tab:gew_Ktau_first_sig0p1_by_sim} \small Z statistic from Geweke test for Kendall's tau for the first ranker, scenario 2, $\sigma = 0.1$, 10 repetitions.}
\begin{tabular}{r|rrrrrrrrrr}
 Repetition  & 1 & 2 & 3 & 4 & 5 & 6 & 7 & 8 & 9 & 10 \\ 
  \hline
ARROBART & -0.11 & -0.98 & -0.93 & -1.00 & -0.67 & 0.02 & -0.81 & 1.17 & 0.19 & -0.81 \\ 
  ARROLinear Lag & -0.09 & -2.01 & 0.05 & 0.31 & 0.81 & 0.38 & 0.31 & -0.23 & 0.38 & -1.12 \\ 
  ARROLinear & 0.44 & -0.61 & 0.67 & 0.08 & 0.69 & 0.45 & 1.69 & -0.96 & 1.06 & 0.95 \\ 
  ROLinear & -0.46 & 1.65 & 0.26 & -0.03 & 0.49 & 0.71 & -0.11 & -0.97 & -0.32 & -0.75 \\ 
  ROBART & -0.06 & -0.38 & 0.39 & 0.90 & 1.65 & 1.65 & -0.10 & 1.92 & -1.07 & -3.79 \\ 
\end{tabular}
\begin{tablenotes}
\footnotesize
\item \!\!\!\! Notes: The variable to which the test is applied is the Kendall's tau distance of the first ranker's period $T+1$ predicted ranks from the true ranks.
\end{tablenotes}
\end{threeparttable}
\end{table}

Similarly, we provide the Geweke test for the first ranker's first item latent $Z_{ijt}$ in Table~\ref{tab:gew_firstZijt_avg_sig0p1_by_sim}, while Table~\ref{tab:MVgew_5Ktaus_sig0p1_by_sim} provides the p-values for the multivariate Geweke test.

\begin{table}[t!h]
\centering
\begin{threeparttable}
\captionsetup{width=0.93\linewidth}
\caption{\label{tab:gew_firstZijt_avg_sig0p1_by_sim} \small Z statistic from Geweke test for the first ranker's first item latent $Z_{ijt}$, scenario 2, $\sigma = 0.1$, 10 repetitions.}
\begin{tabular}{r|rrrrrrrrrr}
 Repetition  & 1 & 2 & 3 & 4 & 5 & 6 & 7 & 8 & 9 & 10 \\ 
  \hline
ARROBART & -0.11 & -0.98 & -0.93 & -1.00 & -0.67 & 0.02 & -0.81 & 1.17 & 0.19 & -0.81 \\ 
  ARROLinear Lag & -0.09 & -2.01 & 0.05 & 0.31 & 0.81 & 0.38 & 0.31 & -0.23 & 0.38 & -1.12 \\ 
  ARROLinear & 0.44 & -0.61 & 0.67 & 0.08 & 0.69 & 0.45 & 1.69 & -0.96 & 1.06 & 0.95 \\ 
  ROLinear & -0.46 & 1.65 & 0.26 & -0.03 & 0.49 & 0.71 & -0.11 & -0.97 & -0.32 & -0.75 \\ 
  ROBART & -0.06 & -0.38 & 0.39 & 0.90 & 1.65 & 1.65 & -0.10 & 1.92 & -1.07 & -3.79 \\ 
\end{tabular}
\begin{tablenotes}
\footnotesize
\item \!\!\!\! Notes: The variable to which the test is applied is the ranker- and item- specific latent outcome $Z_{ijt}$ for the first ranker and first item.
\end{tablenotes}
\end{threeparttable}
\end{table}

\begin{table}[t!h]
\centering
\begin{threeparttable}
\captionsetup{width=0.93\linewidth}
\caption{\label{tab:MVgew_5Ktaus_sig0p1_by_sim} \small P-values for multivariate Geweke test, scenario 2, $\sigma = 0.1$, 10 repetitions.}
\begin{tabular}{r|rrrrrrrrrr}
 Repetition  & 1 & 2 & 3 & 4 & 5 & 6 & 7 & 8 & 9 & 10 \\ 
  \hline
ARROBART & 0.86 & 0.72 & 0.46 & 0.34 & 0.38 & 0.54 & 0.09 & 0.01 & 0.86 & 0.87 \\ 
  ARROLinear Lag & 0.22 & 0.13 & 0.55 & 0.04 & 0.99 & 0.86 & 0.79 & 0.74 & 0.77 & 0.73 \\ 
  ARROLinear & 0.55 & 0.59 & 0.48 & 0.40 & 0.67 & 0.94 & 0.69 & 0.77 & 0.44 & 0.56 \\ 
  ROLinear & 0.32 & 0.36 & 0.92 & 0.97 & 0.12 & 0.66 & 0.19 & 0.11 & 0.70 & 0.32 \\ 
  ROBART & 0.25 & 0.50 & 0.30 & 0.01 & 0.40 & 0.01 & 0.11 & 0.00 & 0.23 & 0.00 \\ 
\end{tabular}
\begin{tablenotes}
\footnotesize
\item \!\!\!\! Notes: The 5 variables in the multivariate Geweke test are ranker-specific Kendall's tau values calculated for each of 5 ranker's predicted rankings in period $T+1$ in each MCMC iteration.
\end{tablenotes}
\end{threeparttable}
\end{table}

Other MCMC diagnostic presented are the stable Gelman-Rubin test statistic for average Kendall's tau, first ranker's Kendall's tau, and first ranker's first item latent $Z_{ijt}$ in Tables~\ref{tab:StableGR_Ktau_avg_sig0p1_by_sim}, \ref{tab:StableGR_Ktau_first_sig0p1_by_sim}, and \ref{tab:StableGR_firstZijt_sig0p1_by_sim}, respectively.

\begin{table}[t!h]
\centering
\begin{threeparttable}
\captionsetup{width=0.93\linewidth}
\caption{\label{tab:StableGR_Ktau_avg_sig0p1_by_sim} \small Stable Gelman-Rubin test statistic for average Kendall's tau, scenario 2, $\sigma = 0.1$, 10 repetitions.}
\begin{tabular}{r|rrrrrrrrrr}
 Repetition  & 1 & 2 & 3 & 4 & 5 & 6 & 7 & 8 & 9 & 10 \\ 
  \hline
ARROBART & 1.00 & 1.00 & 1.00 & 1.00 & 1.00 & 1.00 & 1.00 & 1.00 & 1.00 & 1.00 \\ 
  ARROLinear Lag & 1.00 & 1.00 & 1.00 & 1.00 & 1.00 & 1.00 & 1.00 & 1.00 & 1.00 & 1.00 \\ 
  ARROLinear & 1.00 & 1.00 & 1.00 & 1.00 & 1.00 & 1.00 & 1.00 & 1.00 & 1.00 & 1.00 \\ 
  ROLinear & 1.00 & 1.00 & 1.00 & 1.00 & 1.00 & 1.00 & 1.00 & 1.00 & 1.00 & 1.00 \\ 
  ROBART & 1.00 & 1.00 & 1.00 & 1.00 & 1.00 & 1.00 & 1.00 & 1.00 & 1.00 & 1.00 \\  
\end{tabular}
\begin{tablenotes}
\footnotesize
\item \!\!\!\! Notes: The table contains the Stable Gelman-Rubin statistic for a single chain as introduced by \cite{vats2021revisiting}. The variable to which the test is applied is the average of ranker-specific Kendall's tau values calculated for each of 5 ranker's predicted rankings in period $T+1$ in each MCMC iteration.
\end{tablenotes}
\end{threeparttable}
\end{table}

\begin{table}[t!h]
\centering
\begin{threeparttable}
\captionsetup{width=0.93\linewidth}
\caption{\label{tab:StableGR_Ktau_first_sig0p1_by_sim} \small Stable Gelman-Rubin test statistic for first ranker's Kendall's tau, scenario 2, $\sigma = 0.1$, 10 repetitions.}
\begin{tabular}{r|rrrrrrrrrr}
 Repetition  & 1 & 2 & 3 & 4 & 5 & 6 & 7 & 8 & 9 & 10 \\ 
  \hline
ARROBART & 1.00 & 1.00 & 1.00 & 1.00 & 1.00 & 1.00 & 1.00 & 1.00 & 1.00 & 1.00 \\ 
  ARROLinear Lag & 1.00 & 1.00 & 1.00 & 1.00 & 1.00 & 1.00 & 1.00 & 1.00 & 1.00 & 1.00 \\ 
  ARROLinear & 1.00 & 1.00 & 1.00 & 1.00 & 1.00 & 1.00 & 1.00 & 1.00 & 1.00 & 1.00 \\ 
  ROLinear & 1.00 & 1.00 & 1.00 & 1.00 & 1.00 & 1.00 & 1.00 & 1.00 & 1.00 & 1.00 \\ 
  ROBART & 1.00 & 1.00 & 1.00 & 1.00 & 1.00 & 1.00 & 1.00 & 1.00 & 1.00 & 1.00 \\ 
\end{tabular}
\begin{tablenotes}
\footnotesize
\item \!\!\!\! Notes: The table contains the Stable Gelman-Rubin statistic for a single chain as introduced by \cite{vats2021revisiting}. The variable to which the test is applied is the ranker-specific Kendall's tau values calculated for the predicted rankings of the first ranker in period $T+1$ in each MCMC iteration.
\end{tablenotes}
\end{threeparttable}
\end{table}

\begin{table}[t!h]
\centering
\begin{threeparttable}
\captionsetup{width=0.93\linewidth}
\caption{\label{tab:StableGR_firstZijt_sig0p1_by_sim} \small Stable Gelman-Rubin test statistic for first ranker's first item latent $Z_{ijt}$, scenario 2, $\sigma = 0.1$, 10 repetitions.}
\begin{tabular}{r|rrrrrrrrrr}
 Repetition  & 1 & 2 & 3 & 4 & 5 & 6 & 7 & 8 & 9 & 10 \\ 
  \hline
ARROBART & 1.00 & 1.00 & 1.00 & 1.00 & 1.00 & 1.00 & 1.00 & 1.00 & 1.00 & 1.00 \\ 
  ARROLinear Lag & 1.00 & 1.00 & 1.00 & 1.00 & 1.00 & 1.00 & 1.00 & 1.00 & 1.00 & 1.00 \\ 
  ARROLinear & 1.00 & 1.00 & 1.00 & 1.00 & 1.00 & 1.00 & 1.00 & 1.00 & 1.00 & 1.00 \\ 
  ROLinear & 1.00 & 1.00 & 1.00 & 1.00 & 1.00 & 1.00 & 1.00 & 1.00 & 1.00 & 1.00 \\ 
  ROBART & 1.00 & 1.00 & 1.00 & 1.00 & 1.01 & 1.00 & 1.01 & 1.00 & 1.00 & 1.00 \\ 
\end{tabular}
\begin{tablenotes}
\footnotesize
\item \!\!\!\! Notes: The table contains the Stable Gelman-Rubin statistic for a single chain as introduced by \cite{vats2021revisiting}. The variable to which the test is applied is the ranker- and item- specific latent outcome $Z_{ijt}$ for the first ranker and first item.
\end{tablenotes}
\end{threeparttable}
\end{table}

\begin{figure}[t!h]
\centering
\captionsetup{width=0.93\linewidth}
\hspace*{-4.5ex}
\begin{tabular}{cc}
\includegraphics[trim=0mm 0mm 0mm 0mm,clip,height=6cm,width=8cm
]{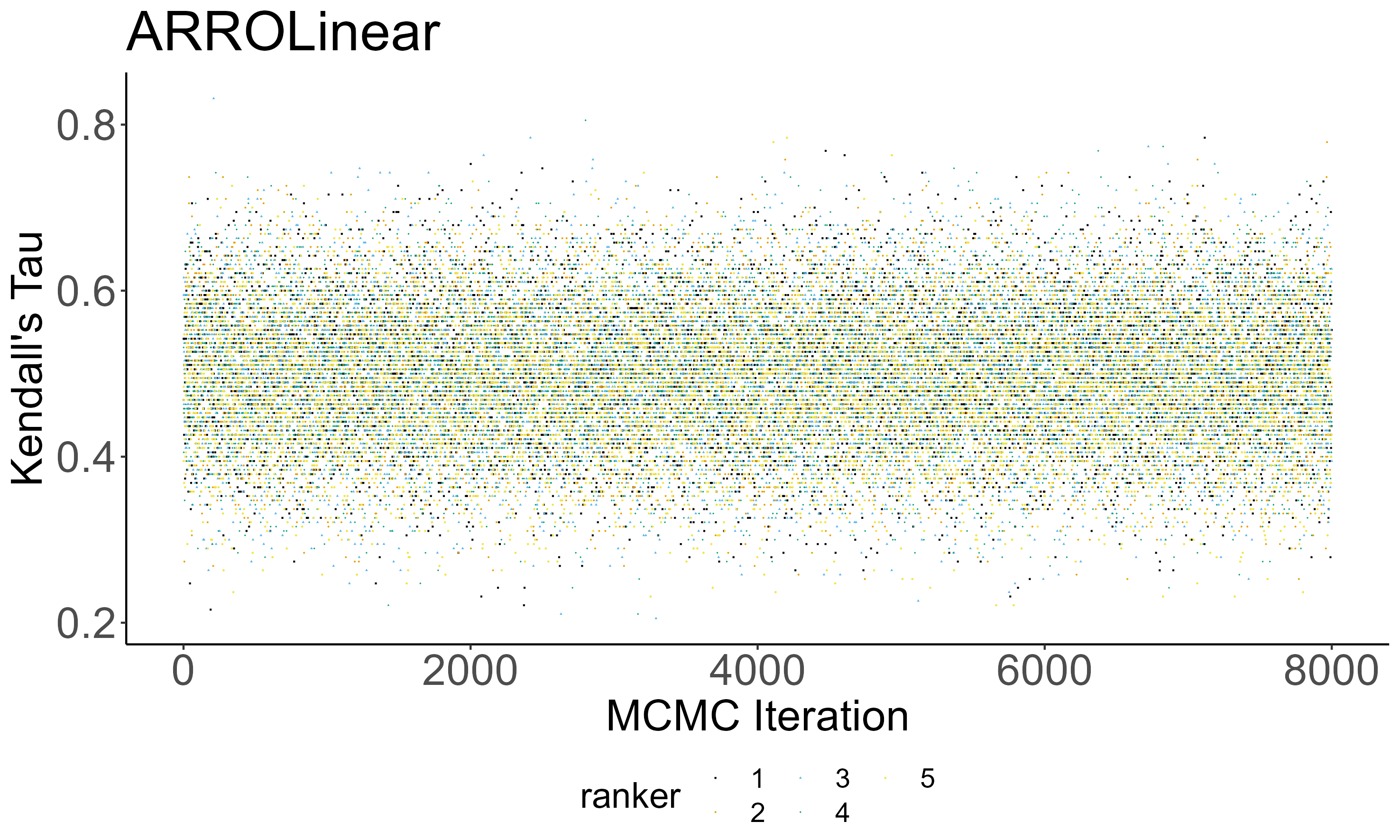} &
\includegraphics[trim=0mm 0mm 0mm 0mm,clip,height=6cm,width=8cm
]{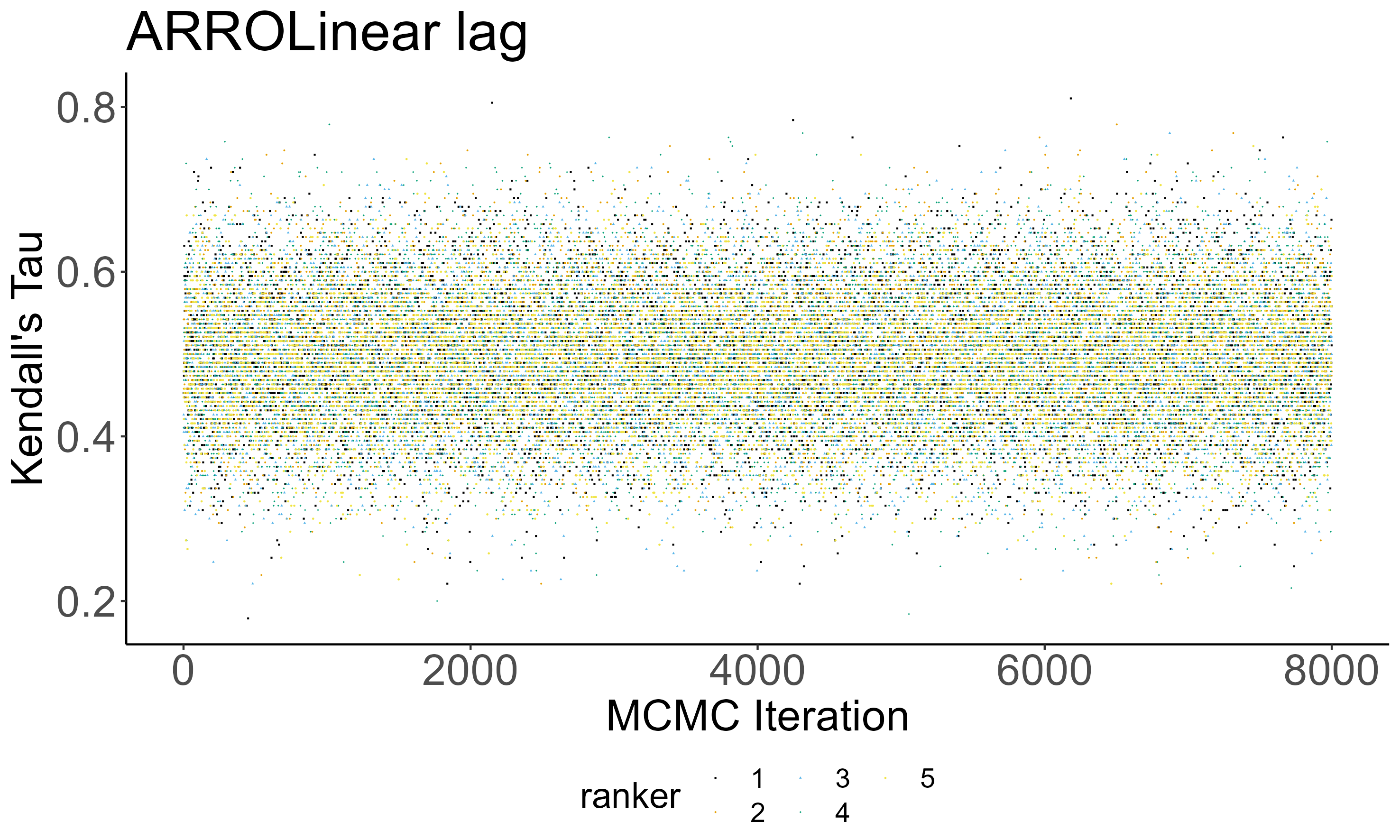}
\end{tabular}
\caption{Kendall's tau distance of each ranker's predicted ranks from true ranks by MCMC iteration for ARROLinear (left) and ARROLinear lag (right). }
\label{fig:Ktau_by_ranker_ARROLinear_ARROLinearlag}
\end{figure}

\begin{figure}[t!h]
\centering
\captionsetup{width=0.93\linewidth}
\hspace*{-4.5ex}
\begin{tabular}{cc}
\includegraphics[trim=0mm 0mm 0mm 0mm,clip,height=6cm,width=8cm
]{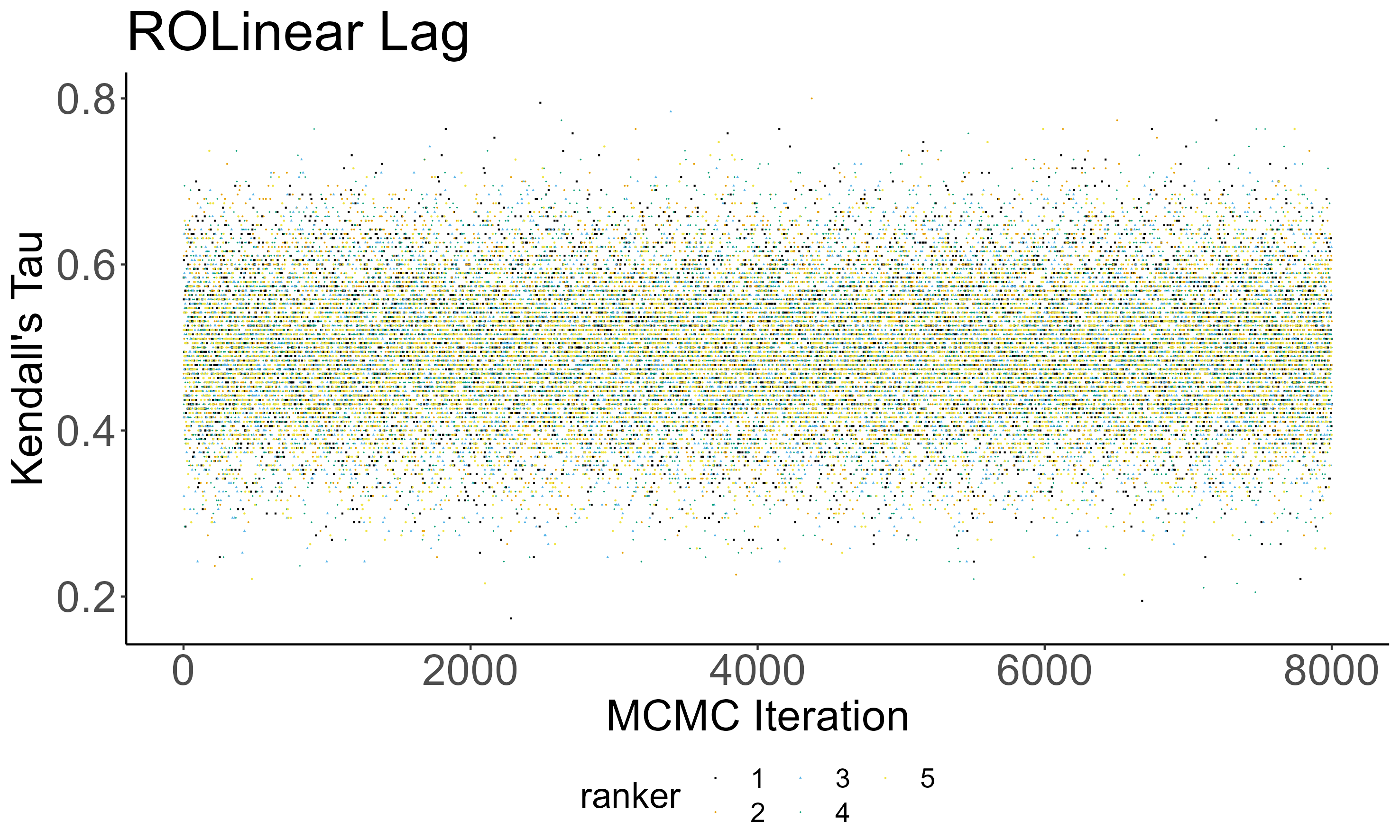} &
\includegraphics[trim=0mm 0mm 0mm 0mm,clip,height=6cm,width=8cm
]{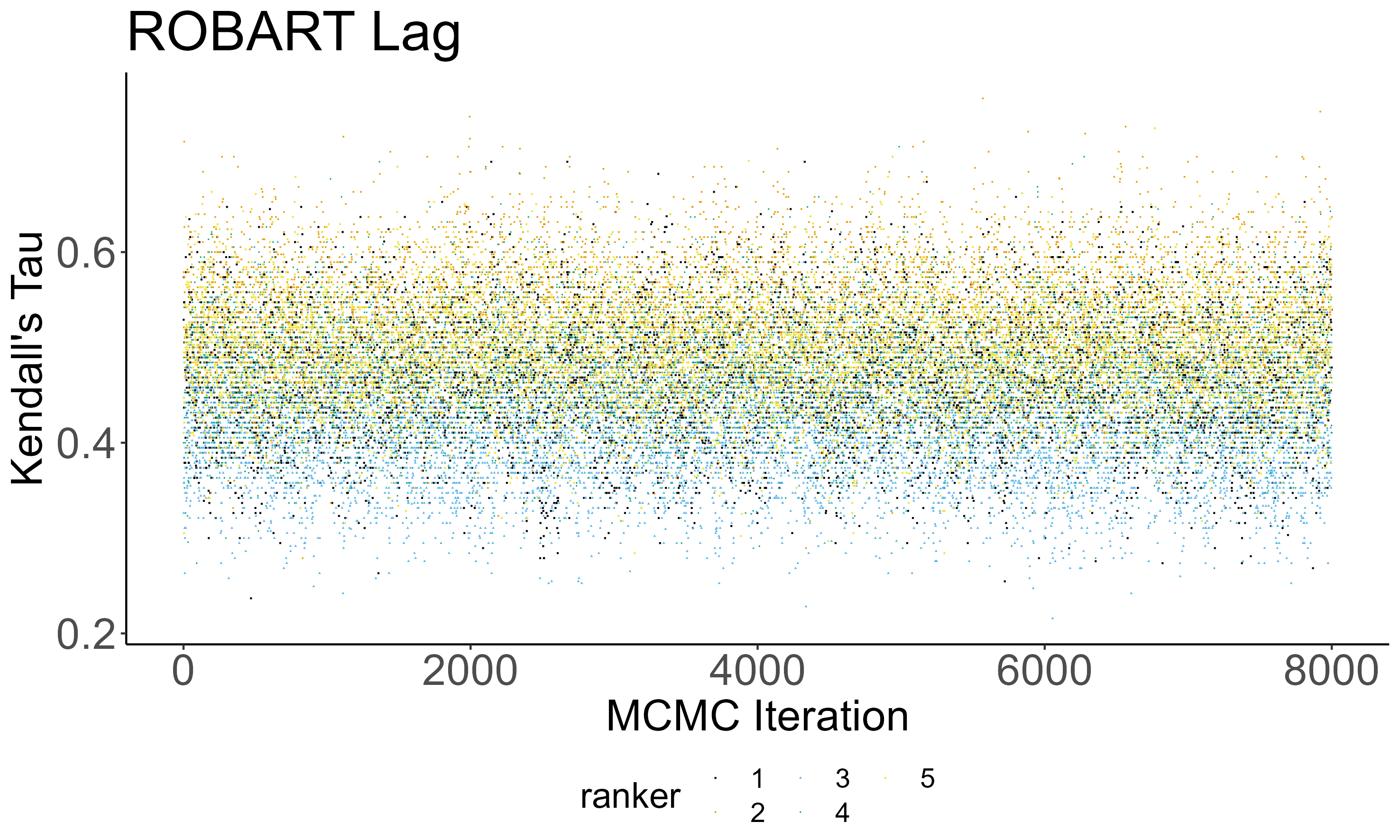}
\end{tabular}
\caption{Kendall's tau distance of each ranker's predicted ranks from true ranks by MCMC iteration for ROLinear lag (left) and ROBART lag (right). }
\label{fig:Ktau_by_ranker_ROLinearlag_ROBARTlag}
\end{figure}

\begin{figure}[t!h]
\centering
\captionsetup{width=0.93\linewidth}
\hspace*{-4.5ex}
\begin{tabular}{cc}
\includegraphics[trim=0mm 0mm 0mm 0mm,clip,height=6cm,width=8cm
]{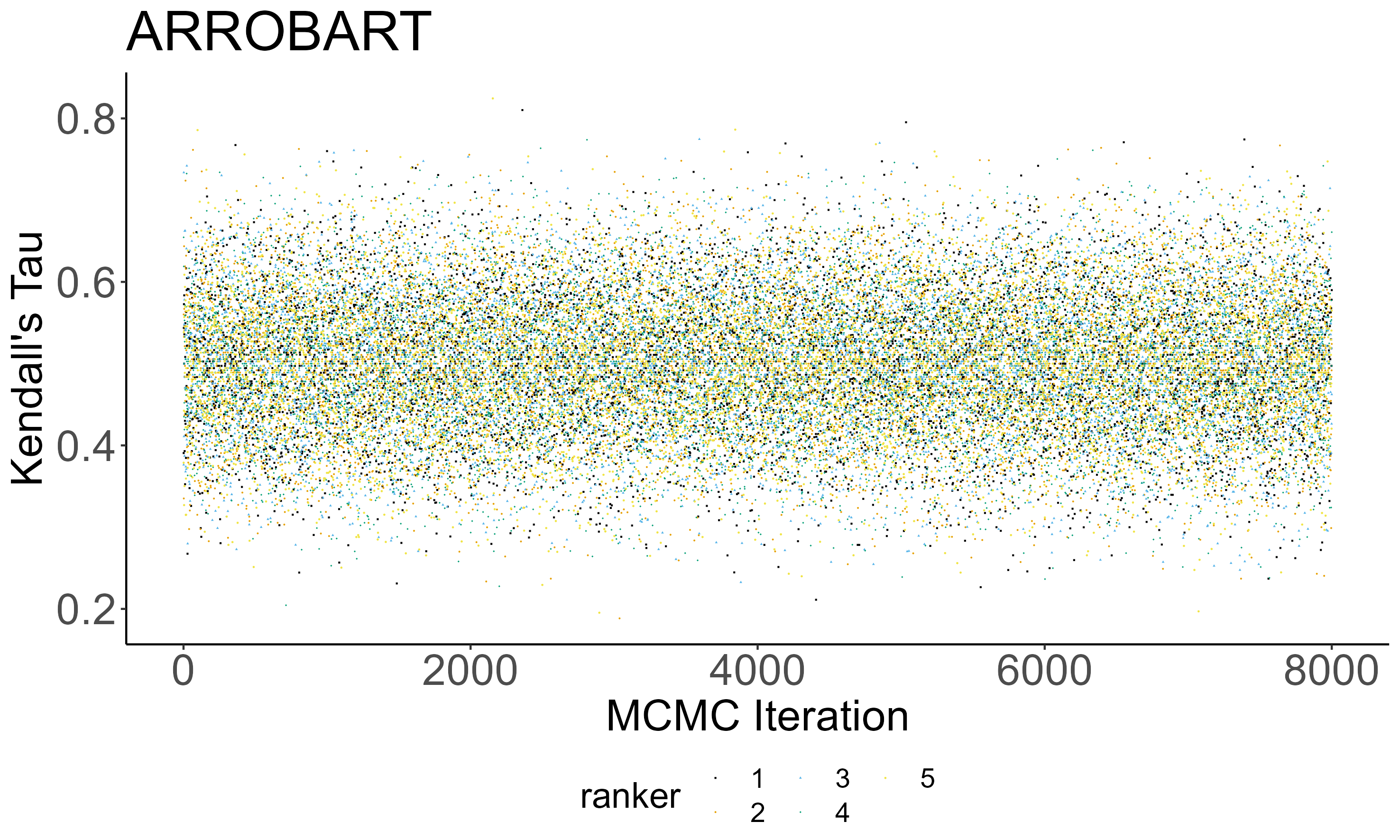} &
\end{tabular}
\caption{Kendall's tau distance of each ranker's predicted ranks from true ranks by MCMC iteration for ARROBART  (left) . }
\label{fig:Ktau_by_ranker_ARROBART}
\end{figure}

Graphically, we highlight the Kendall's tau distance of each ranker's predicted ranks from true ranks by MCMC iteration for ARROLINEAR (Fig.~\ref{fig:Ktau_by_ranker_ARROLinear_ARROLinearlag}), ROLinear and ROBART (Fig.~\ref{fig:Ktau_by_ranker_ROLinearlag_ROBARTlag}) and ARROBART (Fig~\ref{fig:Ktau_by_ranker_ARROBART}).

\begin{figure}[t!h]
\centering
\captionsetup{width=0.93\linewidth}
\hspace*{-4.5ex}
\begin{tabular}{cc}
\includegraphics[trim=0mm 0mm 0mm 0mm,clip,height=6cm,width=8cm
]{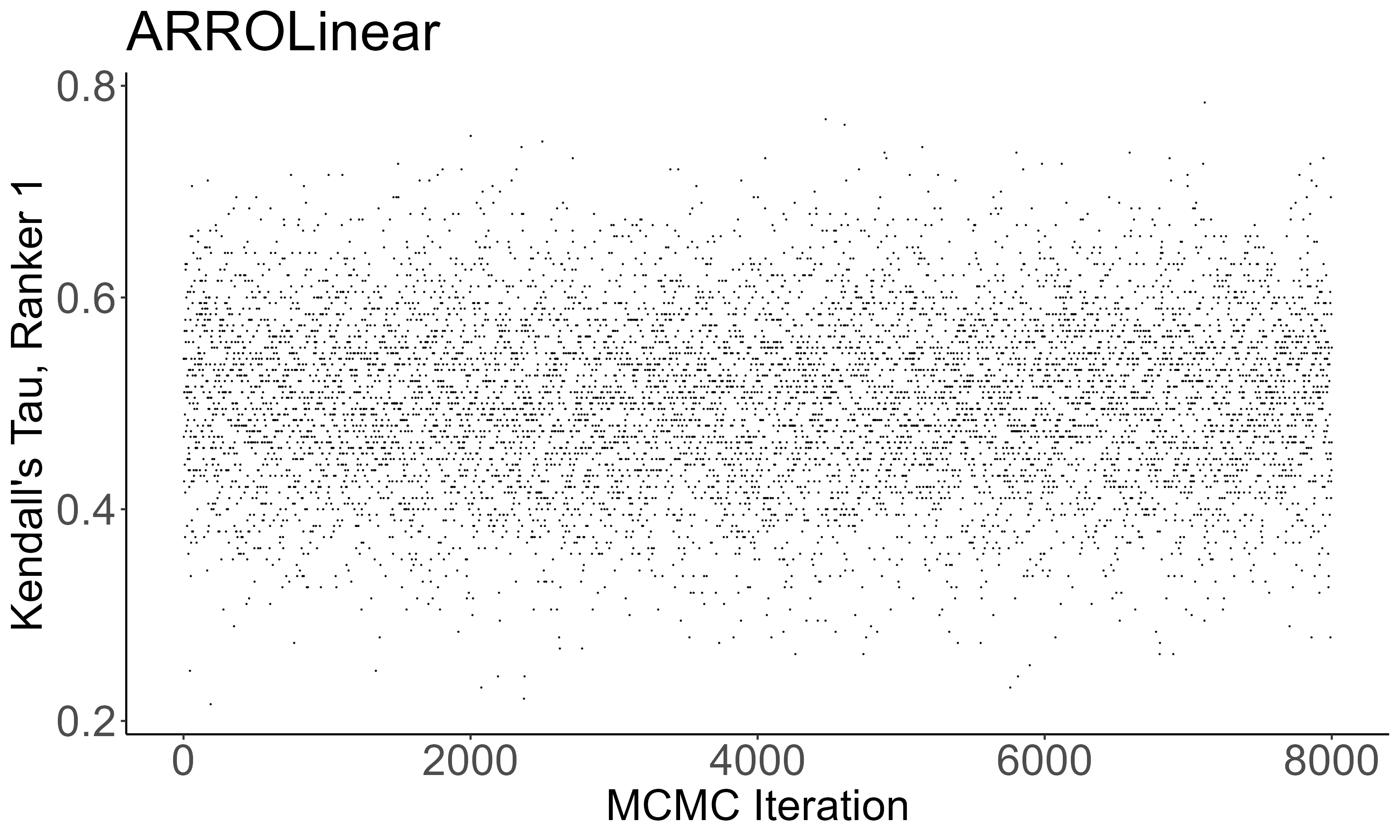} &
\includegraphics[trim=0mm 0mm 0mm 0mm,clip,height=6cm,width=8cm
]{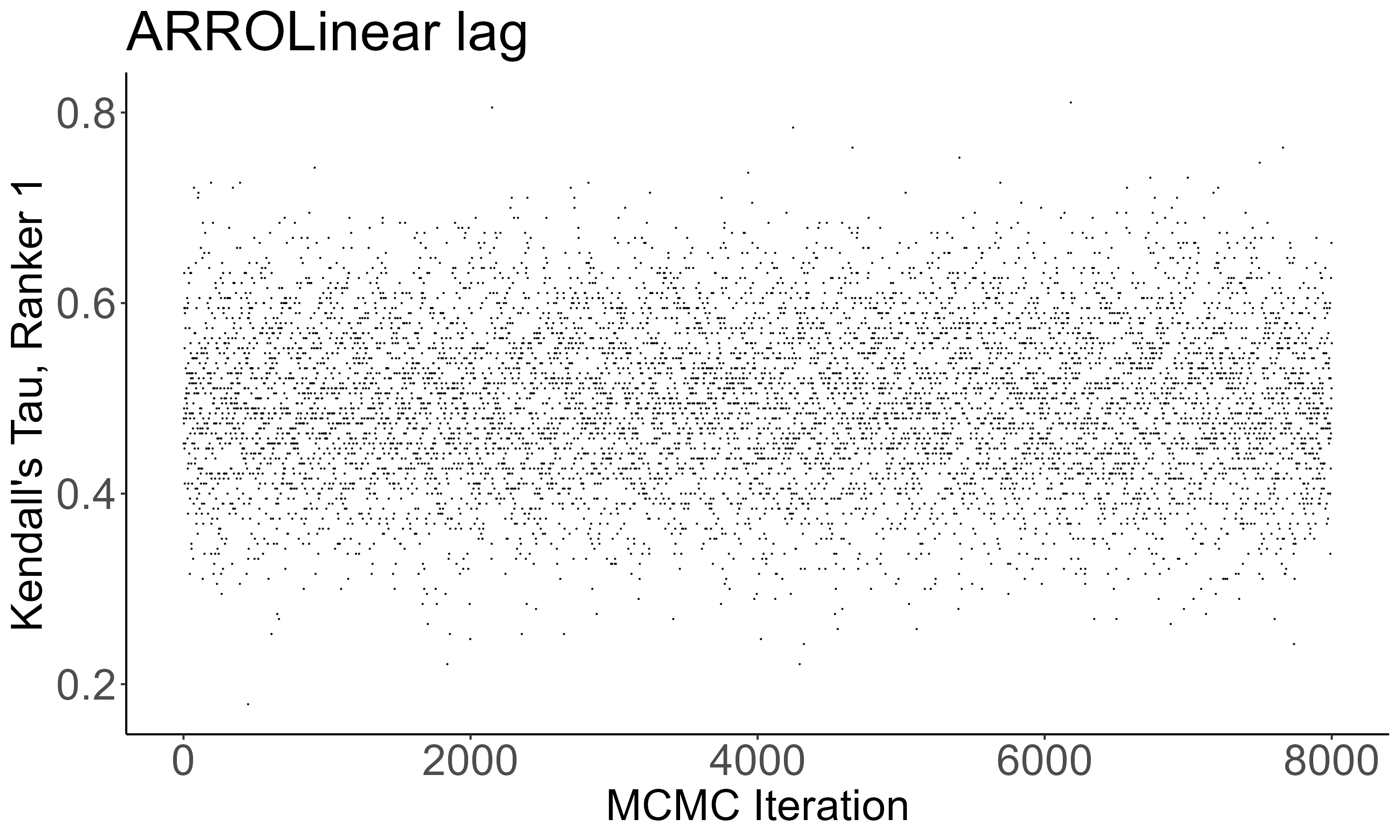}
\end{tabular}
\caption{Kendall's tau distance of first ranker's predicted ranks from true ranks by MCMC iteration for ARROLinear (left) and ARROLinear lag (right). }
\label{fig:Ktau_first_ranker_ARROLinear_ARROLinearlag}
\end{figure}

\begin{figure}[t!h]
\centering
\captionsetup{width=0.93\linewidth}
\hspace*{-4.5ex}
\begin{tabular}{cc}
\includegraphics[trim=0mm 0mm 0mm 0mm,clip,height=6cm,width=8cm
]{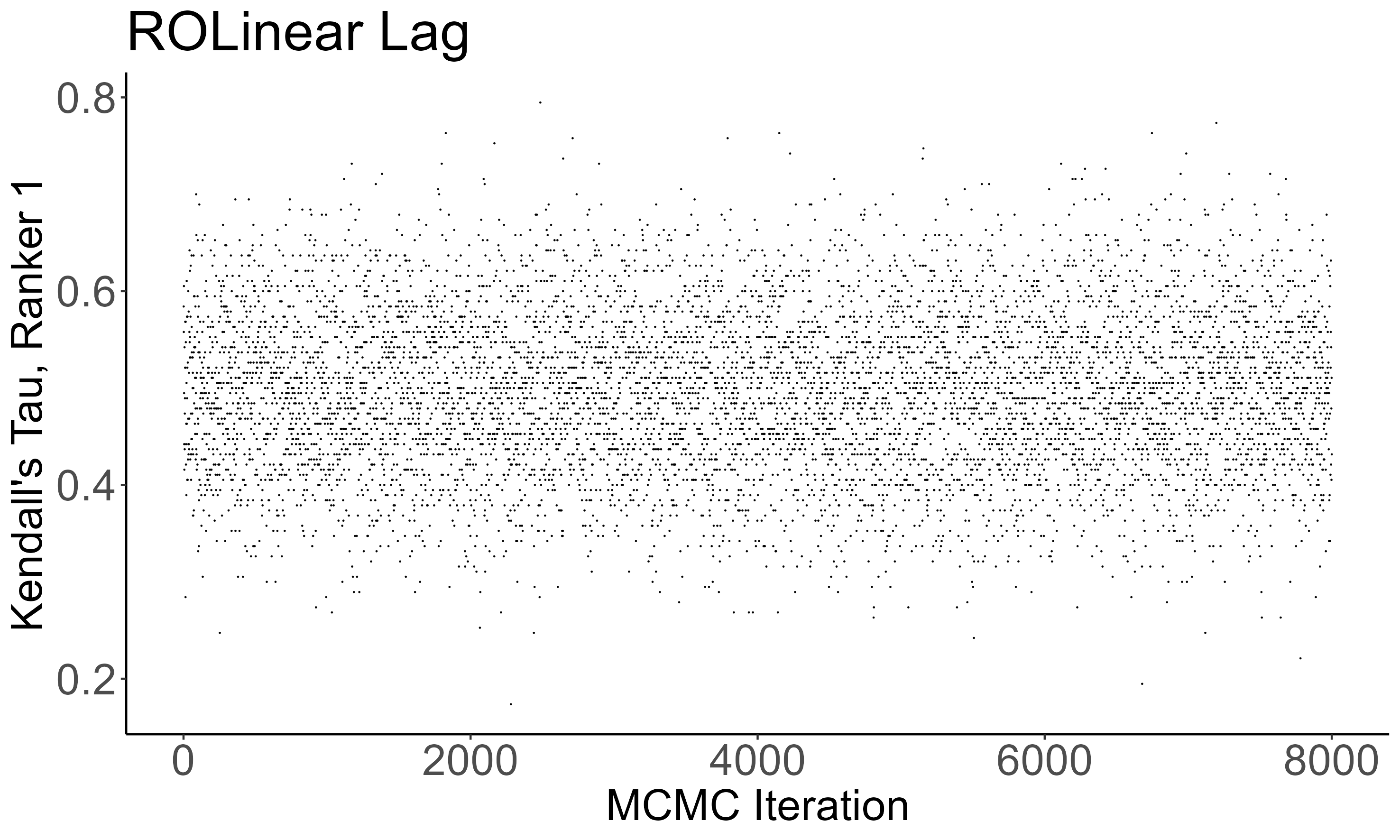} &
\includegraphics[trim=0mm 0mm 0mm 0mm,clip,height=6cm,width=8cm
]{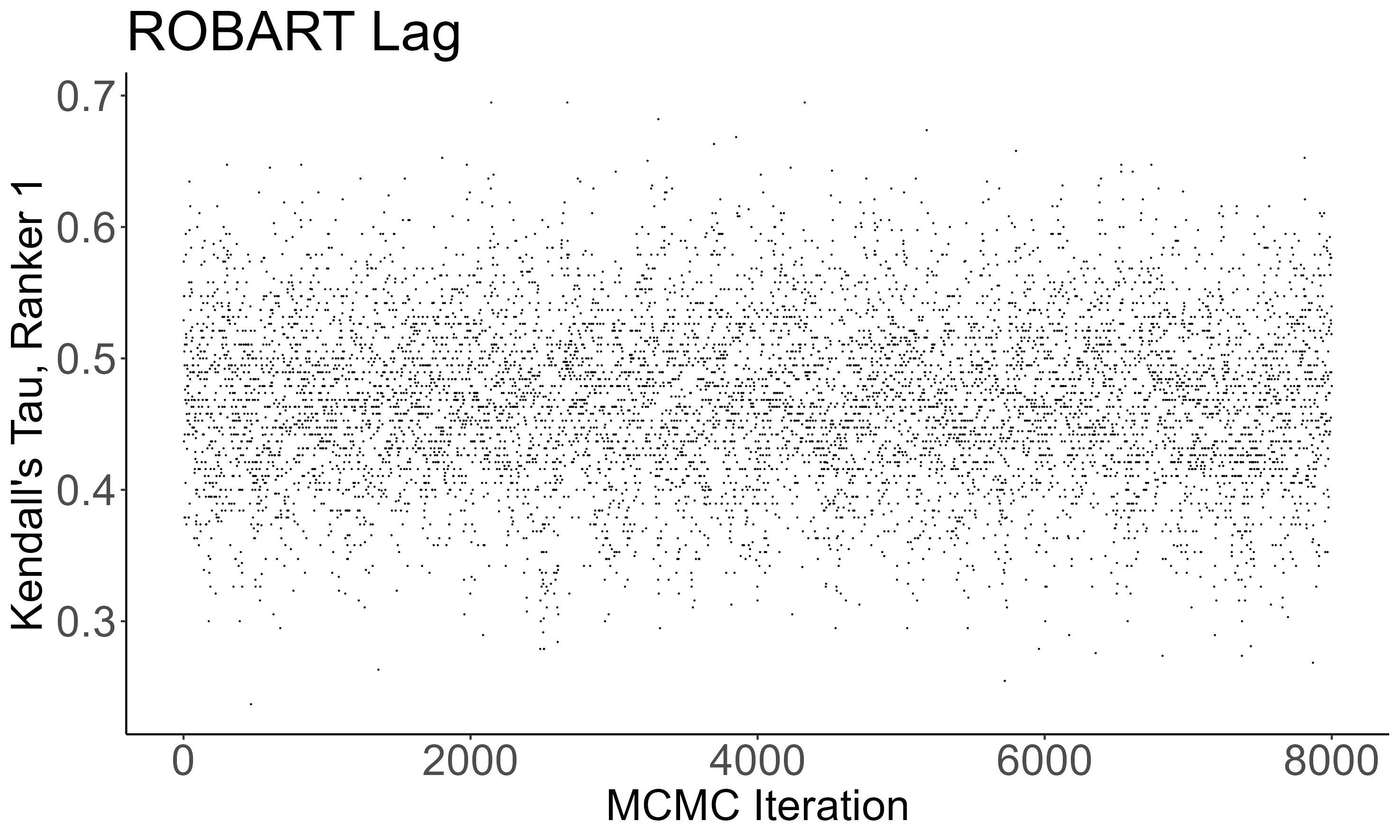}
\end{tabular}
\caption{Kendall's tau distance of first ranker's predicted ranks from true ranks by MCMC iteration for ROLinear lag (left) and ROBART lag (right). }
\label{fig:Ktau_first_ranker_ROLinearlag_ROBARTlag}
\end{figure}

\begin{figure}[t!h]
\centering
\captionsetup{width=0.93\linewidth}
\hspace*{-4.5ex}
\begin{tabular}{cc}
\includegraphics[trim=0mm 0mm 0mm 0mm,clip,height=6cm,width=8cm
]{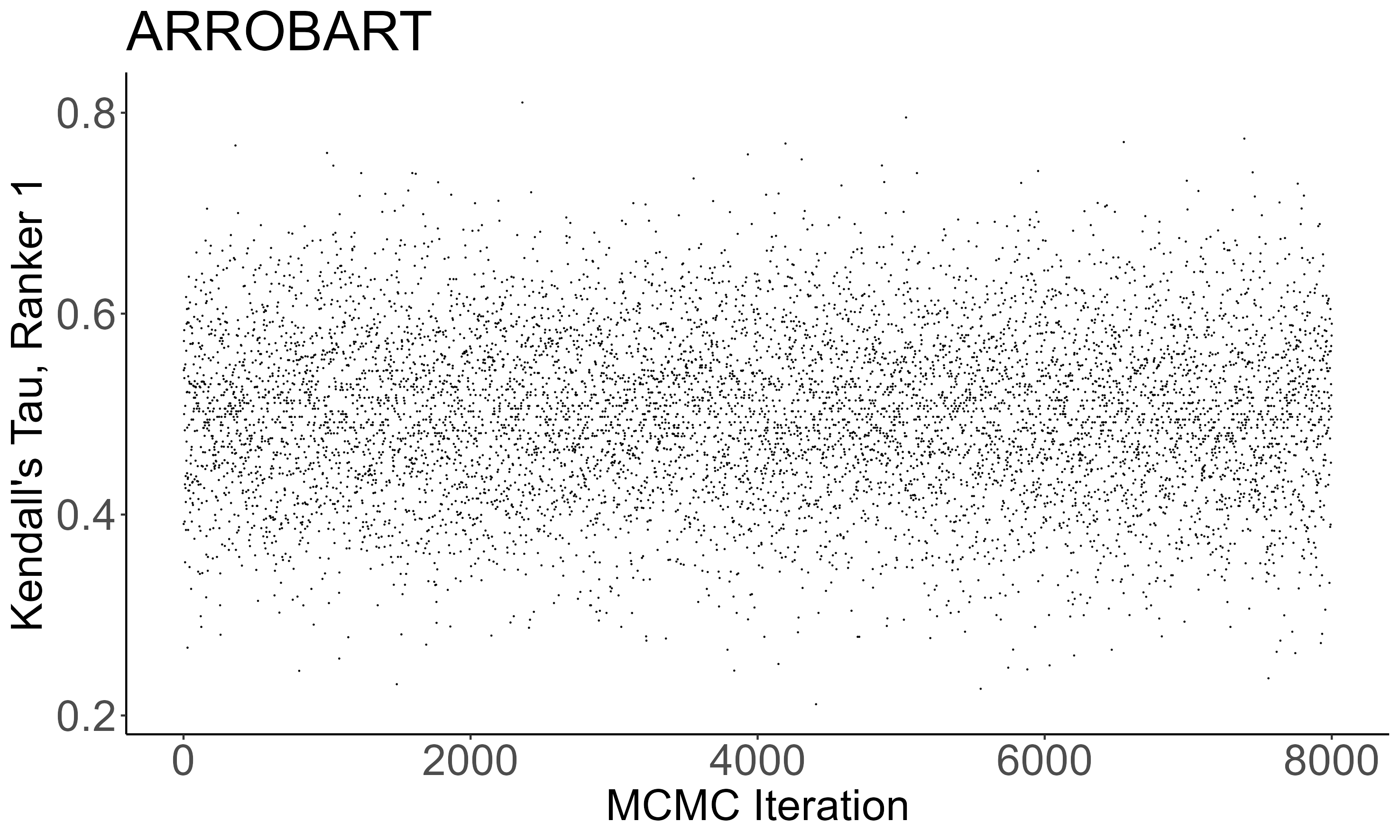} &
\end{tabular}
\caption{Kendall's tau distance of first ranker's predicted ranks from true ranks by MCMC iteration for ARROBART  (left) . }
\label{fig:Ktau_first_ranker_ARROBART}
\end{figure}

In conclusion, the Kendall's tau distance of first ranker's predicted ranks from true ranks by MCMC iteration for ARROLINEAR (Fig.~\ref{fig:Ktau_first_ranker_ARROLinear_ARROLinearlag}), ROLinear and ROBART (Fig.~\ref{fig:Ktau_first_ranker_ROLinearlag_ROBARTlag}) and ARROBART (Fig~\ref{fig:Ktau_first_ranker_ARROBART}).

\FloatBarrier
\subsection{Computational Times}

Table~\ref{tab:Comptimes} provides the computational times (in minutes) for different methods and values of $\sigma$ for the scenario 2 in the simulation studies. 
As noticed the choice of $\sigma$ does not affect the computational timing, while the main differences arise between methods. 
As pointed out in the literature, using a BART specification with autoregressive components lead to biggest computational timing as expected with respect to linear model such as ARROLINEAR or ROLINEAR. Viceversa if we do not include an autoregressive component, the ROBART specification as expected is the fastest methods among the five considered.

\begin{table}[t!h]
\centering
\begin{threeparttable}
\captionsetup{width=0.93\linewidth}
\caption{\label{tab:Comptimes} \small Computational times (in minutes) by method and $\sigma$ for scenario 2.}
\begin{tabular}{l|ccccc}
  \toprule
   & \multicolumn{5}{c}{$\sigma$} \\
  \multicolumn{1}{c}{Model} & 0.1 & 0.5 & 1.0 & 1.5 & 2.0 \\
  \midrule
  ARROBART & 77.61 & 75.98 & 76.66 & 76.03 & 76.63 \\ 
  ARROLinear Lag & 23.88 & 22.73 & 21.04 & 24.48 & 23.02 \\ 
  ARROLinear & 23.84 & 24.32 & 23.16 & 24.11 & 23.61 \\ 
  ROLinear lag & 22.89 & 24.35 & 21.53 & 24.23 & 22.51 \\ 
  ROBART lag & 15.10 & 15.75 & 15.27 & 15.19 & 14.93 \\ 
  \bottomrule
\end{tabular}
\begin{tablenotes}
\footnotesize
\item \!\!\!\! Notes: Times are averages over 10 replications.
\end{tablenotes}
\end{threeparttable}
\end{table}

\FloatBarrier

\section{Dynamic Label Ranking Application with multiple rankers: NCAA data}
\label{sec:apdx_NCAA_application}

\subsection{Data Description}

\begin{figure}[t!h]
\centering
\captionsetup{width=0.93\linewidth}
   \includegraphics[scale = 0.55]{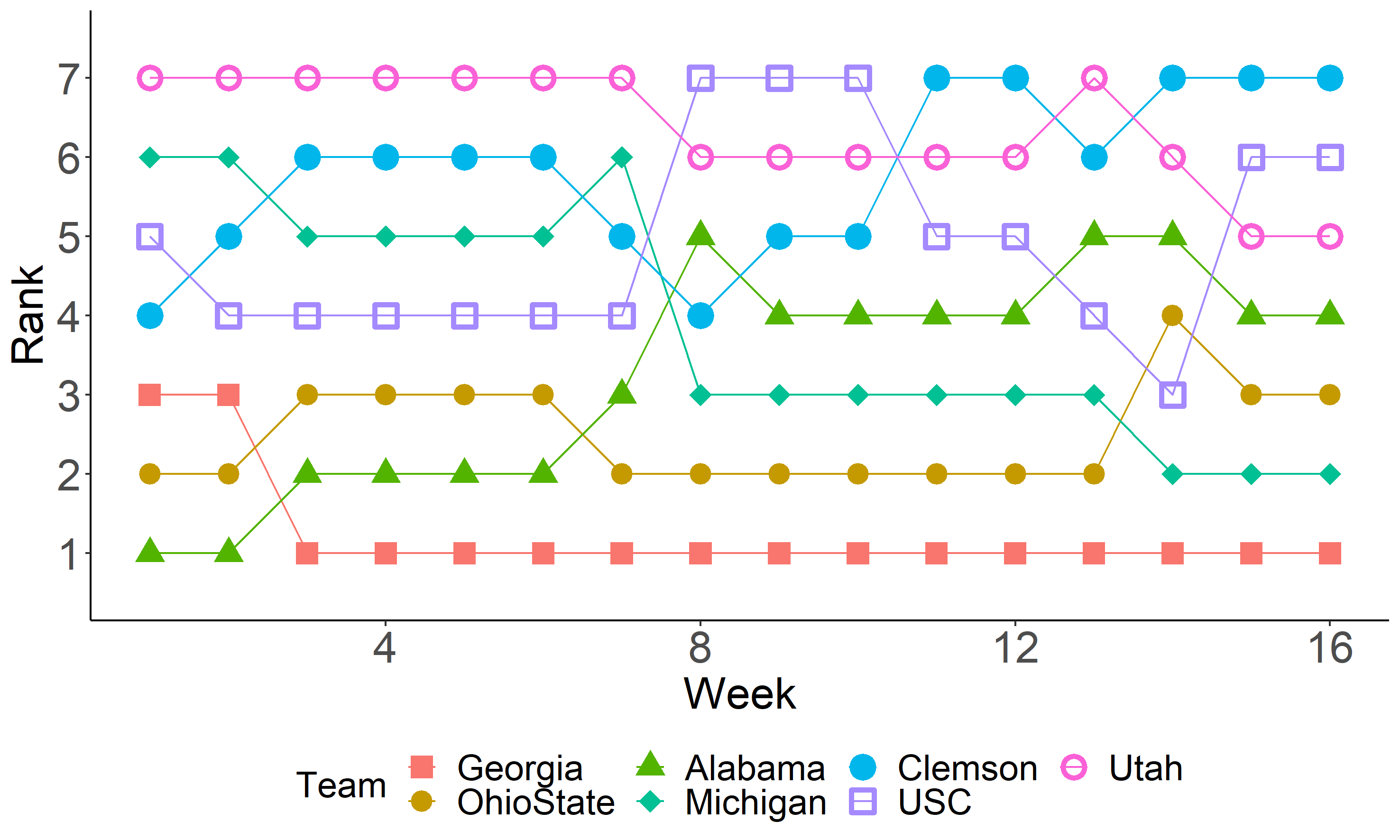}
\caption{AP Pollster Brian Howell: Observed relative ranks for $N=7$ teams of the 2022 NCAA Division I Football season.}
\label{fig:NCAA_data}
\end{figure}

We apply the proposed methods to a real dataset of pollster-specific rankings of American football teams from the Associated Press weekly poll for the 2022 NCAA Division I Football season.\footnote{Data were collected from \url{https://collegepolltracker.com}.}
The dataset includes the ranking lists of $M=15$ pollsters across $T=16$ weeks. As many teams are not listed in the top 20 by at least one pollster in at least one week, we consider a subset of $N=7$ teams to obtain a time series of full rankings.
Figure~\ref{fig:NCAA_data} shows the entire time series of ranking lists for a single pollster.

To account for external factors driving each pollster's evaluation, we include the following covariates, updated each week before pollsters submit their rankings: the win percentage, the average margin of victory, the margin of victory of the last game, and a binary variable equal to 1 if the team won the previous game.\footnote{Data were collected from \url{https://www.teamrankings.com/}.}

The purpose of this application is to assess the forecasting performance of the proposed methods. By comparing the static ROBART and dynamic ARROBART together with their linear counterparts, we aim to determine the relative importance of accounting for nonlinearities (via the BART) and temporal persistence in time series.
Specifically, we perform an out-of-sample forecasting exercise where we train the models using an expanding window with test periods from week 12 to week 16 of the poll.

Table~\ref{tab:NCAA_data_3pollsters} shows the observed time series of ranking lists from three selected AP pollsters.

\begin{table}[h!t]
\centering
\begin{tabular}{r | C{0.25cm} C{0.25cm} C{0.25cm} C{0.25cm} C{0.25cm} C{0.5cm} C{0.35cm} | C{0.25cm} C{0.25cm} C{0.25cm} C{0.25cm} C{0.25cm} C{0.5cm} C{0.35cm} | C{0.25cm} C{0.25cm} C{0.25cm} C{0.25cm} C{0.25cm} C{0.5cm} C{0.35cm}}
  \toprule
   & \multicolumn{7}{c|}{Aaron McMann} & \multicolumn{7}{c|}{Adam Grosbard} & \multicolumn{7}{c}{Adam Zucker} \\
 t & Ga & OS & Al & Mi & Cl & USC & Ut  & Ga & OS & Al & Mi & Cl & USC & Ut  & Ga & OS & Al & Mi & Cl & USC & Ut  \\ 
  \midrule
  1 & 3 & 2 & 1 & 5 & 4 & 7 & 6 & 3 & 2 & 1 & 6 & 4 & 7 & 5& 3 & 2 & 1 & 4 & 6 & 7 & 5 \\ 
  2 & 2 & 3 & 1 & 5 & 4 & 7 & 6 & 3 & 1 & 2 & 4 & 5 & 7 & 6 & 2 & 3 & 1 & 4 & 5 & 6 & 7 \\ 
  3 & 1 & 3 & 2 & 4 & 5 & 6 & 7 & 1 & 2 & 3 & 4 & 6 & 5 & 7  & 1 & 2 & 4 & 3 & 5 & 6 & 7 \\ 
  4 & 1 & 3 & 2 & 4 & 5 & 6 & 7 & 1 & 2 & 3 & 4 & 6 & 5 & 7  & 1 & 2 & 3 & 4 & 5 & 6 & 7 \\ 
  5 & 1 & 3 & 2 & 4 & 5 & 6 & 7 & 1 & 2 & 3 & 4 & 6 & 5 & 7 & 1 & 2 & 3 & 4 & 5 & 6 & 7 \\ 
  6 & 1 & 3 & 2 & 4 & 5 & 6 & 7 & 2 & 1 & 3 & 4 & 5 & 6 & 7 & 3 & 2 & 1 & 4 & 5 & 6 & 7 \\ 
  7 & 1 & 3 & 2 & 4 & 5 & 6 & 7 & 2 & 1 & 3 & 4 & 5 & 6 & 7 & 2 & 1 & 3 & 5 & 4 & 6 & 7 \\ 
  8 & 1 & 2 & 5 & 3 & 4 & 6 & 7 & 2 & 1 & 5 & 3 & 4 & 7 & 6  & 2 & 1 & 5 & 3 & 4 & 6 & 7 \\ 
  9 & 1 & 2 & 5 & 3 & 4 & 6 & 7 & 1 & 2 & 5 & 3 & 4 & 7 & 6 & 2 & 1 & 5 & 3 & 4 & 6 & 7 \\ 
  10 & 1 & 2 & 5 & 3 & 4 & 6 & 7 & 1 & 2 & 5 & 3 & 4 & 6 & 7 & 2 & 1 & 5 & 3 & 4 & 6 & 7 \\ 
  11 & 1 & 2 & 6 & 3 & 4 & 5 & 7 & 1 & 2 & 5 & 3 & 6 & 4 & 7 & 1 & 2 & 4 & 3 & 6 & 5 & 7 \\ 
  12 & 1 & 2 & 6 & 3 & 5 & 4 & 7 & 1 & 2 & 5 & 3 & 6 & 4 & 7  & 1 & 2 & 4 & 3 & 6 & 5 & 7 \\ 
  13 & 1 & 2 & 6 & 3 & 5 & 4 & 7 & 1 & 2 & 6 & 3 & 5 & 4 & 7 & 1 & 2 & 6 & 3 & 5 & 4 & 7 \\ 
  14 & 1 & 4 & 5 & 2 & 6 & 3 & 7 & 1 & 4 & 5 & 2 & 7 & 3 & 6 & 1 & 4 & 5 & 2 & 6 & 3 & 7 \\ 
  15 & 1 & 4 & 5 & 2 & 6 & 3 & 7 & 1 & 3 & 4 & 2 & 7 & 6 & 5 & 1 & 3 & 4 & 2 & 5 & 6 & 7 \\  
  16 & 1 & 3 & 4 & 2 & 6 & 5 & 7  & 1 & 3 & 4 & 2 & 7 & 6 & 5  & 1 & 3 & 4 & 2 & 7 & 6 & 5 \\   
\bottomrule
\end{tabular}
\caption{Observed ranks submitted by AP Pollsters Aaron McMann, Adam Grosbard, and Adam Zucker for the NCAA Football Division 1 2022-23 Season. Ga = Georgia, OS - Ohio State, Al= Alabama, Mi = Michigan, Cl = Clemson, USC = University of Southern California, Ut = Utah.}
\label{tab:NCAA_data_3pollsters}
\end{table}

\subsection{Main Results}
The models under investigation are the proposed ROBART, ARROBART, and ARROBARTX, together with their linear counterparts ROLinear, ARROLinear, and ARROLinearX. The ARROBARTX and ARROLinearX models include exogenous covariates in addition to the lagged latent variable, whereas ROBART and ROLinear, being static, only consider the former.
ARROBART, ARROBARTX, and ROBART were implemented with 25, 25, and 50 trees, respectively, and Table~\ref{tab:football_tau_time_models_extra} contains a robustness check for ARROBART with different numbers of trees (25, 50, 75).

\begin{table}[t!h]
\centering
\begin{threeparttable}
\captionsetup{width=0.93\linewidth}
\caption{\label{tab:football_tau_time_models} \small Kendall's tau distance for forecasts of the AP Football Poll 2022 rankings.}
\begin{tabular}{l c c c c c c}
\hline
  Model & t=12 & t=13 & t=14 & t=15 & t=16 & Average \\ 
  \hline
  ARROBART  & (\textbf{0.04}) & (0.07) & (\textbf{0.09}) & (0.16) & (\textbf{0.07}) & (\textbf{0.09}) \\ 
  ARROBARTX  & 2.00 & \textbf{0.52} & 1.72 & 1.40 & 2.48 & 1.58 \\ 
  ROLinear & 3.43 & 2.62 & 2.76 & \textbf{0.60} & 2.74 & 2.01 \\ 
  ARROLinearX  & 2.79 & 1.05 & 2.24 & 1.40 & 3.57 & 2.03 \\ 
  ARROLinear & 2.79 & 1.05 & 2.24 & 1.40 & 3.57 & 2.03 \\ 
  ROBART & 2.79 & 1.05 & 2.24 & 1.42 & 3.57 & 2.04 \\ 
\hline
\end{tabular}
\begin{tablenotes}
\footnotesize
\item \!\!\!\! Notes: average across pollsters by week (columns 2 to 6) and global average across weeks (column 7). Models are sorted in ascending order according to the global average across weeks. The first row contains the Kendall's tau distances between ARROABRT and the true ranks. For all other models, the ratios to the ARROBART Kendall's tau distances are presented.
\end{tablenotes}
\end{threeparttable}
\end{table}

Table~\ref{tab:football_tau_time_models} displays averages across pollsters of Kendall's tau distances for forecasts of team rankings. Columns 2 to 6 contain week-specific average tau distances, and the last column gives the average over the last 5 weeks.
The results highlight a slight heterogeneity across horizons and models, and we rely on the total average across rankers and time horizons to determine the overall forecasting performance of each model.

Overall, we find strong evidence in favor of the ARROBART class of models, both with and without covariates. 
This aligns with our simulation study, where the ARROBART generally outperforms ARROLinear models.
Moreover, the total average indicates that the dynamic models (ARROBART and ARROLinear) outperform the static framework (ROBART and ROLinear), thus suggesting the importance of accounting for temporal dependence in forecasting this dataset.
More precisely, the ARROLinear specifications outperform the ROLinear in periods $t=12,13,14$ while having slightly worse overall average performance.
Finally, the nonparametric (BART) specification is found to yield the best performance in the autoregressive models and very similar outcomes to the linear counterpart in the static case.
The better average performance of ROLinear is essentially due to a very good forecast at $t=15$, whereas in all the previous periods, it is under-performing compared to the competitors.

\begin{figure}[t!h]
\centering
\captionsetup{width=0.93\linewidth}
\includegraphics[scale = 0.8]{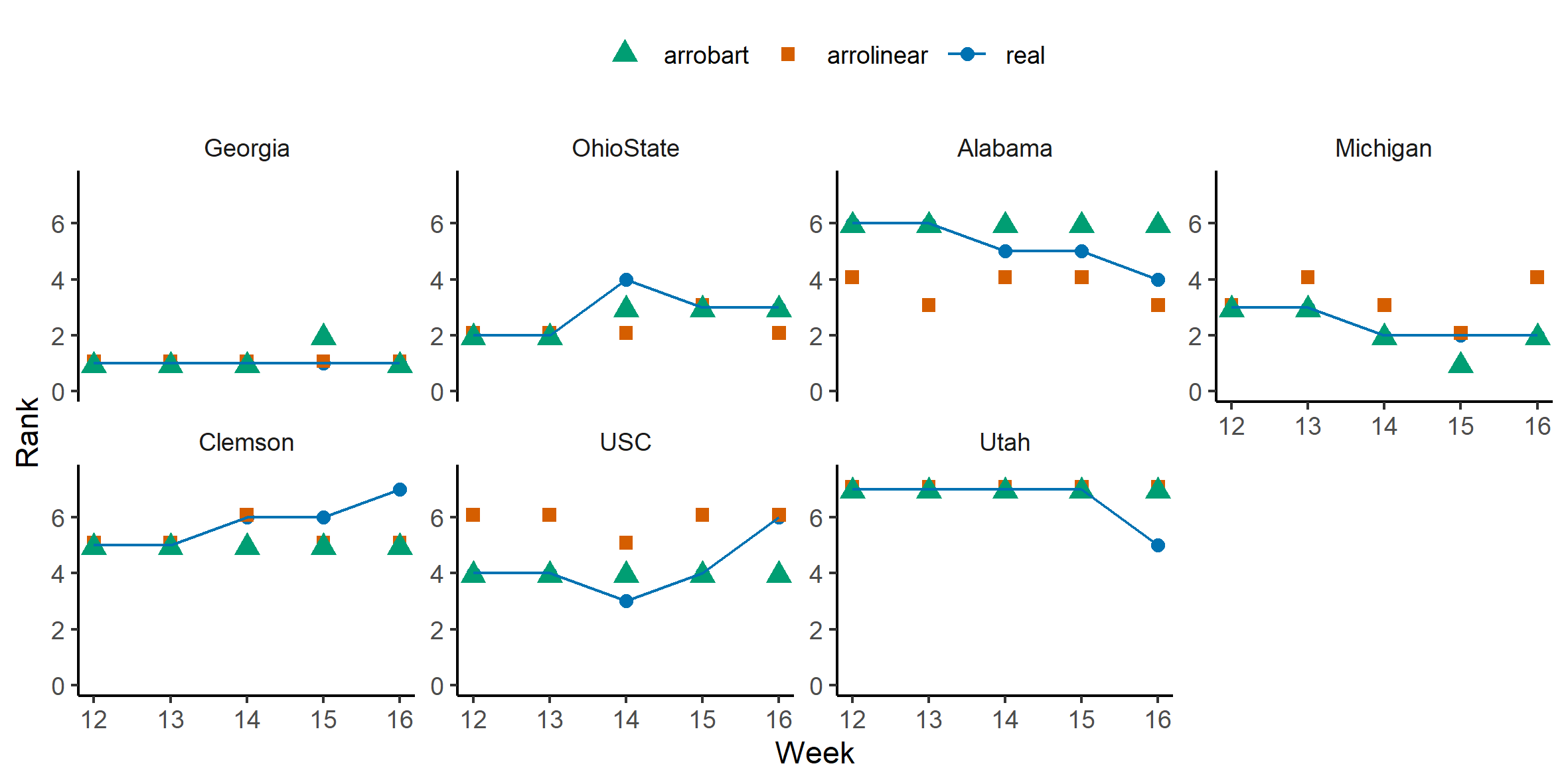}
\caption{AP Pollster Brian Howell: observed poll ranks (blue dots and line) and ranks forecasted using the ARROBART (green triangles) and ARROLinear (orange squares) models.}
\label{fig:realranks_onegraph}
\end{figure}

To inspect in more detail the advantages of the BART specification compared to the linear one, we report the point forecasts from the best ARROBART and ARROLinear models in Fig.~\ref{fig:realranks_onegraph} against the observed ranks, for $N=7$ teams and $h=5$ test periods. As the rankers are assumed to be independent, we present the results for a single pollster.
The ARROBART model is able to correctly predict the rank for several teams, with minor deviations for Alabama and Clemson. Conversely, the ARROLinear forecasts are generally more distant from the actual ranks. 
In particular, it can be observed that while both methods often produce similar forecasts, ARROBART produces more accurate forecasts for Alabama, USC, and Michigan in weeks 12 to 14.

The point forecast falls short of quantifying the uncertainty about the predicted value. To tackle this issue, we investigate in deeper detail the posterior predictive distribution of the best-performing model.
Figure~\ref{fig:postpredictive_michigan_ranker11} reports the posterior predictive distribution of the rank for Alabama and Michigan, whose observed ranks have changed multiple times during the testing sample (see Fig.~\ref{fig:realranks_onegraph}).
For both teams, the predictive density of the rank is unimodal and changes over time in terms of its mode and dispersion, suggesting a horizon-specific uncertainty of predictions. Furthermore, the comparison of Fig.~\ref{fig:realranks_onegraph} and Fig.~\ref{fig:postpredictive_michigan_ranker11}  highlights that in those periods where the point forecast differs from the observed one, the latter falls within the second-most frequent rank of the posterior predictive distribution.

\begin{figure}[t!h]
\centering
\setlength{\abovecaptionskip}{-2pt}
\captionsetup{width=0.93\linewidth}
\hspace*{-4.5ex}
\begin{tabular}{cc}
\includegraphics[trim=0mm 0mm 0mm 0mm,clip,height=6cm,width=8cm
]{figures/rankprob_Arrobart_team_Alabama_ranker_11.png} &
\includegraphics[trim=0mm 0mm 0mm 0mm,clip,height=6cm,width=8cm
]{figures/rankprob_Arrobart_team_Michigan_ranker_11.png}
\end{tabular}
\caption{ARROBART posterior predictive distribution (grey balls, with size proportional to the probability mass) against the observed ranks (black dots and line) for Alabama (left) and Michigan (right) ranks of AP Pollster Brian Howell.}
\label{fig:postpredictive_michigan_ranker11}
\end{figure}

\subsection{Additional Results}

\begin{figure}[t!h]
\centering
\captionsetup{width=0.93\linewidth}
\hspace*{-4.5ex}
\begin{tabular}{cc}
\includegraphics[trim=0mm 0mm 0mm 0mm,clip,height=6cm,width=8cm
]{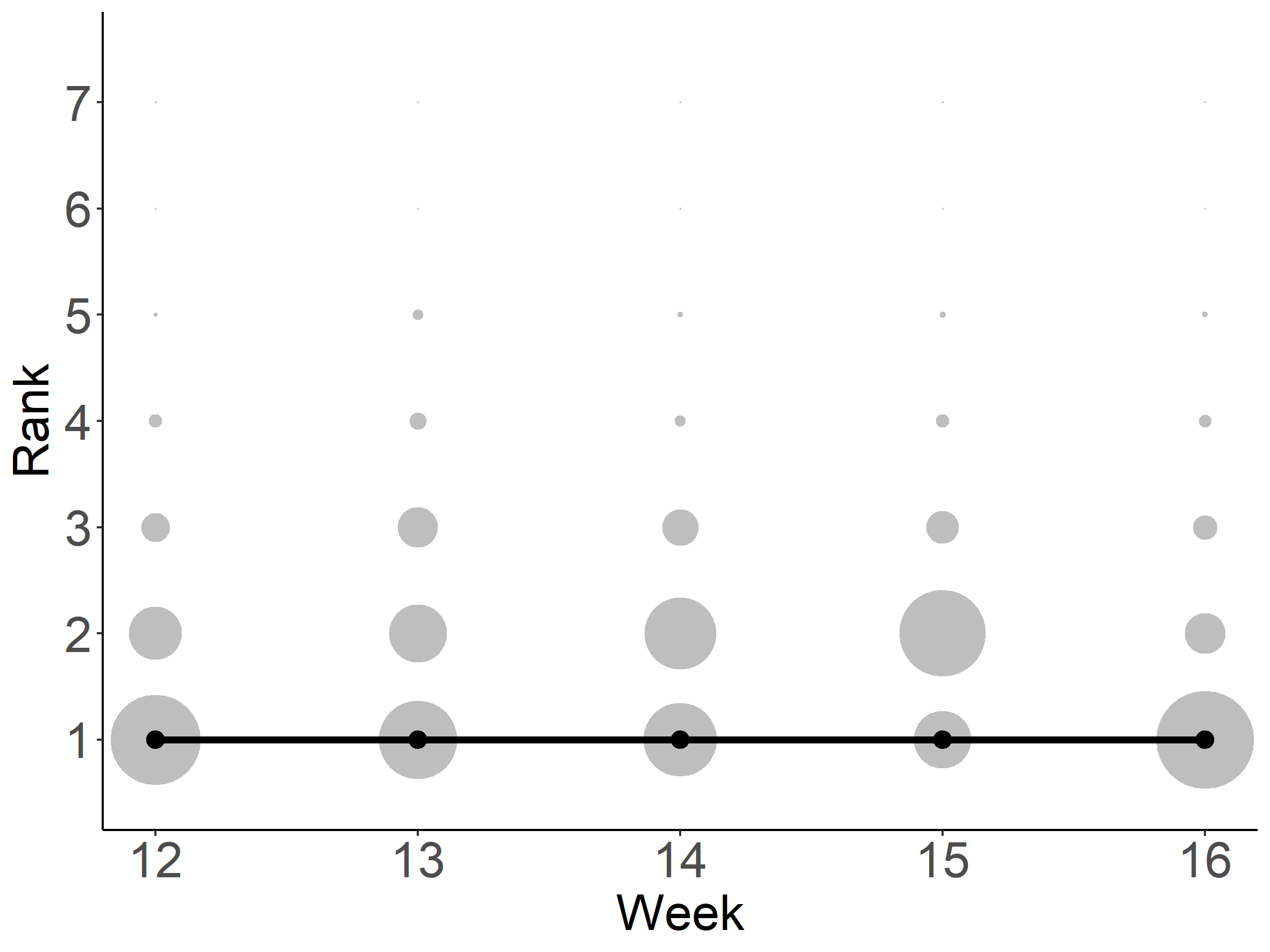} &
\includegraphics[trim=0mm 0mm 0mm 0mm,clip,height=6cm,width=8cm
]{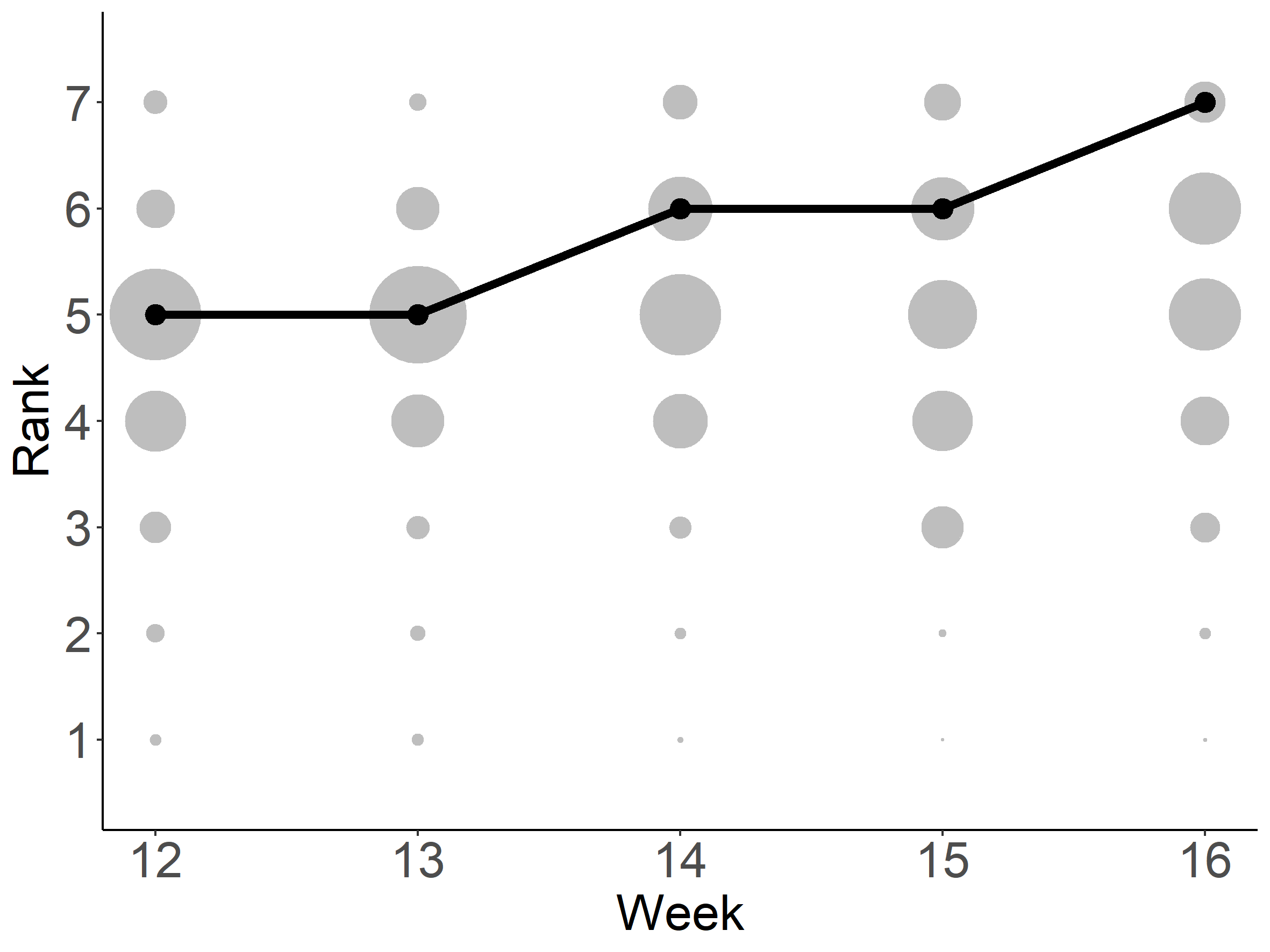}
\end{tabular}
\caption{ARROBART posterior predictive distribution (grey balls, with size proportional to the probability mass) against the observed ranks (black dots and line) for Georgia (left) and Clemson (right) ranks of AP Pollster Brian Howell.}
\label{fig:postpredictive_georgia_clemson_ranker11}
\end{figure}

\begin{figure}[t!h]
\centering
\captionsetup{width=0.93\linewidth}
\hspace*{-4.5ex}
\begin{tabular}{cc}
\includegraphics[trim=0mm 0mm 0mm 0mm,clip,height=6cm,width=8cm
]{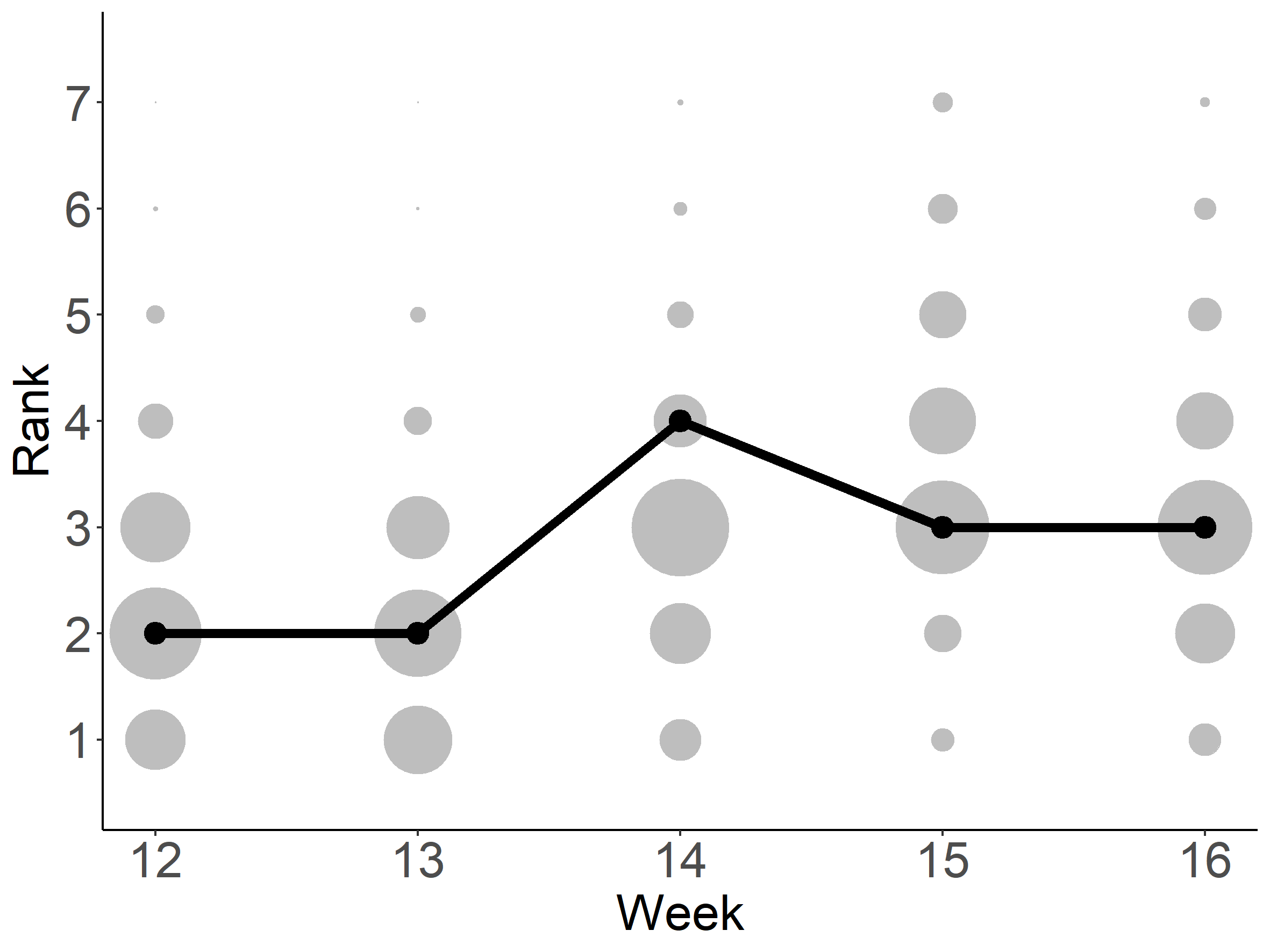} &
\includegraphics[trim=0mm 0mm 0mm 0mm,clip,height=6cm,width=8cm
]{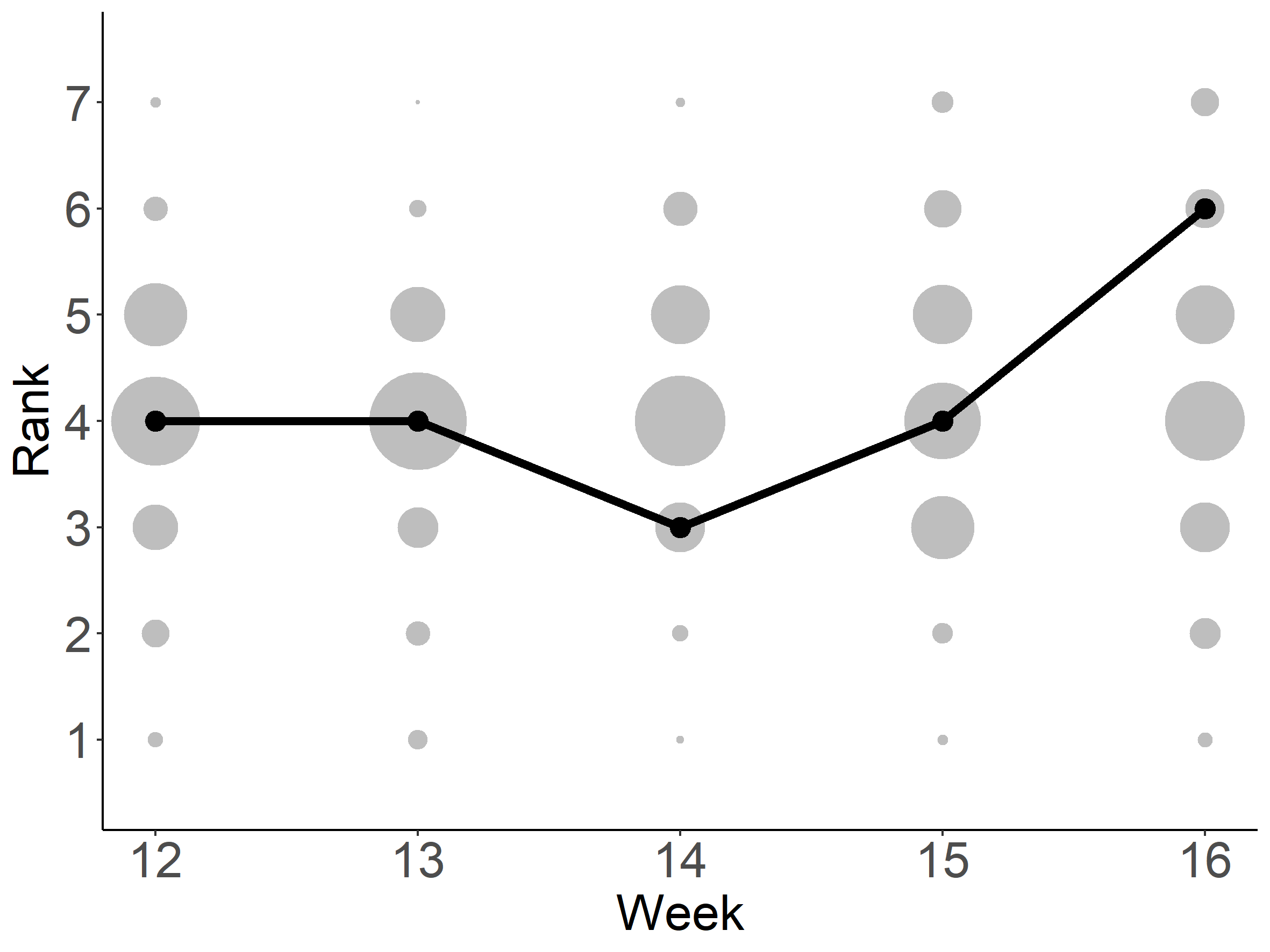}
\end{tabular}
\caption{ARROBART posterior predictive distribution (grey balls, with size proportional to the probability mass) against the observed ranks (black dots and line) for Ohio State (left) and USC (right) ranks of AP Pollster Brian Howell.}
\label{fig:postpredictive_OS_USC_ranker11}
\end{figure}

\begin{figure}[t!h]
\centering
\captionsetup{width=0.93\linewidth}
\hspace*{-4.5ex}
\begin{tabular}{cc}
\includegraphics[trim=0mm 0mm 0mm 0mm,clip,height=6cm,width=8cm
]{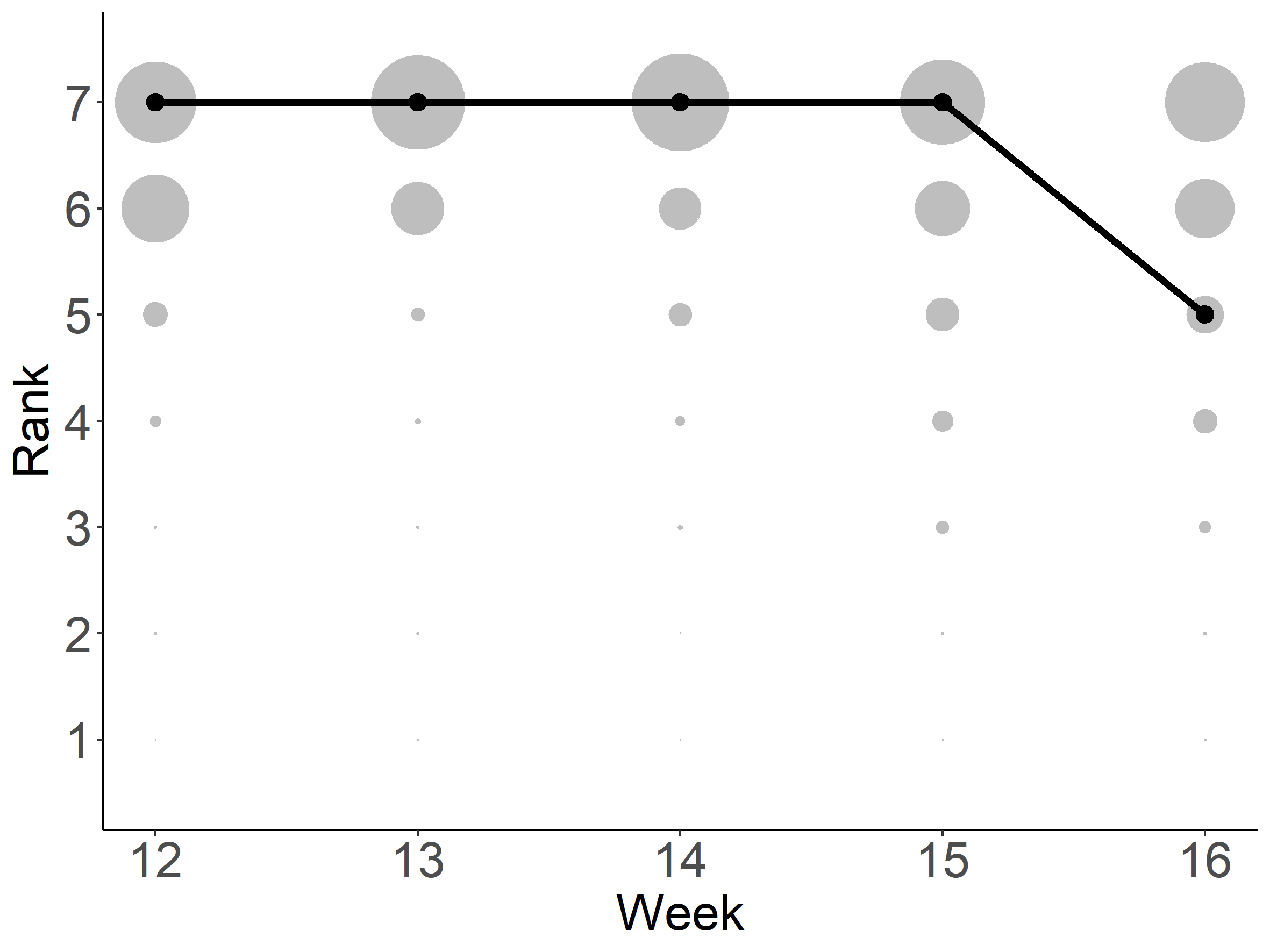} &
\end{tabular}
\caption{ARROBART posterior predictive distribution (grey balls, with size proportional to the probability mass) against the observed ranks (black dots and line) for Utah ranks of AP Pollster Brian Howell.}
\label{fig:postpredictive_utah_ranker11}
\end{figure}

Figures~\ref{fig:postpredictive_alabama_michigan_ranker11_app}-\ref{fig:postpredictive_utah_ranker11} plot the ARROBART posterior predictive distributions for each team against the observed ranks, for a single pollster (Brian Howell).

\begin{figure}[t!h]
\centering
\captionsetup{width=0.93\linewidth}
\hspace*{-4.5ex}
\begin{tabular}{cc}
\includegraphics[trim=0mm 0mm 0mm 0mm,clip,height=6cm,width=8cm
]{figures/rankprob_Arrobart_team_Alabama_ranker_11.png} &
\includegraphics[trim=0mm 0mm 0mm 0mm,clip,height=6cm,width=8cm
]{figures/rankprob_Arrobart_team_Michigan_ranker_11.png}
\end{tabular}
\caption{ARROBART posterior predictive distribution (grey balls, with size proportional to the probability mass) against the observed ranks (black dots and line) Alabama (left) and Michigan (right) ranks of AP Pollster Brian Howell.}
\label{fig:postpredictive_alabama_michigan_ranker11_app}
\end{figure}

Table~\ref{tab:football_tau_time_models_extra} provides the results of the robustness check, where the ARROBART models have been estimated using different numbers of trees (25, 50, 75). For each model, the results do not vary with the number of trees.

\begin{table}[t!h]
\centering
\begin{threeparttable}
\captionsetup{width=0.93\linewidth}
\caption{\label{tab:football_tau_time_models_extra} \small Kendall's tau distance for forecasts of the AP Football Poll 2022 rankings.}
\begin{tabular}{l c c c c c c}
\hline
  Model & t=12 & t=13 & t=14 & t=15 & t=16 & Average \\ 
  \hline
  ARROBART 25 trees & 0.044 & 0.067 & 0.092 & 0.159 & 0.073 & 0.087 \\ 
  ARROBART 50 trees & 0.044 & 0.067 & 0.137 & 0.159 & 0.073 & 0.096 \\ 
  ARROBART 75 trees & 0.044 & 0.067 & 0.137 & 0.159 & 0.073 & 0.096 \\ 
  ARROBARTX 75 trees & 0.089 & 0.032 & 0.095 & 0.222 & 0.175 & 0.123 \\
  ARROBARTX 50 trees & 0.089 & 0.035 & 0.095 & 0.222 & 0.181 & 0.124 \\ 
  ARROBARTX 25 trees & 0.089 & 0.035 & 0.159 & 0.222 & 0.181 & 0.137 \\ 
  ROLinear & 0.152 & 0.175 & 0.254 & 0.095 & 0.200 & 0.175 \\ 
  ARROLinearX  & 0.124 & 0.070 & 0.206 & 0.222 & 0.260 & 0.177 \\ 
  ARROLinear & 0.124 & 0.070 & 0.206 & 0.222 & 0.260 & 0.177 \\ 
  ROBART & 0.124 & 0.070 & 0.206 & 0.225 & 0.260 & 0.177 \\
\hline
\end{tabular}
\begin{tablenotes}
\footnotesize
\item \!\!\!\! Notes: average across pollsters by week (columns 2 to 6) and global average across weeks (column 7). Models are sorted in ascending order according to the global average across weeks.
\end{tablenotes}
\end{threeparttable}
\end{table}

\FloatBarrier
\section{Dynamic Label Ranking Application, One Ranker: Economic Complexity Index, additional details}
\label{ECI_sec}

\subsection{Data sources}

Economic Complexity Index Rankings:\\ \url{https://ourworldindata.org/grapher/economic-complexity-rankings.csv?v=1&csvType=full&useColumnShortNames=true}.

GDP Per Capita was sourced from the Maddison dataset \citep{INZBF2_2024, bolt2025maddison} via Our World in Data: \\\url{https://ourworldindata.org/grapher/gdp-per-capita-maddison-project-database.csv?v=1&csvType=full&useColumnShortNames=true}.

Population data (from which population growth was calculated) was sourced from our Word in Data, which in turn used several sources:\\ \url{https://ourworldindata.org/grapher/population.csv?v=1&csvType=full&useColumnShortNames=true}.

Cereal production data was sourced from Our World in Data, which in turn sourced the data from the Food and Agriculture Organization of the United Nations:\\ \url{https://ourworldindata.org/grapher/cereal-production.csv?v=1&csvType=full&useColumnShortNames=true}.

The age dependency ratio (from which dependency growth was calculated) data was sourced from Our World in Data, which in turn sourced the data from the United Nations:\\ \url{https://ourworldindata.org/grapher/age-dependency-ratio-of-working-age-population.csv?v=1&csvType=full&useColumnShortNames=true}.

\subsection{Results by Window}

Table \ref{tab:ECIkTaubyWindow} contains the ECI ranking application rolling window results by test year. The windows were of length 41 years. It can be observed that BART-based methods generally outperform their linear counterparts. The implementations with sparse splitting, RODART and ARRODART, give the best predictions across almost all windows, with ARRODART generally outperforming RODART.

\begin{table}[t!h]
\centering
\begin{adjustbox}{max width=\textwidth}
\begin{threeparttable}
\captionsetup{width=0.93\linewidth}
\caption{\label{tab:ECIkTaubyWindow} \small Kendall's tau distance for ranking forecasts.}
\begin{tabular}{c cccccc}
  \toprule
Test year & ROLinear & ARROLinear & ROBART & RODART & ARROBART & ARRODART \\ 
  \midrule
  2006 & 1.00 & 0.88 & 1.09 & 0.91 & 1.07 & \textbf{0.87} \\ 
  2007 & 1.00 & 1.50 & 1.02 & 0.93 & 0.97 & \textbf{0.69} \\ 
  2008 & 1.00 & 1.23 & 1.11 & 1.02 & 0.97 & \textbf{0.93} \\ 
  2009 & 1.00 & 1.36 & 0.93 & \textbf{0.70} & 1.10 & 0.73 \\ 
  2010 & 1.00 & 1.60 & \textbf{0.83} & 0.87 & 1.33 & 0.89 \\ 
  2011 & 1.00 & 1.33 & 1.09 & 0.96 & 1.10 & \textbf{0.90} \\ 
  2012 & 1.00 & 1.37 & 0.88 & 0.88 & 0.96 & \textbf{0.77} \\ 
  2013 & 1.00 & 1.44 & 1.01 & 0.99 & 0.92 & \textbf{0.81} \\ 
  2014 & 1.00 & 1.49 & 1.07 & 1.14 & 0.86 & \textbf{0.82} \\ 
  2015 & 1.00 & 1.85 & 0.81 & 0.74 & 0.75 & \textbf{0.70} \\ 
  \bottomrule
\end{tabular}
\begin{tablenotes}
\footnotesize
\item \!\!\!\! Notes: Kendall's tau distances between the one period ahead forecasted rankings and true rankings. All results except the ROLinear results are relative to ROLinear. Bold denotes the best performing model.
\end{tablenotes}
\end{threeparttable}
\end{adjustbox}
\end{table}

\FloatBarrier

\section{Temporal Dependence}
\label{sec:apdx_tempdepend_sec}

This section contains a study of the extent to which the posterior conditional means of latent outcomes and ranks have similar temporal dependence to the true rankings. 
All results in this section are for an ARROBART model trained with the lag of the latent variable and the lag of observed rankings as potential splitting variables for scenario 2 of the dynamic simulation study described in the main text.

\subsection{Pearson Autocorrelation of Latent Outcome Mean}

\begin{table}[t!h]
\centering
\begin{threeparttable}
\captionsetup{width=0.93\linewidth}
\caption{\label{tab:Ktau} \small First order Pearson autocorrelation of conditional mean of latent outcome.}
\begin{tabular}{ccc}
  \toprule
$\sigma$ & True & ARROBART lag \\ 
  \midrule
  0.1 & 0.030 & 0.023\\ 
  0.5 & 0.036 & 0.010\\ 
  1.0 & 0.058 & 0.059\\ 
  1.5 & 0.118 & 0.129\\ 
   \bottomrule
\end{tabular}
\begin{tablenotes}
\footnotesize
\item \!\!\!\! Notes: Averaged across simulation repetitions, time, items, and rankers.
\end{tablenotes}
\end{threeparttable}
\end{table}

\begin{table}[t!h]
\centering
\begin{threeparttable}
\captionsetup{width=0.93\linewidth}
\caption{\label{tab:Ktau} \small Pearson first order autocorrelation of true conditional mean of latent outcome, by ranker.}
\begin{tabular}{cccccc}
  \toprule
$\sigma$  & 1 & 2 & 3 & 4 & 5 \\ 
  \midrule
  0.1 & 0.034 & 0.019 & 0.027 & 0.040 & 0.029 \\ 
  0.5 & 0.006 & 0.046 & 0.045 & 0.043 & 0.037 \\ 
  1.0 & 0.068 & 0.044 & 0.073 & 0.048 & 0.055 \\ 
  1.5 & 0.121 & 0.106 & 0.134 & 0.116 & 0.112 \\ 
   \bottomrule
\end{tabular}
\begin{tablenotes}
\footnotesize
\item \!\!\!\! Averaged across simulation repetitions, time, and items, by ranker.
\end{tablenotes}
\end{threeparttable}
\end{table}

\begin{table}[t!h]
\centering
\begin{threeparttable}
\captionsetup{width=0.93\linewidth}
\caption{\label{tab:Ktau} \small Pearson first order autocorrelation of ARROBART posterior conditional mean of latent outcomes, by ranker.}
\begin{tabular}{cccccc}
  \toprule
$\sigma$  & 1 & 2 & 3 & 4 & 5 \\ 
  \midrule
  0.1 & 0.023 & 0.023 & 0.018 & 0.030 & 0.022 \\ 
  0.5 & -0.011 & 0.009 & 0.019 & 0.011 & 0.020 \\ 
  1.0 & 0.076 & 0.038 & 0.060 & 0.061 & 0.060 \\ 
  1.5 & 0.139 & 0.124 & 0.143 & 0.118 & 0.119 \\ 
   \bottomrule
\end{tabular}
\begin{tablenotes}
\footnotesize
\item \!\!\!\! Notes: Averaged across simulation repetitions, time, and items, by ranker.
\end{tablenotes}
\end{threeparttable}
\end{table}

\begin{table}[t!h]
\centering
\begin{threeparttable}
\captionsetup{width=0.93\linewidth}
\caption{\label{tab:Ktau} \small Pearson first order autocorrelation of true conditional mean of latent outcome, by ranker, one simulation repetition.}
\begin{tabular}{cccccc}
  \toprule
$\sigma$  & 1 & 2 & 3 & 4 & 5 \\ 
  \midrule
  0.1 & -0.001 & -0.004 & 0.040 & 0.111 & 0.029 \\ 
  0.5 & 0.040 & 0.004 & 0.067 & 0.002 & 0.058 \\ 
  1.0 & -0.008 & 0.058 & 0.062 & 0.047 & 0.120 \\ 
  1.5 & 0.131 & 0.108 & 0.105 & 0.079 & 0.150 \\ 
  2.0 & 0.152 & 0.197 & 0.154 & 0.228 & 0.176 \\ 
   \bottomrule
\end{tabular}
\begin{tablenotes}
\footnotesize
\item \!\!\!\! Notes: Averaged across time and items for one simulation repetition, by ranker.
\end{tablenotes}
\end{threeparttable}
\end{table}

\begin{table}[t!h]
\centering
\begin{threeparttable}
\captionsetup{width=0.93\linewidth}
\caption{\label{tab:Ktau} \small Pearson first order autocorrelation of ARROBART posterior conditional mean of latent outcome, by ranker, one simulation repetition.}
\begin{tabular}{cccccc}
  \toprule
$\sigma$  & 1 & 2 & 3 & 4 & 5 \\ 
  \midrule
  0.1 & 0.155 & 0.119 & 0.129 & 0.202 & 0.101 \\ 
  0.5 & 0.042 & 0.058 & 0.077 & 0.012 & 0.076 \\ 
  1.0 & 0.111 & 0.119 & 0.118 & 0.123 & 0.163 \\ 
  1.5 & 0.174 & 0.201 & 0.175 & 0.156 & 0.179 \\ 
  2.0 & 0.249 & 0.198 & 0.228 & 0.267 & 0.256 \\ 
   \bottomrule
\end{tabular}
\begin{tablenotes}
\footnotesize
\item \!\!\!\! Notes: Averaged across time and items for one simulation repetition, by ranker.
\end{tablenotes}
\end{threeparttable}
\end{table}

\FloatBarrier

\begin{table}[t!h]
\centering
\begin{adjustbox}{max width=\textwidth}
\begin{threeparttable}
\captionsetup{width=0.93\linewidth}
\caption{\label{tab:Ktau} \small Pearson first order autocorrelation of true  conditional mean of latent outcome, by item.}
\begin{tabular}{rrrrrrrrrrrrrrrrrrrrr}
  \toprule
$\sigma$  & 1 & 2 & 3 & 4 & 5 & 6 & 7 & 8 & 9 & 10 & 11 & 12 & 13 & 14 & 15 & 16 & 17 & 18 & 19 & 20 \\ 
  \midrule
0.1 & 0.053 & 0.014 & -0.001 & 0.049 & 0.033 & 0.051 & 0.019 & 0.014 & 0.021 & 0.053 & 0.008 & 0.059 & 0.013 & 0.040 & 0.031 & 0.033 & 0.047 & 0.040 & 0.018 & 0.001 \\ 
  0.5 & 0.024 & 0.056 & 0.061 & 0.037 & 0.036 & 0.010 & 0.037 & 0.039 & 0.061 & 0.038 & 0.003 & 0.032 & 0.022 & 0.039 & 0.053 & 0.024 & 0.061 & 0.035 & -0.006 & 0.047 \\ 
  1.0 & 0.035 & 0.074 & 0.045 & 0.045 & 0.073 & 0.074 & 0.088 & 0.052 & 0.092 & 0.069 & 0.078 & 0.033 & 0.038 & 0.046 & 0.058 & 0.069 & 0.035 & 0.047 & 0.054 & 0.048 \\ 
  1.5 & 0.126 & 0.163 & 0.120 & 0.123 & 0.136 & 0.147 & 0.105 & 0.087 & 0.094 & 0.080 & 0.095 & 0.155 & 0.138 & 0.133 & 0.148 & 0.075 & 0.088 & 0.117 & 0.127 & 0.101 \\ 
   \bottomrule
\end{tabular}
\begin{tablenotes}
\footnotesize
\item \!\!\!\! Notes: Averaged across time, rankers, and simulation repetitions, by item.
\end{tablenotes}
\end{threeparttable}
\end{adjustbox}
\end{table}

\begin{table}[t!h]
\centering
\begin{adjustbox}{max width=\textwidth}
\begin{threeparttable}
\captionsetup{width=0.93\linewidth}
\caption{\label{tab:Ktau} \small Pearson first order autocorrelation of ARROBART posterior conditional mean of latent outcome, by item.}
\begin{tabular}{rrrrrrrrrrrrrrrrrrrrr}
  \toprule
$\sigma$  & 1 & 2 & 3 & 4 & 5 & 6 & 7 & 8 & 9 & 10 & 11 & 12 & 13 & 14 & 15 & 16 & 17 & 18 & 19 & 20 \\ 
  \midrule
0.1 & 0.030 & 0.048 & 0.042 & 0.054 & 0.013 & 0.032 & 0.037 & 0.015 & 0.015 & 0.033 & -0.000 & 0.031 & 0.027 & 0.020 & 0.030 & -0.016 & 0.039 & -0.001 & 0.004 & 0.012 \\ 
  0.5 & 0.023 & -0.026 & 0.018 & 0.004 & 0.023 & -0.002 & -0.015 & 0.043 & 0.023 & 0.009 & -0.003 & -0.004 & -0.001 & 0.037 & -0.003 & 0.023 & 0.011 & 0.001 & 0.012 & 0.021 \\ 
  1.0 & 0.053 & 0.085 & 0.037 & 0.050 & 0.090 & 0.084 & 0.067 & 0.056 & 0.083 & 0.041 & 0.057 & 0.033 & 0.034 & 0.036 & 0.083 & 0.051 & 0.068 & 0.037 & 0.056 & 0.085 \\ 
  1.5 & 0.134 & 0.178 & 0.135 & 0.158 & 0.124 & 0.123 & 0.090 & 0.096 & 0.121 & 0.114 & 0.134 & 0.205 & 0.104 & 0.159 & 0.145 & 0.069 & 0.115 & 0.122 & 0.135 & 0.111 \\ 
   \bottomrule
\end{tabular}
\begin{tablenotes}
\footnotesize
\item \!\!\!\! Notes: Averaged across time, rankers, and simulation repetitions, by item.
\end{tablenotes}
\end{threeparttable}
\end{adjustbox}
\end{table}

\begin{table}[t!h]
\centering
\begin{adjustbox}{max width=\textwidth}
\begin{threeparttable}
\captionsetup{width=0.93\linewidth}
\caption{\label{tab:Ktau} \small Pearson first order autocorrelation of true posterior mean of latent outcome, by item, one simulation repetition.}
\begin{tabular}{rrrrrrrrrrrrrrrrrrrrr}
  \toprule
$\sigma$  & 1 & 2 & 3 & 4 & 5 & 6 & 7 & 8 & 9 & 10 & 11 & 12 & 13 & 14 & 15 & 16 & 17 & 18 & 19 & 20 \\ 
  \midrule
0.1 & 0.145 & 0.086 & 0.079 & -0.016 & 0.052 & 0.073 & 0.038 & -0.056 & -0.019 & 0.066 & -0.038 & 0.052 & 0.071 & 0.024 & 0.001 & 0.091 & 0.071 & 0.063 & -0.081 & -0.001 \\ 
  0.5 & 0.018 & 0.084 & 0.050 & 0.119 & -0.009 & -0.026 & 0.036 & 0.063 & 0.225 & 0.066 & -0.017 & -0.019 & -0.002 & 0.044 & 0.030 & -0.036 & -0.050 & 0.090 & 0.028 & -0.006 \\ 
  1.0 & 0.034 & 0.111 & 0.042 & 0.005 & 0.028 & 0.037 & 0.247 & -0.070 & -0.035 & 0.203 & 0.050 & 0.193 & 0.046 & 0.079 & -0.050 & 0.131 & 0.007 & -0.050 & 0.033 & 0.075 \\ 
  1.5 & 0.043 & 0.099 & -0.001 & 0.127 & 0.211 & 0.134 & 0.182 & 0.113 & 0.161 & 0.027 & 0.157 & 0.223 & 0.078 & 0.127 & 0.118 & -0.009 & 0.162 & 0.107 & 0.114 & 0.115 \\ 
  2 & 0.263 & 0.000 & 0.221 & 0.210 & 0.316 & 0.257 & 0.128 & 0.163 & 0.340 & 0.096 & 0.169 & 0.105 & 0.153 & 0.143 & 0.027 & 0.085 & 0.174 & 0.421 & 0.062 & 0.293 \\
   \bottomrule
\end{tabular} 
\begin{tablenotes}
\footnotesize
\item \!\!\!\! Notes: Averaged across time and rankers for one simulation repetition, by item.
\end{tablenotes}
\end{threeparttable}
\end{adjustbox}
\end{table}

\begin{table}[t!h]
\centering
\begin{adjustbox}{max width=\textwidth}
\begin{threeparttable}
\captionsetup{width=0.93\linewidth}
\caption{\label{tab:Ktau} \small Pearson first order autocorrelation of ARROBART posterior mean of latent outcome, by item, one simulation repetition.}
\begin{tabular}{rrrrrrrrrrrrrrrrrrrrr}
  \toprule
$\sigma$  & 1 & 2 & 3 & 4 & 5 & 6 & 7 & 8 & 9 & 10 & 11 & 12 & 13 & 14 & 15 & 16 & 17 & 18 & 19 & 20 \\ 
  \midrule
0.1 & 0.071 & 0.221 & 0.227 & 0.126 & 0.162 & 0.153 & 0.166 & 0.133 & 0.086 & 0.159 & 0.151 & 0.110 & 0.140 & 0.069 & 0.204 & 0.166 & 0.180 & 0.156 & -0.010 & 0.154 \\ 
  0.5 & 0.026 & 0.009 & 0.024 & -0.026 & 0.085 & 0.013 & 0.019 & -0.034 & 0.100 & 0.142 & 0.079 & 0.043 & -0.051 & 0.081 & 0.138 & 0.067 & -0.037 & 0.158 & 0.100 & 0.121 \\ 
  1.0 & 0.104 & 0.191 & 0.155 & 0.127 & 0.191 & 0.102 & 0.254 & -0.037 & -0.000 & 0.237 & 0.093 & 0.143 & 0.241 & 0.104 & 0.156 & 0.011 & 0.083 & 0.089 & 0.101 & 0.194 \\ 
  1.5 & 0.089 & 0.211 & 0.076 & 0.224 & 0.249 & 0.124 & 0.170 & 0.073 & 0.173 & 0.122 & 0.191 & 0.380 & 0.208 & 0.119 & 0.178 & 0.120 & 0.104 & 0.279 & 0.219 & 0.233 \\ 
  2 & 0.331 & 0.208 & 0.235 & 0.286 & 0.238 & 0.219 & 0.127 & 0.223 & 0.419 & 0.214 & 0.250 & 0.218 & 0.190 & 0.163 & 0.169 & 0.174 & 0.197 & 0.354 & 0.235 & 0.339 \\ 
  \bottomrule
\end{tabular}
\begin{tablenotes}
\footnotesize
\item \!\!\!\! Notes: Averaged across time and rankers for one simulation repetition, by item.
\end{tablenotes}
\end{threeparttable}
\end{adjustbox}
\end{table}

\FloatBarrier

\subsection{Pearson Autocorrelation of Latent Outcome Mean}

\begin{table}[t!h]
\centering
\begin{threeparttable}
\captionsetup{width=0.93\linewidth}
\caption{\label{tab:Ktau} \small Pearson first order autocorrelation of true rankings, by ranker, one simulation repetition.}
\begin{tabular}{cccccc}
  \toprule
$\sigma$  & 1 & 2 & 3 & 4 & 5 \\ 
  \midrule
  0.1 & 0.017 & -0.009 & 0.054 & 0.116 & 0.002 \\ 
  0.5 & -0.008 & -0.016 & 0.014 & 0.034 & 0.011 \\ 
  1.0 & -0.023 & 0.047 & 0.039 & -0.020 & 0.092 \\ 
  1.5 & 0.044 & 0.057 & 0.054 & 0.061 & 0.029 \\ 
  2.0 & 0.082 & 0.093 & 0.082 & 0.095 & 0.067 \\ 
   \bottomrule
\end{tabular}
\begin{tablenotes}
\footnotesize
\item \!\!\!\! Notes: Averaged across time and items for one simulation repetition, by ranker.
\end{tablenotes}
\end{threeparttable}
\end{table}

\begin{table}[t!h]
\centering
\begin{threeparttable}
\captionsetup{width=0.93\linewidth}
\caption{\label{tab:Ktau} \small Kendall first order autocorrelation of true rankings, by ranker, one simulation repetition.}
\begin{tabular}{cccccc}
  \toprule
$\sigma$  & 1 & 2 & 3 & 4 & 5 \\ 
  \midrule
  0.1 & 0.014 & -0.009 & 0.037 & 0.085 & 0.003 \\ 
  0.5 & -0.002 & -0.016 & 0.004 & 0.023 & 0.005 \\ 
  1.0 & -0.017 & 0.036 & 0.025 & -0.016 & 0.065 \\ 
  1.5 & 0.031 & 0.041 & 0.039 & 0.038 & 0.021 \\ 
  2.0 & 0.059 & 0.075 & 0.054 & 0.067 & 0.047 \\ 
   \bottomrule
\end{tabular}
\begin{tablenotes}
\footnotesize
\item \!\!\!\! Notes: Averaged across time and items for one simulation repetition, by ranker.
\end{tablenotes}
\end{threeparttable}
\end{table}

\begin{table}[t!h]
\centering
\begin{threeparttable}
\captionsetup{width=0.93\linewidth}
\caption{\label{tab:Ktau} \small Spearman first order autocorrelation of true rankings, by ranker, one simulation repetition.}
\begin{tabular}{cccccc}
  \toprule
$\sigma$  & 1 & 2 & 3 & 4 & 5 \\ 
  \midrule
0.1 & 0.021 & -0.014 & 0.053 & 0.117 & 0.001 \\ 
  0.5 & -0.001 & -0.021 & 0.005 & 0.035 & 0.008 \\ 
  1 & -0.022 & 0.051 & 0.037 & -0.021 & 0.094 \\ 
  1.5 & 0.041 & 0.061 & 0.055 & 0.057 & 0.030 \\ 
  2 & 0.082 & 0.102 & 0.077 & 0.091 & 0.068 \\ 
   \bottomrule
\end{tabular}
\begin{tablenotes}
\footnotesize
\item \!\!\!\! Notes: Averaged across time and items for one simulation repetition, by ranker.
\end{tablenotes}
\end{threeparttable}
\end{table}

\begin{table}[t!h]
\centering
\begin{threeparttable}
\captionsetup{width=0.93\linewidth}
\caption{\label{tab:Ktau} \small Pearson first order autocorrelation of predicted rankings, by ranker, one simulation repetition.}
\begin{tabular}{cccccc}
  \toprule
$\sigma$  & 1 & 2 & 3 & 4 & 5 \\ 
  \midrule
0.1 & 0.156 & 0.130 & 0.167 & 0.219 & 0.105 \\ 
  0.5 & 0.041 & 0.038 & 0.086 & 0.002 & 0.083 \\ 
  1.0 & 0.086 & 0.090 & 0.105 & 0.096 & 0.108 \\ 
  1.5 & 0.163 & 0.169 & 0.185 & 0.144 & 0.152 \\ 
  2.0 & 0.175 & 0.173 & 0.194 & 0.171 & 0.200 \\ 
   \bottomrule
\end{tabular}
\begin{tablenotes}
\footnotesize
\item \!\!\!\! Notes: Averaged across time and items for one simulation repetition, by ranker.
\end{tablenotes}
\end{threeparttable}
\end{table}

\begin{table}[t!h]
\centering
\begin{threeparttable}
\captionsetup{width=0.93\linewidth}
\caption{\label{tab:Ktau} \small Kendall first order autocorrelation of predicted rankings, by ranker, one simulation repetition.}
\begin{tabular}{cccccc}
  \toprule
$\sigma$  & 1 & 2 & 3 & 4 & 5 \\ 
  \midrule
0.1 & 0.112 & 0.087 & 0.117 & 0.151 & 0.069 \\ 
  0.5 & 0.035 & 0.023 & 0.058 & 0.005 & 0.054 \\ 
  1.0 & 0.055 & 0.070 & 0.077 & 0.055 & 0.077 \\ 
  1.5 & 0.109 & 0.119 & 0.136 & 0.098 & 0.115 \\ 
  2.0 & 0.121 & 0.119 & 0.133 & 0.119 & 0.138 \\ 
   \bottomrule
\end{tabular}
\begin{tablenotes}
\footnotesize
\item \!\!\!\! Notes: Averaged across time and items for one simulation repetition, by ranker.
\end{tablenotes}
\end{threeparttable}
\end{table}

\begin{table}[t!h]
\centering
\begin{threeparttable}
\captionsetup{width=0.93\linewidth}
\caption{\label{tab:Ktau} \small Spearman first order autocorrelation of predicted rankings, by ranker, one simulation repetition.}
\begin{tabular}{cccccc}
  \toprule
$\sigma$  & 1 & 2 & 3 & 4 & 5 \\ 
  \midrule
0.1 & 0.155 & 0.128 & 0.161 & 0.210 & 0.096 \\ 
  0.5 & 0.048 & 0.033 & 0.080 & 0.002 & 0.074 \\ 
  1.0 & 0.076 & 0.099 & 0.107 & 0.081 & 0.107 \\ 
  1.5 & 0.156 & 0.168 & 0.189 & 0.140 & 0.160 \\ 
  2.0 & 0.169 & 0.171 & 0.187 & 0.166 & 0.192 \\ 
   \bottomrule
\end{tabular}
\begin{tablenotes}
\footnotesize
\item \!\!\!\! Notes: Averaged across time and items for one simulation repetition, by ranker.
\end{tablenotes}
\end{threeparttable}
\end{table}

\begin{table}[t!h]
\centering
\begin{adjustbox}{max width=\textwidth}
\begin{threeparttable}
\captionsetup{width=0.93\linewidth}
\caption{\label{tab:Ktau} \small Pearson first order autocorrelation of true rankings, by item, one simulation repetition.}
\begin{tabular}{rrrrrrrrrrrrrrrrrrrrr}
  \toprule
$\sigma$  & 1 & 2 & 3 & 4 & 5 & 6 & 7 & 8 & 9 & 10 & 11 & 12 & 13 & 14 & 15 & 16 & 17 & 18 & 19 & 20 \\ 
  \midrule
0.1 & 0.065 & 0.026 & 0.172 & 0.055 & 0.063 & 0.036 & 0.038 & -0.032 & -0.031 & 0.103 & -0.035 & 0.046 & 0.032 & 0.034 & 0.026 & 0.111 & 0.047 & 0.057 & -0.102 & 0.015 \\ 
  0.5 & -0.084 & 0.063 & -0.005 & 0.084 & 0.028 & -0.122 & -0.007 & -0.032 & 0.076 & -0.037 & 0.044 & 0.038 & 0.046 & 0.079 & -0.017 & -0.025 & -0.070 & 0.079 & 0.055 & -0.051 \\ 
  1.0 & 0.037 & 0.077 & 0.060 & 0.005 & 0.029 & -0.030 & 0.155 & -0.089 & -0.051 & 0.121 & 0.057 & -0.010 & 0.142 & 0.009 & -0.040 & 0.112 & 0.016 & -0.066 & -0.050 & 0.053 \\ 
  1.5 & 0.017 & 0.042 & -0.052 & 0.047 & 0.077 & 0.023 & 0.094 & -0.006 & 0.168 & -0.028 & 0.094 & 0.053 & 0.036 & 0.034 & 0.084 & -0.053 & 0.022 & 0.170 & 0.100 & 0.057 \\ 
  2.0 & 0.167 & 0.012 & 0.082 & 0.098 & 0.091 & 0.053 & 0.032 & 0.082 & 0.264 & -0.006 & 0.086 & -0.012 & 0.097 & 0.099 & 0.030 & 0.049 & 0.033 & 0.166 & 0.103 & 0.151 \\
   \bottomrule
\end{tabular} 
\begin{tablenotes}
\footnotesize
\item \!\!\!\! Notes: Averaged across time and rankers for one simulation repetition, by item.
\end{tablenotes}
\end{threeparttable}
\end{adjustbox}
\end{table}

\begin{table}[t!h]
\centering
\begin{adjustbox}{max width=\textwidth}
\begin{threeparttable}
\captionsetup{width=0.93\linewidth}
\caption{\label{tab:Ktau} \small Kendall first order autocorrelation of true rankings, by item, one simulation repetition.}
\begin{tabular}{rrrrrrrrrrrrrrrrrrrrr}
  \toprule
$\sigma$  & 1 & 2 & 3 & 4 & 5 & 6 & 7 & 8 & 9 & 10 & 11 & 12 & 13 & 14 & 15 & 16 & 17 & 18 & 19 & 20 \\ 
  \midrule
0.1 & 0.033 & 0.006 & 0.125 & 0.050 & 0.054 & 0.022 & 0.037 & -0.017 & -0.031 & 0.067 & -0.017 & 0.033 & 0.036 & 0.024 & 0.005 & 0.086 & 0.030 & 0.042 & -0.075 & 0.011 \\ 
  0.5 & -0.067 & 0.028 & 0.007 & 0.066 & 0.017 & -0.090 & -0.006 & -0.024 & 0.030 & -0.034 & 0.036 & 0.034 & 0.044 & 0.046 & -0.026 & -0.008 & -0.042 & 0.056 & 0.031 & -0.045 \\ 
  1.0 & 0.016 & 0.058 & 0.048 & -0.002 & 0.020 & -0.024 & 0.115 & -0.063 & -0.033 & 0.095 & 0.043 & -0.002 & 0.103 & 0.012 & -0.030 & 0.065 & 0.000 & -0.050 & -0.029 & 0.035 \\ 
  1.5 & 0.017 & 0.043 & -0.038 & 0.050 & 0.056 & 0.023 & 0.062 & -0.012 & 0.125 & -0.038 & 0.058 & 0.045 & 0.021 & 0.020 & 0.057 & -0.037 & 0.007 & 0.117 & 0.066 & 0.037 \\ 
  2.0 & 0.136 & 0.003 & 0.061 & 0.070 & 0.082 & 0.041 & 0.017 & 0.055 & 0.193 & -0.009 & 0.056 & -0.023 & 0.068 & 0.064 & 0.033 & 0.043 & 0.019 & 0.126 & 0.062 & 0.108 \\
   \bottomrule
\end{tabular}
\begin{tablenotes}
\footnotesize
\item \!\!\!\! Notes: Averaged across time and rankers for one simulation repetition, by item.
\end{tablenotes}
\end{threeparttable}
\end{adjustbox}
\end{table}

\begin{table}[t!h]
\centering
\begin{adjustbox}{max width=\textwidth}
\begin{threeparttable}
\captionsetup{width=0.93\linewidth}
\caption{\label{tab:Ktau} \small Spearman first order autocorrelation of true rankings, by item, one simulation repetition.}
\begin{tabular}{rrrrrrrrrrrrrrrrrrrrr}
  \toprule
$\sigma$  & 1 & 2 & 3 & 4 & 5 & 6 & 7 & 8 & 9 & 10 & 11 & 12 & 13 & 14 & 15 & 16 & 17 & 18 & 19 & 20 \\ 
  \midrule
0.1 & 0.047 & 0.010 & 0.170 & 0.068 & 0.068 & 0.036 & 0.049 & -0.023 & -0.039 & 0.099 & -0.027 & 0.052 & 0.042 & 0.032 & 0.002 & 0.117 & 0.044 & 0.047 & -0.103 & 0.019 \\ 
  0.5 & -0.090 & 0.041 & 0.009 & 0.085 & 0.026 & -0.125 & -0.008 & -0.029 & 0.049 & -0.038 & 0.047 & 0.044 & 0.060 & 0.067 & -0.041 & -0.011 & -0.051 & 0.083 & 0.043 & -0.060 \\ 
  1.0 & 0.035 & 0.081 & 0.069 & -0.000 & 0.034 & -0.038 & 0.157 & -0.084 & -0.042 & 0.116 & 0.071 & -0.003 & 0.149 & 0.012 & -0.042 & 0.097 & -0.004 & -0.069 & -0.039 & 0.056 \\ 
  1.5 & 0.018 & 0.060 & -0.053 & 0.067 & 0.092 & 0.029 & 0.083 & -0.018 & 0.176 & -0.046 & 0.069 & 0.061 & 0.045 & 0.039 & 0.085 & -0.053 & 0.010 & 0.165 & 0.098 & 0.046 \\ 
  2.0 & 0.186 & 0.002 & 0.080 & 0.106 & 0.104 & 0.064 & 0.029 & 0.078 & 0.268 & -0.012 & 0.076 & -0.021 & 0.094 & 0.085 & 0.038 & 0.059 & 0.029 & 0.181 & 0.086 & 0.148 \\
   \bottomrule
\end{tabular}
\begin{tablenotes}
\footnotesize
\item \!\!\!\! Notes: Averaged across time and rankers for one simulation repetition, by item.
\end{tablenotes}
\end{threeparttable}
\end{adjustbox}
\end{table}

\begin{table}[t!h]
\centering
\begin{adjustbox}{max width=\textwidth}
\begin{threeparttable}
\captionsetup{width=0.93\linewidth}
\caption{\label{tab:Ktau} \small Pearson first order autocorrelation of predicted rankings, by item, one simulation repetition.}
\begin{tabular}{rrrrrrrrrrrrrrrrrrrrr}
  \toprule
$\sigma$  & 1 & 2 & 3 & 4 & 5 & 6 & 7 & 8 & 9 & 10 & 11 & 12 & 13 & 14 & 15 & 16 & 17 & 18 & 19 & 20 \\ 
  \midrule
0.1 & 0.064 & 0.260 & 0.285 & 0.158 & 0.161 & 0.170 & 0.150 & 0.161 & 0.077 & 0.126 & 0.150 & 0.117 & 0.214 & 0.072 & 0.168 & 0.213 & 0.216 & 0.169 & 0.014 & 0.166 \\ 
  0.5 & -0.018 & -0.000 & 0.003 & -0.021 & 0.096 & 0.039 & 0.017 & -0.027 & 0.135 & 0.154 & 0.113 & 0.039 & -0.012 & 0.073 & 0.093 & 0.079 & -0.076 & 0.129 & 0.039 & 0.148 \\ 
  1.0 & 0.082 & 0.086 & 0.118 & 0.132 & 0.169 & 0.103 & 0.175 & -0.059 & -0.036 & 0.203 & 0.041 & 0.127 & 0.231 & 0.033 & 0.181 & -0.011 & 0.074 & 0.045 & 0.086 & 0.161 \\ 
  1.5 & 0.166 & 0.205 & 0.072 & 0.258 & 0.177 & 0.103 & 0.101 & 0.076 & 0.102 & 0.174 & 0.183 & 0.234 & 0.180 & 0.115 & 0.220 & 0.114 & 0.074 & 0.363 & 0.157 & 0.180 \\ 
  2.0 & 0.277 & 0.205 & 0.235 & 0.263 & 0.176 & 0.257 & 0.119 & 0.182 & 0.228 & 0.158 & 0.191 & 0.132 & 0.142 & 0.150 & 0.182 & 0.153 & 0.111 & 0.132 & 0.131 & 0.226 \\ 
   \bottomrule
\end{tabular} 
\begin{tablenotes}
\footnotesize
\item \!\!\!\! Notes: Averaged across time and rankers for one simulation repetition, by item.
\end{tablenotes}
\end{threeparttable}
\end{adjustbox}
\end{table}

\begin{table}[t!h]
\centering
\begin{adjustbox}{max width=\textwidth}
\begin{threeparttable}
\captionsetup{width=0.93\linewidth}
\caption{\label{tab:Ktau} \small Kendall first order autocorrelation of predicted rankings, by item, one simulation repetition.}
\begin{tabular}{rrrrrrrrrrrrrrrrrrrrr}
  \toprule
$\sigma$  & 1 & 2 & 3 & 4 & 5 & 6 & 7 & 8 & 9 & 10 & 11 & 12 & 13 & 14 & 15 & 16 & 17 & 18 & 19 & 20 \\ 
  \midrule
0.1 & 0.032 & 0.183 & 0.193 & 0.104 & 0.119 & 0.138 & 0.104 & 0.118 & 0.046 & 0.091 & 0.114 & 0.069 & 0.138 & 0.048 & 0.121 & 0.142 & 0.145 & 0.129 & -0.009 & 0.121 \\ 
  0.5 & -0.017 & 0.017 & 0.009 & -0.016 & 0.062 & 0.027 & 0.008 & -0.013 & 0.076 & 0.102 & 0.091 & 0.029 & -0.009 & 0.063 & 0.066 & 0.049 & -0.055 & 0.078 & 0.031 & 0.101 \\ 
  1.0 & 0.059 & 0.067 & 0.072 & 0.081 & 0.102 & 0.056 & 0.122 & -0.032 & -0.035 & 0.155 & 0.036 & 0.080 & 0.161 & 0.035 & 0.121 & 0.002 & 0.037 & 0.041 & 0.056 & 0.119 \\ 
  1.5 & 0.121 & 0.138 & 0.067 & 0.191 & 0.131 & 0.073 & 0.053 & 0.064 & 0.077 & 0.123 & 0.116 & 0.167 & 0.123 & 0.066 & 0.162 & 0.082 & 0.055 & 0.250 & 0.131 & 0.122 \\ 
  2.0 & 0.212 & 0.120 & 0.161 & 0.191 & 0.130 & 0.162 & 0.082 & 0.135 & 0.192 & 0.090 & 0.136 & 0.099 & 0.094 & 0.115 & 0.117 & 0.104 & 0.068 & 0.104 & 0.064 & 0.146 \\ 
   \bottomrule
\end{tabular} 
\begin{tablenotes}
\footnotesize
\item \!\!\!\! Notes: Averaged across time and rankers for one simulation repetition, by item.
\end{tablenotes}
\end{threeparttable}
\end{adjustbox}
\end{table}

\begin{table}[t!h]
\centering
\begin{adjustbox}{max width=\textwidth}
\begin{threeparttable}
\captionsetup{width=0.93\linewidth}
\caption{\label{tab:Ktau} \small Spearman first order autocorrelation of predicted rankings, by item, one simulation repetition.}
\begin{tabular}{rrrrrrrrrrrrrrrrrrrrr}
  \toprule
$\sigma$  & 1 & 2 & 3 & 4 & 5 & 6 & 7 & 8 & 9 & 10 & 11 & 12 & 13 & 14 & 15 & 16 & 17 & 18 & 19 & 20 \\ 
  \midrule
0.1 & 0.048 & 0.253 & 0.282 & 0.146 & 0.160 & 0.193 & 0.138 & 0.157 & 0.068 & 0.128 & 0.156 & 0.099 & 0.199 & 0.064 & 0.175 & 0.202 & 0.195 & 0.178 & -0.003 & 0.161 \\ 
  0.5 & -0.016 & 0.019 & 0.007 & -0.022 & 0.086 & 0.049 & 0.003 & -0.023 & 0.108 & 0.139 & 0.116 & 0.031 & -0.011 & 0.082 & 0.095 & 0.072 & -0.085 & 0.113 & 0.040 & 0.145 \\ 
  1.0 & 0.077 & 0.098 & 0.108 & 0.120 & 0.145 & 0.074 & 0.167 & -0.056 & -0.040 & 0.224 & 0.041 & 0.118 & 0.223 & 0.049 & 0.172 & -0.001 & 0.054 & 0.059 & 0.083 & 0.166 \\ 
  1.5 & 0.174 & 0.206 & 0.091 & 0.264 & 0.183 & 0.103 & 0.078 & 0.076 & 0.109 & 0.173 & 0.167 & 0.232 & 0.173 & 0.089 & 0.224 & 0.122 & 0.081 & 0.353 & 0.175 & 0.174 \\ 
  2.0 & 0.296 & 0.174 & 0.231 & 0.268 & 0.184 & 0.234 & 0.110 & 0.192 & 0.267 & 0.136 & 0.189 & 0.131 & 0.120 & 0.162 & 0.164 & 0.146 & 0.093 & 0.149 & 0.092 & 0.203 \\ 
   \bottomrule
\end{tabular} 
\begin{tablenotes}
\footnotesize
\item \!\!\!\! Notes: Averaged across time and rankers for one simulation repetition, by item.
\end{tablenotes}
\end{threeparttable}
\end{adjustbox}
\end{table}

\begin{table}[t!h]
\centering
\begin{threeparttable}
\captionsetup{width=0.93\linewidth}
\caption{\label{tab:Ktau} \small Pearson first order autocorrelation of true rankings, by simulation repetition.}
\begin{tabular}{rrrrrrrrrrr}
  \toprule
 & 1 & 2 & 3 & 4 & 5 & 6 & 7 & 8 & 9 & 10 \\ 
  \midrule
0.1 & 0.036 & 0.027 & 0.004 & 0.034 & 0.049 & 0.057 & 0.011 & 0.035 & -0.012 & 0.115 \\ 
  0.5 & 0.007 & 0.033 & 0.045 & 0.010 & 0.024 & 0.021 & 0.031 & 0.026 & 0.014 & 0.005 \\ 
  1.0 & 0.027 & 0.041 & 0.057 & 0.013 & 0.028 & 0.024 & 0.027 & 0.050 & 0.040 & 0.033 \\ 
  1.5 & 0.049 & 0.076 & 0.081 & 0.056 & 0.060 & 0.049 & 0.057 & 0.053 & 0.058 & 0.075 \\ 
  2.0 & 0.084 & 0.106 & 0.113 & 0.108 & 0.140 & 0.091 & 0.118 & 0.123 & 0.110 & 0.115 \\ 
   \bottomrule
\end{tabular}
\begin{tablenotes}
\footnotesize
\item \!\!\!\! Notes: Averaged across time, ranker, and items, by simulation repetition.
\end{tablenotes}
\end{threeparttable}
\end{table}

\begin{table}[t!h]
\centering
\begin{threeparttable}
\captionsetup{width=0.93\linewidth}
\caption{\label{tab:Ktau} \small Kendall first order autocorrelation of true rankings, by simulation repetition.}
\begin{tabular}{rrrrrrrrrrr}
  \toprule
 & 1 & 2 & 3 & 4 & 5 & 6 & 7 & 8 & 9 & 10 \\ 
  \midrule
0.1 & 0.026 & 0.017 & 0.004 & 0.025 & 0.032 & 0.038 & 0.009 & 0.026 & -0.010 & 0.082 \\ 
  0.5 & 0.003 & 0.023 & 0.033 & 0.004 & 0.016 & 0.018 & 0.021 & 0.019 & 0.009 & 0.002 \\ 
  1.0 & 0.019 & 0.032 & 0.039 & 0.012 & 0.017 & 0.016 & 0.017 & 0.035 & 0.029 & 0.026 \\ 
  1.5 & 0.034 & 0.055 & 0.061 & 0.041 & 0.042 & 0.034 & 0.041 & 0.033 & 0.042 & 0.053 \\ 
  2.0 & 0.060 & 0.080 & 0.086 & 0.080 & 0.101 & 0.069 & 0.083 & 0.089 & 0.081 & 0.082 \\ 
   \bottomrule
\end{tabular}
\begin{tablenotes}
\footnotesize
\item \!\!\!\! Notes: Averaged across time, ranker, and items, by simulation repetition.
\end{tablenotes}
\end{threeparttable}
\end{table}

\begin{table}[t!h]
\centering
\begin{threeparttable}
\captionsetup{width=0.93\linewidth}
\caption{\label{tab:Ktau} \small Spearman first order autocorrelation of true rankings, by simulation repetition.}
\begin{tabular}{rrrrrrrrrrr}
  \toprule
 & 1 & 2 & 3 & 4 & 5 & 6 & 7 & 8 & 9 & 10 \\ 
  \midrule
0.1 & 0.035 & 0.024 & 0.006 & 0.035 & 0.045 & 0.053 & 0.013 & 0.036 & -0.016 & 0.114 \\ 
  0.5 & 0.005 & 0.029 & 0.046 & 0.007 & 0.021 & 0.022 & 0.030 & 0.026 & 0.014 & 0.005 \\ 
  1.0 & 0.028 & 0.045 & 0.056 & 0.015 & 0.023 & 0.023 & 0.025 & 0.050 & 0.040 & 0.034 \\ 
  1.5 & 0.049 & 0.077 & 0.086 & 0.059 & 0.059 & 0.047 & 0.058 & 0.048 & 0.059 & 0.073 \\ 
  2.0 & 0.084 & 0.113 & 0.116 & 0.113 & 0.140 & 0.095 & 0.117 & 0.122 & 0.112 & 0.114 \\ 
    \bottomrule
\end{tabular}
\begin{tablenotes}
\footnotesize
\item \!\!\!\! Notes: Averaged across time, ranker, and items, by simulation repetition.
\end{tablenotes}
\end{threeparttable}
\end{table}

\begin{table}[t!h]
\centering
\begin{threeparttable}
\captionsetup{width=0.93\linewidth}
\caption{\label{tab:Ktau} \small Pearson first order autocorrelation of predicted rankings, by simulation repetition.}
\begin{tabular}{rrrrrrrrrrr}
  \toprule
 & 1 & 2 & 3 & 4 & 5 & 6 & 7 & 8 & 9 & 10 \\ 
  \midrule
0.1 & 0.156 & 0.071 & -0.089 & 0.113 & 0.170 & -0.024 & 0.018 & -0.049 & -0.079 & 0.169 \\ 
  0.5 & 0.050 & 0.005 & -0.036 & -0.026 & 0.051 & 0.034 & 0.077 & 0.021 & -0.116 & -0.033 \\ 
  1.0 & 0.097 & 0.145 & -0.076 & 0.052 & 0.021 & 0.022 & 0.041 & 0.040 & -0.028 & 0.103 \\ 
  1.5 & 0.163 & 0.055 & 0.020 & 0.126 & 0.161 & 0.129 & -0.004 & 0.132 & -0.029 & 0.224 \\ 
  2.0 & 0.183 & 0.200 & 0.135 & 0.087 & 0.158 & 0.050 & 0.111 & 0.164 & 0.127 & 0.169 \\ 
  \bottomrule
\end{tabular}
\begin{tablenotes}
\footnotesize
\item \!\!\!\! Notes: Averaged across time, ranker, and items, by simulation repetition.
\end{tablenotes}
\end{threeparttable}
\end{table}

\begin{table}[t!h]
\centering
\begin{threeparttable}
\captionsetup{width=0.93\linewidth}
\caption{\label{tab:Ktau} \small Kendall first order autocorrelation of predicted rankings, by simulation repetition.}
\begin{tabular}{rrrrrrrrrrr}
  \toprule
 & 1 & 2 & 3 & 4 & 5 & 6 & 7 & 8 & 9 & 10 \\ 
  \midrule
0.1 & 0.107 & 0.050 & -0.064 & 0.083 & 0.120 & -0.016 & 0.016 & -0.032 & -0.055 & 0.124 \\ 
  0.5 & 0.035 & 0.005 & -0.023 & -0.018 & 0.035 & 0.022 & 0.054 & 0.011 & -0.085 & -0.021 \\ 
  1.0 & 0.067 & 0.103 & -0.051 & 0.037 & 0.014 & 0.018 & 0.031 & 0.030 & -0.020 & 0.074 \\ 
  1.5 & 0.116 & 0.040 & 0.017 & 0.089 & 0.115 & 0.091 & -0.005 & 0.092 & -0.019 & 0.161 \\ 
  2.0 & 0.126 & 0.145 & 0.099 & 0.064 & 0.118 & 0.044 & 0.081 & 0.116 & 0.092 & 0.124 \\ 
  \bottomrule
\end{tabular}
\begin{tablenotes}
\footnotesize
\item \!\!\!\! Notes: Averaged across time, ranker, and items, by simulation repetition.
\end{tablenotes}
\end{threeparttable}
\end{table}

\begin{table}[t!h]
\centering
\begin{threeparttable}
\captionsetup{width=0.93\linewidth}
\caption{\label{tab:Ktau} \small Spearman first order autocorrelation of predicted rankings, by simulation repetition.}
\begin{tabular}{rrrrrrrrrrr}
  \toprule
 & 1 & 2 & 3 & 4 & 5 & 6 & 7 & 8 & 9 & 10 \\ 
  \midrule
0.1 & 0.150 & 0.068 & -0.089 & 0.117 & 0.167 & -0.022 & 0.022 & -0.044 & -0.077 & 0.172 \\ 
  0.5 & 0.047 & 0.008 & -0.034 & -0.026 & 0.048 & 0.030 & 0.076 & 0.021 & -0.121 & -0.032 \\ 
  1 & 0.094 & 0.147 & -0.073 & 0.054 & 0.019 & 0.026 & 0.042 & 0.043 & -0.031 & 0.104 \\ 
  1.5 & 0.162 & 0.055 & 0.024 & 0.123 & 0.161 & 0.126 & -0.006 & 0.131 & -0.026 & 0.222 \\ 
  2 & 0.177 & 0.201 & 0.137 & 0.091 & 0.160 & 0.064 & 0.113 & 0.163 & 0.130 & 0.172 \\ 
  \bottomrule
\end{tabular}
\begin{tablenotes}
\footnotesize
\item \!\!\!\! Notes: Averaged across time, ranker, and items, by simulation repetition.
\end{tablenotes}
\end{threeparttable}
\end{table}

\FloatBarrier

\subsection{Autocorrelation Function Plots}

\begin{figure}[t!h]
\centering
\captionsetup{width=0.93\linewidth}
\hspace*{-4.5ex}
\begin{tabular}{cc}
\includegraphics[trim=0mm 0mm 0mm 0mm,clip,height=6cm,width=8cm
]{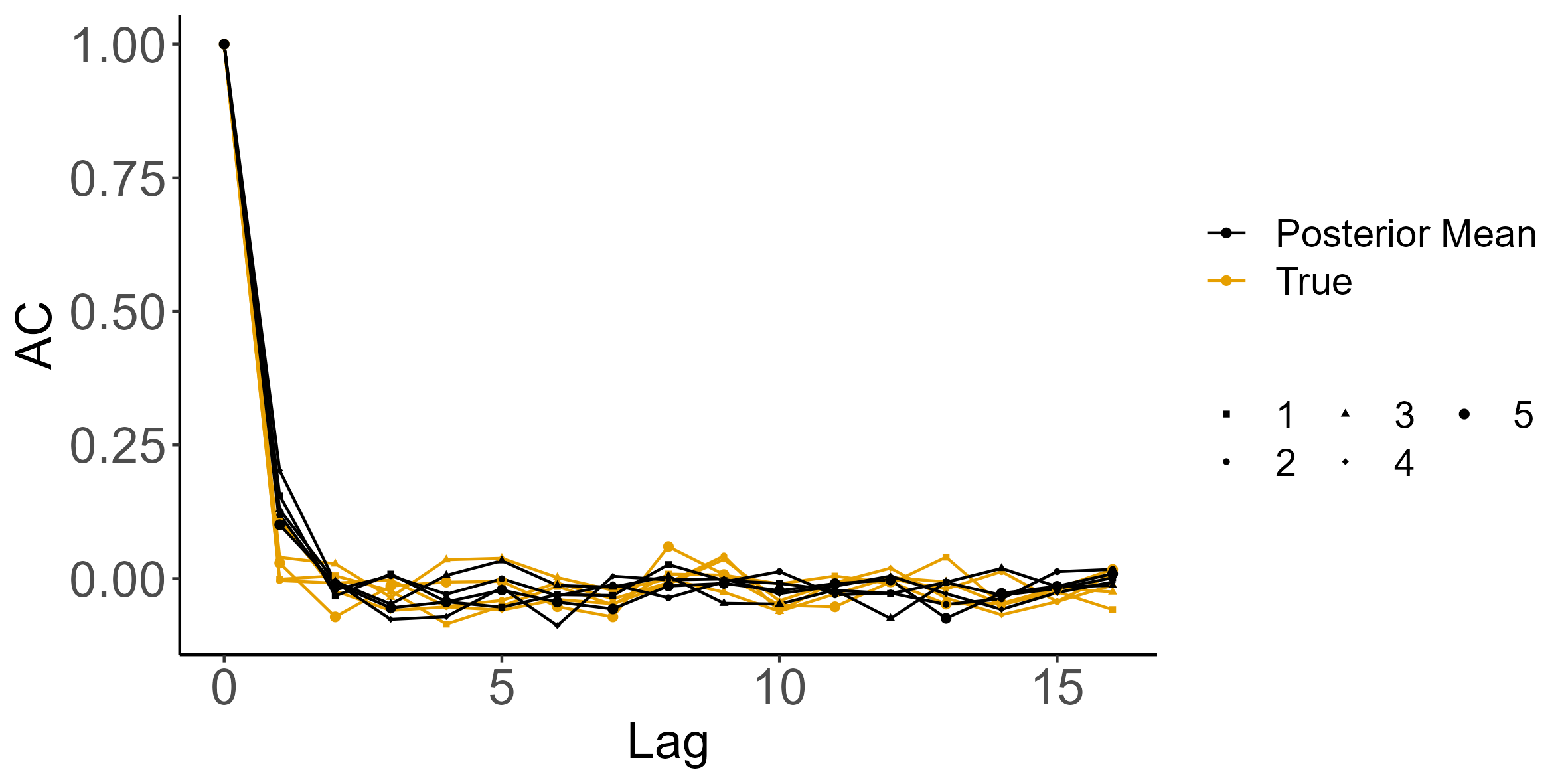} &
\includegraphics[trim=0mm 0mm 0mm 0mm,clip,height=6cm,width=8cm
]{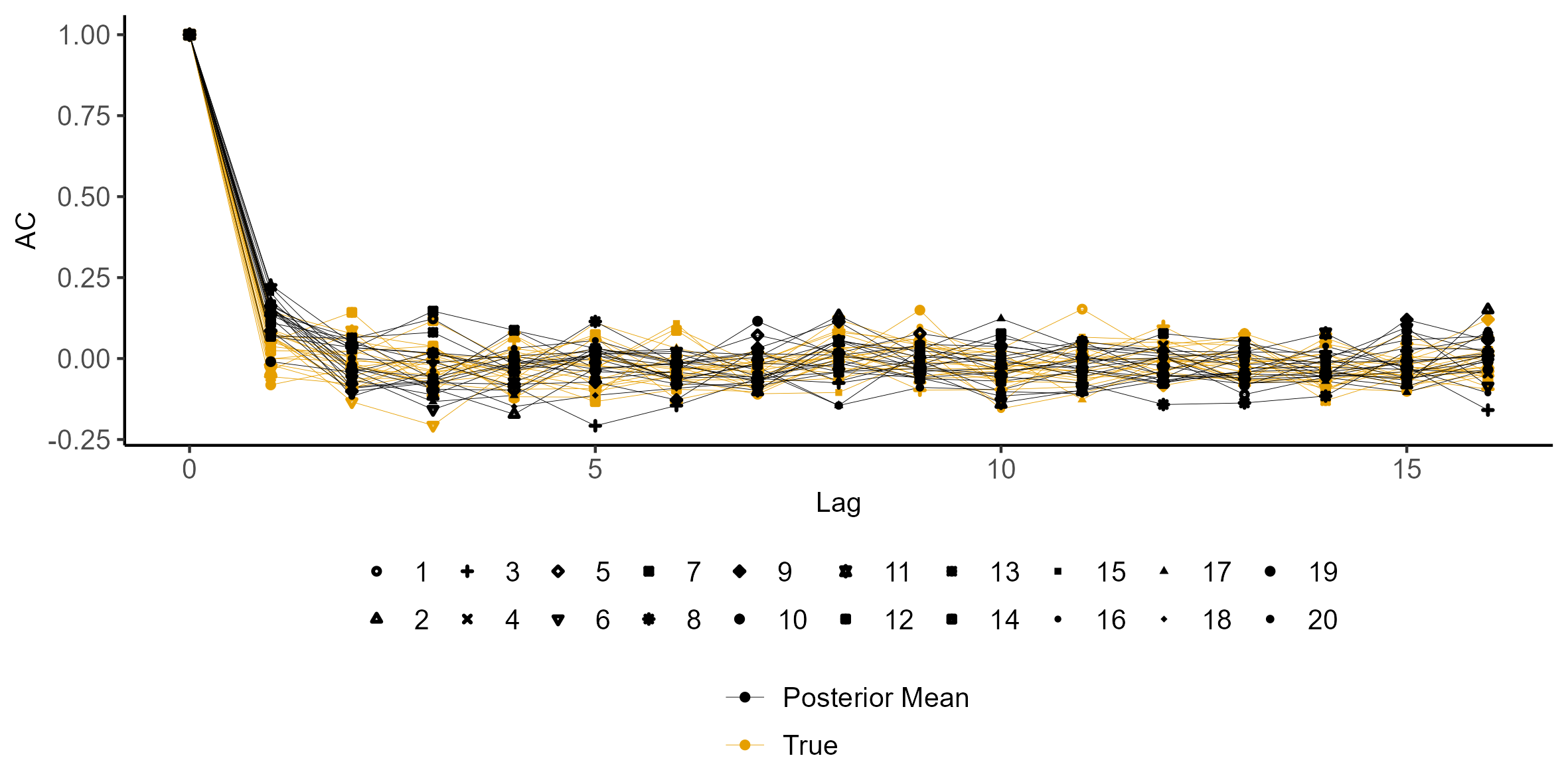}
\end{tabular}
\caption{Pearson sample autocorrelation functions of true and posterior conditional means of latent outcomes for one simulation by ranker averaged over items (left) and by item averaged over rankers (right).  }
\label{fig:true_pred_ACF_by_ranker_by_item_onesim}
\end{figure}

\begin{figure}[t!h]
\centering
\captionsetup{width=0.93\linewidth}
\hspace*{-4.5ex}
\begin{tabular}{cc}
\includegraphics[trim=0mm 0mm 0mm 0mm,clip,height=6cm,width=8cm
]{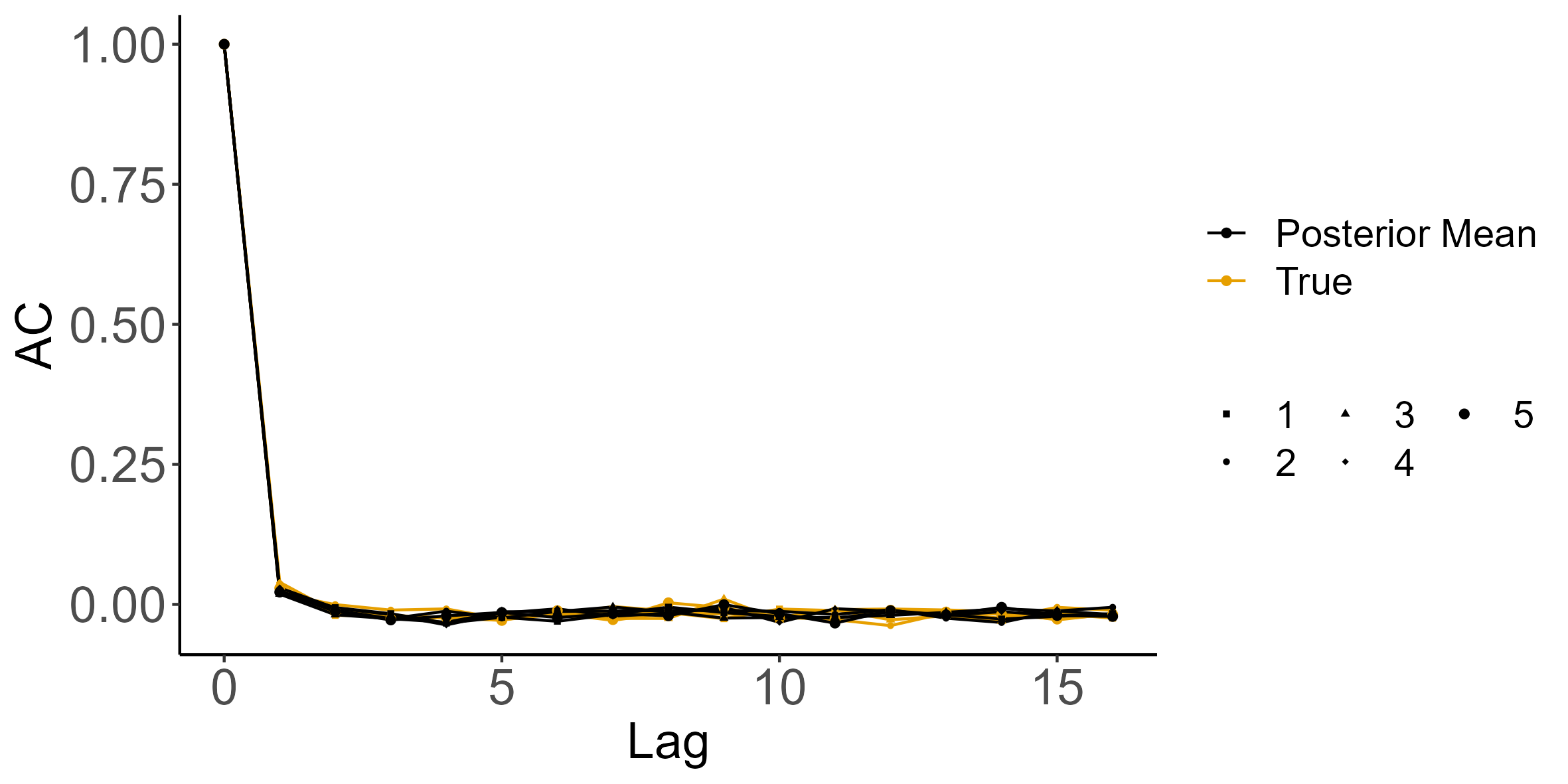} &
\includegraphics[trim=0mm 0mm 0mm 0mm,clip,height=6cm,width=8cm
]{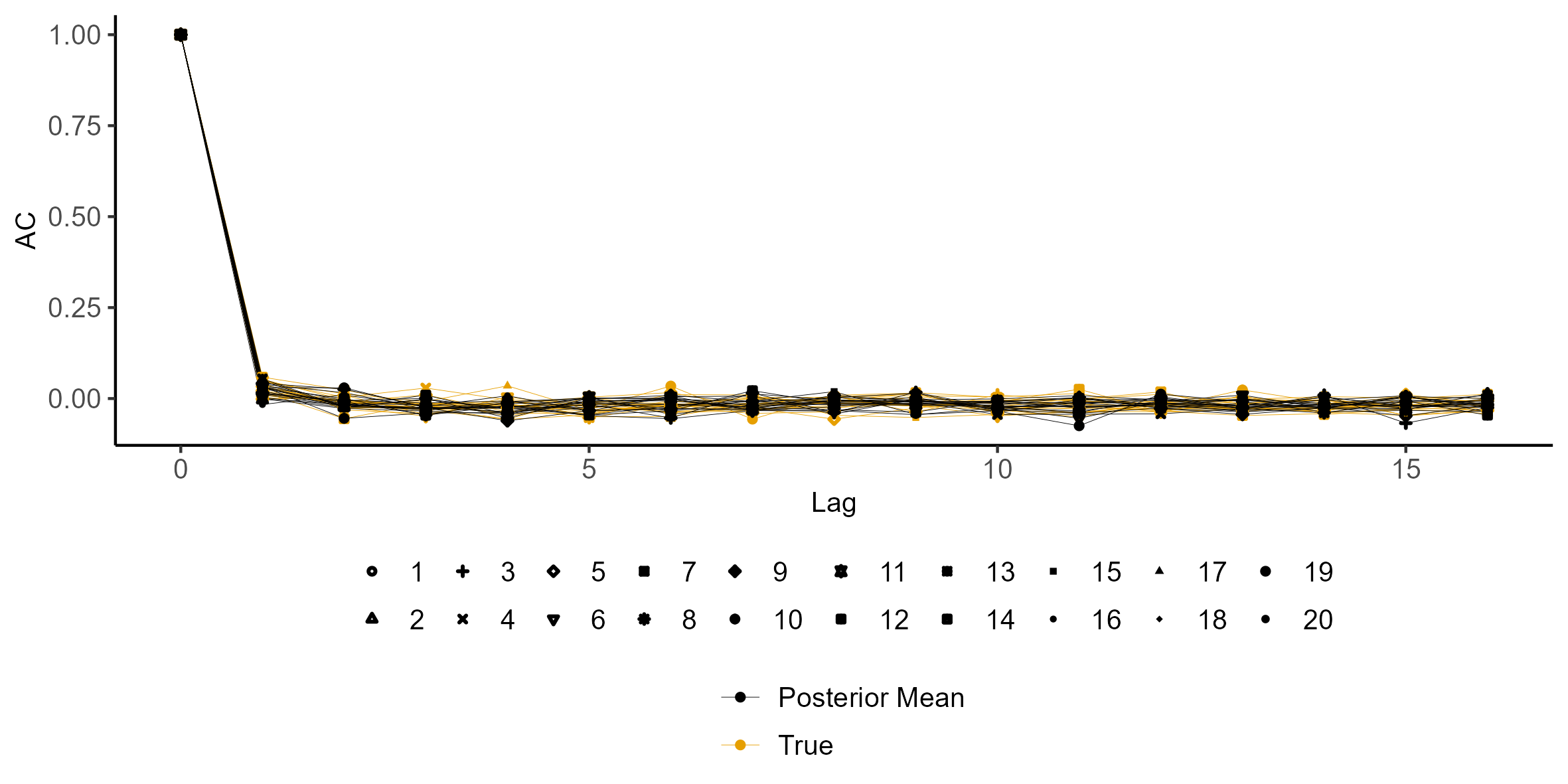}
\end{tabular}
\caption{Pearson sample autocorrelation functions of true and posterior conditional means of latent outcomes by ranker (left) and by item (right), averaged over simulation repetitions.  }
\label{fig:true_pred_ACF_by_ranker_by_item_avg}
\end{figure}

\begin{figure}[t!h]
\centering
\captionsetup{width=0.93\linewidth}
\hspace*{-4.5ex}
\begin{tabular}{cc}
\includegraphics[trim=0mm 0mm 0mm 0mm,clip,height=6cm,width=8cm]{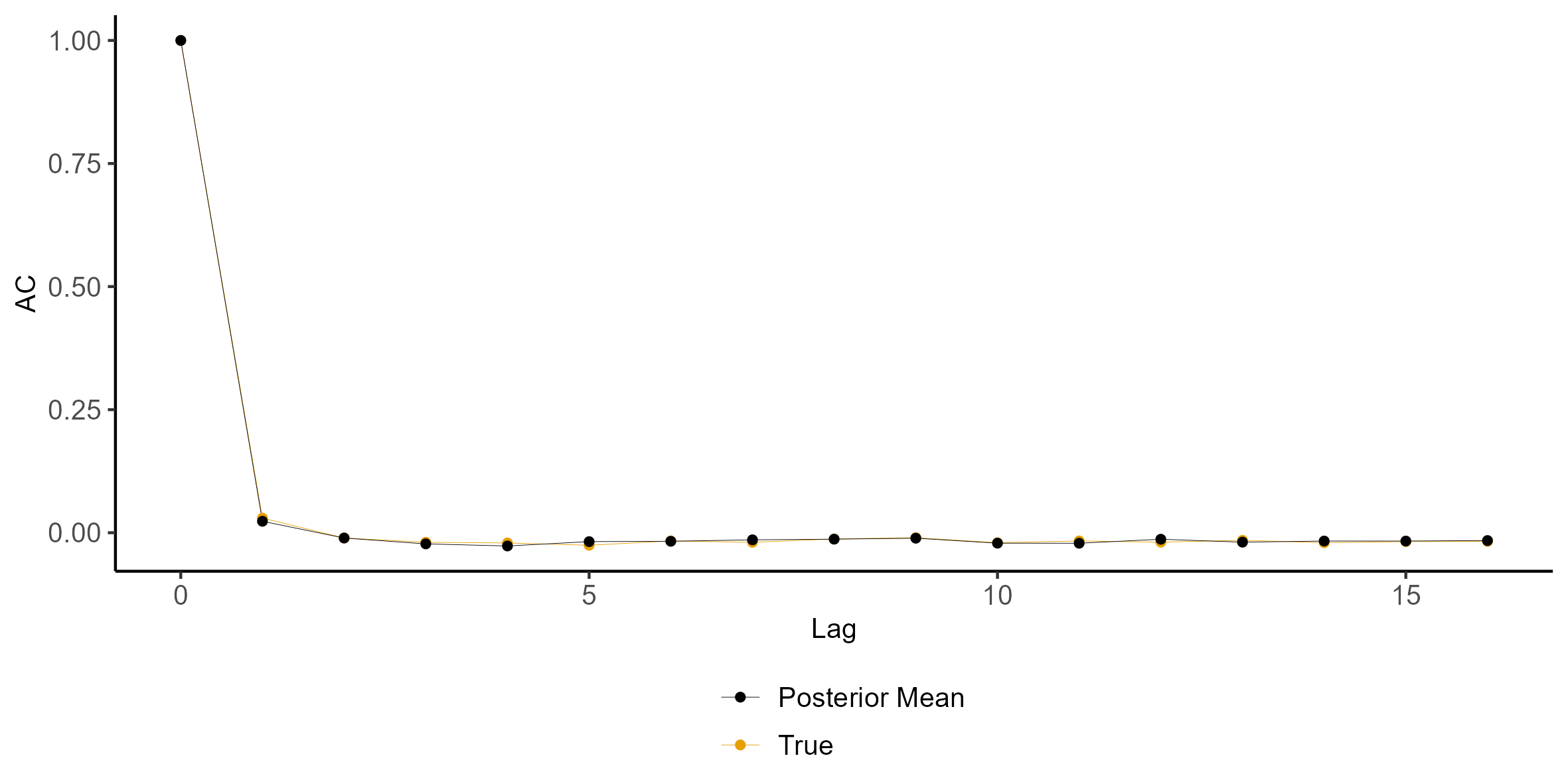} &
\end{tabular}
\caption{Pearson sample autocorrelation functions of true and posterior conditional means of latent outcomes, averaged over simulation repetitions, items, and rankers.  }
\label{fig:true_pred_ACF_overall}
\end{figure}

\end{document}